\documentclass[11pt,a4paper]{book}
\usepackage{hyperref}
\usepackage{makeidx} % Generation de l'index
\usepackage{fancyheadings,minitoc-hyper}
\usepackage{graphicx}
\usepackage{color}
\usepackage{bm} % bold math
\usepackage{amssymb} % symboles AMS (-> \mathbb)
\hypersetup{pdftitle={3+1 formalism and bases of numerical relativity},
  pdfsubject={Lecture Notes},
  pdfauthor={Eric Gourgoulhon <eric.gourgoulhon@obspm.fr>},
  pdfkeywords={LaTeX},
  colorlinks=true
}

\newcommand{\encadre}[1]{\fbox{$\displaystyle #1$}}
\newcommand{\der}[2]{\frac{\partial #1}{\partial #2}}
\newcommand{\dert}[2]{{\partial #1}/{\partial #2}}
\newcommand{\dder}[2]{\frac{\partial^2 #1}{\partial {#2}^2}}
\newcommand{\dderp}[3]{\frac{\partial^2 #1}{\partial #2\partial #3}}
\newcommand{\w}[1]{\bm{#1}}
\newcommand{\uu}[1]{\underline{\w{#1}}}
\newcommand{\Lie}[1]{\bm{\mathcal{L}}_{\w{#1}}\,}
\newcommand{\Liec}[1]{{\mathcal{L}}_{\w{#1}}\,}
\newcommand{\wpar}{\w{\partial}}

\newcommand{\wnab}{\w{\nabla}}
\newcommand{\dd}{\mathbf{d}}
\newcommand{\be}{\begin{equation}}
\newcommand{\ee}{\end{equation}}
\newcommand{\bea}{\begin{eqnarray}}
\newcommand{\eea}{\end{eqnarray}}

\newcommand{\R}{\mathbb{R}}
\newcommand{\M}{\mathcal{M}}
\newcommand{\Sp}{\mathcal{S}}

\newcommand{\T}{\mathcal{T}}
\newcommand{\Df}{\mathcal{D}}
\newcommand{\wDf}{\w{\Df}}
\newcommand{\gam}{\gamma}
\newcommand{\gm}{\gamma}
\newcommand{\wgm}{\w{\gm}}
\newcommand{\wD}{\w{D}}
\newcommand{\vg}{\vec{\w{\gam}}}
\newcommand{\vgs}{\vg^*}
\newcommand{\tgm}{{\tilde\gm}}
\newcommand{\wtgm}{\w{\tgm}}
\newcommand{\tD}{{\tilde D}}
\newcommand{\wtD}{\w{\tD}}
\newcommand{\tA}{{\tilde A}}
\newcommand{\hA}{{\hat A}}

\newcommand{\defin}[1]{\textbf{\slshape #1}\index{#1}}

\newenvironment{remark}%
{\begin{description} \item[\emph{Remark :}]\em}%
{\end{description}}

\newenvironment{example}[1][]%
{\begin{description} \item[\emph{Example #1 :}]\em}%
{\end{description}}

\makeindex

\begin{document}

%\frontmatter

\begin{titlepage}
\begin{center}
\ \\[4cm]
\textcolor{blue}{\textbf{\Huge 3+1 Formalism}} \\[1cm]
\textcolor{blue}{\textbf{\Huge and}} \\[1cm]
\textcolor{blue}{\textbf{\Huge Bases of Numerical Relativity}} \\[2cm]
{\LARGE \slshape Lecture notes} \\[2cm]
{\Large  \'Eric Gourgoulhon} \\[0.5cm]
{Laboratoire Univers et Th\'eories, \\
UMR 8102 du C.N.R.S., Observatoire de Paris, \\
Universit\'e Paris 7 \\
F-92195 Meudon Cedex,  France} \\
\href{mailto:eric.gourgoulhon@obspm.fr}{\texttt{eric.gourgoulhon@obspm.fr}}\\[3cm]
6 March 2007 
\end{center}
\end{titlepage}

\dominitoc

\tableofcontents

%  
%    Preface
%
% $Date: 2007-03-05 22:39:07 +0100 (lun, 05 mar 2007) $
% $Rev: 182 $
% $Author: e_gourgoulhon $
%%%%%%%%%%%%%%%%%%%%%%%%%%%%%

\chapter*{Preface} 
%\chapter{Preface} 

%\verb$Date: 2007-03-05 22:39:07 +0100 (lun, 05 mar 2007) $

\vspace{1cm}

%%%%%%%%%%%%%%%%%%%%%%%%%%%%%%%%%%%%%%%%%%%%%%%%%%%%%%%%%%%%%%%%%%%%%%%%%%%%

These notes are the written version of lectures given in the fall of 2006 at the General Relativity
Trimester at the Institut Henri Poincar\'e in Paris \cite{IHP06} and (for
Chap.~\ref{s:ini}) at the
VII Mexican School on Gravitation and Mathematical Physics in Playa del Carmen 
(Mexico) \cite{Mexico06}.
\medskip

The prerequisites are those of a general relativity course, at the
undergraduate or graduate level, like the textbooks 
by Hartle \cite{Hartl03} or Carroll \cite{Carro04}, of part I of Wald's
book \cite{Wald84}, as well as track 1 of Misner, Thorne and Wheeler 
book \cite{MisneTW73}. 
\medskip

The fact that this is \emph{lecture notes} and not a \emph{review article} implies 
two things:
\begin{itemize}
\item the calculations are rather detailed
(the experienced reader might say \emph{too detailed}), 
with an attempt to made them 
self-consistent and complete, trying to use as less as possible the famous
sentences ``as shown in paper XXX'' or ``see paper XXX for details''; 
\item the bibliographical references do not constitute an extensive survey of 
the mathematical or numerical relativity literature: 
articles have been cited in so far as they
have a direct connection with the main text. 
\end{itemize}

I thank Thibault Damour and Nathalie Deruelle --- the organizers of the IHP 
Trimester, as well as Miguel Alcubierre, Hugo Garcia-Compean and Luis Urena 
--- the organizers of the VII Mexican school, for their invitation to give these lectures. I also warmly thank Marcelo Salgado for the perfect organization 
of my stay in Mexico. 
I am indebted to Nicolas Vasset for his careful reading of the manuscript. 
Finally, I acknowledge the hospitality of the Centre \'Emile Borel of
the Institut Henri Poincar\'e, where a part of these notes 
has been written.
\medskip

Corrections and suggestions for improvement are welcome at
\href{mailto:eric.gourgoulhon@obspm.fr}{\texttt{eric.gourgoulhon@obspm.fr}}. 

%\mainmatter
%  
%    Preface
%
% $Date: 2007-03-06 11:59:03 +0100 (mar, 06 mar 2007) $
% $Rev: 183 $
% $Author: e_gourgoulhon $
%%%%%%%%%%%%%%%%%%%%%%%%%%%%%

\chapter{Introduction} 

%\verb$Date: 2007-03-06 11:59:03 +0100 (mar, 06 mar 2007) $

\vspace{1cm}

%%%%%%%%%%%%%%%%%%%%%%%%%%%%%%%%%%%%%%%%%%%%%%%%%%%%%%%%%%%%%%%%%%%%%%%%%%%%

The \defin{3+1 formalism} is an approach to general relativity and to
Einstein equations that relies on the slicing of the four-dimensional spacetime 
by three-dimensional surfaces \emph{(hypersurfaces)}. 
These hypersurfaces have to be spacelike, so that the metric induced on them
by the \emph{Lorentzian} spacetime metric [signature $(-,+,+,+)$] is
\emph{Riemannian} [signature $(+,+,+)$]. From the mathematical point of view,
this procedure allows to formulate the problem of resolution of Einstein equations 
as a \emph{Cauchy problem} with constraints. From the pedestrian point
of view, it amounts to a decomposition of spacetime 
into ``space'' + ``time'', so that one manipulates only time-varying tensor fields
in the ``ordinary'' three-dimensional space, where the standard scalar product
is Riemannian. Notice that this space + time splitting is not an a priori 
structure of general relativity  but relies on the somewhat arbitrary 
choice of a time coordinate.
The 3+1 formalism should not be confused with the \emph{1+3 formalism}, where the
basic structure is a congruence of one-dimensional curves (mostly timelike curves,
i.e. worldlines), instead of a family of three-dimensional surfaces. 

The 3+1 formalism originates from works by Georges Darmois in the 1920's 
\cite{Darmo27}, Andr\'e Lichnerowicz in the 1930-40's \cite{Lichn39,Lichn44,Lichn52}
and Yvonne Choquet-Bruhat
(at that time Yvonne Four\`es-Bruhat) in the 1950's \cite{Foure52,Foure56}
\footnote{These three persons have some direct filiation: Georges Darmois
was the thesis adviser of Andr\'e Lichnerowicz, who was himself the
thesis adviser of Yvonne Choquet-Bruhat}. 
Notably, in 1952, Yvonne Choquet-Bruhat 
was able to show that the Cauchy problem arising from the 3+1 decomposition has 
locally a unique solution \cite{Foure52}. 
In the late 1950's and early 1960's, 
the 3+1 formalism received a considerable impulse, serving as foundation of Hamiltonian formulations
of general relativity by Paul A.M. Dirac \cite{Dirac58,Dirac59}, and 
Richard Arnowitt, Stanley Deser and Charles W. Misner (ADM) \cite{ArnowDM62}. 
It was also during this time that John A. Wheeler 
put forward the concept of \emph{geometrodynamics} and coined 
the names \emph{lapse} and \emph{shift} \cite{Wheel64}. 
In the 1970's, the 3+1 formalism became the basic tool for the nascent 
numerical relativity. A primordial role has then been played by James W. York,
who developed a general method to solve the initial data problem 
\cite{York73} and who put the 3+1 equations in the shape used afterwards by 
the numerical community \cite{York79}. 
In the 1980's and 1990's, numerical computations increased in
complexity, from 1D (spherical symmetry) to 3D (no symmetry at all).
In parallel, a lot of studies have been devoted to formulating
the 3+1 equations in a form suitable for numerical implementation. 
The authors who participated to this effort are too numerous to be cited
here but it is certainly worth to mention Takashi Nakamura and his school, 
who among other things initiated the formulation which would become the popular \emph{BSSN} 
scheme \cite{NakamOK87,Nakam94,ShibaN95}. 
Needless to say, a strong motivation for the expansion of numerical
relativity has been the development of gravitational wave detectors,
either ground-based (LIGO, VIRGO, GEO600, TAMA) or 
in space (LISA project). 

Today, most numerical codes for solving Einstein equations are based 
on the 3+1 formalism. Other approaches are the  
2+2 formalism or characteristic formulation, as reviewed by 
Winicour \cite{Winic05}, the
conformal field equations by Friedrich \cite{Fried02}
as reviewed by Frauendiener \cite{Fraue04}, 
or the generalized harmonic decomposition used by 
Pretorius \cite{Preto05a,Preto05b,Preto06} for
his recent successful computations of binary black hole merger. 

These lectures are devoted to the 3+1 formalism and theoretical foundations
for numerical relativity. They are not covering numerical techniques, which 
mostly belong to two families: finite difference methods 
and spectral methods. For a pedagogical introduction to these techniques, 
we recommend the lectures by Choptuik \cite{Chopt06} (finite differences)
and the review article by Grandcl\'ement and Novak \cite{GrandN07}
(spectral methods). 

We shall start by two purely geometrical\footnote{by \emph{geometrical} it is
meant \emph{independent of the Einstein equation}} chapters devoted to the study of 
a single hypersurface embedded in spacetime (Chap.~\ref{s:hyp}) and 
to the foliation (or slicing)
of spacetime by a family of spacelike hypersurfaces (Chap.~\ref{s:fol}).
The presentation is divided in two chapters to distinguish clearly
between concepts which are meaningful for a single hypersurface and those 
who rely on a foliation. In some presentations, these notions are blurred; 
for instance the extrinsic curvature is defined as the time derivative of the 
induced metric, giving the impression that it requires a foliation, whereas
it is perfectly well defined for a single hypersurface.
The decomposition of the Einstein equation relative to the foliation
is given in Chap.~\ref{s:dec}, giving rise to the Cauchy problem with constraints,
which constitutes the core of the 3+1 formalism. 
The ADM Hamiltonian formulation of general relativity is also introduced 
in this chapter. 
Chapter~\ref{s:mat} is devoted to the decomposition of the matter and
electromagnetic field equations, focusing on the astrophysically relevant
cases of a perfect fluid and a perfect conductor (MHD). 
An important technical chapter occurs then: Chap.~\ref{s:cfd} introduces some
conformal transformation of the 3-metric on each hypersurface and the 
corresponding rewriting of the 3+1 Einstein equations.
As a byproduct, we also discuss the Isenberg-Wilson-Mathews (or conformally flat)
approximation to general relativity. 
Chapter~\ref{s:glo} details the various global quantities associated with
asymptotic flatness (ADM mass and ADM linear momentum, angular momentum)
or with some symmetries (Komar mass and Komar angular momentum). 
In Chap.~\ref{s:ini}, we study the initial data problem, presenting with some
examples two classical methods: the conformal transverse-traceless method
and the conformal thin sandwich one. Both methods rely on the conformal decomposition
that has been introduced in Chap.~\ref{s:cfd}.
The choice of spacetime coordinates within the 3+1 framework is discussed in Chap.~\ref{s:evo}, starting from the choice of foliation before discussing
the choice of the three coordinates in each leaf of the foliation. 
The major coordinate families used in modern numerical relativity are reviewed.
Finally Chap.~\ref{s:sch} presents various schemes for the time integration of 
the 3+1 Einstein equations,  putting some emphasis on the most successful scheme 
to date, the BSSN one. 
Two appendices are devoted to basic tools of the 3+1 formalism:
the Lie derivative (Appendix~\ref{s:lie})
and the conformal Killing operator and the related vector Laplacian
(Appendix~\ref{s:cko}).

%  
%    Chapitre : Geometry of hypersurfaces
%
% $Date: 2007-03-05 22:39:07 +0100 (lun, 05 mar 2007) $
% $Rev: 182 $
% $Author: e_gourgoulhon $
%%%%%%%%%%%%%%%%%%%%%%%%%%%%%

\chapter{Geometry of hypersurfaces} \label{s:hyp}

%\verb$Date: 2007-03-05 22:39:07 +0100 (lun, 05 mar 2007) $

\minitoc
\vspace{1cm}

%%%%%%%%%%%%%%%%%%%%%%%%%%%%%%%%%%%%%%%%%%%%%%%%%%%%%%%%%%%%%%%%%%%%%%%%%%%%

\section{Introduction}

The notion of hypersurface is the basis of the 3+1 formalism
of general relativity. 
This first chapter is thus devoted to hypersurfaces. 
It is fully independent of the Einstein equation, i.e. 
all results are valid for any spacetime endowed with a
Lorentzian metric, whether the latter is a solution or not 
of Einstein equation.
Otherwise stated, the properties discussed below are purely
geometric, hence the title of this chapter.

Elementary presentations of hypersurfaces are given in numerous textbooks.
To mention a few in the physics literature, let us quote Chap.~3 of 
Poisson's book \cite{Poiss04}, 
Appendix~D of Carroll's one \cite{Carro04}
and Appendix~A of Straumann's one \cite{Strau04}.
The presentation performed here is relatively self-contained
and requires only some elementary knowledge of differential
geometry, at the level of an introductory course in general
relativity (e.g. \cite{Derue06}).
 
\section{Framework and notations}

\subsection{Spacetime and tensor fields}

We consider a spacetime $(\M,\w{g})$ 
where $\M$ is a real smooth (i.e. $\mathcal C^\infty$) manifold
of dimension 4 and $\w{g}$ 
a Lorentzian metric on $\M$, of signature $(-,+,+,+)$.
We assume that $(\M,\w{g})$ is \defin{time orientable}, that is, 
it is possible to divide \emph{continuously over $\M$} each light cone
of the metric $\w{g}$ in two parts, \defin{past}
and \defin{future} \cite{HawkiE73,Wald84}.
We denote by $\wnab$ the affine connection associated
with $\w{g}$, and call it the \defin{spacetime connection}
to distinguish it from other connections introduced in the text.

At a given point $p\in \M$, we denote by $\T_p(\M)$ the \defin{tangent
space}, i.e. the (4-dimensional) space of vectors at $p$.
Its dual space (also called \defin{cotangent space})
is denoted by $\T_p^*(\M)$ and is constituted 
by all linear forms at $p$. 
We denote by $\T(\M)$ (resp. $\T^*(\M)$) the space of smooth
vector fields (resp. 1-forms) on $\M$ \footnote{
The experienced reader is warned that $\T(\M)$ does not stand 
for the tangent bundle of $\M$ (it rather corresponds to the 
space of smooth cross-sections of that bundle). No confusion may arise since 
we shall not use the notion of bundle.}. 

When dealing with indices, we adopt the following conventions:
all Greek indices run in $\{0,1,2,3\}$.
We will use letters from the
beginning of the alphabet ($\alpha$, $\beta$, $\gamma$, ...) for free indices,
and letters starting from $\mu$ ($\mu$, $\nu$, $\rho$, ...) as dumb indices
for contraction (in this way the tensorial degree (valence) of any 
equation is immediately apparent). 
Lower case Latin indices starting from the letter $i$ ($i$, $j$, $k$, ...) 
run in $\{1,2,3\}$, while those starting from the
beginning of the alphabet ($a$, $b$, $c$, ...) run in $\{2,3\}$ only. 

For the sake of clarity, let us recall that if $(\w{e}_\alpha)$
is a vector basis of the tangent space $\T_p(\M)$ and 
$(\w{e}^\alpha)$ is the associate
dual basis, i.e. the basis of $\T_p^*(\M)$ such that 
$\w{e}^\alpha(\w{e}_\beta)=\delta^\alpha_{\ \, \beta}$,
the components 
$T^{\alpha_1\ldots\alpha_p}_{\qquad\ \; \beta_1\ldots\beta_q}$
of a tensor $\w{T}$ of type $\left({p \atop q}\right)$ with 
respect to the bases $(\w{e}_\alpha)$ and $(\w{e}^\alpha)$ 
are given by the expansion
\be \label{e:hyp:comp_tens}
	\w{T} = T^{\alpha_1\ldots\alpha_p}_{\qquad\ \; \beta_1\ldots\beta_q}
		\; \w{e}_{\alpha_1} \otimes \ldots \otimes \w{e}_{\alpha_p} 
                \otimes
		\w{e}^{\beta_1} \otimes \ldots \otimes \w{e}^{\beta_q} .
\ee
The components
$\nabla_{\gamma}  T^{\alpha_1\ldots\alpha_p}_{\qquad\ \; \beta_1\ldots\beta_q}$
of the covariant derivative $\w{\nabla}\w{T}$ are defined by the expansion
\be \label{e:hyp:cov_der_comp}
	\w{\nabla}\w{T} = 
	\nabla_{\gamma} \, 
        T^{\alpha_1\ldots\alpha_p}_{\qquad\ \; \beta_1\ldots\beta_q}
		\; \w{e}_{\alpha_1} \otimes \ldots \otimes \w{e}_{\alpha_p} 
                \otimes
		\w{e}^{\beta_1} \otimes \ldots \otimes \w{e}^{\beta_q} 
		\otimes \w{e}^\gamma  .
\ee
Note the position of the ``derivative index'' $\gamma$ : 
$\w{e}^\gamma$ is the
{\em last} 1-form of the tensorial product on the right-hand side. In this
respect, the notation 
$T^{\alpha_1\ldots\alpha_p}_{\qquad\ \; \beta_1\ldots\beta_q;\gamma}$ instead of 
$\nabla_{\gamma} \, 
T^{\alpha_1\ldots\alpha_p}_{\qquad\ \; \beta_1\ldots\beta_q}$
would have been more appropriate .
This index convention agrees with that 
of MTW \cite{MisneTW73} [cf. their Eq.~(10.17)].
As a result, the covariant derivative of the tensor $\w{T}$ along any
vector field $\w{u}$ is related to $\w{\nabla}\w{T}$ by
\be \label{e:hyp:directional_der}
    \w{\nabla}_{\w{u}}\w{T} = \w{\nabla}\w{T}
        (\underbrace{.,\ldots,.}_{p+q\ {\rm slots}},\w{u}) . 
\ee
The components of $\w{\nabla}_{\w{u}}\w{T}$ are then 
$u^\mu \nabla_{\mu} 
T^{\alpha_1\ldots\alpha_p}_{\qquad\ \; \beta_1\ldots\beta_q}$. 

\subsection{Scalar products and metric duality} \label{s:hyp:metric_dual}

We denote the scalar product of two vectors with respect to the
metric $\w{g}$ by a dot:
\be
  \forall (\w{u},\w{v}) \in \T_p(\M)\times\T_p(\M),\quad 
  \w{u}\cdot\w{v} := \w{g}(\w{u},\w{v}) = g_{\mu\nu} u^\mu v^\nu.
\ee
We also use a dot for the contraction of two tensors $\w{A}$
and $\w{B}$ on the last index of $\w{A}$ and the first index of 
$\w{B}$ (provided of course that these indices are of opposite types). 
For instance if $\w{A}$ is a bilinear form and $\w{B}$ a vector, 
$\w{A}\cdot\w{B}$ is the linear form which components are
\be
    (A\cdot B)_\alpha = A_{\alpha\mu} B^\mu . 
\ee
However, to denote the action of linear forms on vectors, we will use 
brackets instead of a dot: 
\be \label{e:hyp:brackets}
	\forall (\w{\omega},\w{v}) \in \T_p^*(\M)\times\T_p(\M),\quad 
        \langle \w{\omega},\w{v} \rangle = 
        \w{\omega} \cdot \w{v} = \omega_\mu \, v^\mu .
\ee
Given a 1-form $\w{\omega}$ and a vector field $\w{u}$, the directional
covariant derivative $\w{\nabla}_{\w{u}} \, \w{\omega}$ is a 1-form and
we have [combining the notations (\ref{e:hyp:brackets}) and 
(\ref{e:hyp:directional_der})]
\be \label{e:hyp:direc_deriv_1form}
	\forall (\w{\omega},\w{u},\w{v}) \in \T^*(\M)\times\T(\M)\times\T(\M),\quad 
        \langle \w{\nabla}_{\w{u}} \, \w{\omega},\w{v} \rangle = 
        \w{\nabla}\w{\omega} (\w{v},\w{u}).
\ee
Again, notice the ordering in the arguments of the bilinear form $\w{\nabla}\w{\omega}$.
Taking the risk of insisting outrageously, let us stress that this is equivalent
to say that the components 
$(\nabla\omega)_{\alpha\beta}$ of $\w{\nabla}\w{\omega}$ with respect to 
a given basis $(\w{e}^\alpha\otimes\w{e}^\beta)$ of $\T^*(\M)\otimes\T^*(\M)$ are
$\nabla_\beta\omega_\alpha$:
\be \label{e:hyp:grad_1form}
    \w{\nabla}\w{\omega} = \nabla_\beta\omega_\alpha \; 
    \w{e}^\alpha\otimes\w{e}^\beta ,
\ee
this relation constituting a particular case of Eq.~(\ref{e:hyp:cov_der_comp}).

The metric $\w{g}$ induces an isomorphism between
$\T_p(\M)$ (vectors) and $\T_p^*(\M)$ (linear forms) which, in the index notation, 
corresponds to the lowering or raising of the index by contraction
with $g_{\alpha\beta}$ or $g^{\alpha\beta}$. 
In the present lecture, an index-free symbol will always denote
a tensor with a fixed covariance type (e.g. a vector, a 1-form,
a bilinear form, etc...). We will therefore use a different symbol
to denote its image under the metric isomorphism. 
In particular, we denote by an underbar the 
isomorphism $\T_p(\M) \rightarrow \T_p^*(\M)$
and by an arrow the reverse isomorphism $\T_p^*(\M) \rightarrow \T_p(\M)$:
\begin{enumerate}
\item for any vector $\w{u}$ in $\T_p(\M)$, $\underline{\w{u}}$ stands for 
the unique linear form such that 
\be \label{e:hyp:underbar}
	\forall \w{v} \in \T_p(\M),\quad \langle \underline{\w{u}}, \w{v}
		\rangle = \w{g}(\w{u},\w{v}) .
\ee
However, we will omit the underlining on the components
of $\underline{\w{u}}$, since
the position of the index allows to distinguish between vectors
and  linear forms, following the standard usage:
if the components of 
$\w{u}$ in a given basis $(\w{e}_\alpha)$ are denoted by $u^\alpha$,
the components of $\underline{\w{u}}$ in the dual basis $(\w{e}^\alpha)$
are then denoted by $u_\alpha$
[in agreement with Eq.~(\ref{e:hyp:comp_tens})].
\item for any linear form $\w{\omega}$ in $\T_p^*(\M)$, $\vec{\w{\omega}}$
stands for the unique vector of $\T_p(\M)$ such that
\be \label{e:hyp:arrow_form}
	\forall \w{v} \in \T_p(\M),\quad 
        \w{g}(\vec{\w{\omega}},\w{v}) = 
        \langle \w{\omega}, \w{v} \rangle .
\ee
As for the underbar, we will omit the arrow over the components
of $\vec{\w{\omega}}$ by denoting them $\omega^\alpha$. 
\item we extend the arrow notation to {\em bilinear} forms on $\T_p(\M)$:
for any bilinear form $\w{T}\, : \, \T_p(\M)\times\T_p(\M) \rightarrow \R$,
we denote by $\vec{\w{T}}$ the (unique) endomorphism 
$T(\M) \rightarrow T(\M)$ which satisfies 
\be \label{e:hyp:arrow_endo}
    \forall (\w{u},\w{v}) \in \T_p(\M)\times\T_p(\M), \quad 
    \w{T}(\w{u},\w{v}) = \w{u} \cdot \vec{\w{T}}(\w{v}) . 
\ee
If $T_{\alpha\beta}$ are the components of the bilinear form $\w{T}$ 
in some basis $\w{e}^\alpha\otimes\w{e}^\beta$, the matrix of
the endomorphism $\vec{\w{T}}$ with respect to the vector basis 
$\w{e}_\alpha$ (dual to $\w{e}^\alpha$) is $T^\alpha_{\ \, \beta}$.
\end{enumerate}

\subsection{Curvature tensor} \label{s:hyp:curvat}

We follow the MTW convention \cite{MisneTW73} and define the
\defin{Riemann curvature tensor} of the spacetime connection 
$\w{\nabla}$ by\footnote{the superscript `4' stands for the four dimensions
of $\M$ and is used to distinguish from Riemann tensors that will be
defined on submanifolds of $\M$}
\be \label{e:hyp:def_Riemann}
	 \begin{array}{cccc}
	{}^{4}\mathrm{\bf Riem} \ : & \T^*(\M)\times\T(\M)^3 & 
	\longrightarrow & \mathcal{C}^\infty(\M,\R) \\
		& (\w{\omega},\w{w},\w{u},\w{v}) 
		& \longmapsto & \bigg\langle \w{\omega} , \ 
                \w{\nabla}_{\w{u}} \w{\nabla}_{\w{v}} \w{w}
		-  \w{\nabla}_{\w{v}} \w{\nabla}_{\w{u}} \w{w} \\
                & & &
		- \w{\nabla}_{[\w{u},\w{v}]} \w{w} \bigg\rangle ,
	\end{array}  
\ee
where $\mathcal{C}^\infty(\M,\R)$ denotes the space of
smooth scalar fields on $\M$. As it is well known, the above
formula does define a tensor field on $\M$, i.e. the value
of ${}^{4}\mathrm{\bf Riem}(\w{\omega},\w{w},\w{u},\w{v})$ at a given
point $p\in\M$ depends only upon the values of the fields 
$\w{\omega}$, $\w{w}$, $\w{u}$ and $\w{v}$ at $p$ and not
upon their behaviors away from $p$, as the gradients in 
Eq.~(\ref{e:hyp:def_Riemann}) might suggest. 
We denote the components of this tensor in 
a given basis $(\w{e}_\alpha)$, not by 
${}^{4}{\rm Riem}^\gamma_{\ \, \delta \alpha\beta}$, but by
${}^{4}\!R^\gamma_{\ \, \delta \alpha\beta}$. 
The definition (\ref{e:hyp:def_Riemann}) leads then to the
following writing 
(called \defin{Ricci identity}):
\be \label{e:hyp:Ricci_ident}
    \forall\w{w}\in\T(\M),\quad 
        \left(\nabla_\alpha\nabla_\beta  
        - \nabla_\beta\nabla_\alpha\right) w^\gamma
        = {}^{4}\!R^\gamma_{\ \, \mu \alpha\beta} \, w^\mu ,  
\ee
From the definition (\ref{e:hyp:def_Riemann}), the Riemann tensor is
clearly antisymmetric with respect to its last two arguments $(\w{u},\w{v})$.
The fact that the connection $\w{\nabla}$ is associated with a metric
(i.e. $\w{g}$) implies the additional well-known 
antisymmetry:
\be \label{e:hyp:Riemann_antisym12}
    \forall (\w{\omega},\w{w})\in \T^*(\M)\times\T(\M),\
    {}^{4}\mathrm{\bf Riem}(\w{\omega},\w{w},\cdot,\cdot)
    = - {}^{4}\mathrm{\bf Riem}(\underline{\w{w}},\vec{\w{\omega}},\cdot,\cdot) .
\ee
In addition, the Riemann tensor satisfies the cyclic property
\bea   
& & \forall (\w{u},\w{v},\w{w})\in \T(\M)^3, \nonumber \\
& & \ \quad
{}^{4}\mathrm{\bf Riem}(\cdot,\w{u},\w{v},\w{w}) 
+{}^{4}\mathrm{\bf Riem}(\cdot,\w{w},\w{u},\w{v})
+{}^{4}\mathrm{\bf Riem}(\cdot,\w{v},\w{w},\w{u}) = 0 \ . \label{e:hyp:Riemann_cyclic}
\eea

The \defin{Ricci tensor} of the spacetime connection $\w{\nabla}$ is
the bilinear form ${}^{4}\!\w{R}$ defined by 
\be \label{e:hyp:def_Ricci}
	 \begin{array}{cccc}
	{}^{4}\!\w{R} \ : & \T(\M)\times\T(\M) & 
	\longrightarrow & \mathcal{C}^\infty(\M,\R) \\
		& (\w{u},\w{v}) 
		& \longmapsto & 
                {}^{4}\mathrm{\bf Riem}(\w{e}^\mu,\w{u},\w{e}_\mu,\w{v}) .
	\end{array}  
\ee
This definition is independent of the choice of the basis $(\w{e}_\alpha)$
and its dual counterpart $(\w{e}^\alpha)$. Moreover the bilinear form
${}^{4}\!\w{R}$ is symmetric. 
In terms of components:
\be \label{e:hyp:def_Ricci_comp}
    {}^{4}\!R_{\alpha\beta} = {}^{4}\!R^\mu_{\ \, \alpha\mu\beta}.
\ee
Note that, following the standard usage, we are denoting the components
of both the Riemann and Ricci tensors by the same letter $R$, the 
number of indices allowing to distinguish between the two tensors.
On the contrary we are using different symbols, ${}^{4}\mathrm{\bf Riem}$ and
${}^{4}\!\w{R}$, when dealing with the `intrinsic' notation.

Finally, the Riemann tensor can 
be split into (i) a ``trace-trace'' part, represented
by the \defin{Ricci scalar} ${}^{4}\!R:=g^{\mu\nu} {}^{4}\!R_{\mu\nu}$
(also called \defin{scalar curvature}), 
(ii) a ``trace'' part, 
represented by the Ricci tensor ${}^{4}\!\w{R}$
[cf. Eq.~(\ref{e:hyp:def_Ricci_comp})], and (iii) a ``traceless'' part,
which is constituted by the \defin{Weyl conformal curvature tensor}, ${}^{4}\w{C}$:
\bea
	{}^{4}\!R^\gamma_{\ \; \delta\alpha\beta}   & = & 
        {}^{4}C^\gamma_{\ \; \delta\alpha\beta}
	+ \frac{1}{2} \left( {}^{4}\!R^\gamma_{\ \, \alpha} \, g_{\delta\beta}
	   - {}^{4}\!R^\gamma_{\ \, \beta}\,  g_{\delta\alpha}
	   + {}^{4}\!R_{\delta\beta} \, \delta^\gamma_{\ \, \alpha}
	   - {}^{4}\!R_{\delta\alpha} \, \delta^\gamma_{\ \, \beta} \right) 
                            \nonumber \\
	 &&   + \frac{1}{6} {}^{4}\!R \left( g_{\delta\alpha} \, 
         \delta^\gamma_{\ \, \beta}
	   - g_{\delta\beta} \, \delta^\gamma_{\ \, \alpha} \right) . \label{e:hyp:Weyl}
\eea
The above relation can be taken as the definition of ${}^{4}\w{C}$. 
It implies that ${}^{4}\w{C}$ is traceless: 
\bea
\label{e:hyp:Weyl_traceless}
{}^{4}C^\mu_{\ \, \alpha\mu\beta}=0 \ .
\eea
The other possible traces are zero thanks to the symmetry properties of 
the Riemann tensor. 
It is well known that the $20$ independent components
of the Riemann tensor distribute in the $10$ components in the 
Ricci tensor, which are fixed by Einstein equation, and $10$
independent components in the Weyl tensor.

%%%%%%%%%%%%%%%%%%%%%%%%%%%%%%%%%%%%%%%%%%%%%%%%%%%%%%%%%%%%%%%%%%%%%%%%%%%%

\section{Hypersurface embedded in spacetime} 

\subsection{Definition} \label{s:hyp:def_hyp}

\begin{figure}
\centerline{\includegraphics[width=0.8\textwidth]{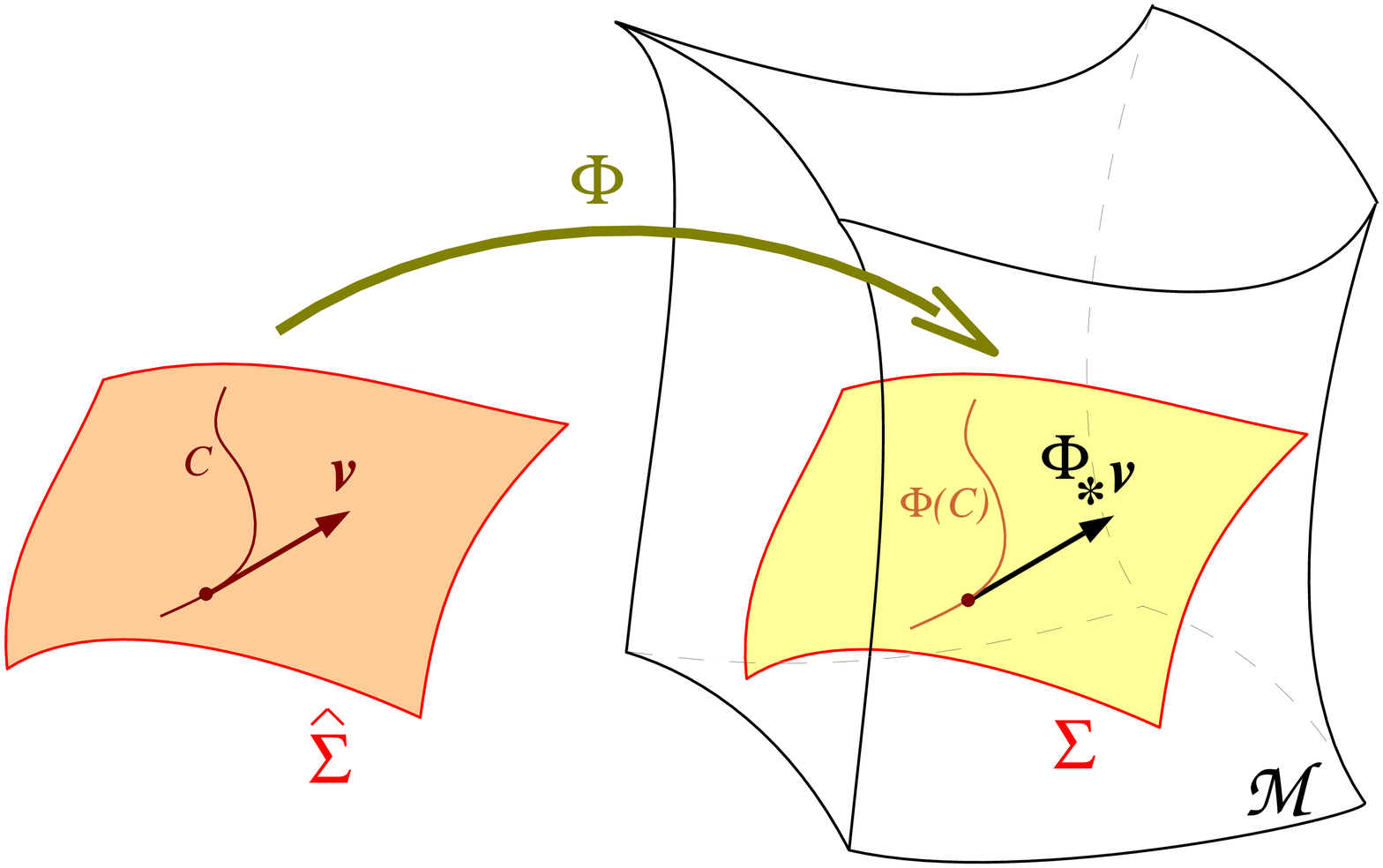}}
\caption[]{\label{f:hyp:embed} \footnotesize
Embedding $\Phi$ of the 3-dimensional manifold $\hat\Sigma$ into the 
4-dimensional manifold $\M$, 
defining the hypersurface $\Sigma = \Phi(\hat\Sigma)$. The push-forward $\Phi_*\w{v}$
of a vector $\w{v}$ tangent to some curve $C$ in $\hat\Sigma$ is a vector
tangent to $\Phi(C)$ in $\M$.}
\end{figure}

A \defin{hypersurface} $\Sigma$ of $\M$ is the image of a 3-dimensional
manifold $\hat\Sigma$ by an embedding $\Phi\,:\, \hat\Sigma \rightarrow \M$
(Fig.~\ref{f:hyp:embed}) :
\be
	\Sigma = \Phi(\hat\Sigma) . 
\ee
Let us recall that \defin{embedding} means that 
$\Phi\,:\, \hat\Sigma \rightarrow \Sigma$ is a homeomorphism, 
i.e. a one-to-one mapping such that both $\Phi$ and $\Phi^{-1}$ are continuous.
The one-to-one character guarantees that $\Sigma$ does not ``intersect itself''.
A hypersurface can be defined locally as the set of points for
which a scalar field on $\M$, $t$ let say, is constant:
\be \label{e:hyp:r_const}
	\forall p \in \M,\quad p \in \Sigma \iff t(p) = 0 .
\ee
For instance, let us assume that $\Sigma$ is a connected submanifold
of $\M$ with topology $\R^3$.
Then we may
introduce locally a coordinate system of $\M$, $x^\alpha=(t,x,y,z)$,
such that $t$ spans $\R$ and $(x,y,z)$ are Cartesian
coordinates spanning $\R^3$. $\Sigma$ is then defined by
the coordinate condition $t=0$ [Eq.~(\ref{e:hyp:r_const})] and 
an explicit form of the mapping $\Phi$ can be obtained by 
considering $x^i=(x,y,z)$ as coordinates on the 3-manifold $\hat\Sigma$ :
\be
	\begin{array}{cccc}
	\Phi \ : & \hat\Sigma & \longrightarrow & \M \\
		& (x,y,z) & \longmapsto & (0,x,y,z) .
	\end{array} 
\ee

The embedding $\Phi$ ``carries along'' curves in $\hat\Sigma$ to
curves in $\M$. Consequently it also ``carries along'' vectors on $\hat\Sigma$
to vectors on $\M$ (cf. Fig.~\ref{f:hyp:embed}).
In other words, it defines a mapping between $\T_p(\hat\Sigma)$ and $\T_p(\M)$. 
This mapping is denoted by $\Phi_*$ and is called the \defin{push-forward mapping};
thanks to the adapted 
coordinate systems $x^\alpha=(t,x,y,z)$, it can be  explicited as follows
\be \label{e:hyp:push_forward}
	\begin{array}{cccc}
	\Phi_* \ : & \T_p(\hat\Sigma) & \longrightarrow & \T_p(\M) \\
		& \w{v} = (v^x,v^y,v^z) & \longmapsto & 
			\Phi_*\w{v} = (0,v^x,v^y,v^z) ,
	\end{array} 
\ee
where $v^i=(v^x,v^y,v^z)$ denotes the components of the vector $\w{v}$ with
respect to the natural basis $\partial/\partial x^i$ of $\T_p(\Sigma)$
associated  with the coordinates $(x^i)$.

Conversely, the embedding $\Phi$ induces a mapping, called the \defin{pull-back mapping}
and denoted $\Phi^*$, between the linear forms on $\T_p(\M)$ and those on $\T_p(\hat\Sigma)$
as follows
\be \label{e:hyp:pull_back}
	\begin{array}{ccccccc}
	\Phi^* \ : & \T_p^*(\M) & \longrightarrow & \T_p^*(\hat\Sigma) & & & \\
		& \w{\omega}  & \longmapsto & 
			\Phi^*\w{\omega} \ :
				& \T_p(\hat\Sigma) & \rightarrow & \R \\
			& & & 	 & \w{v} & \mapsto & 
				\langle \w{\omega} , \Phi_*\w{v}\rangle .
	\end{array} 
\ee
Taking into account (\ref{e:hyp:push_forward}), the pull-back mapping
can be explicited:
\be \label{e:hyp:pull_back_comp}
	\begin{array}{cccc}
	\Phi^* \ : & \T_p^*(\M) & \longrightarrow & \T_p^*(\hat\Sigma)  \\
		& \w{\omega} = (\omega_t,\omega_x,\omega_y,\omega_z)  
			& \longmapsto &
		 \Phi^*\w{\omega} = (\omega_x,\omega_y,\omega_z) ,
	\end{array} 
\ee
where $\omega_\alpha$ denotes the components of the 1-form $\w{\omega}$ with
respect to the basis $\dd x^\alpha$ associated 
with the coordinates $(x^\alpha)$.

In what follows, we identify $\hat\Sigma$ and $\Sigma=\Phi(\hat\Sigma)$.
In particular, we identify any vector on $\hat\Sigma$ with its push-forward
image in $\M$, writing simply $\w{v}$ instead of $\Phi_*\w{v}$.

The pull-back operation can be extended to the multi-linear forms
on $\T_p(\M)$ in an obvious way: if $\w{T}$ is a $n$-linear form
on $\T_p(\M)$, $\Phi^*\w{T}$ is the $n$-linear form on $\T_p(\Sigma)$ defined
by 
\be \label{e:hyp:def_pull-back_multi}
	\forall (\w{v}_1,\ldots,\w{v}_n) \in \T_p(\Sigma)^n,\quad
	\Phi^*\w{T}(\w{v}_1,\ldots,\w{v}_n) = 
	\w{T} (\Phi_* \w{v}_1,\ldots,\Phi_* \w{v}_n) .
\ee

\begin{remark} \label{rem:hyp:emb_map}
By itself, the embedding $\Phi$ induces a mapping
from vectors on $\Sigma$ to vectors on $\M$ (push-forward mapping $\Phi_*$)
and a mapping from 1-forms on $\M$ to 1-forms on $\Sigma$
(pull-back mapping $\Phi^*$),
but not in the reverse way. For instance, one may define ``naively'' 
a reverse mapping $F:\; \T_p(\M) \longrightarrow \T_p(\Sigma)$ by
$\w{v} = (v^t,v^x,v^y,v^z) \longmapsto 
F\w{v} = (v^x,v^y,v^z)$, but it would then depend on 
the choice of coordinates $(t,x,y,z)$, which is not the 
case of the push-forward mapping defined by Eq.~(\ref{e:hyp:push_forward}).
As we shall see below, if $\Sigma$ is a spacelike hypersurface, 
a coordinate-independent reverse mapping is provided by the {\em orthogonal projector}
(with respect to the ambient metric $\w{g}$) onto $\Sigma$. 
\end{remark}

A very important case of pull-back operation is that of the bilinear
form $\w{g}$ (i.e. the spacetime metric), which defines the 
\defin{induced metric on $\Sigma$} : 
\be \label{e:hyp:def_q}
	\encadre{\w{\gam} := \Phi^* \w{g} }
\ee
$\w{\gam}$ is also called the \defin{first fundamental form of $\Sigma$}.
We shall also use the short-hand name \defin{3-metric} to design it.  
Notice that
\be
    \forall (\w{u},\w{v}) \in \T_p(\Sigma)\times\T_p(\Sigma),\quad 
  \w{u}\cdot\w{v} = \w{g}(\w{u},\w{v}) = \w{\gam}(\w{u},\w{v}).
\ee
In terms of the coordinate system\footnote{Let us recall that by
convention Latin indices run in $\{1,2,3\}$.}
$x^i=(x,y,z)$ of $\Sigma$, the components of $\w{\gam}$ are
deduced from (\ref{e:hyp:pull_back_comp}):
\be \label{e:hyp:qAB}
	\encadre{ \gam_{ij} = g_{ij} }.
\ee
The hypersurface is said to be 
\begin{itemize}
\item \defin{spacelike} iff the metric $\w{\gam}$ is definite positive,
i.e. has signature $(+,+,+)$;
\item \defin{timelike} iff the metric $\w{\gam}$ is Lorentzian, 
i.e. has signature $(-,+,+)$;
\item \defin{null} iff the metric $\w{\gam}$ is degenerate,
i.e. has signature $(0,+,+)$.
\end{itemize}

\subsection{Normal vector} \label{s:hyp:norm_vect}

Given a scalar field $t$ on $\M$ such that the hypersurface 
$\Sigma$ is defined
as a level surface of $t$ [cf. Eq.~(\ref{e:hyp:r_const})], the
gradient 1-form $\dd t$ is normal to $\Sigma$, in the 
sense that for every vector $\w{v}$ tangent to $\Sigma$,
$\langle \dd t, \w{v} \rangle = 0$. The metric dual to $\dd t$, 
i.e. the vector $\vec{\wnab} t$ (the component of which are
$\nabla^\alpha t= g^{\alpha\mu}\nabla_\mu t = g^{\alpha\mu}(\dd t)_\mu$)
is a vector normal to $\Sigma$ and satisfies to the following
properties
\begin{itemize}
\item $\vec{\wnab} t$ is timelike iff $\Sigma$ is spacelike;
\item $\vec{\wnab} t$ is spacelike iff $\Sigma$ is timelike;
\item $\vec{\wnab} t$ is null iff $\Sigma$ is null.
\end{itemize}
The vector $\vec{\wnab} t$ defines the unique direction normal
to $\Sigma$. In other words, any other vector $\w{v}$ normal
to $\Sigma$ must be collinear to $\vec{\wnab} t$:
$\w{v} = \lambda \vec{\wnab} t$. Notice a characteristic property
of null hypersurfaces: a vector normal to them is also tangent
to them. This is because null vectors are orthogonal to themselves.

In the case where $\Sigma$ is not null, we can re-normalize
$\vec{\wnab} t$ to make it a unit vector, by setting
\be \label{e:hyp:n_dt_gal}
    \w{n} := \left( \pm \vec{\wnab} t \cdot \vec{\wnab} t \right) ^{-1/2}
            \, \vec{\wnab} t ,
\ee
with the sign $+$ for a timelike hypersurface and the sign $-$ for a
spacelike one. The vector $\w{n}$ is by construction a unit vector:
\bea
    \w{n}\cdot\w{n} = -1 & \quad & \mbox{if $\Sigma$ is spacelike,} \\
    \w{n}\cdot\w{n} = 1 & \quad & \mbox{if $\Sigma$ is timelike.}
\eea
$\w{n}$ is one of the two unit vectors normal to $\Sigma$, the other
one being $\w{n'} = -\w{n}$. 
In the case where $\Sigma$ is a null hypersurface, such a construction
is not possible since $\vec{\wnab} t \cdot \vec{\wnab} t=0$.
Therefore there is no natural way to pick a privileged normal vector 
in this case. Actually, given a null normal $\w{n}$, any vector
$\w{n'} = \lambda \w{n}$, with $\lambda\in\R^*$, is a perfectly valid
alternative to $\w{n}$. 

\subsection{Intrinsic curvature} \label{s:hyp:intrins_curv}

If $\Sigma$ is a spacelike or timelike hypersurface, then the induced
metric $\w{\gam}$ is not degenerate. This implies that there is a unique
connection (or covariant derivative) $\w{D}$ on the manifold $\Sigma$
that is torsion-free and satisfies
\be
	\encadre{ \w{D} \, \w{\gam} = 0 }. 
\ee
$\w{D}$ is the so-called \defin{Levi-Civita connection} associated with the
metric $\w{\gam}$ (see Sec.~2.IV.2 of N.~Deruelle's lectures \cite{Derue06}).
The Riemann tensor associated with this connection represents what can
be called the \defin{intrinsic curvature} of $(\Sigma,\w{\gam})$.
We shall denote it by $\mathrm{\bf Riem}$ (without any superscript `4'),
and its components by the letter $R$, as $R^k_{\ \, lij}$.
$\mathrm{\bf Riem}$ measures the non-commutativity of two successive 
covariant derivatives $\w{D}$, as expressed by 
the Ricci identity, similar to Eq.~(\ref{e:hyp:Ricci_ident})
but at three dimensions:
\be \label{e:hyp:Ricci_ident_3D}
	\forall \w{v}\in\T(\Sigma), \ 
	(D_i D_j - D_j D_i) v^k = R^k_{\ \, lij} \, v^l  .
\ee
The corresponding Ricci tensor is denoted $\w{R}$: $R_{ij} = R^k_{\ \, ikj}$
and the Ricci scalar (scalar curvature) is denoted $R$: $R = \gam^{ij} R_{ij}$.  
$R$ is also called the \defin{Gaussian curvature} of $(\Sigma,\w{\gam})$. 

Let us remind that in dimension 3, the Riemann tensor can be
fully determined from the knowledge of the Ricci tensor,
according to the formula
\be \label{e:hyp:Riem_R_dim3}
	R^i_{\ \, jkl} = \delta^i_{\ \, k} R_{jl}
	- \delta^i_{\ \, l} R_{jk}
	+ \gam_{jl} R^i_{\ \, k} -\gam_{jk} R^i_{\ \, l}
	+ \frac{1}{2} R (\delta^i_{\ \, l} \gam_{jk}
	- \delta^i_{\ \, k} \gam_{jl}) .  
\ee
In other words, the Weyl tensor vanishes identically in dimension 3
[compare Eq.~(\ref{e:hyp:Riem_R_dim3}) with Eq.~(\ref{e:hyp:Weyl})].

\subsection{Extrinsic curvature} \label{s:hyp:extr_curv}

Beside the intrinsic curvature discussed above, one may consider
another type of ``curvature'' regarding hypersurfaces,
namely that related to the ``bending'' of $\Sigma$ in $\M$.
This ``bending'' corresponds to the change of direction of the normal $\w{n}$ 
as one moves on $\Sigma$. 
More precisely, one defines the \defin{Weingarten map} (sometimes called
the \defin{shape operator}) as the endomorphism
of $\T_p(\Sigma)$ which associates with each vector tangent to $\Sigma$
the variation of the normal along that vector, 
the variation being evaluated via 
the spacetime connection $\w{\nabla}$:
\be \label{e:hyp:Weingarten_def} \encadre{
	\begin{array}{cccc}
	\w{\chi}: & \T_p(\Sigma) & \longrightarrow & \T_p(\Sigma) \\
		& \w{v} & \longmapsto & \w{\nabla}_{\w{v}} \, \w{n}
	\end{array} }
\ee
This application is well defined (i.e. its image is in $\T_p(\Sigma)$) since
\be
	\w{n}\cdot \w{\chi}(\w{v}) = \w{n}\cdot \w{\nabla}_{\w{v}} \, \w{n} 
		= {1\over 2} \w{\nabla}_{\w{v}} (\w{n}\cdot\w{n}) = 0 ,
\ee
which shows that $\w{\chi}(\w{v}) \in \T_p(\Sigma)$. 
If $\Sigma$ is not a null hypersurface, the Weingarten map is uniquely defined (modulo the
choice $+\w{n}$ or $-\w{n}$ for the unit normal), whereas if $\Sigma$ is null, 
the definition of $\w{\chi}$ depends upon the choice of the null normal $\w{n}$.

The fundamental property of the Weingarten map is to be {\em self-adjoint}
with respect to the induced metric $\w{\gam}$ :
\be
	\encadre{ \forall (\w{u},\w{v}) \in \T_p(\Sigma)\times\T_p(\Sigma),\quad
	\w{u} \cdot \w{\chi}(\w{v}) = \w{\chi}(\w{u})\cdot \w{v} },
\ee
where the dot means the scalar product with respect to $\w{\gam}$ [considering
$\w{u}$ and $\w{v}$ as vectors of $\T_p(\Sigma)$] or
$\w{g}$ [considering
$\w{u}$ and $\w{v}$ as vectors of $\T_p(\M)$].
Indeed, one obtains from the definition of $\w{\chi}$
\bea
	\w{u}\cdot\w{\chi}(\w{v}) & = & 
		\w{u}\cdot \w{\nabla}_{\w{v}} \, \w{n} = 
                 \w{\nabla}_{\w{v}}\, (\underbrace{\w{u}\cdot\w{n}}_{=0}) -
			\w{n}\cdot \w{\nabla}_{\w{v}} \, \w{u} =
             - \w{n}\cdot \left(   \w{\nabla}_{\w{u}} \, \w{v}
			- [\w{u},\w{v}] \right) \nonumber \\
		& = & - \w{\nabla}_{\w{u}} \, (\underbrace{\w{n}\cdot \w{v}}_{=0})
			+ \w{v} \cdot \w{\nabla}_{\w{u}} \, \w{n}
			+ \w{n} \cdot [\w{u},\w{v}] \nonumber \\
            &= & \w{v}\cdot \w{\chi}(\w{u}) 
		+ \w{n} \cdot [\w{u},\w{v}] . \label{e:hyp:u_dot_chi_v}
\eea
Now the Frobenius theorem states that the commutator $[\w{u},\w{v}]$
of two vectors of the hyperplane $\T(\Sigma)$ belongs to $\T(\Sigma)$
since $\T(\Sigma)$ is surface-forming (see e.g. Theorem B.3.1 in Wald's textbook
\cite{Wald84}). It is straightforward to establish it: 
\bea
	\vec{\wnab}t \cdot [\w{u},\w{v}] & = & \langle \dd t , [\w{u},\w{v}] \rangle
			= \nabla_\mu t \, u^\nu \nabla_\nu v^\mu
			- \nabla_\mu t \,  v^\nu \nabla_\nu u^\mu \nonumber \\
     & = & u^\nu [ \nabla_\nu (\underbrace{\nabla_\mu t \, v^\mu}_{=0})
    - v^\mu \nabla_\nu \nabla_\mu t ]
    - v^\nu [ \nabla_\nu (\underbrace{\nabla_\mu t \, u^\mu}_{=0})
    - u^\mu \nabla_\nu \nabla_\mu t ] \nonumber \\
                       & = & u^\mu v^\nu \left( \nabla_\nu \nabla_\mu t - \nabla_\mu \nabla_\nu t
    \right) = 0 , 
\eea
where the last equality results from the lack of torsion of the connection $\wnab$:
$\nabla_\nu \nabla_\mu t = \nabla_\mu \nabla_\nu t$.
Since $\w{n}$ is collinear to $\vec{\wnab}t$, we have as well 
$\w{n}\cdot[\w{u},\w{v}] = 0$. Once inserted into Eq.~(\ref{e:hyp:u_dot_chi_v}), this 
establishes that the Weingarten map is self-adjoint.

The eigenvalues of the Weingarten map, which are all real numbers
since $\w{\chi}$ is self-adjoint,  are called the 
\defin{principal curvatures} of the hypersurface $\Sigma$ and
the corresponding eigenvectors define the so-called \defin{principal
directions} of $\Sigma$. The \defin{mean curvature} of the hypersurface
$\Sigma$ is the arithmetic mean of the principal curvature:
\be
	H := \frac{1}{3} \left( \kappa_1 + \kappa_2 + \kappa_3 \right)
\ee
where the $\kappa_i$ are the three eigenvalues of $\w{\chi}$.

\begin{remark}
The curvatures defined above are not to be confused with the Gaussian
curvature introduced in Sec.~\ref{s:hyp:intrins_curv}. The latter
is an \emph{intrinsic} quantity, independent of the way the 
manifold $(\Sigma,\w{\gam})$ is embedded in $(\M,\w{g})$.
On the contrary the principal curvatures and mean curvature depend on 
the embedding. For this reason, they are qualified of \emph{extrinsic}.
\end{remark}

The self-adjointness of $\w{\chi}$ implies that the 
bilinear form defined on $\Sigma$'s tangent space by
\be \label{e:hyp:2ndform_def}
	\encadre{
	\begin{array}{cccc}
	\w{K}: & \T_p(\Sigma)\times\T_p(\Sigma) & \longrightarrow & \mathbb{R} \\
		& (\w{u},\w{v}) & \longmapsto & - \w{u} \cdot \w{\chi}(\w{v})
	\end{array} }
\ee
is symmetric. 
It is called
the \defin{second fundamental form} of
the hypersurface $\Sigma$. It is also called the \defin{extrinsic curvature tensor}
of $\Sigma$ (cf. the remark above regarding the qualifier 'extrinsic'). 
$\w{K}$ contains the same information as the Weingarten map.
\begin{remark}
The minus sign in the definition (\ref{e:hyp:2ndform_def}) is chosen so
that $\w{K}$ agrees with the convention used in the numerical relativity
community, as well as in the MTW book \cite{MisneTW73}. Some other authors
(e.g. Carroll \cite{Carro04}, Poisson \cite{Poiss04}, Wald \cite{Wald84}) 
choose the opposite convention.
\end{remark} 
If we make explicit the value of $\w{\chi}$ in the 
definition (\ref{e:hyp:2ndform_def}), we get 
[see Eq.~(\ref{e:hyp:direc_deriv_1form})]
\be
  \forall (\w{u},\w{v}) \in \T_p(\Sigma)\times\T_p(\Sigma),\quad
  \encadre{ \w{K}(\w{u},\w{v}) =  - \w{u} \cdot\w{\nabla}_{\w{v}}\w{n} }
			. \label{e:hyp:Kuv}
\ee 
We shall denote by $K$ the trace of the bilinear form $\w{K}$ with respect to the metric $\w{\gam}$; it 
is the opposite of the trace of the endomorphism $\w{\chi}$ and is 
equal to $-3$ times the mean curvature of $\Sigma$:
\be \label{e:hyp:K_mean_curvature}
	K := \gam^{ij} K_{ij} = - 3 H . 
\ee

\begin{figure}
\centerline{\includegraphics[width=0.6\textwidth]{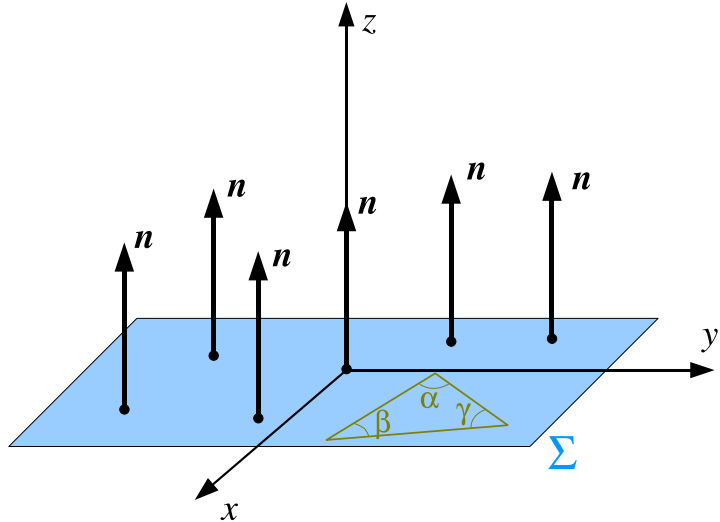}}
\caption[]{\label{f:hyp:plan} \footnotesize
Plane $\Sigma$ as a hypersurface of the Euclidean space $\R^3$. 
Notice that the unit normal vector $\w{n}$ stays constant along
$\Sigma$; this implies that the extrinsic curvature of $\Sigma$
vanishes identically. Besides, the sum of angles of any triangle
lying in $\Sigma$ is $\alpha+\beta+\gamma=\pi$, which shows that
the \emph{intrinsic} curvature of $(\Sigma,\w{\gam})$ vanishes as well.}
\end{figure}

\subsection{Examples: surfaces embedded in the Euclidean space $\R^3$}
\label{s:hyp:examples}

Let us illustrate the previous definitions with some hypersurfaces
of a space which we are very familiar with, namely $\R^3$
endowed with the standard Euclidean metric. In this case, the dimension is reduced by one unit with 
respect to the spacetime $\M$ and the ambient metric $\w{g}$ is Riemannian
(signature $(+,+,+)$) instead of Lorentzian. The hypersurfaces are
2-dimensional submanifolds of $\R^3$, namely they are \emph{surfaces}
by the ordinary meaning of this word.

In this section, and in this section only, we change our index convention
to take into account that the base manifold is of dimension 3 and not 4:
until the next section, the Greek indices run in $\{1,2,3\}$ and the Latin
indices run in $\{1,2\}$.

\begin{example}[1]
\textbf{a plane in $\R^3$} 

Let us take for $\Sigma$ the simplest surface
one may think of: a plane (cf. Fig.~\ref{f:hyp:plan}). 
Let us consider Cartesian coordinates $(X^\alpha)=(x,y,z)$ on $\R^3$, such that
$\Sigma$ is the $z=0$ plane. The scalar function $t$ defining
$\Sigma$ according to Eq.~(\ref{e:hyp:r_const}) is then simply $t=z$.
$(x^i) = (x,y)$ constitutes a coordinate system on $\Sigma$ and the metric
$\w{\gam}$ induced by $\w{g}$ on $\Sigma$ has the components $\gam_{ij} = \mathrm{diag}(1,1)$
with respect to these coordinates.
It is obvious that this metric is flat: $\w{\mathrm{Riem}}(\w{\gam})=0$. 
The unit normal $\w{n}$ has components $n^\alpha = (0,0,1)$ with respect to 
the coordinates $(X^\alpha)$. 
The components of
the gradient $\wnab\w{n}$ being simply given by the partial derivatives
$\nabla_\beta n^\alpha = \partial n^\alpha/\partial X^\beta$ [the Christoffel symbols vanishes for
the coordinates $(X^\alpha)$], we get immediately
$\wnab\w{n}=0$. Consequently, the Weingarten map and the extrinsic curvature vanish identically: $\w{\chi}=0$ and $\w{K}=0$. 
\end{example}

\begin{figure}
\centerline{\includegraphics[width=0.6\textwidth]{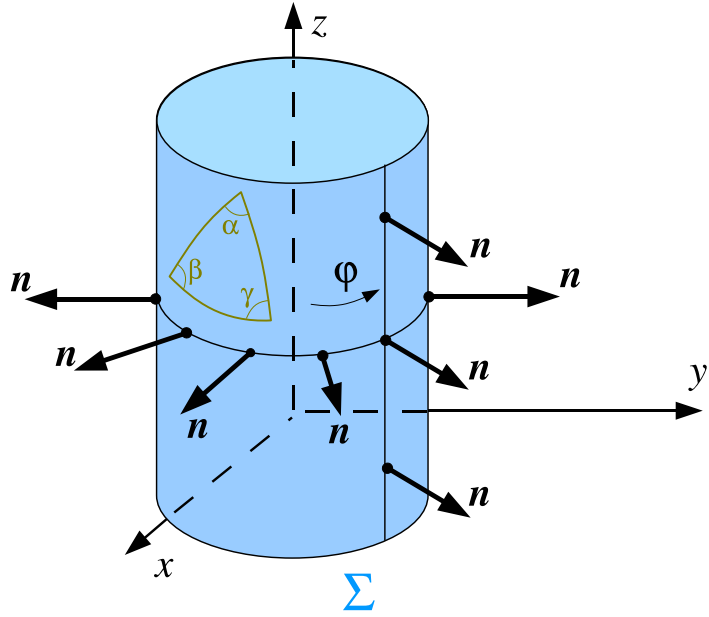}}
\caption[]{\label{f:hyp:cylindre} \footnotesize
Cylinder $\Sigma$ as a hypersurface of the Euclidean space $\R^3$. 
Notice that the unit normal vector $\w{n}$ stays constant when $z$
varies at fixed $\varphi$, whereas its direction changes as $\varphi$ varies
at fixed $z$. Consequently the extrinsic curvature of $\Sigma$ vanishes in the $z$
direction, but is non zero in the $\varphi$ direction.
Besides, the sum of angles of any triangle
lying in $\Sigma$ is $\alpha+\beta+\gamma=\pi$, which shows that
the \emph{intrinsic} curvature of $(\Sigma,\w{\gam})$ is identically zero.}
\end{figure}

\begin{example}[2]
\textbf{a cylinder in $\R^3$} 

Let us at present consider for $\Sigma$ the cylinder defined by
the equation $t:=\rho - R = 0$, where $\rho := \sqrt{x^2+y^2}$ and
$R$ is a positive constant --- the radius of the cylinder (cf Fig.~\ref{f:hyp:cylindre}). 
Let us introduce the cylindrical coordinates $(x^\alpha)=(\rho,\varphi,z)$, 
such that $\varphi\in[0,2\pi)$, $x=r\cos\varphi$ and $y=r\sin\varphi$. 
Then $(x^i)=(\varphi,z)$ constitutes a coordinate system on $\Sigma$.  
The components of the induced metric in this coordinate system are
given by 
\be \label{e:hyp:metr_cylind}
	\gam_{ij} \, dx^i \, dx^j = R^2 d\varphi^2 + dz^2 . 
\ee
It appears that this metric is flat, as for the plane considered
above. Indeed, the change of coordinate $\eta := R \, \varphi$ 
(remember $R$ is a constant !) transforms the metric components into
\be
	\gam_{i'j'} \, dx^{i'} \, dx^{j'} = d\eta^2 + dz^2 , 
\ee
which exhibits the standard Cartesian shape. 

To evaluate the extrinsic curvature of $\Sigma$, let us consider the
unit normal $\w{n}$ to $\Sigma$. Its components with respect to the
Cartesian coordinates $(X^\alpha)=(x,y,z)$ are
\be
	n^\alpha = \left( \frac{x}{\sqrt{x^2+y^2}},\ 
	\frac{y}{\sqrt{x^2+y^2}},\ 0 \right) . 
\ee
It is then easy to compute $\nabla_\beta n^\alpha = \partial n^\alpha/\partial X^\beta$.
We get 
\be \label{e:hyp:nab_n_cylind}
	\nabla_\beta n^\alpha = (x^2+y^2)^{-3/2} \left( 
	\begin{array}{ccc}
	y^2 & - x y & 0 \\
	- x y & x^2 & 0 \\
	0 & 0 & 0
	\end{array}
	\right) .
\ee
From Eq.~(\ref{e:hyp:Kuv}), the components of the extrinsic curvature $\w{K}$
with respect to the basis $(x^i)=(\varphi,z)$ are
\be \label{e:hyp:Kab_ij}
    K_{ij} = \w{K}(\wpar_i,\wpar_j)
    = - \nabla_\beta n_\alpha \, (\partial_i)^\alpha \, (\partial_j)^\beta ,
\ee
where $(\wpar_i)=(\wpar_\varphi,\wpar_z)=(\dert{}{\varphi},\dert{}{z})$ denotes the natural
basis associated with the coordinates $(\varphi,z)$ and $(\partial_i)^\alpha$
the components of the vector $\wpar_i$ with respect to the
natural basis $(\wpar_\alpha)=(\wpar_x,\wpar_y,\wpar_z)$ associated 
with the Cartesian coordinates $(X^\alpha)=(x,y,z)$. 
Specifically, since $\wpar_\varphi = -y \wpar_x + x \wpar_y$, one has
$(\partial_\varphi)^\alpha=(-y,x,0)$ and $(\partial_z)^\alpha = (0,0,1)$. 
From Eq.~(\ref{e:hyp:nab_n_cylind}) and (\ref{e:hyp:Kab_ij}), 
we then obtain
\be \label{e:hyp:Kij_cylinder}
    K_{ij} = \left( \begin{array}{cc}
	K_{\varphi\varphi} & K_{\varphi z} \\
        K_{z\varphi} & K_{zz}
	\end{array} \right)
    = \left(\begin{array}{cc}
	-R & 0 \\
        0 & 0 
	\end{array} \right) .
\ee
From Eq.~(\ref{e:hyp:metr_cylind}), 
$\gam^{ij} = \mathrm{diag}(R^{-2}, 1)$, so that the 
trace of $\w{K}$ is 
\be \label{e:hyp:K_cylinder}
	K = - \frac{1}{R} . 
\ee
\end{example}

\begin{figure}
\centerline{\includegraphics[width=0.5\textwidth]{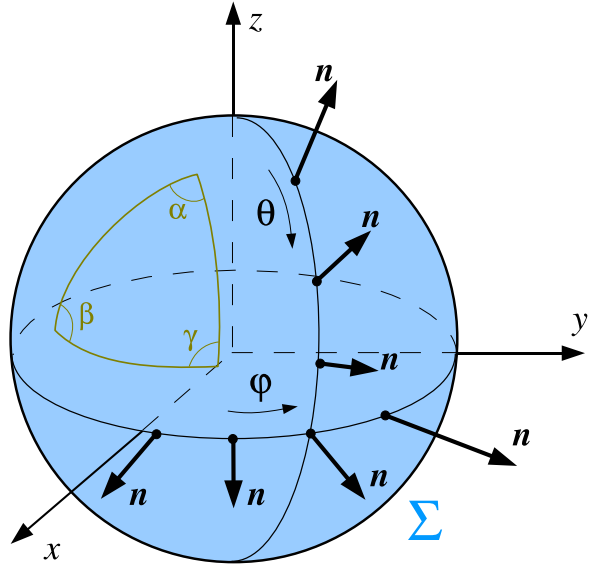}}
\caption[]{\label{f:hyp:sphere} \footnotesize
Sphere $\Sigma$ as a hypersurface of the Euclidean space $\R^3$. 
Notice that the unit normal vector $\w{n}$ changes its direction 
when displaced on $\Sigma$. This shows that the extrinsic curvature 
of $\Sigma$ does not vanish.
Moreover all directions being equivalent at the surface of the
sphere, $\w{K}$ is necessarily proportional to the induced metric
$\w{\gam}$, as found by the explicit calculation leading to 
Eq.~(\ref{e:hyp:Kab_sphere}).
Besides, the sum of angles of any triangle
lying in $\Sigma$ is $\alpha+\beta+\gamma>\pi$, which shows that
the \emph{intrinsic} curvature of $(\Sigma,\w{\gam})$ does not vanish either.}
\end{figure}

\begin{example}[3]
\textbf{a sphere in $\R^3$} 
Our final simple example is constituted by the sphere of radius $R$
(cf. Fig.~\ref{f:hyp:sphere}), 
the equation of which is
$t:=r-R=0$, with $r=\sqrt{x^2+y^2+z^2}$. Introducing the
spherical coordinates $(x^\alpha)=(r,\theta,\varphi)$ such that
$x=r\sin\theta\cos\varphi$, $y=r\sin\theta\sin\varphi$ and
$z=r\cos\theta$, $(x^i)=(\theta,\varphi)$ constitutes a
coordinate system on $\Sigma$. The components of the induced
metric $\w{\gam}$ in this coordinate system are given by
\be 
	\gam_{ij} \, dx^i \, dx^j = R^2 \left( d\theta^2 + \sin^2\theta d\varphi^2
	\right) .  
\ee
Contrary to the previous two examples, this metric is not flat:
the Ricci scalar, Ricci tensor and Riemann tensor of $(\Sigma,\w{\gam})$
are respectively\footnote{the superscript $\Sigma$ has been put on the Ricci
scalar to distinguish it from the sphere's radius $R$.}
\be \label{e:hyp:curv_sphere}
	{}^\Sigma R = \frac{2}{R^2},\quad
	R_{ij} = \frac{1}{R^2} \, \gam_{ij},\quad
	R^i_{\ \, jkl} = \frac{1}{R^2} \left( \delta^i_{\ \, k} \gam_{jl}
	- \delta^i_{\ \, l} \gam_{jk} \right) .
\ee
The non vanishing of the Riemann tensor is reflected by the well-known property
that the sum of angles of any triangle drawn at the surface of a sphere
is larger than $\pi$ (cf. Fig.~\ref{f:hyp:sphere}). 

The unit vector $\w{n}$ normal to $\Sigma$ (and oriented towards the exterior of the sphere)
has the following components with respect to the coordinates $(X^\alpha)=(x,y,z)$:
\be
	n^\alpha = \left( \frac{x}{\sqrt{x^2+y^2+z^2}},\ 
	\frac{y}{\sqrt{x^2+y^2+z^2}},\ \frac{z}{\sqrt{x^2+y^2+z^2}} \right) . 
\ee
It is then easy to compute $\nabla_\beta n^\alpha = \partial n^\alpha/\partial X^\beta$
to get
\be
	\nabla_\beta n^\alpha = (x^2+y^2+z^2)^{-3/2} \left( 
	\begin{array}{ccc}
	y^2+z^2 & - xy & - x z \\
	- x y & x^2 + z^2 & - y z \\
	- x z & - y z & x^2 + y^2 
	\end{array} \right) .
\ee
The natural basis associated with the coordinates $(x^i)=(\theta,\varphi)$ on
$\Sigma$ is 
\bea
	\wpar_\theta & = & (x^2+y^2)^{-1/2} \left[ x z\, \wpar_x 
	+ y z \, \wpar_y  - (x^2+y^2) \, \wpar_z \right]
	\label{e:CG:partial_theta} \\
	\wpar_\varphi & = & - y \, \wpar_x + x \, \wpar_y  .
\eea
The components of the extrinsic curvature tensor in this basis are
obtained from $K_{ij} = \w{K}(\wpar_i,\wpar_j)
    = - \nabla_\beta n_\alpha \, (\partial_i)^\alpha \, (\partial_j)^\beta$.
We get
\be \label{e:hyp:Kab_sphere}
    K_{ij} = \left( \begin{array}{cc}
	K_{\theta\theta} & K_{\theta\varphi} \\
        K_{\varphi\theta} & K_{\varphi\varphi}
	\end{array} \right)
    = \left(\begin{array}{cc}
	-R & 0 \\
        0 & - R \sin^2 \theta
	\end{array} \right) = - \frac{1}{R} \, \gam_{ij}.
\ee
The trace of $\w{K}$ with respect to $\w{\gam}$ is then 
\be \label{e:hyp:K_sphere}
	K = - \frac{2}{R} . 
\ee
\end{example}

With these examples, we have encountered hypersurfaces with intrinsic and extrinsic curvature
both vanishing (the plane), the intrinsic curvature vanishing but not the extrinsic one
(the cylinder), and with both curvatures non vanishing (the sphere). 
As we shall see in Sec.~\ref{s:hyp:Gauss_Codaz}, the extrinsic curvature is not fully
independent from the intrinsic one: they are related by the Gauss equation. 

%%%%%%%%%%%%%%%%%%%%%%%%%%%%%%%%%%%%%%%%%%%%%%%%%%%%%%%%%%%%%%%%%%%%%%%%%%%%%%%%

\section{Spacelike hypersurface}

From now on, we focus on spacelike hypersurfaces, i.e. hypersurfaces $\Sigma$ such
that the induced metric $\w{\gam}$ is definite positive (Riemannian), or equivalently
such that the unit normal vector $\w{n}$ is timelike (cf. Secs.~\ref{s:hyp:def_hyp}
and \ref{s:hyp:norm_vect}).

\subsection{The orthogonal projector}

At each point $p\in\Sigma$, the space of all spacetime vectors can be
orthogonally decomposed as 
\be \label{e:hyp:TM_ortho}
	\encadre{ \T_p(\M) = \T_p(\Sigma) \oplus \mathrm{Vect}(\w{n}) }, 
\ee
where $\mathrm{Vect}(\w{n})$ stands for the 1-dimensional subspace of $\T_p(\M)$ generated
by the vector $\w{n}$. 
\begin{remark}
The orthogonal decomposition (\ref{e:hyp:TM_ortho}) holds for spacelike and timelike hypersurfaces,
but not for the null ones. Indeed for any normal $\w{n}$ to a null hypersurface $\Sigma$, $\mathrm{Vect}(\w{n})\subset\T_p(\Sigma)$.
\end{remark}
The \defin{orthogonal projector onto $\Sigma$} is the operator 
$\vg$ associated with the decomposition (\ref{e:hyp:TM_ortho})
according to 
\be \label{e:hyp:def_vg}
	\encadre{
	\begin{array}{cccc}
	\vg: & \T_p(\M) & \longrightarrow & \T_p(\Sigma) \\
		& \w{v} & \longmapsto & \w{v} + (\w{n}\cdot\w{v})\,  \w{n} .
	\end{array} }
\ee
In particular, as a direct consequence of $\w{n}\cdot\w{n}=-1$, $\vg$ satisfies
\be
	\vg(\w{n}) = 0 .
\ee
Besides, it reduces to the identity operator for any vector tangent to $\Sigma$:
\be
	\forall \w{v}\in\T_p(\Sigma),\quad \vg(\w{v}) = \w{v} . 
\ee
According to Eq.~(\ref{e:hyp:def_vg}), the components of $\vg$ with respect to any
basis $(\w{e}_\alpha)$ of $\T_p(\M)$ are
\be \label{e:hyp:vecgam_ab}
	\encadre{\gam^\alpha_{\ \, \beta} = \delta^\alpha_{\ \, \beta} 
	+ n^\alpha n_\beta }.
\ee

We have noticed in Sec.~\ref{s:hyp:def_hyp} that the embedding $\Phi$ of $\Sigma$
in $\M$ induces a mapping $\T_p(\Sigma)\rightarrow\T_p(\M)$ (push-forward)
and a mapping $\T_p^*(\M)\rightarrow\T_p^*(\Sigma)$ (pull-back), but does not
provide any mapping in the reverse ways, i.e. from $\T_p(\M)$ to $\T_p(\Sigma)$
and from $\T_p^*(\Sigma)$ to $\T_p^*(\M)$. The orthogonal projector naturally provides
these reverse mappings: from its very definition, it is a mapping $\T_p(\M)\rightarrow\T_p(\Sigma)$
and we can construct from it a mapping $\vgs_{\scriptscriptstyle \M}:\ \T_p^*(\Sigma)\rightarrow\T_p^*(\M)$ by setting,
for any linear form $\w{\omega} \in \T_p^*(\Sigma)$,
\be 
	\begin{array}{cccc}
	\vgs_{\scriptscriptstyle \M}\w{\omega}: & \T_p(\M) & \longrightarrow & \R \\
		& \w{v} & \longmapsto & \w{\omega}(\vg(\w{v})) .
	\end{array} 
\ee
This clearly defines a linear form belonging to $\T_p^*(\M)$.
Obviously, we can extend the operation $\vgs_{\scriptscriptstyle \M}$ to any multilinear form $\w{A}$ 
acting on $\T_p(\Sigma)$,
by setting 
\be \label{e:hyp:def_vgs_mform}
	\begin{array}{cccc}
	\vgs_{\scriptscriptstyle \M}\w{A} : & \T_p(\M)^n & \longrightarrow & \mathbb{R} \\
		& (\w{v}_1,\ldots,\w{v}_n) & \longmapsto & 
		\w{A} \left( \vg(\w{v}_1),\ldots,
                \vg(\w{v}_n) \right) .
	\end{array}  
\ee
Let us apply this definition to the bilinear form on $\Sigma$ constituted by the induced
metric $\w{\gam}$: $\vgs_{\scriptscriptstyle \M}\w{\gam}$ is then a bilinear form on $\M$, which 
coincides with $\w{\gam}$ if its two arguments are vectors tangent to $\Sigma$ and 
which gives zero if any of its argument is a vector orthogonal to $\Sigma$, 
i.e. parallel to $\w{n}$. Since it constitutes an ``extension'' of $\w{\gam}$
to all vectors in $\T_p(\M)$, we shall denote it by the same symbol:
\be 
	\encadre{ \w{\gam} := \vgs_{\scriptscriptstyle \M} \w{\gam} } .
\ee
This extended $\w{\gam}$ can be expressed in terms of the metric tensor $\w{g}$
and the linear form $\uu{n}$ dual to the normal vector $\w{n}$ according to
\be \label{e:hyp:gam_g_nn}
	\encadre{ \w{\gam} = \w{g} + \uu{n}\otimes\uu{n} } . 
\ee
In components:
\be \label{e:hyp:gam_ab}
	\gam_{\alpha\beta} =  g_{\alpha\beta} + n_\alpha \, n_\beta . 
\ee
Indeed, if $\w{v}$ and $\w{u}$ are vectors both tangent to $\Sigma$, 
$\w{\gam}(\w{u},\w{v})=\w{g}(\w{u},\w{v}) + \langle\uu{n},\w{u}\rangle
	\langle\uu{n},\w{v}\rangle 
	= \w{g}(\w{u},\w{v}) + 0 =\w{g}(\w{u},\w{v}) $, 
and if $\w{u}=\lambda\w{n}$, then, for any 
$\w{v}\in\T_p(\M)$,  $\w{\gam}(\w{u},\w{v})= \lambda \w{g}(\w{n},\w{v})+ \lambda 
	\langle\uu{n},\w{n}\rangle \langle\uu{n},\w{v}\rangle
	= \lambda [\w{g}(\w{n},\w{v}) - \langle\uu{n},\w{v}\rangle]=0 $. 
This establishes Eq.~(\ref{e:hyp:gam_g_nn}).
Comparing Eq.~(\ref{e:hyp:gam_ab}) with Eq.~(\ref{e:hyp:vecgam_ab}) justifies the notation
$\vg$ employed for the orthogonal projector onto $\Sigma$, according to the 
convention set in Sec.~\ref{s:hyp:metric_dual} [see Eq.~(\ref{e:hyp:arrow_endo})]: 
$\vg$ is nothing but the 
''extended'' induced metric $\w{\gam}$ with the first index raised by the metric $\w{g}$.

Similarly, we may use the $\vgs_{\scriptscriptstyle \M}$ operation to extend 
the extrinsic curvature tensor
$\w{K}$, defined a priori as a bilinear form on $\Sigma$ [Eq.~(\ref{e:hyp:2ndform_def})], 
to a bilinear form on $\M$, 
and we shall use the same symbol to denote this extension:
\be \label{e:hyp:def_K_extend}
	\encadre{\w{K} := \vgs_{\scriptscriptstyle \M} \w{K}}.
\ee

\begin{remark}
In this lecture, we will very often use such a ``four-dimensional point of view'', 
i.e. we shall treat tensor fields defined on $\Sigma$ as if they were defined
on $\M$. For covariant tensors (multilinear forms), if not mentioned explicitly, 
the four-dimensional extension is performed via the $\vgs_{\scriptscriptstyle \M}$ operator, 
as above for $\w{\gam}$ and $\w{K}$. For contravariant tensors, 
the identification is provided by the push-forward mapping $\Phi_*$ discussed in Sec.~\ref{s:hyp:def_hyp}.
This four-dimensional point of view has been advocated by
Carter \cite{Carte92a,Carte92b,Carte97} and results in an easier manipulation 
of tensors defined in $\Sigma$, by treating them as ordinary tensors on $\M$.
In particular this avoids the introduction of special coordinate systems and complicated notations. 
\end{remark}

In addition to the extension of three dimensional tensors to four dimensional
ones, we use the orthogonal projector $\vg$ to define an ``orthogonal projection 
operation'' for \emph{all tensors on $\M$} in the following way.
Given a tensor $\w{T}$ of type $\left({p \atop q}\right)$ on $\M$, we 
denote by $\vgs\w{T}$ another tensor on $\M$, of the same type and such that its components in any basis 
$(\w{e}_\alpha)$ of $\T_p(\M)$ are expressed in terms of those of $\w{T}$ by
\be \label{e:hyp:vgs_comp}
	(\vgs \w{T})^{\alpha_1\ldots\alpha_p}_{\qquad\ \; \beta_1\ldots\beta_q}
	= \gam^{\alpha_1}_{\ \  \mu_1} \ldots \gam^{\alpha_p}_{\ \ \mu_p}
	\gam^{\nu_1}_{\ \ \beta_1} \ldots \gam^{\nu_q}_{\ \ \beta_q}
	\, T^{\mu_1\ldots\mu_p}_{\qquad\ \; \nu_1\ldots\nu_q} . 
\ee
Notice that for any multilinear form $\w{A}$ on $\Sigma$, 
$\vgs(\vgs_{\scriptscriptstyle \M} \w{A}) = \vgs_{\scriptscriptstyle \M} \w{A}$, for
a vector $\w{v}\in\T_p(\M)$, $\vgs\w{v} = \vg(\w{v})$,
for a linear form $\w{\omega}\in\T^*_p(\M)$, $\vgs\w{\omega} = \w{\omega} \circ\vg$,
and for any tensor $\w{T}$, $\vgs \w{T}$ is \defin{tangent to $\Sigma$},
in the sense that $\vgs \w{T}$ results in zero if one of its arguments is
$\w{n}$ or $\uu{n}$.

\subsection{Relation between $K$ and $\nabla n$} \label{s:hyp:rel_K_nabn}

A priori the unit vector $\w{n}$ normal to $\Sigma$ is defined only at points
belonging to $\Sigma$. Let us consider some extension of $\w{n}$ in an open
neighbourhood of $\Sigma$. If $\Sigma$ is a level surface of some scalar field
$t$, such a natural extension is provided by the gradient of $t$, according
to Eq.~(\ref{e:hyp:n_dt_gal}). Then the tensor fields $\wnab\w{n}$
and $\wnab{\uu{n}}$ are well defined quantities. 
In particular, we can introduce the vector 
\be
	\w{a} := \wnab_{\w{n}} \w{n} . 
\ee
Since $\w{n}$ is a timelike unit vector, it can be regarded as the 4-velocity
of some observer, and $\w{a}$ is then the corresponding 4-acceleration.
$\w{a}$ is orthogonal to $\w{n}$ and hence tangent to $\Sigma$,
since $\w{n}\cdot\w{a} = \w{n} \cdot \wnab_{\w{n}} \w{n} =
1/2 \, \wnab_{\w{n}} (\w{n}\cdot\w{n}) = 1/2 \, \wnab_{\w{n}} (-1) = 0$.

Let us make explicit the definition of the tensor $\w{K}$
extend to $\M$ by Eq.~(\ref{e:hyp:def_K_extend}). From the definition 
of the operator $\vgs_{\scriptscriptstyle \M}$ [Eq.~(\ref{e:hyp:def_vgs_mform})]
and the original definition of $\w{K}$ [Eq.~(\ref{e:hyp:Kuv})], we have
\bea
	\forall (\w{u},\w{v})\in\T_p(\M)^2,\quad
	\w{K}(\w{u},\w{v}) & = & \w{K}(\vg(\w{u}),\vg(\w{v})) 
	= - \vg(\w{u}) \cdot \wnab_{\vg(\w{v})} \w{n} \nonumber \\
  & = & -\vg(\w{u}) \cdot \wnab_{\w{v} + (\w{n}\cdot\w{v})\w{n}} \, \w{n} 
	\nonumber \\
	& = &  - [\w{u} + (\w{n}\cdot\w{u})\w{n}] \cdot
	[ \wnab_{\w{v}} \w{n} + (\w{n}\cdot\w{v}) \wnab_{\w{n}} \w{n} ]
	\nonumber \\
	& = & - \w{u}\cdot \wnab_{\w{v}} \w{n} 
		- (\w{n}\cdot\w{v})\w{u} \cdot \underbrace{\wnab_{\w{n}} \w{n}}_{=\w{a}}
		- (\w{n}\cdot\w{u})
	\underbrace{\w{n}\cdot\wnab_{\w{v}} \w{n}}_{=0} \nonumber \\
	& & - (\w{n}\cdot\w{u})(\w{n}\cdot\w{v})
	\underbrace{\w{n}\cdot\wnab_{\w{n}} \w{n}}_{=0} \nonumber \\
	& = & - \w{u}\cdot \wnab_{\w{v}} \w{n} 
		- (\w{a}\cdot\w{u})(\w{n}\cdot\w{v}) , \nonumber \\
	& = & - \wnab\uu{n}(\w{u},\w{v}) - \langle\uu{a},\w{u}\rangle 
		\langle \uu{n},\w{v} \rangle ,\label{e:hyp:Kuv_interm}
\eea
where we have used the fact that $\w{n}\cdot\w{n}=-1$ to set
$\w{n}\cdot\wnab_{\w{x}}\w{n}=0$ for any vector $\w{x}$. 
Since Eq.~(\ref{e:hyp:Kuv_interm}) is valid for any pair of vectors $(\w{u},\w{v})$
in $\T_p(\M)$, we conclude that 
\be \label{e:hyp:nab_n_K}
	\encadre{ \wnab \uu{n}  = - \w{K} - \uu{a}\otimes \uu{n} } .
\ee
In components:
\be \label{e:hyp:nab_n_K_comp}
	\encadre{ \nabla_\beta \, n_\alpha = - K_{\alpha\beta} - a_\alpha \, n_\beta } . 
\ee
Notice that Eq.~(\ref{e:hyp:nab_n_K}) implies that the (extended) extrinsic curvature tensor
is nothing but the gradient of the 1-form $\uu{n}$ to which the projector operator
$\vgs$ is applied:
\be \label{e:hyp:K_vgs_nabn}
	\encadre{ \w{K} = - \vgs \wnab\uu{n}} . 
\ee
\begin{remark}
Whereas the bilinear form $\wnab\uu{n}$ is a priori not symmetric, its projected part
$-\w{K}$ is a symmetric bilinear form. 
\end{remark}

Taking the trace of Eq.~(\ref{e:hyp:nab_n_K}) with respect to the metric
$\w{g}$ (i.e. contracting Eq.~(\ref{e:hyp:nab_n_K_comp}) with $g^{\alpha\beta}$)
yields a simple relation between the divergence of the vector $\w{n}$ and
the trace of the extrinsic curvature tensor:
\be \label{e:hyp:K_div_n}
	\encadre{ K = -\wnab\cdot \w{n} }. 
\ee

\subsection{Links between the $\nabla$ and $D$ connections} \label{e:hyp:link_nab_D}

Given a tensor field $\w{T}$ on $\Sigma$, its covariant derivative 
$\w{D}\w{T}$ with respect to the Levi-Civita connection $\w{D}$ of the metric
$\w{\gam}$ (cf. Sec.~\ref{s:hyp:intrins_curv}) is expressible in terms
of the covariant derivative $\wnab\w{T}$ with respect to the spacetime
connection $\wnab$ according to the formula
\be \label{e:hyp:link_D_nab}
	\encadre{ \w{D}\w{T} = \vgs \w{\nabla} \w{T} }  , 
\ee
the component version of which is [cf. Eq.~(\ref{e:hyp:vgs_comp})]:
\be \label{e:hyp:link_D_nab_comp}
\encadre{ D_\rho T^{\alpha_1\ldots\alpha_p}_{\ \qquad\beta_1\ldots\beta_q}
		= \gamma_{\ \ \, \mu_1}^{\alpha_1} \, \cdots 
		 \gamma_{\ \ \, \mu_p}^{\alpha_p} \,
		  \gamma_{\ \ \, \beta_1}^{\nu_1} \, \cdots
		  \gamma_{\ \ \, \beta_q}^{\nu_q} \,
		  \gamma_{\ \ \, \rho}^{\sigma} \, \nabla_\sigma
		  T^{\mu_1\ldots\mu_p}_{\ \qquad\nu_1\ldots\nu_q} } .		  
\ee
Various comments are appropriate: first of all, the $\w{T}$ in the
right-hand side of Eq.~(\ref{e:hyp:link_D_nab}) should be the four-dimensional extension $\vgs_{\scriptscriptstyle \M}\w{T}$
provided by Eq.~(\ref{e:hyp:def_vgs_mform}). Following the remark made
above, we write $\w{T}$ instead of $\vgs_{\scriptscriptstyle \M}\w{T}$. 
Similarly the right-hand side should write $\vgs_{\scriptscriptstyle \M} \w{D}\w{T}$, so that Eq.~(\ref{e:hyp:link_D_nab}) is a equality between tensors on $\M$.
Therefore the rigorous version of Eq.~(\ref{e:hyp:link_D_nab}) is
\be 
	\vgs_{\scriptscriptstyle \M} \w{D}\w{T} = \vgs [\wnab
     (\vgs_{\scriptscriptstyle \M}\w{T}) ].  
\ee
Besides, even if
$\w{T}:=\vgs_{\scriptscriptstyle \M}\w{T}$ is a four-dimensional tensor, 
its suppport (domain of definition)
remains the hypersurface $\Sigma$. In order to define the covariant derivative
$\wnab\w{T}$, the support must be an open set of $\M$, 
which $\Sigma$ is not. Accordingly, one must first construct some extension 
$\w{T}'$ of $\w{T}$ in an open neighbourhood of $\Sigma$
in $\M$ and then compute $\wnab\w{T}'$. The
key point is that thanks to the operator $\vgs$ acting on 
$\wnab\w{T}'$, the result does not depend of the choice of the extension
$\w{T}'$, provided that $\w{T}'=\w{T}$ at every point in $\Sigma$.

The demonstration of the formula (\ref{e:hyp:link_D_nab}) takes two steps.
First, one can show easily that $\vgs\wnab$ (or more precisely
the pull-back of $\vgs\wnab\vgs_{\scriptscriptstyle \M}$) is a torsion-free connection 
on $\Sigma$, for it satisfies all the defining properties of a connection
(linearity, reduction to the gradient for a scalar field,
commutation with contractions and Leibniz' rule) and its torsion vanishes.
Secondly, this connection vanishes when applied to the metric tensor
$\w{\gam}$: indeed, using Eqs.~(\ref{e:hyp:vgs_comp}) and (\ref{e:hyp:gam_ab}),
\bea
  \left( \vgs\wnab\w{\gam}\right) _{\alpha\beta\gamma}
	& = & \gam^\mu_{\ \, \alpha} \gam^\nu_{\ \, \beta}
	\gam^\rho_{\ \, \gamma} \nabla_\rho \gam_{\mu\nu} \nonumber \\
	& = &\gam^\mu_{\ \, \alpha} \gam^\nu_{\ \, \beta}
	\gam^\rho_{\ \, \gamma} ( 
	\underbrace{ \nabla_\rho \, g_{\mu\nu} }_{=0}
	+ \nabla_\rho n_\mu \, n_\nu	
	+ n_\mu \nabla_\rho n_\nu ) \nonumber \\
 	& = & \gam^\rho_{\ \, \gamma}  ( \gam^\mu_{\ \, \alpha} 
	\underbrace{\gam^\nu_{\ \, \beta}  n_\nu}_{=0}
	\nabla_\rho n_\mu +
	\underbrace{\gam^\mu_{\ \, \alpha} n_\mu}_{=0}
	\nabla_\rho n_\nu ) \nonumber \\
	& = & 0 .
\eea 
Invoking the uniqueness of the torsion-free connection associated with a given
non-degenerate metric (the Levi-Civita connection, cf. Sec.~2.IV.2 of N. Deruelle's lecture
\cite{Derue06}), we conclude that necessarily $\vgs\wnab=\w{D}$.

One can deduce from Eq.~(\ref{e:hyp:link_D_nab}) an interesting formula
about the derivative of a vector field $\w{v}$ along 
another vector field $\w{u}$, when both vectors are tangent to $\Sigma$. Indeed, from 
Eq.~(\ref{e:hyp:link_D_nab}),
\bea
	(\w{D}_{\w{u}} \w{v})^\alpha & = & u^\sigma D_\sigma v^\alpha
	= \underbrace{u^\sigma \gam^\nu_{\ \, \sigma}}_{=u^\nu} 
		\gam^\alpha_{\ \, \mu} 
	\nabla_\nu v^\mu 
	= u^\nu \left( \delta^\alpha_{\ \, \mu} + n^\alpha n_\mu \right)
	\nabla_\nu v^\mu  \nonumber \\
	& = & u^\nu \nabla_\nu v^\alpha 
	+ n^\alpha u^\nu 
	\underbrace{n_\mu \nabla_\nu v^\mu}_{=-v^\mu \nabla_\nu n_\mu}  
	= u^\nu \nabla_\nu v^\alpha - n^\alpha u^\nu v^\mu \nabla_\mu n_\nu ,
\eea
where we have used $n_\mu v^\mu =0$ ($\w{v}$ being tangent to $\Sigma$) to
write $n_\mu \nabla_\nu v^\mu = -v^\mu \nabla_\nu n_\mu$.
Now, from Eq.~(\ref{e:hyp:Kuv}), 
$- u^\nu v^\mu \nabla_\mu n_\nu = \w{K}(\w{u},\w{v})$, so that the above formula
becomes
\be \label{e:hyp:Duv_nabuv_K}
  \forall (\w{u},\w{v})\in \T(\Sigma)\times\T(\Sigma),\quad
	\encadre{ \w{D}_{\w{u}} \w{v} = \wnab_{\w{u}} \w{v}
	+ \w{K}(\w{u},\w{v}) \, \w{n} } . 
\ee
This equation provides another interpretation of the extrinsic curvature tensor
$\w{K}$: $\w{K}$ measures the deviation of the derivative of any vector of $\Sigma$
along another vector of $\Sigma$, taken with the intrinsic connection $\w{D}$
of $\Sigma$ from the derivative taken with the spacetime connection $\wnab$. 
Notice from Eq.~(\ref{e:hyp:Duv_nabuv_K}) that this deviation is always 
in the direction of the normal vector $\w{n}$. 

\begin{figure}
\centerline{\includegraphics[width=0.5\textwidth]{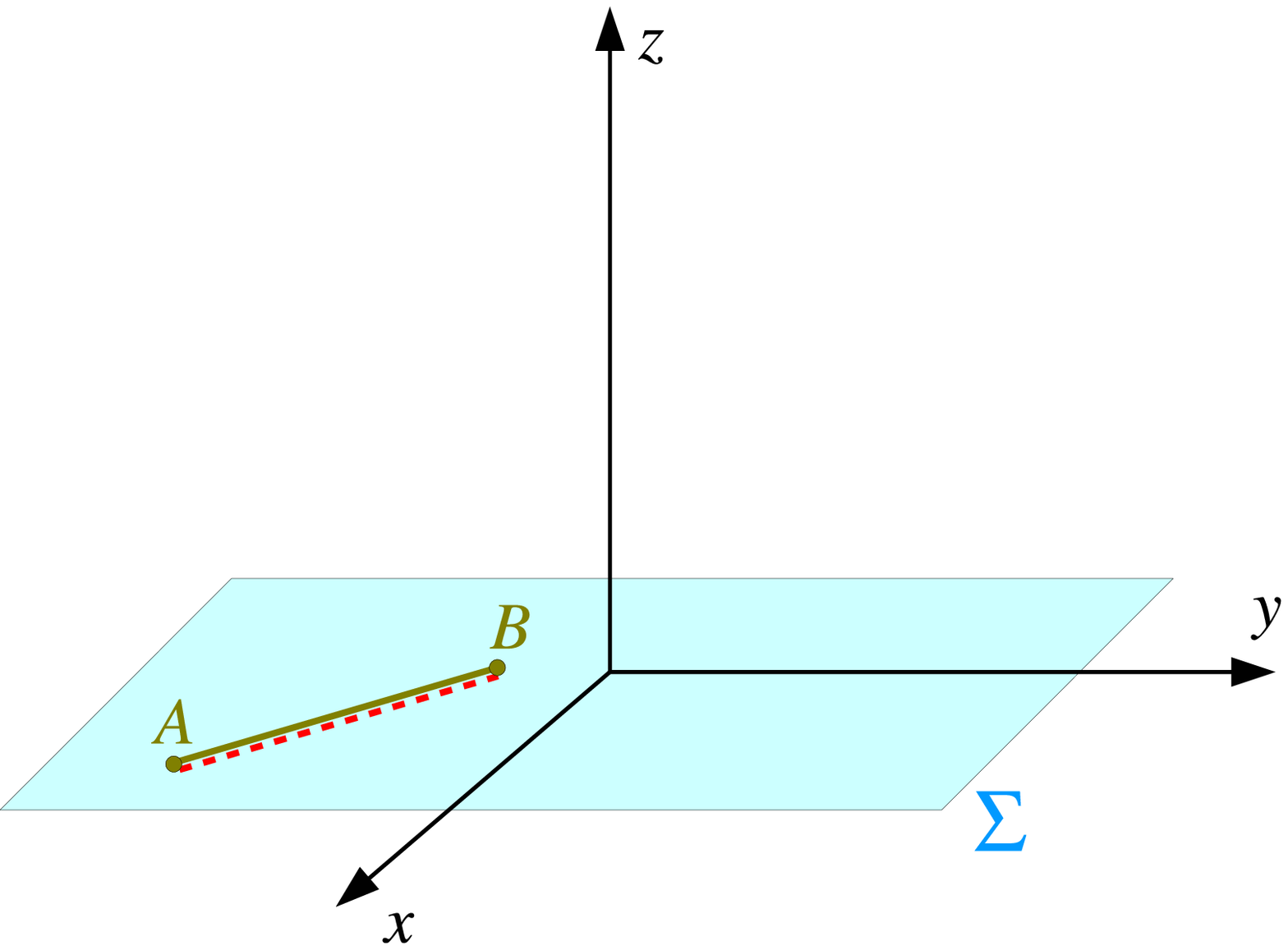}
\quad \includegraphics[width=0.4\textwidth]{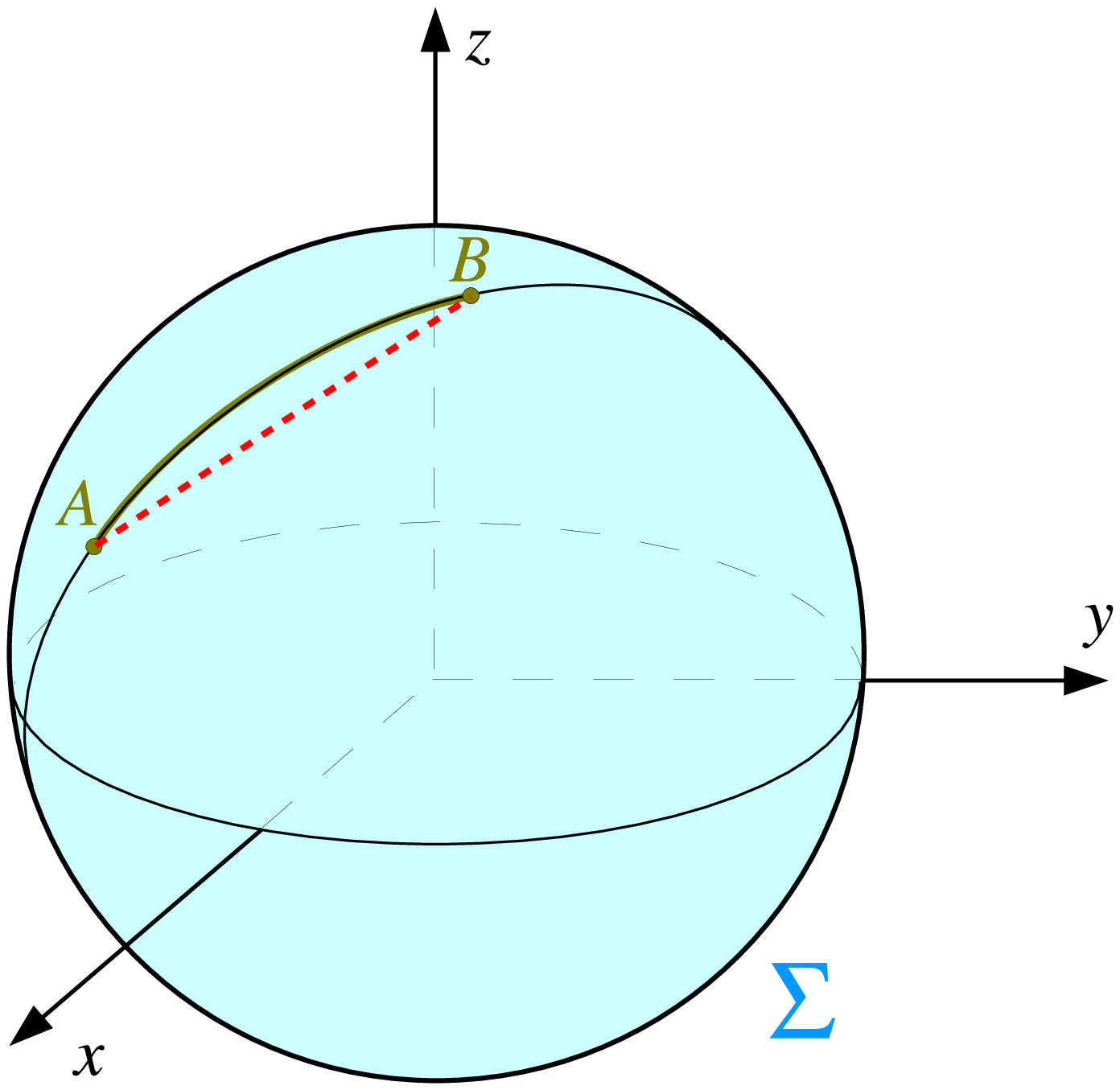}}
\caption[]{\label{f:hyp:plan_sphere_geod} \footnotesize
In the Euclidean space $\R^3$, the plane $\Sigma$ is a totally geodesic hypersurface, for the 
geodesic between two points $A$ and $B$ within $(\Sigma,\w{\gam})$ (solid line) coincides
with the geodesic in the ambient space (dashed line). On the contrary, for the sphere, the
two geodesics are distinct, whatever the position of points $A$ and $B$.}
\end{figure}

Consider a geodesic curve $\cal L$ in $(\Sigma,\w{\gam})$ and the tangent vector 
$\w{u}$ associated with some affine parametrization of $\cal L$. 
Then $\w{D}_{\w{u}} \w{u} = 0$
and Eq.~(\ref{e:hyp:Duv_nabuv_K}) leads to 
$\wnab_{\w{u}} \w{u}=-\w{K}(\w{u},\w{u}) \, \w{n}$.
If $\cal L$ were a geodesic of $(\M,\w{g})$, one should have 
$\wnab_{\w{u}} \w{u} = \kappa \w{u}$, for some non-affinity parameter $\kappa$.
Since $\w{u}$ is never parallel to $\w{n}$, we conclude that the
extrinsic curvature tensor $\w{K}$ measures the failure of a geodesic of
$(\Sigma,\w{\gam})$ to be a geodesic of $(\M,\w{g})$. 
Only in the case where $\w{K}$ vanishes, the two notions of geodesics coincide. 
For this reason, hypersurfaces for which $\w{K}=0$ are called 
\defin{totally geodesic hypersurfaces}. 

\begin{example}
The plane in the Euclidean 
space $\R^3$ discussed as Example~1 in Sec.~\ref{s:hyp:examples}
is a totally geodesic hypersurface: $\w{K}=0$. This is obvious since
the geodesics of the plane are straight lines, which are also geodesics of $\R^3$
(cf. Fig.~\ref{f:hyp:plan_sphere_geod}).
A counter-example is provided by the sphere embedded in $\R^3$ 
(Example~3 in Sec.~\ref{s:hyp:examples}): given two points $A$ and $B$, the geodesic curve
with respect to $(\Sigma,\w{\gam})$ joining them is a portion of a sphere's great circle, whereas from the point of view of $\R^3$, the geodesic from $A$ to $B$ is a 
straight line (cf. Fig.~\ref{f:hyp:plan_sphere_geod}).
\end{example}

%%%%%%%%%%%%%%%%%%%%%%%%%%%%%%%%%%%%%%%%%%%%%%%%%%%%%%%%%%%%%%%%%%%%%%%%%%%%%%%%

\section{Gauss-Codazzi relations} \label{s:hyp:Gauss_Codaz}

We derive here equations that will constitute the basis of the 3+1 formalism for
general relativity. They are
decompositions of the spacetime Riemann tensor, ${}^4\mathbf{Riem}$ 
[Eq.~(\ref{e:hyp:def_Riemann})], in terms of quantities relative to the spacelike 
hypersurface $\Sigma$, namely the Riemann tensor associated with the induced metric $\w{\gam}$,
$\mathbf{Riem}$ [Eq.~(\ref{e:hyp:Ricci_ident_3D})] and the extrinsic curvature tensor 
of $\Sigma$, $\w{K}$.

\subsection{Gauss relation}

Let us consider the Ricci identity (\ref{e:hyp:Ricci_ident_3D}) 
defining the (three-dimensional) Riemann tensor $\mathbf{Riem}$ as measuring
the lack of commutation of two successive covariant derivatives with
respect to the connection $\w{D}$ associated with $\Sigma$'s 
metric $\w{\gam}$. The four-dimensional version of this identity
is
\be \label{e:hyp:Ricci_ident_3D_comp}
	D_\alpha D_\beta v^\gam - D_\beta D_\alpha v^\gam = R^\gam_{\ \, \mu\alpha\beta} \, v^\mu ,
\ee
where $\w{v}$ is a generic vector field tangent to $\Sigma$. 
Let us use formula (\ref{e:hyp:link_D_nab_comp}) which relates the $\w{D}$-derivative to the $\wnab$-derivative, to write
\be
	D_\alpha D_\beta v^\gam  = D_\alpha( D_\beta v^\gam) = 
	\gam^\mu_{\ \, \alpha} \gam^\nu_{\ \, \beta}
	\gam^\gam_{\ \, \rho} \nabla_\mu ( D_\nu v^\rho ) .
\ee
Using again formula (\ref{e:hyp:link_D_nab_comp})  to express $D_\nu v^\rho$ yields
\be
	D_\alpha D_\beta v^\gam  = 
	\gam^\mu_{\ \, \alpha} \gam^\nu_{\ \, \beta}
	\gam^\gam_{\ \, \rho} \nabla_\mu \left( \gam^\sigma_{\ \, \nu} \gam^\rho_{\ \, \lambda}
	\nabla_\sigma v^\lambda \right) . 	
\ee
Let us expand this formula by making use of Eq.~(\ref{e:hyp:vecgam_ab}) to write
$\nabla_\mu \gam^\sigma_{\ \, \nu} = \nabla_\mu \left( \delta^\sigma_{\ \, \nu} + n^\sigma n_\nu
\right) = \nabla_\mu n^\sigma\, n_\nu + n^\sigma \nabla_\mu n_\nu$. Since 
$\gam^\nu_{\ \, \beta} n_\nu = 0$, we get
\bea
	D_\alpha D_\beta v^\gam & = & 
	\gam^\mu_{\ \, \alpha} \gam^\nu_{\ \, \beta}
	\gam^\gam_{\ \, \rho} \bigg( n^\sigma \nabla_\mu n_\nu \, \gam^\rho_{\ \, \lambda}
	\nabla_\sigma v^\lambda 
	+ \gam^\sigma_{\ \, \nu} \nabla_\mu n^\rho\, 
	\underbrace{n_\lambda \nabla_\sigma v^\lambda}_{=-v^\lambda \nabla_\sigma n_\lambda}
	+ \gam^\sigma_{\ \, \nu} \gam^\rho_{\ \, \lambda} \nabla_\mu \nabla_\sigma v^\lambda 
	\bigg)  \nonumber \\
	& = & \gam^\mu_{\ \, \alpha} \gam^\nu_{\ \, \beta} \gam^\gam_{\ \, \lambda}
	\nabla_\mu n_\nu \, n^\sigma \nabla_\sigma v^\lambda
	- \gam^\mu_{\ \, \alpha} \gam^\sigma_{\ \, \beta} \gam^\gam_{\ \, \rho}
	v^\lambda \nabla_\mu n^\rho \, \nabla_\sigma n_\lambda
	+ \gam^\mu_{\ \, \alpha} \gam^\sigma_{\ \, \beta} \gam^\gam_{\ \, \lambda}
	\nabla_\mu \nabla_\sigma v^\lambda \nonumber \\
	& = & - K_{\alpha\beta} \, \gam^\gam_{\ \, \lambda} \, n^\sigma \nabla_\sigma v^\lambda
	- K^\gam_{\ \, \alpha} K_{\beta\lambda} \, v^\lambda
	+ \gam^\mu_{\ \, \alpha} \gam^\sigma_{\ \, \beta} \gam^\gam_{\ \, \lambda}
	\nabla_\mu \nabla_\sigma v^\lambda , 	\label{e:hyp:DaDbvg}
\eea
where we have used the idempotence of the projection operator $\vg$, i.e.
$\gam^\gam_{\ \, \rho} \gam^\rho_{\ \, \lambda} = \gam^\gam_{\ \, \lambda}$ to get the 
second line and $\gam^\mu_{\ \, \alpha} \gam^\nu_{\ \, \beta} \nabla_\mu n_\nu = - K_{\beta\alpha}$
[Eq.~(\ref{e:hyp:K_vgs_nabn})] to get the third one.
When we permute the indices $\alpha$ and $\beta$ and substract from Eq.~(\ref{e:hyp:DaDbvg})
to form $D_\alpha D_\beta v^\gam  - D_\beta D_\gam v^\gam$, the first term vanishes since
$K_{\alpha\beta}$ is symmetric in $(\alpha,\beta)$. There remains
\be
	D_\alpha D_\beta v^\gam  - D_\beta D_\gam v^\gam = \left( K_{\alpha\mu} K^\gam_{\ \, \beta}
	 - K_{\beta\mu} K^\gam_{\ \, \alpha} \right) v^\mu 
	+ \gam^\rho_{\ \, \alpha} \gam^\sigma_{\ \, \beta} \gam^\gam_{\ \, \lambda}
	\left( \nabla_\rho \nabla_\sigma v^\lambda - \nabla_\sigma \nabla_\rho v^\lambda \right) . 
\ee
Now the Ricci identity (\ref{e:hyp:Ricci_ident}) for the connection $\wnab$ gives
$\nabla_\rho \nabla_\sigma v^\lambda - \nabla_\sigma \nabla_\rho v^\lambda  = 
	{}^4\! R^\lambda_{\ \, \mu \rho\sigma} v^\mu$. 
Therefore
\be
	D_\alpha D_\beta v^\gam  - D_\beta D_\gam v^\gam = \left( K_{\alpha\mu} K^\gam_{\ \, \beta}
	 - K_{\beta\mu} K^\gam_{\ \, \alpha} \right) v^\mu 
	+ \gam^\rho_{\ \, \alpha} \gam^\sigma_{\ \, \beta} \gam^\gam_{\ \, \lambda}
	{}^4\! R^\lambda_{\ \, \mu \rho\sigma} v^\mu .
\ee
Substituting this relation for the left-hand side of Eq.~(\ref{e:hyp:Ricci_ident_3D_comp})
results in 
\be
	\left( K_{\alpha\mu} K^\gam_{\ \, \beta}
	 - K_{\beta\mu} K^\gam_{\ \, \alpha} \right) v^\mu 
	+ \gam^\rho_{\ \, \alpha} \gam^\sigma_{\ \, \beta} \gam^\gam_{\ \, \lambda}
	{}^4\! R^\lambda_{\ \, \mu \rho\sigma} v^\mu 
 	= R^\gam_{\ \, \mu\alpha\beta} \, v^\mu ,
\ee
or equivalently, since $v^\mu = \gam^\mu_{\ \, \sigma} v^\sigma$,  
\be
	\gam^\mu_{\ \, \alpha} \gam^\nu_{\ \, \beta} \gam^\gam_{\ \, \rho} \gam^\sigma_{\ \, \lambda}	
	{}^4\! R^\rho_{\ \, \sigma\mu\nu} v^\lambda 
	= R^\gam_{\ \, \lambda\alpha\beta} \, v^\lambda
	+ \left( K^\gam_{\ \, \alpha} K_{\lambda\beta} - 
		K^\gam_{\ \, \beta} K_{\alpha\lambda} \right) v^\lambda .
\ee
In this identity, $\w{v}$ can be replaced by any vector of $\T(\M)$ without changing the
results, thanks to the presence of the projector operator $\vg$ 
and to the fact that both $\w{K}$ and $\mathbf{Riem}$
are tangent to $\Sigma$. Therefore we conclude that 
\be \label{e:hyp:Gauss}
\encadre{ \gam^\mu_{\ \, \alpha} \gam^\nu_{\ \, \beta} \gam^\gam_{\ \, \rho} \gam^\sigma_{\ \, \delta}	
	{}^4\! R^\rho_{\ \, \sigma\mu\nu}  
	= R^\gam_{\ \, \delta\alpha\beta} 
	+ K^\gam_{\ \, \alpha} K_{\delta\beta} - 
		K^\gam_{\ \, \beta} K_{\alpha\delta} } . 
\ee
This is the \defin{Gauss relation}. 

If we contract the Gauss relation on the indices $\gam$ and $\alpha$ and use
$\gam^\mu_{\ \, \alpha} \gam^\alpha_{\ \, \rho} = \gam^\mu_{\ \, \rho} = \delta^\mu_{\ \, \rho}
	+ n^\mu n_\rho$, we obtain an expression that lets appear the Ricci tensors 
${}^4\!\w{R}$ and $\w{R}$ associated with $\w{g}$ and $\w{\gam}$ respectively:
\be \label{e:hyp:Gauss_contracted}
  \encadre{ \gamma^\mu_{\ \, \alpha} \gamma^\nu_{\ \, \beta}  {}^4 \!R_{\mu\nu} 
  + \gamma_{\alpha\mu} n^\nu \gamma^\rho_{\ \, \beta} n^\sigma
   \, {}^4 \!R^\mu_{\ \, \nu\rho\sigma}  = R_{\alpha\beta}
   + K K_{\alpha\beta} - K_{\alpha\mu} K^\mu_{\ \, \beta} } . 
\ee
We call this equation the \defin{contracted Gauss relation}.
Let us take its  trace with respect to $\w{\gam}$, 
taking into account that $K^\mu_{\ \, \mu} = K^i_{\ \, i} = K$,
$K_{\mu\nu} K^{\mu\nu} = K_{ij} K^{ij}$ and
\be
	\gam^{\alpha\beta} \gamma_{\alpha\mu} n^\nu \gamma^\rho_{\ \, \beta} n^\sigma
   \, {}^4 \!R^\mu_{\ \, \nu\rho\sigma} = \gam^\rho_{\ \, \mu} n^\nu n^\sigma
	{}^4 \!R^\mu_{\ \, \nu\rho\sigma} 
	= \underbrace{{}^4 \!R^\mu_{\ \, \nu\mu\sigma}}_{={}^4 \!R_{\nu\sigma}}
	 n^\nu n^\sigma + 
  \underbrace{ {}^4 \!R^\mu_{\ \, \nu\rho\sigma} n^\rho n_\mu n^\nu n^\sigma}_{=0}
	= {}^4 \!R_{\mu\nu}n^\mu n^\nu .
\ee
We obtain 
\be \label{e:hyp:Gauss_scalar}
    \encadre{ {}^4\!R + 2\,  {}^4\!R_{\mu\nu} n^\mu n^\nu = R + K^2 
	- K_{ij} K^{ij} }.
\ee
Let us call this equation the \defin{scalar Gauss relation}.
It constitutes a generalization of Gauss' famous \defin{Theorema Egregium}
(\emph{remarkable theorem}) \cite{Berge03,BergeG87}. 
It relates the intrinsic curvature of $\Sigma$,
represented by the Ricci scalar $R$, to its extrinsic curvature, represented
by $K^2-K_{ij}K^{ij}$.
Actually, the original version of Gauss' theorem was for two-dimensional surfaces
embedded in the Euclidean space $\R^3$. Since the curvature of the latter
is zero, the left-hand side of Eq.~(\ref{e:hyp:Gauss_scalar}) vanishes 
identically in this case. Moreover, the metric $\w{g}$ of the Euclidean space $\R^3$
is Riemannian, not Lorentzian. Consequently the term $K^2-K_{ij}K^{ij}$
has the opposite sign, so that Eq.~(\ref{e:hyp:Gauss_scalar}) becomes
\be \label{e:hyp:TEgr_ori1}
	R - K^2 + K_{ij} K^{ij} = 0  \quad (\w{g}\ \mbox{Euclidean}) .
\ee
This change of sign stems from the fact that for a Riemannian ambient metric,
the unit normal vector $\w{n}$ is spacelike and 
the orthogonal projector is 
$\gam^\alpha_{\ \, \beta} = \delta^\alpha_{\ \, \beta} - n^\alpha n_\beta$
instead of $\gam^\alpha_{\ \, \beta} = \delta^\alpha_{\ \, \beta} + n^\alpha n_\beta$
[the latter form has been used explicitly in the calculation leading to 
Eq.~(\ref{e:hyp:DaDbvg})].
Moreover, in dimension 2, formula (\ref{e:hyp:TEgr_ori1}) can be simplified
by letting appear the principal curvatures $\kappa_1$ and $\kappa_2$ of $\Sigma$
(cf. Sec.~\ref{s:hyp:extr_curv}).
Indeed, $\w{K}$ can be diagonalized in an orthonormal basis (with respect to $\gam$)
so that $K_{ij} = \mathrm{diag}(\kappa_1,\kappa_2)$ and 
$K^{ij} = \mathrm{diag}(\kappa_1,\kappa_2)$. Consequently, $K=\kappa_1+\kappa_2$
and $K_{ij} K^{ij}=\kappa_1^2 + \kappa_2^2$ and Eq.~(\ref{e:hyp:TEgr_ori1})
becomes
\be \label{e:hyp:TEgr_ori2}
	R = 2 \kappa_1 \kappa_2  \quad (\w{g}\ \mbox{Euclidean},
		\ \Sigma\ \mbox{dimension 2}) .
\ee

\begin{example}
We may check the Theorema Egregium (\ref{e:hyp:TEgr_ori1}) for the examples of Sec.~\ref{s:hyp:examples}.
It is trivial for the plane, since each term vanishes separately. 
For the cylinder of radius $r$, $R=0$, $K=-1/r$ [Eq.~(\ref{e:hyp:K_cylinder})],
$K_{ij} K^{ij} = 1/r^2$ [Eq.~(\ref{e:hyp:Kij_cylinder})], so that Eq.~(\ref{e:hyp:TEgr_ori1}) is satisfied. 
For the sphere of radius $r$, $R=2/r^2$ [Eq.~(\ref{e:hyp:curv_sphere})],
$K=-2/r$ [Eq.~(\ref{e:hyp:K_sphere})], $K_{ij}K^{ij}=2/r^2$ 
[Eq.~(\ref{e:hyp:Kab_sphere})], so that Eq.~(\ref{e:hyp:TEgr_ori1}) is satisfied
as well.
\end{example}

\subsection{Codazzi relation}

Let us at present apply the Ricci identity (\ref{e:hyp:Ricci_ident})
to the normal vector $\w{n}$ (or more precisely to any extension of $\w{n}$ around
$\Sigma$, cf. Sec.~\ref{s:hyp:rel_K_nabn}):
\be \label{e:hyp:Ricci_id_n}
        \left(\nabla_\alpha\nabla_\beta  
        - \nabla_\beta\nabla_\alpha\right) n^\gamma
        = {}^{4}\!R^\gamma_{\ \, \mu \alpha\beta} \, n^\mu .  
\ee
If we project this relation onto $\Sigma$, we get
\be
	\gam^\mu_{\ \, \alpha} \gam^\nu_{\ \, \beta} \gam^\gam_{\ \, \rho}
	{}^{4}\!R^\rho_{\ \, \sigma\mu\nu} n^\sigma = 
	\gam^\mu_{\ \, \alpha} \gam^\nu_{\ \, \beta} \gam^\gam_{\ \, \rho}
	\left( \nabla_\mu \nabla_\nu n^\rho -  \nabla_\nu  \nabla_\mu n^\rho
	\right) .	\label{e:hyp:Codaz_interm1}
\ee
Now, from Eq.~(\ref{e:hyp:nab_n_K_comp}), 
\bea
	\gam^\mu_{\ \, \alpha} \gam^\nu_{\ \, \beta} \gam^\gam_{\ \, \rho}
	\nabla_\mu \nabla_\nu n^\rho & =&
	\gam^\mu_{\ \, \alpha} \gam^\nu_{\ \, \beta} \gam^\gam_{\ \, \rho}
	\nabla_\mu \left( - K^\rho_{\ \, \nu} - a^\rho n_\nu \right) \nonumber \\
	& = & - \gam^\mu_{\ \, \alpha} \gam^\nu_{\ \, \beta} \gam^\gam_{\ \, \rho}
		\left( \nabla_\mu K^\rho_{\ \, \nu} 
			+ \nabla_\mu a^\rho \, n_\nu + a^\rho \nabla_\mu n_\nu \right)
				\nonumber \\
	& = & - D_\alpha K^\gam_{\ \, \beta}
		+ a^\gam K_{\alpha\beta} , \label{e:hyp:Codaz_interm2}
\eea
where we have used Eq.~(\ref{e:hyp:link_D_nab_comp}), as well as 
$\gam^\nu_{\ \, \beta} n_\nu = 0$, $\gam^\gam_{\ \, \rho} a^\rho = a^\gam$,
and 
$\gam^\mu_{\ \, \alpha} \gam^\nu_{\ \, \beta} \nabla_\mu n_\nu = - K_{\alpha\beta}$
to get the last line.
After permutation of the indices $\alpha$ and $\beta$ and substraction 
from Eq.~(\ref{e:hyp:Codaz_interm2}), taking into account the symmetry of
$K_{\alpha\beta}$, we see that Eq.~(\ref{e:hyp:Codaz_interm1})
becomes
\be \label{e:hyp:Codazzi}
\encadre{ 
	 \gam^\gam_{\ \, \rho}\, 
	n^\sigma \, \gam^\mu_{\ \, \alpha} \gam^\nu_{\ \, \beta}
	\, {}^{4}\!R^\rho_{\ \, \sigma\mu\nu}  = 
	D_\beta K^\gam_{\ \, \alpha} - D_\alpha K^\gam_{\ \, \beta} } . 
\ee
This is the \defin{Codazzi relation}, also called \defin{Codazzi-Mainardi relation}
in the mathematical litterature \cite{Berge03}.
\begin{remark}
Thanks to the symmetries of the Riemann tensor (cf. Sec.~\ref{s:hyp:curvat}),
changing the index contracted with $\w{n}$ in Eq.~(\ref{e:hyp:Codazzi}) (for
instance considering $n_\rho \gam^{\gam\sigma} \, 
	\gam^\mu_{\ \, \alpha} \gam^\nu_{\ \, \beta}
	\, {}^{4}\!R^\rho_{\ \, \sigma\mu\nu} $ or 
$\gam^\gam_{\ \, \rho}\, \gam^\sigma_{\ \, \alpha}
	\, n^\mu \,  \gam^\nu_{\ \, \beta}
	\, {}^{4}\!R^\rho_{\ \, \sigma\mu\nu}$ 
)
would not give an independent relation: at most it would result in a change of
sign of the right-hand side. 
\end{remark}

Contracting the Codazzi relation on the indices $\alpha$ and $\gamma$ yields
to
\be
	\gam^\mu_{\ \, \rho} \, n^\sigma \gam^\nu_{\ \, \beta}
	\, {}^{4}\!R^\rho_{\ \, \sigma\mu\nu}  = D_\beta K 
	- D_\mu K^\mu_{\ \, \beta} , 
\ee
with $\gam^\mu_{\ \, \rho} \, n^\sigma \gam^\nu_{\ \, \beta}
	\, {}^{4}\!R^\rho_{\ \, \sigma\mu\nu} = (\delta^\mu_{\ \, \rho} + n^\mu
		n_\rho) \, n^\sigma \gam^\nu_{\ \, \beta}
	\, {}^{4}\!R^\rho_{\ \, \sigma\mu\nu} = n^\sigma \gam^\nu_{\ \, \beta}
	\, {}^{4}\!R_{\sigma\nu} 
	+ \gam^\nu_{\ \, \beta} \, {}^{4}\!R^\rho_{\ \, \sigma\mu\nu} 
		n_\rho n^\sigma n^\mu $.
Now, from the antisymmetry of the Riemann tensor with respect to its first two
indices [Eq.~(\ref{e:hyp:Riemann_antisym12}), the last term vanishes, so that
one is left with
\be \label{e:hyp:Codazzi_contract}
	\encadre{ \gam^\mu_{\ \, \alpha} n^\nu \, {}^{4}\!R_{\mu\nu} = 
		D_\alpha K - D_\mu K^\mu_{\ \, \alpha} }. 
\ee
We shall call this equation the \defin{contracted Codazzi relation}.

\begin{example}
The Codazzi relation is trivially satisfied by the three examples of 
Sec.~\ref{s:hyp:examples} because the Riemann tensor vanishes for the Euclidean space
$\R^3$ and for each of the considered surfaces, 
either $\w{K}=0$ (plane) or $\w{K}$ is constant on $\Sigma$, in the sense that
$\w{D}\w{K}=0$.
\end{example}

%  
%    Chapitre : Geometry of foliations
%
% $Date: 2007-03-05 22:39:07 +0100 (lun, 05 mar 2007) $
% $Rev: 182 $
% $Author: e_gourgoulhon $
%%%%%%%%%%%%%%%%%%%%%%%%%%%%%

\chapter{Geometry of foliations} \label{s:fol}

%\verb$Date: 2007-03-05 22:39:07 +0100 (lun, 05 mar 2007) $

\minitoc
\vspace{1cm}

%%%%%%%%%%%%%%%%%%%%%%%%%%%%%%%%%%%%%%%%%%%%%%%%%%%%%%%%%%%%%%%%%%%%%%%%%%%%%%%%

\section{Introduction}

In the previous chapter, we have studied a single hypersurface $\Sigma$
embedded in the spacetime $(\M,\w{g})$. At present, we consider a continuous
set of hypersurfaces $\left( \Sigma_t \right) _{t\in\R}$ that covers the manifold
$\M$. This is possible for a wide class of spacetimes to which we shall restrict
ourselves: the so-called globally hyperbolic spacetimes. Actually the latter
ones cover most of the spacetimes of astrophysical or cosmological interest.
Again the title of this chapter is ``Geometry...'', since as in Chap.~\ref{s:hyp},
all the results are independent of the Einstein equation. 

%%%%%%%%%%%%%%%%%%%%%%%%%%%%%%%%%%%%%%%%%%%%%%%%%%%%%%%%%%%%%%%%%%%%%%%%%%%%%%%%

\section{Globally hyperbolic spacetimes and foliations} 

\subsection{Globally hyperbolic spacetimes} \label{s:fol:glob_hyp}

A \defin{Cauchy surface} is a spacelike
hypersurface $\Sigma$ in $\M$ such that each causal
(i.e. timelike or null) curve without end point intersects $\Sigma$ once and only once
\cite{HawkiE73}. Equivalently, $\Sigma$ is a Cauchy surface iff its domain of 
dependence is the whole spacetime $\M$.
Not all spacetimes admit a Cauchy surface. For instance spacetimes with closed
timelike curves do not. Other examples are provided in Ref.~\cite{Fried04}.
A spacetime $(\M,\w{g})$ that admits a Cauchy surface $\Sigma$
is said to be \defin{globally hyperbolic}.
The name \emph{globally hyperbolic} stems from the fact that the scalar wave equation
is well posed, 

The topology of a globally hyperbolic spacetime $\M$ is 
necessarily $\Sigma\times \R$ (where $\Sigma$ is the Cauchy surface entering in 
the definition of global hyperbolicity). 

\begin{remark}
The original definition of a globally hyperbolic spacetime is actually more technical
that the one given above, but the latter has been shown to be equivalent to the original
one (see e.g. Ref.~\cite{ChoquY80} and references therein).
\end{remark}

\begin{figure}
\centerline{\includegraphics[width=0.5\textwidth]{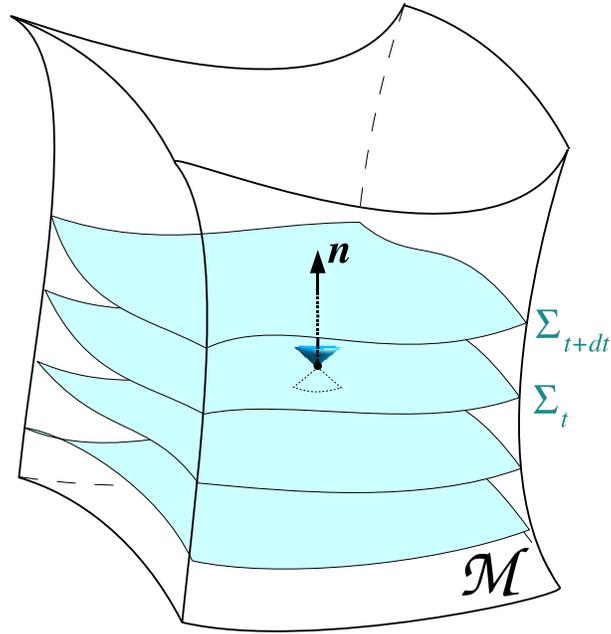}}
\caption[]{\label{f:fol:foliat} \footnotesize
Foliation of the spacetime $\M$ by a family of spacelike hypersurfaces
$(\Sigma_t)_{t\in\R}$.}
\end{figure}

\subsection{Definition of a foliation} \label{s:fol:def_foliat}

Any globally hyperbolic spacetime $(\M,\w{g})$ can be foliated by a family of 
spacelike hypersurfaces $\left( \Sigma_t \right) _{t\in\R}$. 
By \defin{foliation} or \defin{slicing}, 
it is meant that there exists a smooth scalar field $\hat t$ on $\M$,
which is regular (in the sense that its gradient never vanishes), 
such that each hypersurface is a level surface of this scalar field:
\be
	\forall t\in \R,\quad \Sigma_t := \left\{ p\in\M,\ \hat t(p) = t \right\} . 
\ee
Since $\hat t$ is regular, the hypersurfaces $\Sigma_t$ are non-intersecting:
\be \label{e:fol:non_intersect}
	\Sigma_t \cap \Sigma_{t'} = \emptyset \quad \mbox{for}\ t\not=t' .
\ee
In the following, we do no longer distinguish between $t$ and $\hat t$, i.e. we skip
the hat in the name of the scalar field. 
Each hypersurface $\Sigma_t$ is called a \defin{leaf} or a \defin{slice}
of the foliation.
We assume that all $\Sigma_t$'s are spacelike and that the foliation covers $\M$
(cf. Fig.~\ref{f:fol:foliat}):
\be
	\M = \bigcup_{t\in\R} \Sigma_t . 
\ee

%%%%%%%%%%%%%%%%%%%%%%%%%%%%%%%%%%%%%%%%%%%%%%%%%%%%%%%%%%%%%%%%%%%%%%%%%%%%%%%%

\section{Foliation kinematics}

\subsection{Lapse function} \label{s:fol:def_lapse}

As already noticed in Sec.~\ref{s:hyp:norm_vect}, the timelike and future-directed
unit vector $\w{n}$ normal to the slice $\Sigma_t$ is necessarily collinear to the vector
$\vec{\wnab} t$ associated with the gradient 1-form $\dd t$.
Hence we may write 
\be \label{e:fol:n_lapse}
    \encadre{ \w{n} := - N \vec{\wnab} t}
\ee
with
\be \label{e:fol:def_lapse}
    N := \left( - \vec{\wnab} t \cdot \vec{\wnab} t \right) ^{-1/2} 
    =  \left( - \langle \dd t,  \vec{\wnab} t \rangle \right) ^{-1/2} .
\ee
The minus sign in (\ref{e:fol:n_lapse})
is chosen so that the vector $\w{n}$ is future-oriented
if the scalar field $t$ is increasing towards
the future. Notice that the value of $N$ ensures that 
$\w{n}$ is a unit vector: $\w{n}\cdot\w{n}=-1$.
The scalar field $N$ hence defined is called the \defin{lapse function}.
The name \emph{lapse} has been coined by Wheeler in 1964 \cite{Wheel64}.
\begin{remark}
In most of the numerical relativity literature, the lapse function is
denoted $\alpha$ instead of $N$. We follow here the ADM \cite{ArnowDM62}
and MTW \cite{MisneTW73} notation. 
\end{remark}

Notice that by construction [Eq.~(\ref{e:fol:def_lapse})],
\be
	N > 0 . 
\ee
In particular, the lapse function never vanishes for a regular foliation. 
Equation~(\ref{e:fol:n_lapse}) also says that $-N$ is the proportionality
factor between the gradient 1-form $\dd t$ and the 1-form $\uu{n}$
associated to the vector $\w{n}$ by the metric duality: 
\be \label{e:fol:un_lapse}
	\encadre{ \uu{n} = - N \, \dd t } . 
\ee

\begin{figure}
\centerline{\includegraphics[width=0.7\textwidth]{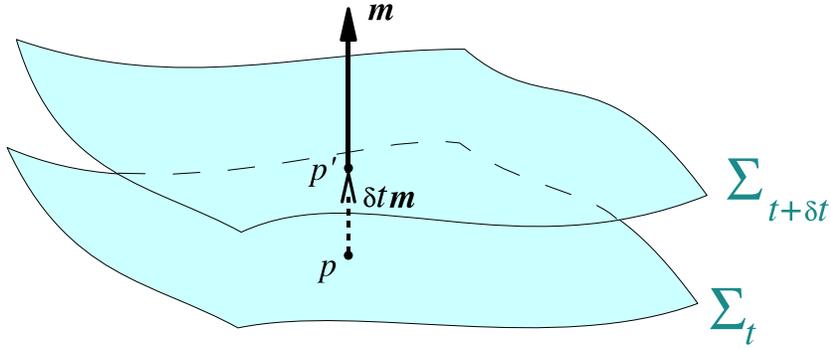}}
\caption[]{\label{f:fol:liem_sigma} \footnotesize
The point $p'$ deduced from $p\in\Sigma_t$ by the displacement $\delta t \, \w{m}$
belongs to $\Sigma_{t+\delta t}$, i.e. the hypersurface $\Sigma_t$ is transformed
to $\Sigma_{t+\delta t}$ by the vector field $\delta t \, \w{m}$ (Lie dragging).}
\end{figure}

\subsection{Normal evolution vector} \label{s:fol:norm_evol}

Let us define the \defin{normal evolution vector} as the timelike vector
normal to $\Sigma_t$ such that
\be \label{e:fol:def_m}
	\encadre{\w{m} := N \w{n}} . 
\ee
Since $\w{n}$ is a unit vector, the scalar square of $\w{m}$ is
\be \label{e:fol:mmN2}
	\w{m}\cdot\w{m} = - N^2 . 
\ee
Besides, we have
\be
	\langle \dd t, \w{m} \rangle = N \langle \dd t, \w{n} \rangle 
	= N^2 \underbrace{(-\langle \dd t, \vec{\wnab} t \rangle)}_{=N^{-2}}
	= 1 , 
\ee
where we have used Eqs.~(\ref{e:fol:n_lapse}) and (\ref{e:fol:def_lapse}). 
Hence 
\be \label{e:fol:dt_m_1}
	\encadre{ 
	\langle \dd t, \w{m} \rangle = \wnab_{\w{m}} \, t = m^\mu \nabla_\mu \, t = 1 } . 
\ee
This relation means that the normal vector $\w{m}$ is ``adapted'' to the scalar field
$t$, contrary to the normal vector $\w{n}$.
A geometrical consequence of this property is that the 
hypersurface $\Sigma_{t+\delta t}$ can be obtained
from the neighbouring hypersurface $\Sigma_t$ by the small displacement 
$\delta t\, \w{m}$ of each point of $\Sigma_t$. 
Indeed consider some point $p$ in $\Sigma_t$
and displace it by the infinitesimal vector $\delta t \, \w{m}$ to the
point $p' = p + \delta t \, \w{m}$ (cf. Fig.~\ref{f:fol:liem_sigma}). 
From the very definition of the gradient
1-form $\dd t$, the value of the scalar field $t$ at $p'$
is
\bea
    t(p') & = & t(p + \delta t \, \w{m}) 
          = t(p) + \langle \dd t, \delta t \, \w{m} \rangle
          = t(p) + \delta t 
          \underbrace{\langle \dd t,  \w{m} \rangle}_{=1} \nonumber \\
          & = & t(p) + \delta t . 
\eea
This last equality shows that $p' \in \Sigma_{t+\delta t}$.
Hence the vector $\delta t \, \w{m}$ carries the hypersurface $\Sigma_t$
into the neighbouring one $\Sigma_{t+\delta t}$. One says equivalently that
the hypersurfaces $(\Sigma_t)$ are \defin{Lie dragged} by the vector $\w{m}$. 
This justifies the name \emph{normal evolution vector} given to $\w{m}$. 

\begin{figure}
\centerline{\includegraphics[width=0.7\textwidth]{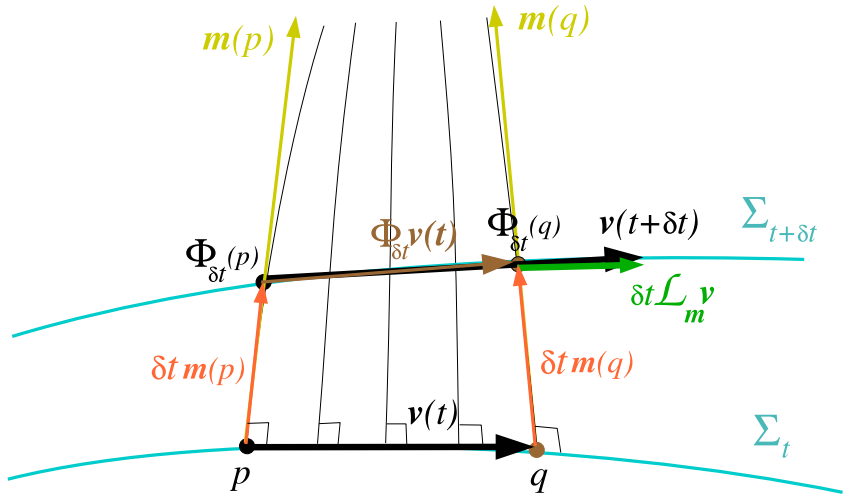}}
\caption[]{\label{f:fol:liem_vect} \footnotesize
Geometrical construction showing that $\Lie{m} \w{v} \in\T(\Sigma_t)$
for any vector $\w{v}$ tangent to the hypersurface $\Sigma_t$: on $\Sigma_t$,
a vector can be identified to a infinitesimal displacement between 
two points, $p$ and $q$ say. These points are transported onto the
neighbouring hypersurface $\Sigma_{t+\delta t}$ along the field lines of the
vector field $\w{m}$ (thin lines on the figure) 
by the diffeomorphism $\Phi_{\delta t}$
associated with $\w{m}$: the displacement between $p$ and $\Phi_{\delta t}(p)$
is the vector $\delta t\, \w{m}$. The couple of points 
$(\Phi_{\delta t}(p),\Phi_{\delta t}(q))$ defines the vector 
$\Phi_{\delta t} \w{v}(t)$, which is tangent to $\Sigma_{t+\delta t}$
since both points $\Phi_{\delta t}(p)$ and $\Phi_{\delta t}(q)$ belong to 
$\Sigma_{t+\delta t}$. The Lie derivative
of $\w{v}$ along $\w{m}$ is then defined by the difference between the
value of the vector field $\w{v}$ at the point $\Phi_{\delta t}(p)$,
i.e. $\w{v}(t+\delta t)$, and the vector transported from $\Sigma_t$ along
$\w{m}$'s field lines, i.e. $\Phi_{\delta t} \w{v}(t)$ :
$\Lie{\w{m}}{\w{v}}(t+\delta t) = \lim_{\delta t\rightarrow 0} [ \w{v}(t+\delta t) -
\Phi_{\delta t} \w{v}(t)]/\delta t$. Since both vectors $ \w{v}(t+\delta t)$
and $\Phi_{\delta t} \w{v}(t)$ are in $\T(\Sigma_{t+\delta t})$, it follows then
that $\Lie{\w{m}}{\w{v}}(t+\delta t) \in\T(\Sigma_{t+\delta t})$.}
\end{figure}

An immediate consequence of the Lie dragging of the hypersurfaces $\Sigma_t$
by the vector $\w{m}$ is that the Lie derivative along $\w{m}$ of any
vector tangent to $\Sigma_t$ is also a vector tangent to $\Sigma_t$:
\be \label{e:fol:Lie_m_v}
	\encadre{ \forall \w{v}\in\T(\Sigma_t),\quad \Lie{m} \w{v} \in\T(\Sigma_t) } .
\ee
This is obvious from the geometric definition of the Lie derivative 
(cf. Fig.~\ref{f:fol:liem_vect}). 
The reader not familiar with the concept of Lie derivative may
consult Appendix~\ref{s:lie}.

\subsection{Eulerian observers} \label{s:fol:Eulerian}

Since $\w{n}$ is a unit timelike vector, it can be regarded as the 4-velocity
of some observer. We call such observer an \defin{Eulerian observer}. 
It follows that the worldlines of the Eulerian observers are orthogonal to the hypersurfaces
$\Sigma_t$. Physically, this means that the hypersurface $\Sigma_t$ is 
\emph{locally} the
set of events that are simultaneous from the point of view of the Eulerian observer, 
according to Einstein's simultaneity convention. 
\begin{remark}
The Eulerian observers are sometimes called \defin{fiducial observers}
(e.g. \cite{ThornM82}). In the special case of axisymmetric and stationary spacetimes,
they are called \defin{locally nonrotating observers} \cite{Barde70}
or \defin{zero-angular-momentum observers} (\defin{ZAMO}) \cite{ThornM82}. 
\end{remark}
Let us consider two close events $p$ and $p'$ on the worldline of some Eulerian
observer. Let $t$ be the ``coordinate time'' of the event $p$ and 
$t+\delta t$ ($\delta t>0$) that of $p'$, in the sense that $p\in\Sigma_t$ and
$p'\in\Sigma_{t+\delta t}$. Then $p' = p + \delta t \, \w{m}$, as above. 
The proper time $\delta \tau$ between the events $p$ and $p'$, as measured the Eulerian
observer, is given by the metric length of the vector linking $p$ and $p'$:
\be
	\delta \tau = \sqrt{ - \w{g}(\delta t \, \w{m}, \delta t \, \w{m})}
	= \sqrt{-\w{g}(\w{m}, \w{m})}\; \delta t . 
\ee
Since $\w{g}(\w{m}, \w{m})=-N^2$ [Eq.~(\ref{e:fol:mmN2})], we get (assuming $N>0$)
\be \label{e:fol:dtau_Ndt}
	\encadre{ \delta\tau = N \, \delta t } .  
\ee
This equality justifies the name \emph{lapse function} given to $N$: $N$ relates
the ``coordinate time'' $t$ labelling the leaves of the foliation to the
physical time $\tau$ measured by the Eulerian observer.

The 4-acceleration of the Eulerian observer is
\be
	\w{a} = \wnab_{\w{n}} \w{n} . 
\ee
As already noticed in Sec.~\ref{s:hyp:rel_K_nabn}, the vector $\w{a}$ is orthogonal
to $\w{n}$ and hence tangent to $\Sigma_t$. Moreover, it can be expressed in 
terms of the spatial gradient of the lapse function. Indeed, 
by means Eq.~(\ref{e:fol:un_lapse}), we have
\bea
	a_\alpha &=& n^\mu \nabla_\mu n_\alpha
	= -n^\mu \nabla_\mu (N\nabla_\alpha t)
	= -n^\mu \nabla_\mu N \nabla_\alpha t - N n^\mu 
	\underbrace{\nabla_\mu \nabla_\alpha t}_{=\nabla_\alpha \nabla_\mu t}
		\nonumber \\
	& = &  \frac{1}{N} n_\alpha  n^\mu \nabla_\mu N
	+ N n^\mu \nabla_\alpha \left( -\frac{1}{N} n_\mu \right)
	 =  \frac{1}{N} n_\alpha  n^\mu \nabla_\mu N
	+ \frac{1}{N} \nabla_\alpha N \, \underbrace{n^\mu n_\mu}_{=-1}
	- \underbrace{n^\mu \nabla_\alpha n_\mu}_{=0} \nonumber \\
	& =& \frac{1}{N} 
	\left( \nabla_\alpha N + n_\alpha  n^\mu \nabla_\mu N \right)
	=  \frac{1}{N} \gam^\mu_{\ \, \alpha} \nabla_\mu N \nonumber \\
	& = & \frac{1}{N}  D_\alpha N = D_\alpha \ln N, 
\eea
where we have used the torsion-free character of the connection $\wnab$
to write $\nabla_\mu \nabla_\alpha t = \nabla_\alpha \nabla_\mu t$,
as well as the expression (\ref{e:hyp:vecgam_ab}) of the orthogonal projector
onto $\Sigma_t$, $\vg$, and the relation (\ref{e:hyp:link_D_nab_comp})
between $\wnab$ and $\w{D}$ derivatives. 
Thus we have
\be \label{e:fol:a_DN}
	\encadre{ \uu{a} = \w{D}\ln N } 
\qquad \mbox{and} \qquad
	\encadre{ \w{a} = \vec{\w{D}} \ln N } . 
\ee
Thus, the 4-acceleration of the Eulerian observer appears to be nothing but
the gradient within $(\Sigma_t,\w{\gam})$ 
of the logarithm of the lapse function. 
Notice that since a spatial gradient is always tangent to $\Sigma_t$, we recover
immediately from formula (\ref{e:fol:a_DN}) that $\w{n}\cdot\w{a}=0$.

\begin{remark} 
Because they are hypersurface-orthogonal, the congruence formed by all the Eulerian
observers' worldlines has a vanishing vorticity, hence the name ``non-rotating''
observer given sometimes to the Eulerian observer. \label{p:fol:nonrot}
\end{remark}

\subsection{Gradients of $n$ and $m$}
 
Substituting Eq.~(\ref{e:fol:a_DN}) for $\uu{a}$ into Eq.~(\ref{e:hyp:nab_n_K})
leads to the following relation between the extrinsic curvature tensor,
the gradient of $\uu{n}$ and the spatial gradient of the lapse function: 
\be \label{e:fol:nab_n_K}
	\encadre{ \wnab \uu{n}  = - \w{K} - \w{D} \ln N\otimes \uu{n} } , 
\ee
or, in components:
\be \label{e:fol:nab_n_K_comp}
	\encadre{ \nabla_\beta \, n_\alpha = 
	- K_{\alpha\beta} - D_\alpha \ln N \; n_\beta } . 
\ee

The covariant derivative of the normal evolution vector is 
deduced from $\wnab\uu{m} = \wnab(N\uu{n}) 
= N \wnab \uu{n} + \uu{n} \otimes \wnab N$. We get
\be
	\encadre{ \wnab \w{m} = - N \vec{\w{K}}
	- \vec{\w{D}} N \otimes \uu{n} 
	+ \w{n} \otimes \wnab N }, 
\ee
or, in components:
\be \label{e:fol:nab_m_comp}
	\encadre{ \nabla_\beta\,  m^\alpha = - N K^\alpha_{\ \, \beta}
	- D^\alpha N \, n_\beta + n^\alpha \nabla_\beta N } . 
\ee

\subsection{Evolution of the 3-metric}

The evolution of $\Sigma_t$'s metric $\w{\gam}$
is naturally given by the Lie derivative of $\w{\gam}$ along the normal 
evolution vector $\w{m}$ (see Appendix~\ref{s:lie}). 
By means of Eqs.~(\ref{e:lie:der_comp_nab}) and (\ref{e:fol:nab_m_comp}), we get
\bea
	\Liec{m} \gam_{\alpha\beta} &=& m^\mu \nabla_\mu
	\gam_{\alpha\beta} + \gam_{\mu\beta} \nabla_\alpha m^\mu
	+ \gam_{\alpha\mu} \nabla_\beta m^\mu \nonumber \\
 	& = & N n^\mu \nabla_\mu(n_\alpha n_\beta)
	- \gam_{\mu\beta} \left( N K^\mu_{\ \, \alpha}
	+ D^\mu N \, n_\alpha - n^\mu \nabla_\alpha N  \right) \nonumber \\
	& & - \gam_{\alpha\mu}  \left( N K^\mu_{\ \, \beta}
	+ D^\mu N \, n_\beta - n^\mu \nabla_\beta N  \right) \nonumber \\
	& = & N \big( \underbrace{n^\mu \nabla_\mu 
	n_\alpha}_{\underbrace{a_\alpha}_{=N^{-1} D_\alpha N}} 
	\; n_\beta + n_\alpha \, 
	\underbrace{n^\mu \nabla_\mu 
		n_\beta}_{\underbrace{a_\beta}_{=N^{-1} D_\beta N}} 
	\big) - N K_{\beta\alpha} - D_\beta N \, n_\alpha
	- N K_{\alpha\beta} - D_\alpha N \, n_\beta \nonumber \\
	& =& -2 N K_{\alpha\beta} . 
\eea
Hence the simple result: 
\be \label{e:fol:Lie_m_gam}
	\encadre{ \Lie{m}\w{\gam} = - 2 N \w{K} } . 
\ee

One can deduce easily from this relation the value of the Lie derivative of 
the 3-metric along the unit normal $\w{n}$. Indeed, since $\w{m}=N\w{n}$, 
\bea
	\Liec{m} \gam_{\alpha\beta} & = & 
	{\mathcal{L}}_{N\w{n}} \gam_{\alpha\beta} \nonumber \\
	& = & N n^\mu \nabla_\mu \gam_{\alpha\beta} 
	+ \gam_{\mu\beta} \nabla_\alpha (N n^\mu)
	+ \gam_{\alpha\mu} \nabla_\beta (N n^\mu) \nonumber \\
	& = & N n^\mu \nabla_\mu \gam_{\alpha\beta} 
	+  \underbrace{\gam_{\mu\beta} n^\mu}_{=0} \nabla_\alpha N  
	+ N \gam_{\mu\beta} \nabla_\alpha n^\mu
	+  \underbrace{\gam_{\alpha\mu} n^\mu}_{=0} \nabla_\beta N  
	+ N \gam_{\alpha\mu} \nabla_\beta n^\mu\nonumber \\
	& = & N \Liec{n} \gam_{\alpha\beta} .
\eea
Hence 
\be \label{e:fol:Lie_n_m_gam}
	\Lie{n}\w{\gam} = \frac{1}{N} \Lie{m}\w{\gam} . 
\ee
Consequently, Eq.~(\ref{e:fol:Lie_m_gam}) leads to
\be \label{e:fol:Lie_n_gam}
	\encadre{\w{K} = -\frac{1}{2} \Lie{n}\w{\gam} } . 
\ee
This equation sheds some new light on the extrinsic curvature tensor $\w{K}$.
In addition to being the projection on $\Sigma_t$ of the gradient of the unit normal
to $\Sigma_t$ [cf. Eq.~(\ref{e:hyp:K_vgs_nabn})], 
\be
	\w{K} = - \vgs \wnab\uu{n} ,
\ee
as well as the measure of the difference between $\w{D}$-derivatives
and $\wnab$-derivatives for vectors tangent to $\Sigma_t$
[cf. Eq.~(\ref{e:hyp:Duv_nabuv_K})],
\be
	 \forall (\w{u},\w{v})\in \T(\Sigma)^2,\ 
	\w{K}(\w{u},\w{v}) \, \w{n} =
	\w{D}_{\w{u}} \w{v} - \wnab_{\w{u}} \w{v},	
\ee
$\w{K}$ is also minus one half the Lie derivative of $\Sigma_t$'s metric along
the unit timelike normal. 
\begin{remark}
In many numerical relativity articles, 
Eq.~(\ref{e:fol:Lie_n_gam}) is used to \emph{define} the extrinsic curvature tensor
of the hypersurface $\Sigma_t$.  
It is worth to keep in mind that this equation has a meaning only 
because $\Sigma_t$ is member of a foliation. 
Indeed the right-hand side is the derivative of the induced metric in a
direction which is not parallel to the hypersurface and therefore this quantity
could not be defined for a single hypersurface, as considered in Chap.~\ref{s:hyp}.
\end{remark}

\subsection{Evolution of the orthogonal projector} \label{s:fol:evol_ortho}

Let us now evaluate the Lie derivative of the orthogonal projector 
onto $\Sigma_t$ along the normal evolution vector. Using 
Eqs.~(\ref{e:lie:der_comp_nab}) and (\ref{e:fol:nab_m_comp}), we have
\bea
	\Liec{m}\gam^\alpha_{\ \, \beta} & = &
	m^\mu \nabla_\mu \gam^\alpha_{\ \, \beta}
	- \gam^\mu_{\ \, \beta} \nabla_\mu m^\alpha
	+ \gam^\alpha_{\ \, \mu} \nabla_\beta m^\mu \nonumber \\
	& = & N n^\mu \nabla_\mu (n^\alpha n_\beta)
	+ \gam^\mu_{\ \, \beta} \left( N K^\alpha_{\ \, \mu}
	+ D^\alpha N \, n_\mu - n^\alpha \nabla_\mu N \right) \nonumber \\
	&&- \gam^\alpha_{\ \, \mu} \left( N K^\mu_{\ \, \beta}
	+ D^\mu N \, n_\beta - n^\mu \nabla_\beta N \right) \nonumber \\
	& = & N \big( \underbrace{n^\mu \nabla_\mu 
	n^\alpha}_{=N^{-1} D^\alpha N} 
	\; n_\beta + n^\alpha \, 
	\underbrace{n^\mu \nabla_\mu 
		n_\beta}_{=N^{-1} D_\beta N} 
	\big)
	+ N K^\alpha_{\ \, \beta}- n^\alpha D_\beta N
	- N K^\alpha_{\ \, \beta} - D^\alpha N \, n_\beta \nonumber \\
	& = & 0 , 
\eea
i.e.
\be \label{e:fol:Lie_m_vg}
	\encadre{ \Lie{m} \vg = 0 } . 
\ee
An important consequence of this is that the Lie derivative along $\w{m}$ of any tensor field $\w{T}$ tangent to $\Sigma_t$ is a tensor field tangent to $\Sigma_t$:
\be \label{e:fol:liem_preserve}
	\encadre{ \w{T} \mbox{\ tangent to\ } \Sigma_t \ \Longrightarrow\ 
	\Lie{m} \w{T} \mbox{\ tangent to\ } \Sigma_t }.
\ee
Indeed a distinctive feature of a tensor field tangent to $\Sigma_t$ is
\be
	\vgs \w{T} = \w{T} . 
\ee
Assume for instance that $\w{T}$ is a tensor field of type $\left( {1\atop 1} \right)$.
Then the above equation writes [cf. Eq.~(\ref{e:hyp:vgs_comp})]
\be
	\gam^\alpha_{\ \, \mu} \gam^\nu_{\ \, \beta} T^\mu_{\ \, \nu}
	= T^\alpha_{\ \, \beta} .
\ee
Taking the Lie derivative along $\w{m}$ of this relation, 
employing the Leibniz rule and making use 
of Eq.~(\ref{e:fol:Lie_m_vg}), leads to
\bea
	& & 
	\Liec{m} \left( \gam^\alpha_{\ \, \mu} \gam^\nu_{\ \, \beta} T^\mu_{\ \, \nu}
	\right)	
	= 	\Liec{m}  T^\alpha_{\ \, \beta} \nonumber \\
	& & \underbrace{\Liec{m} \gam^\alpha_{\ \, \mu}}_{=0} 
	\; \gam^\nu_{\ \, \beta} T^\mu_{\ \, \nu}
	+ 
	\gam^\alpha_{\ \, \mu} 
	\underbrace{\Liec{m} \gam^\nu_{\ \, \beta}}_{=0} \; T^\mu_{\ \, \nu}
	 + \gam^\alpha_{\ \, \mu} \gam^\nu_{\ \, \beta} \, \Liec{m} T^\mu_{\ \, \nu}
	= 	\Liec{m}  T^\alpha_{\ \, \beta} \nonumber \\
	& & \vgs \Lie{m} \w{T} = \Lie{m} \w{T} .	
\eea
This shows that $\Lie{m}\w{T}$ is tangent to $\Sigma_t$.
The proof is readily extended to any type of tensor field tangent to $\Sigma_t$. 
Notice that the property (\ref{e:fol:liem_preserve}) generalizes that obtained for vectors in
Sec.~\ref{s:fol:norm_evol} [cf. Eq.~(\ref{e:fol:Lie_m_v})].

\begin{remark}
An illustration of property (\ref{e:fol:liem_preserve}) is provided by 
Eq.~(\ref{e:fol:Lie_m_gam}), which says that $\Lie{m}\w{\gam}$ is $-2N\w{K}$:
$\w{K}$ being tangent to $\Sigma_t$, we have immediately that $\Lie{m}\w{\gam}$
is tangent to $\Sigma_t$.
\end{remark}

\begin{remark}
Contrary to $\Lie{n}\w{\gam}$ and $\Lie{m}\w{\gam}$, which are related by Eq.~(\ref{e:fol:Lie_n_m_gam}), $\Lie{n}\vg$ and $\Lie{m}\vg$
are not proportional. Indeed a calculation similar to that which lead to 
Eq.~(\ref{e:fol:Lie_n_m_gam}) gives
\be
	\Lie{n}\vg = \frac{1}{N}  \Lie{m}\vg + \w{n}\otimes\w{D} \ln  N . 
\ee
Therefore the property $\Lie{m}\vg=0$ implies 
\be
	\Lie{n}\vg =\w{n}\otimes\w{D} \ln  N \not= 0 .
\ee 
Hence the privileged role played by $\w{m}$ regarding the evolution
of the hypersurfaces $\Sigma_t$ is not shared by $\w{n}$; this merely reflects that
the hypersurfaces are Lie dragged by $\w{m}$, not by $\w{n}$.
\end{remark}

%%%%%%%%%%%%%%%%%%%%%%%%%%%%%%%%%%%%%%%%%%%%%%%%%%%%%%%%%%%%%%%%%%%%%%%%%%%%%%%%

\section{Last part of the 3+1 decomposition of the Riemann tensor}

\subsection{Last non trivial projection of the spacetime Riemann tensor}

In Chap.~\ref{s:hyp}, we have formed the fully projected part of the 
spacetime Riemann tensor, i.e. $\vgs \, {}^4\mathbf{Riem}$, yielding
the Gauss equation [Eq.~(\ref{e:hyp:Gauss})], as well as the part projected three
times onto $\Sigma_t$ and once along the normal $\w{n}$, yielding 
the Codazzi equation [Eq.~(\ref{e:hyp:Codazzi})]. These two decompositions
involve only fields tangents to $\Sigma_t$ and their derivatives in directions
parallel to $\Sigma_t$, namely $\w{\gam}$, $\w{K}$, $\mathbf{Riem}$
and $\w{D}\w{K}$. This is why they could be defined for a single hypersurface. 
In the present section, we form the projection of the spacetime Riemann tensor
twice onto $\Sigma_t$ and twice along $\w{n}$. As we shall see, this involves
a derivative in the direction normal to the hypersurface. 

As for the Codazzi equation, the starting point of the calculation is
the Ricci identity applied to the vector $\w{n}$, i.e. Eq.~(\ref{e:hyp:Ricci_id_n}).
But instead of projecting it totally onto $\Sigma_t$, let us project it
only twice onto $\Sigma_t$ and once along $\w{n}$:
\be
	\gam_{\alpha\mu} n^\sigma \gam^\nu_{\ \, \beta}
	(\nabla_\nu \nabla_\sigma n^\mu - \nabla_\sigma \nabla_\nu n^\mu)
	= \gam_{\alpha\mu} n^\sigma \gam^\nu_{\ \, \beta} \, 
	{}^4\!R^\mu_{\ \, \rho\nu\sigma} n^\rho . 
\ee
By means of Eq.~(\ref{e:fol:nab_n_K_comp}), we get successively
\bea
	\gam_{\alpha\mu} \, n^\rho \gam^\nu_{\ \, \beta} \, n^\sigma \, 
	{}^4\!R^\mu_{\ \, \rho\nu\sigma} & = &
	\gam_{\alpha\mu} n^\sigma \gam^\nu_{\ \, \beta} \left[
	- \nabla_\nu ( K^\mu_{\ \, \sigma} + D^\mu\ln N  \; n_\sigma)
	+ \nabla_\sigma (K^\mu_{\ \, \nu} + D^\mu\ln N  \; n_\nu) \right]
	\nonumber \\
	& = &\gam_{\alpha\mu} n^\sigma \gam^\nu_{\ \, \beta} \big[
	- \nabla_\nu  K^\mu_{\ \, \sigma} -\nabla_\nu n_\sigma \, D^\mu\ln N  
	- n_\sigma \nabla_\nu D^\mu\ln N  \nonumber \\
	& & \qquad\qquad\ \   
	+ \nabla_\sigma  K^\mu_{\ \, \nu} +\nabla_\sigma n_\nu \, D^\mu\ln N  
	+ n_\nu \nabla_\sigma D^\mu\ln N  \; \big]  \nonumber \\
	& = & \gam_{\alpha\mu} \gam^\nu_{\ \, \beta} \left[
	K^\mu_{\ \, \sigma} \nabla_\nu n^\sigma + \nabla_\nu D^\mu\ln N  
	+ n^\sigma \nabla_\sigma K^\mu_{\ \, \nu}
	+ D_\nu\ln N  \, D^\mu \ln N  \right] \nonumber \\
	& = &- K_{\alpha\sigma} K^\sigma_{\ \, \beta}
	+ D_\beta D_\alpha \ln  N  +\gam^\mu_{\ \, \alpha} \gam^\nu_{\ \, \beta}
	\, n^\sigma \nabla_\sigma K_{\mu\nu}
	+ D_\alpha \ln N  D_\beta \ln N   \nonumber \\
	& = & - K_{\alpha\sigma} K^\sigma_{\ \, \beta} 
	+ \frac{1}{N} D_\beta D_\alpha N +\gam^\mu_{\ \, \alpha} \gam^\nu_{\ \, \beta}
	\, n^\sigma \nabla_\sigma K_{\mu\nu} . \label{e:fol:ggRiem_prov}
\eea
Note that we have used $K^\mu_{\ \,\sigma} n^\sigma = 0$, 
$n^\sigma\nabla_\nu n_\sigma=0$, $n_\sigma n^\sigma=-1$,
$n^\sigma \nabla_\sigma n_\nu = D_\nu \ln N $ and $\gam^\nu_{\ \,\beta} n_\nu = 0$
to get the third equality. 
Let us now show that the term $\gam^\mu_{\ \, \alpha} \gam^\nu_{\ \, \beta}
	\, n^\sigma \nabla_\sigma K_{\mu\nu}$
is related to $\Lie{m} \w{K}$. 
Indeed, from the expression (\ref{e:lie:der_comp_nab}) of the Lie derivative:
\be
	\Liec{m} K_{\alpha\beta} = m^\mu \nabla_\mu K_{\alpha\beta}
	+ K_{\mu\beta} \nabla_\alpha m^\mu + K_{\alpha\mu} \nabla_\beta m^\mu . 
\ee
Substituting Eq.~(\ref{e:fol:nab_m_comp}) for $\nabla_\alpha m^\mu$
and $\nabla_\beta m^\mu$ leads to 
\be
	\Liec{m} K_{\alpha\beta} = N n^\mu \nabla_\mu K_{\alpha\beta}
	- 2N K_{\alpha\mu} K^\mu_{\ \, \beta}
	- K_{\alpha\mu} D^\mu N \; n_\beta
	- K_{\beta\mu} D^\mu N \; n_\alpha . 
\ee
Let us project this equation onto $\Sigma_t$, i.e. apply the operator $\vgs$ to both
sides.
Using the property $\vgs \Lie{m} \w{K} =  \Lie{m} \w{K}$, which stems from 
the fact that $\Lie{m} \w{K}$ is tangent to $\Sigma_t$ since $\w{K}$
is [property~(\ref{e:fol:liem_preserve})], we get
\be
	\Liec{m} K_{\alpha\beta} = N \, \gam^\mu_{\ \, \alpha} \gam^\nu_{\ \, \beta}
	\, n^\sigma \nabla_\sigma K_{\mu\nu}
	- 2N K_{\alpha\mu} K^\mu_{\ \, \beta} . 
\ee
Extracting $\gam^\mu_{\ \, \alpha} \gam^\nu_{\ \, \beta}
	\, n^\sigma \nabla_\sigma K_{\mu\nu}$ from this relation and
plugging it into Eq.~(\ref{e:fol:ggRiem_prov}) results in 
\be \label{e:fol:ggRiem}
	\encadre{ \gam_{\alpha\mu} \, n^\rho \gam^\nu_{\ \, \beta} \, n^\sigma \, 
	{}^4\!R^\mu_{\ \, \rho\nu\sigma} = \frac{1}{N} \Liec{m} K_{\alpha\beta} 
	+ \frac{1}{N} D_\alpha D_\beta N + K_{\alpha\mu} K^\mu_{\ \, \beta} } . 
\ee
Note that we have written $D_\beta D_\alpha N =  D_\alpha D_\beta N$ ($\w{D}$ has
no torsion). Equation~(\ref{e:fol:ggRiem}) is the relation we sought.
It is sometimes called the \defin{Ricci equation}
[not to be confused with the \emph{Ricci identity} (\ref{e:hyp:Ricci_ident})].
Together with the Gauss equation (\ref{e:hyp:Gauss})
and the Codazzi equation (\ref{e:hyp:Codazzi}), it 
completes the 3+1 decomposition of the spacetime Riemann tensor. 
Indeed 
the part projected three times along $\w{n}$ vanish identically, 
since
${}^4\mathbf{Riem}(\uu{n},\w{n},\w{n},.)=0$ and 
${}^4\mathbf{Riem}(.,\w{n},\w{n},\w{n})=0$ thanks to 
the partial antisymmetry of the Riemann tensor.
Accordingly one can project ${}^4\mathbf{Riem}$ at most twice along $\w{n}$ 
to get some non-vanishing result. 

It is worth to note that the left-hand side of the Ricci equation
(\ref{e:fol:ggRiem}) is a term which appears in
the contracted Gauss equation
(\ref{e:hyp:Gauss_contracted}). Therefore, by combining the two equations, 
we get a formula which does no longer contain the spacetime Riemann tensor,
but only the spacetime Ricci tensor: 
\be \label{e:fol:proj_Ricci_comp}
	\encadre{ 
	\gamma^\mu_{\ \, \alpha} \gamma^\nu_{\ \, \beta}  {}^4 \!R_{\mu\nu}
	= - \frac{1}{N} \Liec{m} K_{\alpha\beta} 
	-  \frac{1}{N} D_\alpha D_\beta N + R_{\alpha\beta}
	+ K K_{\alpha\beta} - 2 K_{\alpha\mu} K^\mu_{\ \, \beta}
	} ,
\ee
or in index-free notation:
\be \label{e:fol:proj_Ricci}
	\encadre{
	\vgs \, {}^4 \!\w{R} = - \frac{1}{N} \Lie{m} \w{K}
	-  \frac{1}{N} \w{D} \w{D} N 
	+ \w{R} + K\, \w{K} -2 \w{K}\cdot\vec{\w{K}} } . 
\ee

\subsection{3+1 expression of the spacetime scalar curvature}

Let us take the trace of Eq.~(\ref{e:fol:proj_Ricci}) with respect to the
metric $\w{\gam}$. This amounts to contracting Eq.~(\ref{e:fol:proj_Ricci_comp})
with $\gam^{\alpha\beta}$. In the left-hand side, we have 
$\gam^{\alpha\beta} \gamma^\mu_{\ \, \alpha} \gamma^\nu_{\ \, \beta} = 
\gam^{\mu\nu}$ and in the right-hand we can limit the range of variation
of the indices to $\{1,2,3\}$ since all the involved tensors are spatial ones
[including $\Lie{m}\w{K}$, thanks to the property (\ref{e:fol:liem_preserve})]
Hence
\be \label{e:fol:gam_Riem1}
	\gam^{\mu\nu} {}^4\!R_{\mu\nu}
	= - \frac{1}{N}\gam^{ij}  \Liec{m} K_{ij}
	- \frac{1}{N} D_i D^i N  + R + K^2 - 2K_{ij} K^{ij} .  
\ee 
Now $\gam^{\mu\nu} {}^4\!R_{\mu\nu} = (g^{\mu\nu} + n^\mu n^\nu) {}^4\!R_{\mu\nu}
= {}^4\!R + {}^4\!R_{\mu\nu} n^\mu n^\nu$ and 
\be
	- \gam^{ij}  \Liec{m} K_{ij} = - 
	\Liec{m}(\underbrace{\gam^{ij} K_{ij}}_{=K})
	+ K_{ij}  \Liec{m} \gam^{ij} , \label{e:fol:gamup_LiemK}
\ee
with $\Liec{m} \gam^{ij}$ evaluted from the very definition of the inverse
3-metric:
\bea
	& & \gam_{ik} \gam^{kj} = \delta^j_{\ \, i} \nonumber \\
	& \Rightarrow  & \Liec{m} \gam_{ik} \, \gam^{kj} +
	 \gam_{ik}\, \Liec{m}\gam^{kj} = 0 \nonumber \\
	& \Rightarrow & \gam^{il}  \gam^{kj} 
	\Liec{m} \gam_{lk} + 
	\underbrace{\gam^{il} \gam_{lk}}_{=\delta^i_{\ \, k}}
	 \Liec{m} \gam^{lj}
	= 0 \nonumber \\
	& \Rightarrow & \Liec{m} \gam^{ij} =
	-  \gam^{ik}  \gam^{jl} \Liec{m}\gam_{kl} \nonumber \\
	&  \Rightarrow & \Liec{m} \gam^{ij} = 2N \gam^{ik}  \gam^{kl}
	K_{kl} \nonumber \\
	&  \Rightarrow & 
	\encadre{\Liec{m} \gam^{ij} = 2N K^{ij} }, 	\label{e:fol:Liem_gam_up}
\eea
where we have used Eq.~(\ref{e:fol:Lie_m_gam}).
Pluging Eq.~(\ref{e:fol:Liem_gam_up}) into Eq.~(\ref{e:fol:gamup_LiemK}) gives
\be \label{e:fol:trLiemK}
	- \gam^{ij}  \Liec{m} K_{ij} = - \Liec{m} K + 2 N K_{ij} K^{ij} . 
\ee
Consequently Eq.~(\ref{e:fol:gam_Riem1}) becomes
\be
	\encadre{ {}^4\!R + {}^4\!R_{\mu\nu} n^\mu n^\nu = R + K^2
	- \frac{1}{N} \Liec{m} K 
	- \frac{1}{N} D_i D^i N } .
\ee
It is worth to combine with equation with the scalar Gauss relation
(\ref{e:hyp:Gauss_scalar}) to get rid of the Ricci tensor term
${}^4\!R_{\mu\nu} n^\mu n^\nu$ and obtain an equation which involves
only the spacetime scalar curvature ${}^4\!R$:
\be \label{e:fol:4R_3p1}
	\encadre{ {}^4\!R = R + K^2 + K_{ij} K^{ij}
	- \frac{2}{N} \Liec{m} K 
	- \frac{2}{N} D_i D^i N } . 
\ee

%  
%    Chapitre : 3+1 decomposition of Einstein equation
%
% $Date: 2007-03-06 11:59:03 +0100 (mar, 06 mar 2007) $
% $Rev: 183 $
% $Author: e_gourgoulhon $
%%%%%%%%%%%%%%%%%%%%%%%%%%%%%

\chapter{3+1 decomposition of Einstein equation} \label{s:dec}

%\verb$Date: 2007-03-06 11:59:03 +0100 (mar, 06 mar 2007) $

\minitoc
\vspace{1cm}

%%%%%%%%%%%%%%%%%%%%%%%%%%%%%%%%%%%%%%%%%%%%%%%%%%%%%%%%%%%%%%%%%%%%%%%%%%%%

\section{Einstein equation in 3+1 form}

\subsection{The Einstein equation}

After the first two chapters devoted to the geometry of hypersurfaces
and foliations, we are now back to physics: we consider a
spacetime $(\M,\w{g})$ such that $\w{g}$ obeys to the
Einstein equation (with zero cosmological constant): 
\be \label{e:dec:Einstein}
	\encadre{ {}^4\!\w{R} - \frac{1}{2} {}^4\!R \, \w{g} = 8\pi \w{T} },
\ee
where ${}^4\!\w{R}$ is the Ricci tensor associated with $\w{g}$
[cf. Eq.~(\ref{e:hyp:def_Ricci})],
${}^4\!R$ the corresponding Ricci scalar, and 
$\w{T}$ is the matter stress-energy tensor. 

We shall also use the equivalent form
\be \label{e:dec:Einstein2}
   {}^4\!\w{R} = 8\pi \left( \w{T} - \frac{1}{2} T \, \w{g} \right),
\ee
where $T := g^{\mu\nu} T_{\mu\nu}$ stands for the trace (with respect to $\w{g}$) 
of the stress-energy tensor $\w{T}$.

Let us assume that the spacetime $(\M,\w{g})$ is globally hyperbolic
(cf. Sec.~\ref{s:fol:glob_hyp}) and let be $(\Sigma_t)_{t\in\R}$ by
a foliation of $\M$ by a family of spacelike hypersurfaces. 
The foundation of the 3+1 formalism amounts to projecting the
Einstein equation (\ref{e:dec:Einstein}) onto $\Sigma_t$ and perpendicularly
to $\Sigma_t$. 
To this purpose let us first consider the 3+1 decomposition of
the stress-energy tensor. 

\subsection{3+1 decomposition of the stress-energy tensor} \label{s:dec:T3p1}

From the very definition of a stress-energy tensor, the 
\defin{matter energy density}
as measured by the Eulerian observer introduced in Sec.~\ref{s:fol:Eulerian}
is
\be \label{e:dec:E_def}
	\encadre{ E := \w{T}(\w{n},\w{n}) } . 
\ee
This follows from the fact that the 4-velocity of the Eulerian observer
in the unit normal vector $\w{n}$.

Similarly, also from the very definition of a stress-energy tensor, 
the \defin{matter momentum density} as measured by the Eulerian 
observer is the linear form 
\be \label{e:dec:p_def}
	\encadre{ \w{p} := - \w{T}(\w{n}, \vg( . ) ) }, 
\ee
i.e. the linear form defined by
\be
	\forall \w{v}\in \T_p(\M),\quad
	\langle \w{p}, \w{v} \rangle =  
	- \w{T}(\w{n}, \vg( \w{v} ) ) . 
\ee
In components:
\be
	p_\alpha = - T_{\mu\nu} \, n^\mu \, \gam^\nu_{\ \, \alpha} . 
\ee
Notice that, thanks to the projector $\vg$, 
$\w{p}$ is a linear form tangent to $\Sigma_t$. 
\begin{remark}
The momentum density $\w{p}$ is often denoted $\w{j}$. Here we
reserve the latter for electric current density. 
\end{remark}

Finally, still from the very definition of a stress-energy tensor,
the \defin{matter stress tensor} as measured by the Eulerian observer
is the bilinear form
\be \label{e:dec:S_def}
	\encadre{ \w{S} := \vgs \w{T} } , 
\ee
or, in components,
\be \label{e:dec:S_def_comp}
	S_{\alpha\beta} = 
	T_{\mu\nu} \gamma^\mu_{\ \, \alpha} \gamma^\nu_{\ \, \beta}  
\ee
As for $\w{p}$, $\w{S}$ is a tensor field tangent to $\Sigma_t$.
Let us recall the physical interpretation of the stress tensor $\w{S}$:
given two spacelike unit vectors $\w{e}$ and $\w{e}'$ (possibly equal) in the
rest frame of the Eulerian observer (i.e. two unit vectors orthogonal to
$\w{n}$), $S(\w{e},\w{e}')$ is the force in the direction $\w{e}$
acting on the unit surface whose normal is $\w{e}'$.
Let us denote by $S$ the trace of $\w{S}$ with respect to the metric $\w{\gam}$
(or equivalently with respect to the metric $\w{g}$):
\be
	\encadre{ S := \gam^{ij} S_{ij} = g^{\mu\nu} S_{\mu\nu} } . 
\ee

The knowledge of $(E,\w{p},\w{S})$ is sufficient to reconstruct $\w{T}$
since 
\be \label{e:dec:T_3p1}
	\encadre{ \w{T} = \w{S} + \uu{n}\otimes\w{p} + \w{p}\otimes\uu{n}
	+ E \, \uu{n}\otimes\uu{n} } . 
\ee
This formula is easily established by substituting Eq.~(\ref{e:hyp:vecgam_ab})
for $\gamma^\alpha_{\ \, \beta}$ into Eq.~(\ref{e:dec:S_def_comp}) and 
expanding the result. 
Taking the trace of Eq.~(\ref{e:dec:T_3p1}) with respect to the metric $\w{g}$ yields
\be
	T = S + 2 \underbrace{\langle \w{p},\w{n} \rangle}_{=0}
	+ E \underbrace{ \langle \uu{n},\w{n} \rangle}_{=-1},
\ee
hence
\be \label{e:dec:trT_SmE}
	T = S - E . 
\ee

\subsection{Projection of the Einstein equation} \label{s:dec:project_Einstein}

With the above 3+1 decomposition of the stress-energy tensor
and the 3+1 decompositions of the spacetime Ricci tensor obtained
in Chapters~\ref{s:hyp} and \ref{s:fol}, we are fully equipped to 
perform the projection of the Einstein equation (\ref{e:dec:Einstein})
onto the hypersurface $\Sigma_t$ and along its normal. 
There are only three possibilities:

\subsubsection{(1) Full projection onto $\Sigma_t$}
This amounts to applying the operator $\vgs$ to the Einstein equation.
It is convenient to take the version (\ref{e:dec:Einstein2}) of the latter; 
we get
\be \label{e:dec:EEdyn_proj}
	   \vgs\, {}^4\!\w{R} = 8\pi \left( \vgs \w{T} 
	- \frac{1}{2} T \, \vgs\w{g} \right).
\ee
$\vgs \, {}^4 \!\w{R}$ is given by Eq.~(\ref{e:fol:proj_Ricci})
(combination of the contracted Gauss equation with the Ricci equation), 
$\vgs\w{T}$ is by definition $\w{S}$, $T=S-E$ [Eq.~(\ref{e:dec:trT_SmE})], and 
$\vgs \w{g}$ is simply $\w{\gam}$. Therefore
\be
	- \frac{1}{N} \Lie{m} \w{K}
	-  \frac{1}{N} \w{D} \w{D} N 
	+ \w{R} + K\, \w{K} -2 \w{K}\cdot\vec{\w{K}} =
	8\pi \left[ \w{S} - \frac{1}{2} (S-E) \, \w{\gam} \right], 
\ee
or equivalently
\be \label{e:dec:EEdyn}
   \encadre{ \Lie{m} \w{K} = - \w{D} \w{D} N + N\left\{
	\w{R} + K \w{K} -2 \w{K}\cdot\vec{\w{K}} 
	+ 4\pi \left[ (S-E) \w{\gam} - 2 \w{S} \right]
	\right\} } .    
\ee
In components:
\be \label{e:dec:EEdyn_comp_ab}
  \Liec{m} K_{\alpha\beta} = - D_\alpha D_\beta N + N\left\{
	R_{\alpha\beta} + K K_{\alpha\beta} -2 K_{\alpha\mu} K^\mu_{\ \, \beta}
	+ 4\pi \left[ (S-E) \gam_{\alpha\beta} - 2 S_{\alpha\beta} \right]
	\right\}  .    
\ee
Notice that each term in the above equation is a tensor field tangent to $\Sigma_t$.
For $\Lie{m} \w{K}$, this results from the fundamental property 
(\ref{e:fol:liem_preserve}) of $\Lie{m}$. Consequently, we may restrict 
to spatial indices without any loss of generality and write Eq.~(\ref{e:dec:EEdyn_comp_ab})
as
\be \label{e:dec:EEdyn_comp}
   \encadre{ \Liec{m} K_{ij} = - D_i D_j N + N\left\{
	R_{ij} + K K_{ij} -2 K_{ik} K^k_{\ \, j}
	+ 4\pi \left[ (S-E) \gam_{ij} - 2 S_{ij} \right]
	\right\} } .    
\ee

\subsubsection{(2) Full projection perpendicular to $\Sigma_t$}

This amounts to applying
the Einstein equation (\ref{e:dec:Einstein}), which is an identity
between bilinear forms, to the couple $(\w{n},\w{n})$; we get, since $\w{g}(\w{n},\w{n})=-1$,
\be
	{}^4\!\w{R}(\w{n},\w{n}) + \frac{1}{2} {}^4\! R = 8\pi \w{T}(\w{n},\w{n}) .
\ee
Using the scalar Gauss equation (\ref{e:hyp:Gauss_scalar}), and noticing that
$\w{T}(\w{n},\w{n})=E$ [Eq.~(\ref{e:dec:E_def})] yields
\be \label{e:dec:Ham_constr}
	\encadre{ R + K^2 - K_{ij} K^{ij} = 16\pi E} .
\ee
This equation is called the \defin{Hamiltonian constraint}. The word \emph{`constraint'}
will be justified in Sec.~\ref{s:dec:constraints} and the qualifier \emph{`Hamiltonian'}
in Sec.~\ref{s:dec:ADM_Ham}. 

\subsubsection{(3) Mixed projection}

Finally, let us project the Einstein equation (\ref{e:dec:Einstein}) 
once onto $\Sigma_t$ and once
along the normal $\w{n}$:
\be
	{}^4\!\w{R}(\w{n}, \vg(.)) -\frac{1}{2} {}^4\! R 
	\underbrace{\w{g}(\w{n},\vg(.))}_{=0} = 8\pi \w{T}(\w{n}, \vg(.) ) .
\ee
By means of the contracted Codazzi equation (\ref{e:hyp:Codazzi_contract}) and
$\w{T}(\w{n}, \vg(.) )=-\w{p}$ [Eq.~(\ref{e:dec:p_def})], we
get
\be \label{e:dec:mom_constr}
	\encadre{ \w{D} \cdot \vec{\w{K}} - \w{D} K = 8\pi \w{p} } ,
\ee
or, in components,
\be \label{e:dec:mom_constr_comp}
	\encadre{ D_j K^j_{\ \, i} - D_i K = 8\pi  p_i     } . 
\ee
This equation is called the \defin{momentum constraint}. Again, the word 
\emph{`constraint'} will be justified in Sec.~\ref{s:dec:Cauchy}.

\subsubsection{Summary}

The Einstein equation is equivalent to the system of three equations:
(\ref{e:dec:EEdyn}), (\ref{e:dec:Ham_constr}) and (\ref{e:dec:mom_constr}).
Equation~(\ref{e:dec:EEdyn}) is a rank 2 tensorial (bilinear forms) equation 
within $\Sigma_t$, involving
only symmetric tensors: it has therefore 6 independent components.
Equation~(\ref{e:dec:Ham_constr}) is a scalar equation and Eq.~(\ref{e:dec:mom_constr})
is a rank 1 tensorial (linear forms) within $\Sigma_t$: it has therefore 3 independent
components. The total number of independent components 
is thus $6+1+3=10$, i.e. the same as the
original Einstein equation (\ref{e:dec:Einstein}).

%%%%%%%%%%%%%%%%%%%%%%%%%%%%%%%%%%%%%%%%%%%%%%%%%%%%%%%%%%%%%%%%%%%%%%%%%%%%

\section{Coordinates adapted to the foliation}

\subsection{Definition of the adapted coordinates}

The system (\ref{e:dec:EEdyn})+(\ref{e:dec:Ham_constr})+(\ref{e:dec:mom_constr})
is a system of tensorial equations. In order to transform it into a
system of partial differential equations (PDE), one must introduce coordinates
on the spacetime manifold $\M$, which we have not done yet.
Coordinates adapted to the foliation $(\Sigma_t)_{t\in\R}$ are set in the following
way. On each hypersurface $\Sigma_t$ one introduces some coordinate system
$(x^i)=(x^1,x^2,x^3)$. If this coordinate system varies smoothly between
neighbouring hypersurfaces, then $(x^\alpha)=(t,x^1,x^2,x^3)$ constitutes a well-behaved
coordinate system on $\M$. We shall call $(x^i)=(x^1,x^2,x^3)$ the \defin{spatial
coordinates}. 

\begin{figure}
\centerline{\includegraphics[width=0.7\textwidth]{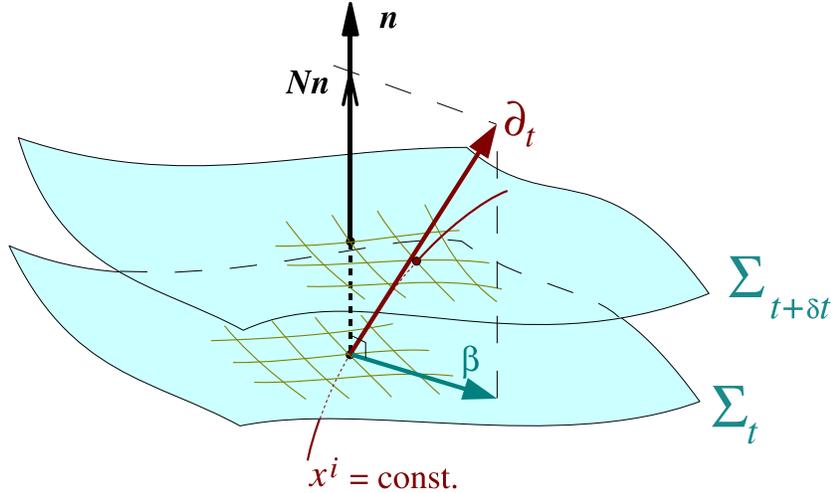}}
\caption[]{\label{f:dec:shift} \footnotesize
Coordinates $(x^i)$ on the hypersurfaces $\Sigma_t$:
each line $x^i={\rm const}$ cuts across the foliation 
$(\Sigma_t)_{t\in\R}$ and defines the time vector $\wpar_t$ 
and the shift vector $\w{\beta}$ of the
spacetime coordinate system $(x^\alpha)=(t,x^i)$.
}
\end{figure}

Let us denote by $(\wpar_\alpha)=(\wpar_t,\wpar_i)$ the natural basis 
of $\T_p(\M)$ associated with the coordinates $(x^\alpha)$, i.e. the set of 
vectors
\bea
		& & \wpar_t := \der{}{t} \\
		& & \wpar_i := \der{}{x^i}, \quad i\in\{1,2,3\} .
\eea
Notice that the vector $\wpar_t$ is tangent to the lines of constant 
spatial coordinates, i.e. the curves of $\M$ defined by
 $(x^1=K^1,x^2=K^2,x^3=K^3)$, where $K^1$, 
$K^2$ and $K^3$ are three constants (cf. Fig.~\ref{f:dec:shift}).
We shall call $\wpar_t$ the \defin{time vector}.
\begin{remark}
$\wpar_t$ is not necessarily a timelike vector.
This will be discussed further below 
[Eqs.~(\ref{e:dec:part_time})-(\ref{e:dec:part_space})].
\end{remark}

For any $i\in\{1,2,3\}$, the vector $\wpar_i$ is tangent to the lines
$t=K^0$, $x^j=K^j$ ($j\not=i$), where $K^0$ and $K^j$ ($j\not=i$) are
three constants. Having $t$ constant, these lines belong to the hypersurfaces $\Sigma_t$.
This implies that $\wpar_i$ is tangent to $\Sigma_t$:
\be \label{e:dec:wpi_tang_Sigma}
	\wpar_i \in \T_p(\Sigma_t) , \quad i\in\{1,2,3\} .
\ee

\subsection{Shift vector} \label{s:dec:shift}

The dual basis associated with $(\wpar_\alpha)$ is the gradient 1-form basis 
$(\dd x^\alpha)$, which is a basis of the space of linear forms $\T_p^*(\M)$:
\be
	\langle \dd x^\alpha, \wpar_\beta \rangle = \delta^\alpha_{\ \, \beta}.
\ee
In particular, the 1-form $\dd t$ is dual to the vector $\wpar_t$:
\be \label{e:dec:dt_part_1}
	\langle \dd t , \wpar_t \rangle = 1 .
\ee
Hence the time vector $\wpar_t$ obeys to the same property as the 
normal evolution vector $\w{m}$, since $\langle \dd t , \w{m} \rangle = 1$
[Eq.~(\ref{e:fol:dt_m_1})]. In particular, $\wpar_t$ Lie drags the hypersurfaces
$\Sigma_t$, as $\w{m}$ does (cf. Sec.~\ref{s:fol:norm_evol}).
In general the two vectors $\wpar_t$ and $\w{m}$ differ.
They coincide only if the coordinates $(x^i)$ are such that the lines
$x^i={\rm const}$ are orthogonal to the hypersurfaces $\Sigma_t$ 
(cf. Fig.~\ref{f:dec:shift}). 
The difference between $\wpar_t$ and $\w{m}$  is called the \defin{shift vector}
and is denoted $\w{\beta}$:
\be \label{e:dec:t_m_b}
	\encadre{ \wpar_t =: \w{m} + \w{\beta} } . 
\ee
As for the lapse, the name \emph{shift vector} has been coined by Wheeler (1964) \cite{Wheel64}. 
By combining Eqs.~(\ref{e:dec:dt_part_1}) and (\ref{e:fol:dt_m_1}), we get
\be
	\langle \dd t, \w{\beta} \rangle = \langle \dd t, \wpar_t \rangle
	- \langle \dd t, \w{m} \rangle = 1-1=0 ,
\ee
or equivalently, since $\dd t = -N^{-1} \uu{n}$ [Eq.~(\ref{e:fol:un_lapse})],
\be \label{e:dec:n_beta}
	\encadre{\w{n}\cdot\w{\beta} = 0} . 
\ee
Hence the vector $\w{\beta}$ is tangent to the hypersurfaces $\Sigma_t$.

The lapse function and the shift vector have been introduced for the first time
explicitly, although without their present names,
by Y. Choquet-Bruhat in 1956 \cite{Foure56}. 

It usefull to rewrite Eq.~(\ref{e:dec:t_m_b}) by means of the relation
$\w{m}=N\w{n}$ [Eq.~(\ref{e:fol:def_m})]:
\be \label{e:dec:t_Nn_b}
	\encadre{ \wpar_t = N\w{n} + \w{\beta} } . 
\ee
Since the vector $\w{n}$ is normal to $\Sigma_t$ and $\w{\beta}$ tangent to $\Sigma_t$, 
Eq.~(\ref{e:dec:t_Nn_b}) can be seen as a 3+1 decomposition of the time vector
$\wpar_t$.

The scalar square of $\wpar_t$ is deduced immediately from Eq.~(\ref{e:dec:t_Nn_b}),
taking into account $\w{n}\cdot\w{n}=-1$ and Eq.~(\ref{e:dec:n_beta}):
\be \label{e:dec:wt_square} 
	\wpar_t\cdot\wpar_t = -N^2 + \w{\beta}\cdot\w{\beta} .
\ee
Hence we have the following:
\bea
	\wpar_t \mbox{ is timelike} & \Longleftrightarrow & 
	\w{\beta}\cdot\w{\beta} < N^2 , \label{e:dec:part_time} \\
	\wpar_t \mbox{ is null} & \Longleftrightarrow & 
	\w{\beta}\cdot\w{\beta} = N^2 ,\\
	\wpar_t \mbox{ is spacelike} & \Longleftrightarrow & 
	\w{\beta}\cdot\w{\beta} > N^2 .\label{e:dec:part_space} 
\eea

\begin{remark}
A shift vector that fulfills the condition (\ref{e:dec:part_space}) is
sometimes called a \defin{superluminal shift}. Notice that, since a priori
the time vector $\wpar_t$ is a pure coordinate quantity and is not associated
with the 4-velocity of some observer (contrary to $\w{m}$, which is proportional
to the 4-velocity of the Eulerian observer), there is nothing unphysical in
having $\wpar_t$ spacelike.
\end{remark}

Since $\w{\beta}$ is tangent to $\Sigma_t$, let us introduce the components
of $\w{\beta}$ and the metric dual form $\uu{\beta}$ with respect to
the spatial coordinates $(x^i)$ according to 
\be
	\w{\beta} =: \beta^i \, \wpar_i
	\qquad \mbox{and} \qquad 
	\uu{\beta} =: \beta_i \, \dd x^i . 
\ee
Equation~(\ref{e:dec:t_Nn_b}) then shows that  
the components of the unit normal vector $\w{n}$ with respect to the
natural basis $(\wpar_\alpha)$ are expressible in terms of $N$ and $(\beta^i)$
as
\be
	n^\alpha = \left(\frac{1}{N},-\frac{\beta^1}{N},
	-\frac{\beta^2}{N},-\frac{\beta^3}{N} \right) .
\ee
Notice that the covariant components (i.e. the components of $\uu{n}$
with respect to the basis $(\dd x^\alpha)$ of $\T_p^*(\M)$) are 
immediately deduced from the relation $\uu{n} = - N \dd t$
[Eq.~(\ref{e:fol:un_lapse})] : 
\be \label{e:dec:uun_comp}
	n_\alpha = (-N, 0,0,0) . 
\ee

\subsection{3+1 writing of the metric components}

Let us introduce the components $\gam_{ij}$ of the 3-metric $\w{\gam}$
with respect to the coordinates $(x^i)$
\be
	\w{\gam} =: \gam_{ij} \, \dd x^i \otimes \dd x^j . 
\ee
From the definition of $\uu{\beta}$, we have
\be \label{e:dec:beta_i}
	\beta_i = \gam_{ij} \, \beta^j . 
\ee

The components $g_{\alpha\beta}$ of the metric $\w{g}$ with respect to
the coordinates $(x^\alpha)$ are defined by
\be
	\w{g} =: g_{\alpha\beta} \, \dd x^\alpha \otimes \dd x^\beta .
\ee
Each component can be computed as
\be
	g_{\alpha\beta} = \w{g}(\wpar_\alpha,\wpar_\beta) . 
\ee
Accordingly, thanks to Eq.~(\ref{e:dec:wt_square}),
\be
	g_{00} = \w{g}(\wpar_t,\wpar_t) = \wpar_t \cdot\wpar_t
	= -N^2 + \w{\beta}\cdot\w{\beta}
	= -N^2 + \beta_i \beta^i 	\label{e:dec:g00}
\ee
and, thanks to Eq.~(\ref{e:dec:t_m_b})
\be
	g_{0i} = \w{g}(\wpar_t,\wpar_i) = (\w{m}+\w{\beta}) \cdot \wpar_i . 
\ee
Now, as noticed above [cf. Eq.~(\ref{e:dec:wpi_tang_Sigma})], the vector
$\wpar_i$ is tangent to $\Sigma_t$, so that $\w{m}\cdot\wpar_i=0$.
Hence
\be
	g_{0i} = \w{\beta}\cdot\wpar_i = \langle\uu{\beta},\wpar_i\rangle
	= \langle \beta_j \, \dd x^j, \wpar_i \rangle 
	= \beta_j \underbrace{\langle \dd x^j, \wpar_i \rangle}_{=\delta^j_{\ \, i}}
	= \beta_i . \label{e:dec:g0i}
\ee
Besides, since $\wpar_i$ and $\wpar_j$ are tangent to $\Sigma_t$,
\be
	g_{ij} = \w{g}(\wpar_i,\wpar_j) = \w{\gam}(\wpar_i,\wpar_j) = \gam_{ij}.
		\label{e:dec:gij}
\ee
Collecting Eqs.~(\ref{e:dec:g00}), (\ref{e:dec:g0i})
and (\ref{e:dec:gij}), we get the following expression of the metric components
in terms of 3+1 quantities:
\be \label{e:dec:g_cov}
	\encadre{g_{\alpha\beta} = \left( \begin{array}{cc}
		g_{00} & g_{0j} \\
		g_{i0} & g_{ij}
		\end{array} \right) =
		\left( \begin{array}{cc}
		-N^2 + \beta_k \beta^k & \beta_j \\
		\beta_i & \gamma_{ij}
		\end{array} \right) } ,
\ee
or, in terms of line elements [using Eq.~(\ref{e:dec:beta_i})],
\be \label{e:dec:g_gam_N_beta}
	\encadre{	g_{\mu\nu} \, dx^\mu\, dx^\nu
	= - N^2 dt^2 + \gamma_{ij} (dx^i + \beta^i dt)
		(dx^j + \beta^j dt) } .  
\ee
The components of the inverse metric are given by the matrix inverse of
(\ref{e:dec:g_cov}):
\be \label{e:dec:g_con}
	\encadre{ g^{\alpha\beta} = \left( \begin{array}{cc}
		g^{00} & g^{0j} \\
		g^{i0} & g^{ij}
		\end{array} \right) =
		\left( \begin{array}{cc}
		-\frac{1}{N^2} & \frac{\beta^j}{N^2} \\
		\frac{\beta^i}{N^2} & \gamma^{ij} 
		- \frac{\beta^i\beta^j}{N^2}
		\end{array} \right) }.
\ee
Indeed, it is easily checked that the matrix product 
$g^{\alpha\mu} g_{\mu\beta}$ is equal to the identity matrix 
$\delta^\alpha_{\ \, \beta}$. 
\begin{remark}
Notice that $g_{ij}=\gam_{ij}$ but that in general 
$g^{ij}\not=\gam^{ij}$.
\end{remark}

One can deduce from the above formul\ae\ a simple relation between the 
determinants of $\w{g}$ and $\w{\gam}$. Let us first define the latter ones by
\be
	\encadre{ g := \det (g_{\alpha\beta}) },
\ee
\be
	\encadre{ \gam := \det (\gam_{ij}) } . 
\ee
Notice that $g$ and $\gam$ depend upon the choice of the coordinates 
$(x^\alpha)$. They are not scalar quantities, but scalar \emph{densities}. 
Using Cramer's rule for expressing the inverse $(g^{\alpha\beta})$
of the matrix $(g_{\alpha\beta})$, we have
\be \label{e:dec:g00_Cramer}
	g^{00} = \frac{C_{00}}{\det (g_{\alpha\beta})} = \frac{C_{00}}{g}, 
\ee
where $C_{00}$ is the element $(0,0)$ of the cofactor matrix associated
with $(g_{\alpha\beta})$. It is given by $C_{00}=(-1)^0 M_{00}=M_{00}$, where
$M_{00}$ is the minor $(0,0)$ of the matrix $(g_{\alpha\beta})$, i.e. 
the determinant of the $3\times3$ matrix deduced from $(g_{\alpha\beta})$
by suppressing the first line and the first column. 
From Eq.~(\ref{e:dec:g_cov}), we read
\be
	M_{00} = \det (\gam_{ij}) = \gam .
\ee
Hence Eq.~(\ref{e:dec:g00_Cramer}) becomes
\be
	g^{00} = \frac{\gam}{g} . 
\ee
Expressing $g^{00}$ from Eq.~(\ref{e:dec:g_con}) yields then
$g = -N^2 \gam$, or equivalently, 
\be \label{e:fol:detg_Ngetgam}
	\encadre{\sqrt{-g} = N \sqrt{\gam} } . 
\ee

\subsection{Choice of coordinates via the lapse and the shift}
\label{s:dec:coord_lapse_shift}

We have seen above that giving a coordinate system $(x^\alpha)$ on $\M$
such that the hypersurfaces $x^0={\rm const.}$ are spacelike determines
uniquely a lapse function $N$ and a shift vector $\w{\beta}$. 
The converse is true in the following sense: setting on some hypersurface $\Sigma_0$
a scalar field $N$, a vector field $\w{\beta}$ and
a coordinate system $(x^i)$
uniquely specifies a coordinate system $(x^\alpha)$ in some neighbourhood of $\Sigma_0$,
such that the hypersurface $x^0=0$ is $\Sigma_0$. Indeed, the knowledge of the
lapse function a each point of $\Sigma_0$ determines a unique vector $\w{m}=N\w{n}$
and consequently the location of the ``next'' hypersurface $\Sigma_{\delta t}$
by Lie transport along $\w{m}$ (cf. Sec.~\ref{s:fol:norm_evol}). Graphically, we may
also say that for each point of $\Sigma_0$ the lapse function specifies how far is the
point of $\Sigma_{\delta t}$ located ``above'' it 
(``above'' meaning perpendicularly to $\Sigma_0$, cf. Fig.~\ref{f:fol:liem_sigma}).
Then the shift vector tells how to propagate the
coordinate system $(x^i)$ from $\Sigma_0$ to $\Sigma_{\delta t}$
(cf. Fig.~\ref{f:dec:shift}).

This way of choosing coordinates via the lapse function and the shift vector
is one of the main topics in 3+1 numerical relativity and
will be discussed in detail in Chap.~\ref{s:evo}.

%%%%%%%%%%%%%%%%%%%%%%%%%%%%%%%%%%%%%%%%%%%%%%%%%%%%%%%%%%%%%%%%%%%%%%%%%%%%
 
\section{3+1 Einstein equation as a PDE system}

\subsection{Lie derivatives along $m$ as partial derivatives}

Let us consider the term $\Lie{m}\w{K}$ which occurs in the 3+1
Einstein equation (\ref{e:dec:EEdyn}). Thanks to Eq.~(\ref{e:dec:t_m_b}), 
we can write
\be  \label{e:dec:LiemK_t_beta}
	\Lie{m}\w{K} = \bm{\mathcal{L}}_{\wpar_t} \w{K} - \Lie{\beta} \w{K} . 
\ee
This implies that $\bm{\mathcal{L}}_{\wpar_t} \w{K}$ is a tensor field tangent to 
$\Sigma_t$, since both $\Lie{m}\w{K}$ and $\Lie{\beta} \w{K}$
are tangent to $\Sigma_t$, the former by the property (\ref{e:fol:liem_preserve})
and the latter because
$\w{\beta}$ and $\w{K}$ are tangent to $\Sigma_t$. 
Moreover, if one uses tensor components with respect to a coordinate system
$(x^\alpha)=(t,x^i)$ adapted to the foliation, the Lie derivative along
$\wpar_t$ reduces simply to the partial derivative with respect to $t$
[cf. Eq.~(\ref{e:Lie_adapted})]:
\be \label{e:dec:LietK}
	{\mathcal{L}}_{\wpar_t} K_{ij} = \der{K_{ij}}{t}  . 
\ee
By means of formula~(\ref{e:Lie_der_comp}), one can also express
$\Lie{\beta} \w{K}$ in terms of partial derivatives:
\be \label{e:dec:Liebeta_K}
	\Liec{\beta} K_{ij} = \beta^k \der{K_{ij}}{x^k} 
	+ K_{kj} \der{ \beta^k}{x^i} + K_{ik} \der{\beta^k}{x^j} . 
\ee

Similarly, the relation (\ref{e:fol:Lie_m_gam}) between
$\Lie{m}\w{\gam}$ and $\w{K}$ becomes
\be \label{e:dec:Liem_gam_t_beta}
	\bm{\mathcal{L}}_{\wpar_t} \w{\gam} - \Lie{\beta} \w{\gam} =
	- 2N\w{K} , 
\ee
with 
\be  \label{e:dec:Liet_gam}
	{\mathcal{L}}_{\wpar_t} \gam_{ij} = \der{\gam_{ij}}{t}  . 
\ee
and, evaluating the Lie derivative with the connection $\w{D}$
instead of partial derivatives [cf. Eq.~(\ref{e:lie:der_comp_nab})]:
\be
	\Liec{\beta} \gam_{ij} = \beta^k 
	\underbrace{D_k \gam_{ij}}_{=0} 
	+ \gam_{kj} D_i  \beta^k  + \gam_{ik} D_j \beta^k , 
\ee
i.e. 
\be \label{e:dec:Lie_beta_gam}
	\Liec{\beta} \gam_{ij} = D_i \beta_j + D_j \beta_i . 
\ee

\subsection{3+1 Einstein system}

Using Eqs.~(\ref{e:dec:LiemK_t_beta}) and (\ref{e:dec:LietK}), 
as well as (\ref{e:dec:Liem_gam_t_beta}) and (\ref{e:dec:Liet_gam}),
we rewrite the 3+1 Einstein system (\ref{e:dec:EEdyn_comp}), (\ref{e:dec:Ham_constr}) and (\ref{e:dec:mom_constr_comp}) as
\bea 
 & & \encadre{
 \left( \der{}{t} - \Liec{\beta} \right) \gam_{ij} = - 2N K_{ij}    }
 	\label{e:dec:Einstein_PDE1}\\
 & & \encadre{ \left( \der{}{t} - \Liec{\beta} \right) K_{ij} 
	= - D_i D_j N + N\left\{
	R_{ij} + K K_{ij} -2 K_{ik} K^k_{\ \, j} 
	+ 4\pi \left[ (S-E) \gam_{ij} - 2 S_{ij} \right] \right\} }\nonumber \\
	& & \label{e:dec:Einstein_PDE2} \\
 & & \encadre{ R + K^2 - K_{ij} K^{ij} = 16\pi E } \label{e:dec:Einstein_PDE3}\\
 & & \encadre{ D_j K^j_{\ \, i} - D_i K = 8\pi  p_i } . \label{e:dec:Einstein_PDE4}
\eea
In this system, the covariant derivatives $D_i$ can be expressed in terms
of partial derivatives with respect to the spatial coordinates $(x^i)$
by means of the Christoffel symbols $\Gamma^i_{\ \, jk}$
of $\w{D}$ associated with $(x^i)$:
\bea
  & & D_i D_j N = \dderp{N}{x^i}{x^j} - \Gamma^k_{\ \, ij}
	\der{N}{x^k} , \label{e:dec:DDN} \\
  &&	D_j K^j_{\ \, i} = \der{K^j_{\ \, i}}{x^j}
	+ \Gamma^j_{\ \, jk} K^k_{\ \, i} 
	- \Gamma^k_{\ \, ji} K^j_{\ \, k} , \\
  && D_i K = \der{K}{x^i} .
\eea
The Lie derivatives along $\w{\beta}$ can be expressed 
in terms of partial derivatives with respect to the 
spatial coordinates $(x^i)$, via Eqs.~(\ref{e:dec:Liebeta_K})
and (\ref{e:dec:Lie_beta_gam}):
\bea
	& & \Liec{\beta}\gam_{ij} = \der{\beta_i}{x^j} + \der{\beta_j}{x^i}
	- 2 \Gamma^k_{\ \, ij} \beta_k \\
	& & \Liec{\beta} K_{ij} = \beta^k \der{K_{ij}}{x^k} 
	+ K_{kj} \der{ \beta^k}{x^i} + K_{ik} \der{\beta^k}{x^j} . 
\eea
Finally, the Ricci tensor and scalar curvature of $\w{\gam}$ are expressible 
according to the standard expressions:
\bea
  & & R_{ij} = \der{\Gamma^k_{\ \, ij}}{x^k} - \der{\Gamma^k_{\ \, ik}}{x^j}
	+ \Gamma^k_{\ \, ij} \Gamma^l_{\ \, kl}
	- \Gamma^l_{\ \, ik} \Gamma^k_{\ \, lj} \label{e:dec:Ricci_gam} \\
  & & R = \gam^{ij} R_{ij} . 
\eea
For completeness, let us recall the expression of the Christoffel symbols
in terms of partial derivatives of the metric:
\be \label{e:dec:Christoffel_gam}
	\Gamma^k_{\ \, ij} = \frac{1}{2} \gam^{kl} \left(
	\der{\gam_{lj}}{x^i} + \der{\gam_{il}}{x^j} - \der{\gam_{ij}}{x^l}
	\right) . 
\ee

Assuming that matter ``source terms'' $(E, p_i, S_{ij})$ are given, the
system (\ref{e:dec:Einstein_PDE1})-(\ref{e:dec:Einstein_PDE4}), 
with all the terms explicited 
according to Eqs.~(\ref{e:dec:DDN})-(\ref{e:dec:Christoffel_gam}) constitutes
a second-order non-linear PDE system for the unknowns $(\gam_{ij},K_{ij},N,\beta^i)$.
It has been first derived by Darmois, as early as 1927 \cite{Darmo27}, in the
special case $N=1$ and $\w{\beta}=0$ (Gaussian normal coordinates, to be discussed
in Sec.~\ref{s:dec:Gaussian_normal}). The case $N\not=1$, but still with 
$\w{\beta}=0$, has been obtained by Lichnerowicz in 1939 \cite{Lichn39,Lichn44} and
the general case (arbitrary lapse and shift) 
by Choquet-Bruhat in 1948 \cite{Foure48,Foure56}. 
A slightly different form, with $K_{ij}$ replaced by the ``momentum conjugate to
 $\gam_{ij}$'', namely
$\pi^{ij} := \sqrt{\gam} (K \gam^{ij} - K^{ij})$, has been derived by 
Arnowitt, Deser and Misner (1962) \cite{ArnowDM62} from their Hamiltonian
formulation of general relativity (to be discussed in Sec.~\ref{s:dec:ADM}).
\begin{remark}
In the numerical relativity literature, the 3+1 Einstein equations (\ref{e:dec:Einstein_PDE1})-(\ref{e:dec:Einstein_PDE4}) are sometimes called
the \emph{``ADM equations''}, in reference of the above mentioned work 
by Arnowitt, Deser and Misner \cite{ArnowDM62}. However, the major contribution
of ADM is an Hamiltonian formulation of general relativity (which we will
discuss succinctly in Sec.~\ref{s:dec:ADM}).
This Hamiltonian approach is not used in numerical relativity, which proceeds by
integrating the system (\ref{e:dec:Einstein_PDE1})-(\ref{e:dec:Einstein_PDE4}).
The latter was known before ADM work. In particular, the recognition of the
extrinsic curvature $\w{K}$ as a fundamental 3+1 variable was already 
achieved  by Darmois in 1927 \cite{Darmo27}. 
Moreoever, as stressed by York \cite{York04} (see also Ref.~\cite{AnderY98}),  
Eq.~(\ref{e:dec:Einstein_PDE2}) is the spatial projection of the 
spacetime Ricci tensor [i.e. is derived from the Einstein equation in
the form (\ref{e:dec:Einstein2}), cf. Sec.~\ref{s:dec:project_Einstein}] 
whereas the dynamical equation 
in the ADM work \cite{ArnowDM62} is instead the spatial projection of the
Einstein tensor [i.e. is derived from the Einstein equation in
the form (\ref{e:dec:Einstein})].
\end{remark}

%%%%%%%%%%%%%%%%%%%%%%%%%%%%%%%%%%%%%%%%%%%%%%%%%%%%%%%%%%%%%%%%%%%%%%%%%%%%%%%%%%%%%%

\section{The Cauchy problem} \label{s:dec:Cauchy}

\subsection{General relativity as a three-dimensional dynamical system}
\label{s:dec:geometrodynamics}

The system (\ref{e:dec:Einstein_PDE1})-(\ref{e:dec:Christoffel_gam})
involves only three-dimensional quantities, i.e. tensor fields defined 
on the hypersurface $\Sigma_t$, and their time derivatives. 
Consequently one may forget about the four-dimensional origin of
the system and consider that (\ref{e:dec:Einstein_PDE1})-(\ref{e:dec:Christoffel_gam})
describes \emph{time evolving} tensor fields on a \emph{single} three-dimensional
manifold $\Sigma$, without any reference to some ambient four-dimensional spacetime.
This constitutes the \emph{geometrodynamics} point of view developed by Wheeler
\cite{Wheel64} (see also Fischer and Marsden 
\cite{FischM72,FischM79} for a more formal treatment).

It is to be noticed that the system
(\ref{e:dec:Einstein_PDE1})-(\ref{e:dec:Christoffel_gam})
does not contain any time derivative of the lapse function $N$ nor of the
shift vector $\w{\beta}$. This means that $N$ and $\w{\beta}$ are not
dynamical variables. This should not be surprising if one remembers
that they are associated with the choice of coordinates $(t,x^i)$
(cf. Sec.~\ref{s:dec:coord_lapse_shift}).
Actually the coordinate freedom of general relativity implies
that we may choose the lapse and shift freely, without changing
the physical solution $\w{g}$ of the Einstein equation. The only things
to avoid are coordinate singularities, to which a arbitrary choice of
lapse and shift might lead. 

\subsection{Analysis within Gaussian normal coordinates}
\label{s:dec:Gaussian_normal}

To gain some insight in the nature of the system
(\ref{e:dec:Einstein_PDE1})-(\ref{e:dec:Christoffel_gam}), let us simplify
it by using the freedom in the choice of lapse and shift: we set 
\bea
	N & = & 1 	\label{e:dec:lapse_geod} \\
	\w{\beta} & = & 0 , 
\eea
in some neighbourhood a given hypersurface $\Sigma_0$ where the coordinates
$(x^i)$ are specified arbitrarily. 
This means that the lines of constant spatial coordinates are orthogonal
to the hypersurfaces $\Sigma_t$ (see Fig.~\ref{f:dec:shift}). 
Moreover, with $N=1$, the coordinate time $t$ coincides with the proper time measured by the Eulerian
observers between neighbouring hypersurfaces $\Sigma_t$ [cf. Eq.~(\ref{e:fol:dtau_Ndt})].
Such coordinates are called \defin{Gaussian normal coordinates}. 
The foliation away from $\Sigma_0$ selected by the choice (\ref{e:dec:lapse_geod})
of the lapse function is called a \defin{geodesic slicing}.
This name stems from the fact that the worldlines of the Eulerian observers are
geodesics, the parameter $t$ being then an affine parameter along them.
This is immediate from Eq.~(\ref{e:fol:a_DN}), which, for $N=1$, implies 
the vanishing of the 4-accelerations of the Eulerian observers (free fall).

In Gaussian normal coordinates, the spacetime metric tensor takes
a simple form [cf. Eq.~(\ref{e:dec:g_gam_N_beta})]:
\be \label{e:dec:GN_ds2}
	g_{\mu\nu} \, dx^\mu\, dx^\nu
	= - dt^2 + \gamma_{ij} \, dx^i \, dx^j .
\ee
In general it is not possible to get a Gaussian normal coordinate system
that covers all $\M$. This results from the well known tendencies of
timelike geodesics without vorticity (such as the worldlines of the
Eulerian observers) to focus and eventually cross.
This reflects the attractive nature of gravity and is best
seen on the Raychaudhuri equation (cf. Lemma 9.2.1 in \cite{Wald84}).  
However, for the purpose of the present discussion it is sufficient to 
consider Gaussian normal coordinates in some neighbourhood of the hypersurface
$\Sigma_0$; provided that the neighbourhood is small enough, this is always
possible.
The 3+1 Einstein system (\ref{e:dec:Einstein_PDE1})-(\ref{e:dec:Einstein_PDE4}) 
reduces then to
\bea  
	& & \der{\gam_{ij}}{t} = - 2 K_{ij} \label{e:fol:GN_dgam_K} \\
 	& & \der{K_{ij}}{t} 
	= R_{ij} + K K_{ij} -2 K_{ik} K^k_{\ \, j} 
	+ 4\pi \left[ (S-E) \gam_{ij} - 2 S_{ij} \right] \\
	& &  R + K^2 - K_{ij} K^{ij} = 16\pi E \\
 	& & D_j K^j_{\ \, i} - D_i K = 8\pi  p_i .
\eea
Using the short-hand notation
\be \label{e:dec:def_dot_gam}
	{\dot\gam}_{ij} := \der{\gam_{ij}}{t} 
\ee
and replacing everywhere $K_{ij}$ thanks to Eq.~(\ref{e:fol:GN_dgam_K}), we get
\bea
	& & - \dder{\gam_{ij}}{t} = 2 R_{ij} + \frac{1}{2} 
	\gam^{kl} {\dot\gam}_{kl} \, {\dot\gam}_{ij} - 2 \gam^{kl} {\dot\gam}_{ik} 	
	{\dot\gam}_{lj} + 8\pi \left[ (S-E) \gam_{ij} - 2 S_{ij} \right] 
				\label{e:dec:GN_ddgam_R} \\
	& & R + \frac{1}{4} (\gam^{ij} {\dot\gam}_{ij})^2
	- \frac{1}{4} \gam^{ik} \gam^{jl} {\dot\gam}_{ij} {\dot\gam}_{kl}
	= 16\pi E \\
	& & D_j( \gam^{jk} {\dot\gam}_{ki} ) - \der{}{x^i} \left( \gam^{kl} {\dot\gam}_{kl}
	\right) 
	= - 16\pi p_i . \label{e:dec:GN_mom_contr1} 
\eea
As far as the gravitational field is concerned, this equation contains only
the 3-metric $\w{\gam}$. In particular the Ricci tensor can be explicited by 
plugging Eq.~(\ref{e:dec:Christoffel_gam}) into Eq.~(\ref{e:dec:Ricci_gam}).
We need only the \defin{principal part} for our analysis, that is the
part containing the derivative of $\gam_{ij}$ of hightest degree (two in the present
case). We get, denoting by ``$\cdots$'' everything but a second order derivative
of $\gam_{ij}$:
\bea
	R_{ij}& = & 
	\der{\Gamma^k_{\ \, ij}}{x^k} - \der{\Gamma^k_{\ \, ik}}{x^j} + \cdots \nonumber \\
	& = &  \frac{1}{2} \der{}{x^k} \left[
	\gam^{kl} \left(
	\der{\gam_{lj}}{x^i} + \der{\gam_{il}}{x^j} - \der{\gam_{ij}}{x^l}
	\right) \right]
	- \frac{1}{2} \der{}{x^j} \left[
	\gam^{kl} \left(
	\der{\gam_{lk}}{x^i} + \der{\gam_{il}}{x^k} - \der{\gam_{ik}}{x^l}
	\right) \right] + \cdots \nonumber \\
	& = & 
	\frac{1}{2} \gam^{kl} \left(
	\dderp{\gam_{lj}}{x^k}{x^i} + \dderp{\gam_{il}}{x^k}{x^j} 
	- \dderp{\gam_{ij}}{x^k}{x^l} - \dderp{\gam_{lk}}{x^j}{x^i}
	- \dderp{\gam_{il}}{x^j}{x^k} + \dderp{\gam_{ik}}{x^j}{x^l}
	\right) + \cdots \nonumber \\
  R_{ij} & = & - \frac{1}{2} \gam^{kl} \left( \dderp{\gam_{ij}}{x^k}{x^l}
	+ \dderp{\gam_{kl}}{x^i}{x^j}
	- \dderp{\gam_{lj}}{x^i}{x^k} 
	- \dderp{\gam_{il}}{x^j}{x^k} \right) 
	+ \mathcal{Q}_{ij}\left(\gam_{kl},\der{\gam_{kl}}{x^m}\right) , 
		\label{e:dec:Ricci_principal}
\eea
where $\mathcal{Q}_{ij}(\gam_{kl},\dert{\gam_{kl}}{x^m})$ is a (non-linear) expression
containing the components $\w{\gam}_{kl}$ and their first spatial derivatives only.
Taking the trace of (\ref{e:dec:Ricci_principal}) (i.e. contracting with $\gam^{ij}$),
we get 
\be \label{e:dec:Ricci_scal_principal}
	R = \gam^{ik} \gam^{jl} \dderp{\gam_{ij}}{x^k}{x^l}
	- \gam^{ij} \gam^{kl} \dderp{\gam_{ij}}{x^k}{x^l} 
	+ \mathcal{Q}\left(\gam_{kl},\der{\gam_{kl}}{x^m}\right) .
\ee
Besides
\bea
	D_j( \gam^{jk} {\dot\gam}_{ki} ) & =&  \gam^{jk} D_j {\dot\gam}_{ki}
	= \gam^{jk} \left( \der{{\dot\gam}_{ki}}{x^j}
	- \Gamma^l_{\ \, jk} {\dot\gam}_{li}
	- \Gamma^l_{\ \, ji} {\dot\gam}_{kl} \right) \nonumber \\
	& =  & \gam^{jk} \dderp{\gam_{ki}}{x^j}{t} 
	+ \mathcal{Q}_i\left(\gam_{kl},\der{\gam_{kl}}{x^m}, \der{\gam_{kl}}{t}\right) ,
	\label{e:dec:Djgamjk}
\eea
where $\mathcal{Q}_i ( \gam_{kl},\dert{\gam_{kl}}{x^m}, \dert{\gam_{kl}}{t})$ is some
expression that does not contain any second order derivative of
$\gam_{kl}$.
Substituting Eqs.~(\ref{e:dec:Ricci_principal}), (\ref{e:dec:Ricci_scal_principal})
and (\ref{e:dec:Djgamjk})
in Eqs.~(\ref{e:dec:GN_ddgam_R})-(\ref{e:dec:GN_mom_contr1}) gives
\bea
	& & - \dder{\gam_{ij}}{t} 
	+ \gam^{kl} \left( \dderp{\gam_{ij}}{x^k}{x^l}
	+ \dderp{\gam_{kl}}{x^i}{x^j}
	- \dderp{\gam_{lj}}{x^i}{x^k} 
	- \dderp{\gam_{il}}{x^j}{x^k} \right) 
	= 8\pi \left[ (S-E) \gam_{ij} - 2 S_{ij} \right] \nonumber \\
 & & \qquad \qquad \qquad \qquad 
	\qquad \qquad \qquad \qquad \qquad \qquad \qquad \qquad 	
 +  \mathcal{Q}_{ij}\left(\gam_{kl},\der{\gam_{kl}}{x^m},\der{\gam_{kl}}{t}\right) 
		\label{e:dec:GN_Einstein1} \\
 & & \gam^{ik} \gam^{jl} \dderp{\gam_{ij}}{x^k}{x^l}
	- \gam^{ij} \gam^{kl} \dderp{\gam_{ij}}{x^k}{x^l} 
	= 16\pi E 
	+ \mathcal{Q}\left(\gam_{kl},\der{\gam_{kl}}{x^m},\der{\gam_{kl}}{t}\right) 
				\label{e:dec:GN_Einstein2}  \\
 & & \gam^{jk} \dderp{\gam_{ki}}{x^j}{t}
	 - \gam^{kl} \dderp{\gam_{kl}}{x^i}{t} 
	= - 16\pi p_i + \mathcal{Q}_i\left(\gam_{kl},\der{\gam_{kl}}{x^m},
	 \der{\gam_{kl}}{t}\right) . \label{e:dec:GN_Einstein3} 
\eea
Notice that we have incorporated the first order time derivatives into
the $\mathcal{Q}$ terms.

Equations (\ref{e:dec:GN_Einstein1})-(\ref{e:dec:GN_Einstein3}) constitute
a system of PDEs for the unknowns $\gam_{ij}$. This system is of second order
and non linear, but \defin{quasi-linear}, i.e. linear with respect to all the second order
derivatives. Let us recall that, in this system, the $\gam^{ij}$'s are to be considered
as functions of the $\gam_{ij}$'s, these functions being given by expressing
the matrix $(\gam_{ij})$ as the inverse of the matrix $(\gam_{ij})$
(e.g. via Cramer's rule).

A key feature of the system (\ref{e:dec:GN_Einstein1})-(\ref{e:dec:GN_Einstein3})
is that it contains $6+1+3=10$ equations for the $6$ unknowns $\gam_{ij}$. 
Hence it is an over-determined system.
Among the three sub-systems (\ref{e:dec:GN_Einstein1}), (\ref{e:dec:GN_Einstein2})
and (\ref{e:dec:GN_Einstein3}), only the first one involves second-order time 
derivatives. Moreover the sub-system (\ref{e:dec:GN_Einstein1}) contains the 
same numbers of equations than unknowns (six) and it is in a form 
tractable as a \emph{Cauchy problem}, namely one could search for a solution, 
given some initial data. More precisely, the sub-system (\ref{e:dec:GN_Einstein1})
being of second order and in the form 
\be
	\dder{\gam_{ij}}{t} =
	 F_{ij}\left(\gam_{kl},\der{\gam_{kl}}{x^m},\der{\gam_{kl}}{t},
		\dderp{\gam_{kl}}{x^m}{x^n} \right) , 
\ee 
the Cauchy problem 
amounts to finding a solution $\gam_{ij}$ for $t>0$
given the knowledge of $\gam_{ij}$ and $\dert{\gam_{ij}}{t}$ at $t=0$,
i.e. the values of $\gam_{ij}$ and $\dert{\gam_{ij}}{t}$ on the hypersurface
$\Sigma_0$.
Since $F_{ij}$ is a analytical function\footnote{it is polynomial in the derivatives
of $\gam_{kl}$ and involves at most rational fractions in $\gam_{kl}$
(to get the inverse metric $\gam^{kl}$}, we can invoke the 
Cauchy-Kovalevskaya theorem (see e.g. \cite{CouraH62b}) to guarantee the existence and uniqueness of a solution $\gam_{ij}$ in a neighbourhood of $\Sigma_0$, for any initial data $(\gam_{ij},\dert{\gam_{ij}}{t})$
on $\Sigma_0$ that are analytical functions of the coordinates $(x^i)$.

The complication arises because of the extra equations (\ref{e:dec:GN_Einstein2})
and (\ref{e:dec:GN_Einstein3}), which must be fulfilled to ensure
that the metric $\w{g}$ reconstructed from $\gam_{ij}$ via Eq.~(\ref{e:dec:GN_ds2})
is indeed a solution of Einstein equation. Equations (\ref{e:dec:GN_Einstein2})
and (\ref{e:dec:GN_Einstein3}), which cannot be put in the form 
such that the Cauchy-Kovalevskaya theorem applies, constitute \defin{constraints}
for the Cauchy problem (\ref{e:dec:GN_Einstein1}).
In particular one has to make sure that the initial data 
$(\gam_{ij},\dert{\gam_{ij}}{t})$ on $\Sigma_0$ satisfies these constraints.
A natural question which arises is then: suppose that we prepare initial
data $(\gam_{ij},\dert{\gam_{ij}}{t})$ which satisfy the constraints
(\ref{e:dec:GN_Einstein2})-(\ref{e:dec:GN_Einstein3}) and that we get
a solution of the Cauchy problem (\ref{e:dec:GN_Einstein1}) from these
initial data, are the constraints satisfied by the solution for $t>0$ ?
The answer is yes, thanks to the Bianchi identities, as we shall see in
Sec.~\ref{s:sch:prop_constraints}. 

\subsection{Constraint equations} \label{s:dec:constraints}

The main conclusions of the above discussion
remain valid for the general 3+1 Einstein system as given by 
Eqs.~(\ref{e:dec:Einstein_PDE1})-(\ref{e:dec:Einstein_PDE4}): 
Eqs.~(\ref{e:dec:Einstein_PDE1})-(\ref{e:dec:Einstein_PDE2})
constitute a time evolution system tractable as a Cauchy problem,
whereas Eqs.~(\ref{e:dec:Einstein_PDE3})-(\ref{e:dec:Einstein_PDE4})
constitute constraints. This partly justifies the names \emph{Hamiltonian
constraint} and \emph{momentum constraint} given respectively to 
Eq.~(\ref{e:dec:Einstein_PDE3}) and to Eq.~(\ref{e:dec:Einstein_PDE4}).

The existence of constraints is not specific to general relativity.
For instance the Maxwell equations for the electromagnetic field
can be treated as a Cauchy problem subject to the constraints
$\w{D}\cdot\w{B}=0$ and $\w{D}\cdot\w{E}=\rho/\epsilon_0$
(see Ref.~\cite{KnappWB02} or Sec.~2.3 of Ref.~\cite{BaumgS03} for 
details of the electromagnetic analogy).

\subsection{Existence and uniqueness of solutions to the Cauchy problem}
\label{s:dec:existence_uniqueness}

In the general case of arbitrary lapse and shift, 
the time derivative ${\dot\gam}_{ij}$ introduced in 
Sec.~\ref{s:dec:Gaussian_normal} has to be replaced by the extrinsic
curvature $K_{ij}$, so that the initial data on a given hypersurface
$\Sigma_0$ is $(\w{\gam},\w{K})$. The couple $(\w{\gam},\w{K})$
has to satisfy the constraint equations 
(\ref{e:dec:Einstein_PDE3})-(\ref{e:dec:Einstein_PDE4})
on $\Sigma_0$. One may then ask the question: given
a set $(\Sigma_0,\w{\gam},\w{K},E,\w{p})$, where $\Sigma_0$ is a three-dimensional
manifold, $\w{\gam}$ a Riemannian metric on $\Sigma_0$, 
$\w{K}$ a symmetric bilinear form field on $\Sigma_0$, $E$ a scalar field on $\Sigma_0$
and $\w{p}$ a vector field on $\Sigma_0$, which obeys the constraint
equations (\ref{e:dec:Einstein_PDE3})-(\ref{e:dec:Einstein_PDE4}):
\bea
 & &  R + K^2 - K_{ij} K^{ij} = 16\pi E \\
 & &  D_j K^j_{\ \, i} - D_i K = 8\pi  p_i  ,
\eea
does there exist a spacetime $(\M,\w{g},\w{T})$ such that $(\w{g},\w{T})$
fulfills the Einstein equation and $\Sigma_0$ can be embedded as an hypersurface
of $\M$ with induced metric $\w{\gam}$ and extrinsic curvature $\w{K}$ ?

Darmois (1927) \cite{Darmo27} and Lichnerowicz (1939)  \cite{Lichn39}
have shown that the answer is yes for the vacuum case ($E=0$ and $p_i=0$),
when the initial data $(\w{\gam},\w{K})$ are \emph{analytical} functions of
the coordinates $(x^i)$ on $\Sigma_0$. Their analysis is based on the 
Cauchy-Kovalevskaya theorem mentioned in Sec.~\ref{s:dec:Gaussian_normal}
(cf. Chap.~10 of Wald's textbook \cite{Wald84} for details). 
However, on physical grounds, the analytical case is too restricted. 
One would like to deal instead with \emph{smooth (i.e. differentiable)}
initial data. There are at least two reasons for this:
\begin{itemize}
\item The smooth manifold structure of $\M$ imposes only that the change of coordinates
are differentiable, not necessarily analytical. Consequently if $(\w{\gam},\w{K})$ are 
analytical functions of the coordinates, they might not be analytical functions
of another coordinate system $({x'}^i)$.
\item An analytical function is fully determined by its value and those of all its
derivatives at a single point. Equivalently an analytical function is fully 
determined by its value in some small open domain $D$. This fits badly with causality
requirements, because a small change to the initial data, localized in a small
region, should not change the whole solution at all points of $\M$. The change should
take place only in the so-called \emph{domain of dependence} of $D$.  
\end{itemize}

This is why the major breakthrough in the Cauchy problem of general relativity
has been achieved by Choquet-Bruhat in 1952 \cite{Foure52} when she showed 
existence and uniqueness of the solution in a small neighbourhood of $\Sigma_0$
for \emph{smooth} (at least $C^5$) initial data $(\w{\gam},\w{K})$.
We shall not give any sketch on the proof (beside the original publication
\cite{Foure52}, see the review articles \cite{BartnI04} and \cite{ChoquY80})
but simply mentioned that it is based on \emph{harmonic coordinates}. 

A major improvement has been then the \emph{global} existence and uniqueness 
theorem by Choquet-Bruhat and Geroch (1969) \cite{ChoquG69}. The latter tells 
that among all the spacetimes $(\M,\w{g})$ solution of the Einstein equation
and such that $(\Sigma_0,\w{\gam},\w{K})$ is an embedded Cauchy surface, there
exists a maximal spacetime $(\M^*,\w{g}^*)$ and it is unique. 
\emph{Maximal} means that any spacetime $(\M,\w{g})$ solution of the Cauchy
problem is isometric to a subpart of $(\M^*,\w{g}^*)$.
For more details about the existence and uniqueness of solutions to the Cauchy
problem, see the reviews by Choquet-Bruhat and York \cite{ChoquY80}, 
Klainerman and  Nicol\`o \cite{KlainN99}, 
Andersson \cite{Ander04} and Rendall \cite{Renda05}.

\section{ADM Hamiltonian formulation} \label{s:dec:ADM}

Further insight in the 3+1 Einstein equations is provided by the Hamiltonian
formulation of general relativity. Indeed the latter makes use of the 3+1 formalism,
since any Hamiltonian approach involves the concept of a physical state
``at a certain time'', which is translated in general relativity by 
the state on a spacelike hypersurface $\Sigma_t$. The 
Hamiltonian formulation of general relativity has been developed
notably by Dirac in the late fifties \cite{Dirac58,Dirac59} (see also Ref.~\cite{Deser04}), by 
Arnowitt, Deser and Misner (ADM) in the early sixties \cite{ArnowDM62}
and by Regge and Teitelboim in the seventies \cite{ReggeT74}.
Pedagogical presentations are given 
in Chap.~21 of MTW \cite{MisneTW73}, in Chap.~4 of Poisson's book \cite{Poiss04},
in M. Henneaux's lectures \cite{Henne06} and in G. Sch\"afer's ones \cite{Schae06}.
Here we focuss on the ADM approach, which makes a direct use of the lapse function
and shift vector (contrary to Dirac's one). 
For simplicity, we consider only the vacuum Einstein equation in this
section. Also we shall disregard any boundary term in the action integrals. 
Such terms will be restored in Chap.~\ref{s:glo} in order to discuss total energy
and momentum. 

\subsection{3+1 form of the Hilbert action}

Let us consider the standard Hilbert action for 
general relativity (see N. Deruelle's lecture \cite{Derue06}):
\be \label{e:dec:S_Hilbert}
	S =  \int_{\mathcal{V}} {}^4\!R \sqrt{-g} \, d^4 x ,
\ee
where $\mathcal{V}$ is a part of $\M$ delimited by two hypersurfaces 
$\Sigma_{t_1}$ and $\Sigma_{t_2}$ ($t_1<t_2$) of the foliation
$(\Sigma_t)_{t\in\R}$:
\be \label{e:fol:def_V}
	\mathcal{V} := \bigcup_{t=t_1}^{t_2} \Sigma_t . 
\ee
Thanks to the 3+1 decomposition of ${}^4\!R$ provided by Eq.~(\ref{e:fol:4R_3p1})
and to the relation $\sqrt{-g} = N \sqrt{\gam}$ [Eq.~(\ref{e:fol:detg_Ngetgam})]
we can write
\be \label{e:dec:S_Hilbert1}
	S =  
	\int_{\mathcal{V}} \left[ N \left( R + K^2 + K_{ij} K^{ij} \right)
	- 2 \Liec{m} K -2 D_i D^i N \right] \sqrt{\gam} \, d^4 x . 
\ee
Now 
\bea
	\Liec{m} K & = & m^\mu \nabla_\mu K = N n^\mu \nabla_\mu K
	= N [ \nabla_\mu(K n^\mu) - K \underbrace{\nabla_\mu n^\mu}_{=-K} ] \nonumber \\
	& = & N [ \nabla_\mu (K n^\mu) + K^2 ] . 
\eea
Hence Eq.~(\ref{e:dec:S_Hilbert1}) becomes
\be
	S =  
	\int_{\mathcal{V}}  \left[ N \left( R + K_{ij} K^{ij} - K^2 \right)
	- 2 N \nabla_\mu(K n^\mu) -2 D_i D^i N \right] \sqrt{\gam} \, d^4 x . 
\ee
But 
\be
	\int_{\mathcal{V}}  N \nabla_\mu(K n^\mu) \sqrt{\gam} \, d^4 x  
	= \int_{\mathcal{V}}  \nabla_\mu(K n^\mu) \sqrt{-g} \, d^4 x
	= \int_{\mathcal{V}}  \der{}{x^\mu} \left( \sqrt{-g} K n^\mu \right) \, d^4 x
\ee
is the integral of a pure divergence and we can disregard this term in the action.
Accordingly, the latter becomes 
\be
	S =  \int_{t_1}^{t_2} \left\{ \int_{\Sigma_t} \left[ N 
	\left(R+K_{ij}K^{ij} -K^2 \right)
	- 2 D_i D^i N \right] \sqrt{\gam} \, d^3 x \right\} dt , 
\ee
where we have used (\ref{e:fol:def_V}) to split the four-dimensional integral
into a time integral and a three-dimensional one.
Again we have a divergence term:
\be
	\int_{\Sigma_t} D_i D^i N \, \sqrt{\gam} \, d^3x 
	= \int_{\Sigma_t} \der{}{x^i} \left( \sqrt{\gam} D^i N \right) d^3x ,
\ee
which we can disregard. 
Hence the 3+1 writing of the Hilbert action is
\be \label{e:dec:S_Hilbert2}
   \encadre{ S =  \int_{t_1}^{t_2} \left\{ \int_{\Sigma_t} N 
	\left(R+K_{ij}K^{ij} -K^2\right) \sqrt{\gam} \, d^3 x \right\} dt } .
\ee

\subsection{Hamiltonian approach} \label{s:dec:ADM_Ham}

The action (\ref{e:dec:S_Hilbert2}) is to be considered as a functional of
the ``configuration'' variables $q=(\gam_{ij},N,\beta^i)$ [which describe
the full spacetime metric components $g_{\alpha\beta}$, cf. Eq.~(\ref{e:dec:g_cov})] 
and their time derivatives\footnote{we use the same notation as that defined by
Eq.~(\ref{e:dec:def_dot_gam})} $\dot q = ({\dot\gam}_{ij},\dot N, {\dot\beta}^i)$:
$S=S[q,\dot q]$. 
In particular $K_{ij}$ in Eq.~(\ref{e:dec:S_Hilbert2}) is
the function of ${\dot\gam}_{ij}$, $\gam_{ij}$, $N$ and $\beta^i$ given
by Eqs.~(\ref{e:dec:Einstein_PDE1}) and (\ref{e:dec:Lie_beta_gam}):
\be \label{e:dec:Kij_gam_dot}
	K_{ij} = \frac{1}{2N} \left( \gam_{ik} D_j \beta^k + 
	\gam_{jk} D_i \beta^k - {\dot\gam}_{ij} \right) . 
\ee
From Eq.~(\ref{e:dec:S_Hilbert2}), we read that the gravitational field
Lagrangian density is
\be \label{e:dec:Langragian}
	\encadre{ L(q,\dot q) = N \sqrt{\gam} (R + K_{ij} K^{ij} - K^2) =
	N \sqrt{\gam}
	\left[ R + (\gam^{ik} \gam^{jl} - \gam^{ij} \gam^{kl}) K_{ij} K_{kl}
	\right] }, 
\ee
with $K_{ij}$ and $K_{kl}$ expressed as (\ref{e:dec:Kij_gam_dot}).
Notice that this Lagrangian does not depend upon
the time derivatives of $N$ and $\beta^i$: this shows that the lapse
function and the shift vector are not dynamical variables. 
Consequently the only dynamical variable is $\gam_{ij}$.
The momentum canonically conjugate to it is
\be
	\pi^{ij} := \der{L}{{\dot\gam}_{ij}} . 
\ee
From Eqs.~(\ref{e:dec:Langragian}) and (\ref{e:dec:Kij_gam_dot}), we get
\be
	\pi^{ij} = N \sqrt{\gam} \left[
	(\gam^{ik} \gam^{jl} - \gam^{ij} \gam^{kl}) K_{kl}
	+ (\gam^{ki} \gam^{lj} - \gam^{kl} \gam^{ij}) K_{kl} \right]
	\times \left( - \frac{1}{2N} \right) ,
\ee
i.e.
\be \label{e:dec:ADM_momentum}
	\encadre{ \pi^{ij} = \sqrt{\gam} \left( K \gam^{ij} - K^{ij} \right) } .
\ee
The Hamiltonian density is given by the Legendre transform
\be
	\mathcal{H} = \pi^{ij} {\dot\gam}_{ij} - L .
\ee
Using Eqs.~(\ref{e:dec:Kij_gam_dot}), (\ref{e:dec:ADM_momentum}) and
(\ref{e:dec:Langragian}), we have
\bea
	\mathcal{H}  & = & \sqrt{\gam} \left( K \gam^{ij} - K^{ij} \right)
	\left( - 2N K_{ij} + D_i\beta_j + D_j \beta_i \right)
	- N \sqrt{\gam} (R + K_{ij} K^{ij} - K^2) \nonumber \\
	& = & \sqrt{\gam} \left[ - N (R+K^2-K_{ij}K^{ij})
	+ 2 \left( K \gam^j_{\ \, i} - K^j_{\ \, i} \right) D_j \beta^i \right] \nonumber \\
	&= & - \sqrt{\gam} \left[ N (R+K^2-K_{ij}K^{ij}) 
	+ 2 \beta^i \left( D_i K - D_j K^j_{\ \, i} \right) \right] \nonumber \\
	& & 
	+ 2 \sqrt{\gam} D_j\left( K \beta^j - K^j_{\ \, i} \beta^i \right) .
		\label{e:dec:Hamil_dens}
\eea
The corresponding Hamiltonian is
\be
	H = \int_{\Sigma_t} \mathcal{H} \, d^3 x .
\ee
Noticing that the last term in Eq.~(\ref{e:dec:Hamil_dens}) is a divergence
and therefore does not contribute to the integral, we get
\be \label{e:dec:Hamilt}
	\encadre{
	H = - \int_{\Sigma_t} \left( N C_0 - 2 \beta^i C_i \right) \sqrt{\gam} d^3 x 
	},
\ee
where 
\bea
	C_0 &:=& R+K^2-K_{ij}K^{ij} ,\\
	C_i &:=& D_j K^j_{\ \, i} - D_i K 
\eea
are the left-hand sides of the constraint equations (\ref{e:dec:Einstein_PDE3})
and (\ref{e:dec:Einstein_PDE4}) respectively.

The Hamiltonian $H$ is a functional of the configuration variables $(\gam_{ij},N,\beta^i)$
and their conjugate momenta $(\pi^{ij},\pi^N,\pi^{\w{\beta}}_i)$, the last two ones
being identically zero since
\be
	\pi^N := \der{L}{\dot N} = 0 \qquad \mbox{and} \qquad
	\pi^{\w{\beta}}_i :=  \der{L}{{\dot\beta}^i} = 0 .
\ee
The scalar curvature $R$ which appears in $H$ via $C_0$ is a function of 
$\gam_{ij}$ and its spatial derivatives, via 
Eqs.~(\ref{e:dec:Ricci_gam})-(\ref{e:dec:Christoffel_gam}), whereas $K_{ij}$
which appears in both $C_0$ and $C_i$ is a function of $\gam_{ij}$ and $\pi^{ij}$,
obtained by ``inverting'' relation (\ref{e:dec:ADM_momentum}):
\be
	K_{ij} = K_{ij}[\w{\gam},\w{\pi}]
	= \frac{1}{\sqrt{\gam}} \left( \frac{1}{2} \gam_{kl} \pi^{kl} \gam_{ij}
	- \gam_{ik} \gam_{jl} \pi^{kl} \right) . 
\ee
The minimization of the Hilbert action is equivalent to the Hamilton equations
\bea
	& & \frac{\delta H}{\delta \pi^{ij}} = {\dot\gam}_{ij} \\
	& & \frac{\delta H}{\delta \gam_{ij}} = - {\dot\pi}^{ij} \\
	& & \frac{\delta H}{\delta N} = - {\dot\pi}^{N} = 0 \\
	& & \frac{\delta H}{\delta \beta^i} = - {\dot\pi}^{\w{\beta}}_i = 0 . 
\eea
Computing the functional derivatives from the expression (\ref{e:dec:Hamilt})
of $H$ leads the equations
\bea
	& & \frac{\delta H}{\delta \pi^{ij}} = - 2NK_{ij} 
	+ D_i\beta_j + D_j\beta_i = {\dot\gam}_{ij} \label{e:dec:Hamil1} \\
	& & \frac{\delta H}{\delta \gam_{ij}} = - {\dot\pi}^{ij} \label{e:dec:Hamil2}\\
	& & \frac{\delta H}{\delta N} = -C_0 = 0 \label{e:dec:Hamil3}\\
	& & \frac{\delta H}{\delta \beta^i} = 2 C_i = 0 . \label{e:dec:Hamil4}
\eea
Equation~(\ref{e:dec:Hamil1}) is nothing but the first equation of the 3+1 Einstein system
(\ref{e:dec:Einstein_PDE1})-(\ref{e:dec:Einstein_PDE4}). 
We do not perform the computation of the variation
(\ref{e:dec:Hamil2}) but the explicit calculation 
(see e.g. Sec.~4.2.7 of Ref.~\cite{Poiss04}) yields an equation which is equivalent
to the dynamical Einstein equation (\ref{e:dec:Einstein_PDE2}).
Finally, 
Eq.~(\ref{e:dec:Hamil3}) is the Hamiltonian constraint (\ref{e:dec:Einstein_PDE3})
with $E=0$ (vacuum) and Eq.~(\ref{e:dec:Hamil4}) is the momentum constraint (\ref{e:dec:Einstein_PDE4}) with $p_i=0$.

Equations~(\ref{e:dec:Hamil3}) and (\ref{e:dec:Hamil4})
show that in the ADM Hamiltonian approach, the lapse function and the shift vector
turn out to be Lagrange multipliers to enforce respectively the Hamiltonian constraint
and the momentum constraint, the true dynamical variables being $\gam_{ij}$ and
$\pi^{ij}$.

%  
%    Chapitre : 3+1 equations for matter and electromagnetic field
%
% $Date: 2007-03-05 22:39:07 +0100 (lun, 05 mar 2007) $
% $Rev: 182 $
% $Author: e_gourgoulhon $
%%%%%%%%%%%%%%%%%%%%%%%%%%%%%

\chapter{3+1 equations for matter and electromagnetic field} \label{s:mat}

%\verb$Date: 2007-03-05 22:39:07 +0100 (lun, 05 mar 2007) $

\minitoc
\vspace{1cm}

%%%%%%%%%%%%%%%%%%%%%%%%%%%%%%%%%%%%%%%%%%%%%%%%%%%%%%%%%%%%%%%%%%%%%%%%%%%%

\section{Introduction}

After having considered mostly the left-hand side of Einstein equation,
in this chapter we focus on the right-hand side, namely on 
the matter represented by its stress-energy tensor
$\w{T}$. By ``matter'', we actually mean any kind of non-gravitational field, 
which is minimally coupled to gravity. 
This includes the electromagnetic field, which we shall treat in Sec.~\ref{s:mat:em}.
The matter obeys two types of equations. The first one is the vanishing of
the spacetime divergence of the stress-energy tensor:
\be \label{e:mat:divT}
	\encadre{ \vec{\wnab} \cdot \w{T} = 0 }, 
\ee
which, thanks to the contracted Bianchi identities, is a consequence
of Einstein equation (\ref{e:dec:Einstein}) (see N. Deruelle's lectures \cite{Derue06}).
The second type of equations is the field equations that must be
satisfied independently of the Einstein equation, for instance the baryon number
conservation law or
the Maxwell equations for the electromagnetic field.

\section{Energy and momentum conservation}

\subsection{3+1 decomposition of the 4-dimensional equation}

Let us replace $\w{T}$ in Eq.~(\ref{e:mat:divT}) by its 3+1 expression
(\ref{e:dec:T_3p1}) in terms of the energy density $E$, the momentum density $\w{p}$
and the stress tensor $\w{S}$, all of them as measured by the Eulerian observer. 
We get, successively,
\bea
	& & \nabla_\mu T^\mu_{\ \, \alpha} = 0 \nonumber \\
	& & \nabla_\mu \left( S^\mu_{\ \, \alpha} + n^\mu p_\alpha + p^\mu n_\alpha
	+ E n^\mu n_\alpha  \right) = 0  \nonumber \\
	& & \nabla_\mu S^\mu_{\ \, \alpha} - K p_\alpha + n^\mu \nabla_\mu p_\alpha
	+ \nabla_\mu p^\mu \, n_\alpha - p^\mu K_{\mu\alpha}
	- KE n_\alpha + E D_\alpha\ln N \nonumber \\
	& & \qquad \qquad + n^\mu \nabla_\mu E \, n_\alpha = 0 , \label{e:mat:divT3p1}
\eea
where we have used Eq.~(\ref{e:fol:nab_n_K_comp}) to express the $\w{\nabla}\uu{n}$
in terms of $\w{K}$ and $\w{D}\ln N$.

\subsection{Energy conservation}

Let us project Eq.~(\ref{e:mat:divT3p1}) along the normal to the hypersurfaces
$\Sigma_t$, i.e. contract Eq.~(\ref{e:mat:divT3p1}) with $n^\alpha$. We get,
since $\w{p}$, $\w{K}$ and $\w{D}\ln N$ are all orthogonal to $\w{n}$:
\be \label{e:mat:energ_prov0}
	n^\nu \nabla_\mu S^\mu_{\ \, \nu} + n^\mu n^\nu \nabla_\mu p_\nu
	- \nabla_\mu p^\mu + KE - n^\mu \nabla_\mu E = 0 .
\ee
Now, since $\w{n}\cdot\w{S}=0$, 
\be \label{e:mat:energ_prov1}
	n^\nu \nabla_\mu S^\mu_{\ \, \nu} = - S^\mu_{\ \, \nu}\nabla_\mu n^\nu
	= S^\mu_{\ \, \nu} (K^\nu_{\ \, \mu} + D^\nu \ln N \, n_\mu)
	= K_{\mu\nu} S^{\mu\nu} . 
\ee
Similarly
\be \label{e:mat:energ_prov2}
	n^\mu n^\nu \nabla_\mu p_\nu = - p_\nu n^\mu \nabla_\mu n^\nu
	= - p_\nu D^\nu \ln N . 
\ee
Besides, let us express the 4-dimensional divergence $\nabla_\mu p^\mu$ is terms
of the 3-dimensional one, $D_\mu p^\mu$. 
For any vector $\w{v}$ tangent to $\Sigma_t$, like $\vec{\w{p}}$, 
Eq.~(\ref{e:hyp:link_D_nab_comp}) gives
\be
	D_\mu v^\mu = \gam^\rho_{\ \, \mu} \gam^\mu_{\ \, \sigma} \nabla_\rho v^\sigma
	= \gam^\rho_{\ \, \sigma} \nabla_\rho v^\sigma
	= (\delta^\rho_{\ \, \sigma} + n^\rho \, n_\sigma)
	\nabla_\rho v^\sigma
	= \nabla_\rho v^\rho - v^\sigma n^\rho \nabla_\rho n_\sigma
	= \nabla_\rho v^\rho - v^\sigma D_\sigma \ln N
\ee
Hence the usefull relation between the two divergences
\be \label{e:mat:divergence}
	\encadre{ \forall \w{v}\in \T(\Sigma_t),\quad
	\wnab\!\cdot\!\w{v} = \w{D}\!\cdot\!\w{v}
	+ \w{v} \cdot \w{D}\ln N } ,  
\ee
or in terms of components,
\be
\forall \w{v}\in \T(\Sigma_t),\quad \nabla_\mu v^\mu = D_i v^i + v^i D_i\ln N . 
\ee
Applying this relation to $\w{v}=\w{p}$ and taking into account 
Eqs.~(\ref{e:mat:energ_prov1}) and (\ref{e:mat:energ_prov2}), Eq.~(\ref{e:mat:energ_prov0})
becomes
\be
	\Lie{n} E + \w{D}\cdot\vec{\w{p}} + 2 \vec{\w{p}}\cdot \w{D}\ln N 
	- KE - K_{ij} S^{ij}
	= 0 .
\ee
\begin{remark}
We have written the derivative of $E$ along $\w{n}$ as a Lie 
derivative. $E$ being a scalar field, we have of course the alternative expressions
\be
	\Lie{n} E = \w{\nabla}_{\w{n}} E = \w{n}\cdot\wnab E
	= n^\mu \nabla_\mu E = n^\mu \der{E}{x^\mu}
	= \langle \dd E, \w{n} \rangle . 
\ee
\end{remark}
$\Lie{n} E$ is the derivative of $E$ with respect to the proper time of the 
Eulerian observers: $\Lie{n} E= dE/d\tau$, for $\w{n}$ is the 4-velocity
of these observers. It is easy to let appear the derivative with respect to 
the coordinate time $t$ instead, thanks to the relation
$\w{n} = N^{-1} (\wpar_t - \w{\beta})$ [cf. Eq.~(\ref{e:dec:t_Nn_b})]:
\be
	\Lie{n} E = \frac{1}{N} \left( \der{}{t} - \Lie{\beta} \right) E . 
\ee
Then 
\be \label{e:mat:ener_cons}
	\encadre{ \left( \der{}{t} - \Lie{\beta} \right) E 
	+ N \left( \w{D}\cdot\vec{\w{p}} 
	- KE - K_{ij} S^{ij} \right) + 2 \vec{\w{p}}\cdot \w{D} N 
	= 0 },
\ee
in components:
\be
	\left( \der{}{t} - \beta^i \der{}{x^i} \right) E 
	+ N \left( D_i p^i
	- KE - K_{ij} S^{ij} \right) + 2 p^i D_i N 
	= 0 .
\ee
This equation has been obtained by York (1979) in his seminal article \cite{York79}.

\subsection{Newtonian limit} \label{s:mat:Newt_lim}

As a check, let us consider the Newtonian limit of Eq.~(\ref{e:mat:ener_cons}).
For this purpose let us assume that the gravitational field is weak and static.
It is then always possible to find a coordinate system $(x^\alpha)=(x^0=ct,x^i)$ such that the metric components take the form (cf. N. Deruelle's lectures \cite{Derue06})
\be \label{e:mat:gab_weak}
   g_{\mu\nu} dx^\mu dx^\nu =
	- \left( 1 + 2\Phi \right)  \, dt^2
	+ \left( 1 - 2\Phi \right) f_{ij} \, dx^i dx^j,
\ee
where $\Phi$ is the Newtonian gravitational potential (solution of
Poisson equation $\Delta\Phi=4\pi G\rho$) and $f_{ij}$ are the components
the flat Euclidean metric $\w{f}$ in the 3-dimensional space. 
For a weak gravitational field (Newtonian limit), 
$|\Phi| \ll 1$ (in units where the light velocity is not one, this should read
$|\Phi|/c^2 \ll 1$).
Comparing Eq.~(\ref{e:mat:gab_weak}) with (\ref{e:dec:g_gam_N_beta}), we 
get $N=\sqrt{1+2\Phi}\simeq 1+\Phi$, $\w{\beta}=0$ and $\w{\gam}=(1-2\Phi)\w{f}$.
From Eq.~(\ref{e:dec:Einstein_PDE1}), we then obtain immediately that
$\w{K}=0$. To summarize:
\be \label{e:mat:Newt_limit}
	\mbox{Newtonian limit:}\quad  N = 1+\Phi,\quad 
	\w{\beta} = 0, \quad
	\w{\gam} =  \left( 1 - 2\Phi \right) \w{f}, \quad
	\w{K} = 0 ,\quad
	|\Phi| \ll 1.
\ee
Notice that the Eulerian observer becomes a Galilean (inertial)
observer for he is non-rotating (cf. remark page~\pageref{p:fol:nonrot}). 

Taking into account the limits (\ref{e:mat:Newt_limit}), Eq.~(\ref{e:mat:ener_cons})
reduces to
\be \label{e:mat:ener_cons_Newt0}
	\der{E}{t} + \w{D}\cdot\vec{\w{p}} = - 2 \vec{\w{p}}\cdot\w{D}\Phi . 
\ee
Let us denote by $\w{\Df}$ the Levi-Civita connection associated with the
flat metric $\w{f}$. Obviously $\w{D}\Phi=\w{\Df}\Phi$. On the other side,
let us express the divergence $\w{D}\cdot\w{\vec{p}}$ in terms of the divergence
$\w{\Df}\cdot\w{\vec{p}}$. From Eq.~(\ref{e:mat:Newt_limit}), we have
$\gam^{ij} = (1-2\Phi)^{-1} f^{ij} \simeq (1+2\Phi) f^{ij}$ as well as the 
relation
$\sqrt{\gam}=\sqrt{ (1-2\Phi)^3 f} \simeq (1-3\Phi) \sqrt{f}$
between the determinants $\gam$ and $f$ of respectively $(\gam_{ij})$ and $(f_{ij})$.
Therefore
\bea
	\w{D}\cdot\vec{\w{p}} & = & \frac{1}{\sqrt{\gam}} \der{}{x^i}
	\left( \sqrt{\gam} p^i \right)
	=  \frac{1}{\sqrt{\gam}} \der{}{x^i}
	\left( \sqrt{\gam} \gam^{ij} p_j \right) \nonumber \\
	& \simeq & \frac{1}{(1-3\Phi)\sqrt{f}} \der{}{x^i}
	\left[ (1-3\Phi)\sqrt{f}  (1+2\Phi) f^{ij} p_j \right]
	\simeq \frac{1}{\sqrt{f}} \der{}{x^i}
	\left[ (1-\Phi)\sqrt{f} f^{ij} p_j \right] \nonumber \\
	&\simeq& \frac{1}{\sqrt{f}} \der{}{x^i} \left( \sqrt{f} f^{ij} p_j \right)
	- f^{ij} p_j \der{\Phi}{x^i} \nonumber \\
	& \simeq & \w{\Df}\cdot\vec{\w{p}} - \vec{\w{p}} \cdot\w{\Df} \Phi .
			\label{e:mat:Dp_Dfp_Newt} 
\eea
Consequently Eq.~(\ref{e:mat:ener_cons_Newt0}) becomes
\be \label{e:mat:ener_cons_Newt}
	\der{E}{t} + \w{\Df}\cdot\vec{\w{p}} = - \vec{\w{p}}\cdot\w{\Df}\Phi . 
\ee
This is the standard energy conservation relation in a Galilean frame
with the source term $- \vec{\w{p}}\cdot\w{\Df}\Phi$. The latter  constitutes
the density of power provided to the system by the gravitational field
(this will be clear in the perfect fluid case, to be discussed below). 

\begin{remark}   
In the left-hand side of Eq.~(\ref{e:mat:ener_cons_Newt}), the quantity
$\w{p}$ plays the role of an \emph{energy flux}, whereas it had been 
defined in Sec.~\ref{s:dec:T3p1}
as a \emph{momentum density}. 
It is well known that both aspects are 
equivalent (see e.g. Chap.~22 of \cite{Hartl03}).
\end{remark}

\subsection{Momentum conservation}

Let us now project Eq.~(\ref{e:mat:divT3p1}) onto $\Sigma_t$:
\be \label{e:mat:mom_prov0}
\gam^\nu_{\ \, \alpha} \nabla_\mu S^\mu_{\ \, \nu} - K p_\alpha + 
\gam^\nu_{\ \, \alpha} n^\mu \nabla_\mu p_\nu
	 - K_{\alpha\mu} p^\mu 
	+ E D_\alpha\ln N  = 0 .
\ee
Now, from relation (\ref{e:hyp:link_D_nab_comp}),
\bea
	D_\mu S^\mu_{\ \, \alpha} &= & \gam^\rho_{\ \, \mu} \gam^\mu_{\ \, \sigma}
	\gam^\nu_{\ \, \alpha} \nabla_\rho S^\sigma_{\ \, \nu} = 
	\gam^\rho_{ \ \, \sigma} \gam^\nu_{\ \, \alpha} \nabla_\rho S^\sigma_{\ \, \nu}
	\nonumber \\
	& = & 
\gam^\nu_{\ \, \alpha} (\delta^\rho_{ \ \, \sigma} + n^\rho n_\sigma)
	\nabla_\rho S^\sigma_{\ \, \nu} 
= \gam^\nu_{\ \, \alpha} \big( \nabla_\rho S^\rho_{\ \, \nu}
	- S^\sigma_{\ \, \nu} \underbrace{n^\rho \nabla_\rho n_\sigma}_{=D_\sigma \ln N}
	 \big) \nonumber \\
	& = &  \gam^\nu_{\ \, \alpha} \nabla_\mu S^\mu_{\ \, \nu} - 
	S^\mu_{\ \, \alpha} D_\mu \ln N .  \label{e:mat:mom_prov1}
\eea
Besides
\bea
	\gam^\nu_{\ \, \alpha} n^\mu \nabla_\mu p_\nu & = &
	N^{-1} \gam^\nu_{\ \, \alpha} m^\mu \nabla_\mu p_\nu 
	= N^{-1} \gam^\nu_{\ \, \alpha} \left( \Liec{m} p_\nu - p_\mu \nabla_\nu m^\mu
	\right) \nonumber \\
	& = & N^{-1} \Lie{m} p_\alpha + K_{\alpha\mu} p^\mu , \label{e:mat:mom_prov2}
\eea
where use has been made of Eqs.~(\ref{e:fol:nab_m_comp}) and
(\ref{e:fol:nab_m_comp}) to get the second line. 
In view of Eqs.~(\ref{e:mat:mom_prov0}) and (\ref{e:mat:mom_prov1}), 
Eq.~(\ref{e:mat:mom_prov2}) becomes
\be
	\frac{1}{N} \Lie{m} p_\alpha + D_\mu S^\mu_{\ \, \alpha}
	+ S^\mu_{\ \, \alpha} D_\mu \ln N 
		- K p_\alpha + E D_\alpha\ln N = 0 
\ee
Writing $\Lie{m} = \dert{}{t} - \Lie{\beta}$, we obtain
\be \label{e:mat:mom_cons}
	\encadre{ \left( \der{}{t} - \Lie{\beta} \right) \w{p}
	+ N \w{D}\cdot\vec{\w{S}}
	+ \w{S} \cdot \vec{\w{D}} N
	- NK \w{p} + E \w{D} N = 0 } ,
\ee
or in components 
\be
	 \left( \der{}{t} - \Liec{\beta} \right) p_i + N D_j S^j_{\ \, i}
	+ S_{ij} D^j N 
		- N K p_i + E D_i N = 0 .
\ee
Again, this equation appears in York's article \cite{York79}.
Actually York's version [his Eq.~(41)] contains an additional term,
for it is written for the vector $\vec{\w{p}}$ dual to the linear form $\w{p}$,
and since $\Liec{m}\gam^{ij}\not=0$, this generates the extra term
$p_j \Liec{m}\gam^{ij} = 2N K^{ij} p_j$.

To take the Newtonian limit of Eq.~(\ref{e:mat:mom_cons}), we shall consider
not only Eq.~(\ref{e:mat:Newt_limit}),
which provides the Newtonian limit of the gravitational field, by in addition
the relation
\be \label{e:mat:SllE_Newt}
	\mbox{Newtonian limit:}\quad |S^i_{\ \, j}| \ll E , 
\ee
which expresses that the matter is not relativistic.
Then the Newtonian limit of (\ref{e:mat:mom_cons}) is
\be \label{e:mat:mom_cons_Newt}
  \der{\w{p}}{t} + \w{\Df}\cdot\vec{\w{S}}
	= -  E \w{\Df}\Phi .
\ee
Note that in relating $\w{D}\cdot\vec{\w{S}}$ to $ \w{\Df}\cdot\vec{\w{S}}$, 
there should appear derivatives of $\Phi$, as in Eq.~(\ref{e:mat:Dp_Dfp_Newt}),
but thanks to property (\ref{e:mat:SllE_Newt}), these terms are negligible 
in front of $E\w{\Df}\Phi$.
Equation (\ref{e:mat:mom_cons_Newt}) is the standard momentum conservation law,
with $-  E \w{\Df}\Phi$ being the gravitational force density.

\section{Perfect fluid}

\subsection{kinematics}

The \defin{perfect fluid} model of matter relies on a vector field $\w{u}$
of 4-velocities, giving at each point the 4-velocity of a fluid
particle. In addition the perfect fluid is characterized by an isotropic 
pressure in the fluid frame. More precisely, the perfect fluid
model is entirely defined by the following stress-energy tensor:
\be \label{e:mat:T_fluid_parfait}
	\encadre{ \w{T} = (\rho  + P)\,  \uu{u}\otimes\uu{u} + P\, \w{g} }, 
\ee
where $\rho$ and $P$ are two scalar fields, representing respectively
the matter energy density and the pressure, both measured
in the fluid frame (i.e. by an observer who is comoving with the fluid), 
and $\uu{u}$ is the 1-form associated to the 4-velocity $\w{u}$ by the metric 
tensor $\w{g}$ [cf. Eq.~(\ref{e:hyp:underbar})].

\begin{figure}
\centerline{\includegraphics[width=0.8\textwidth]{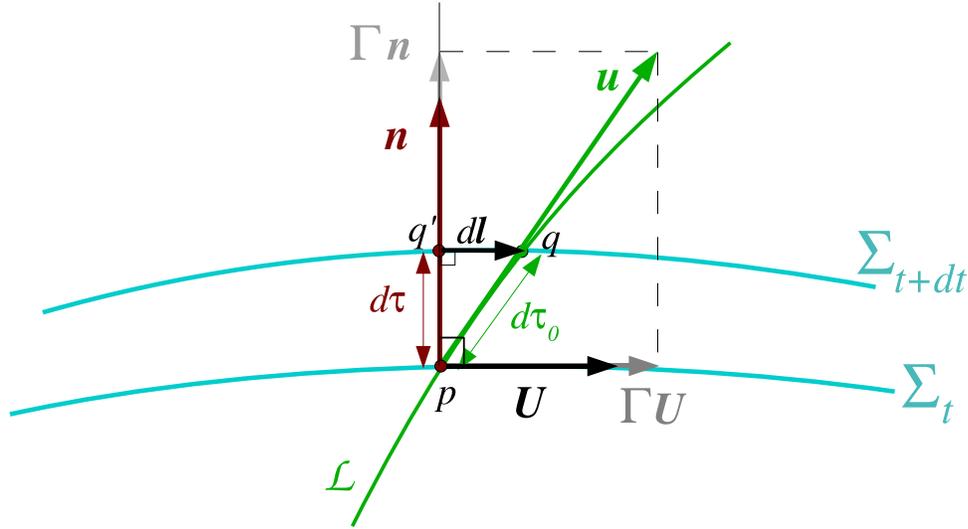}}
\caption[]{\label{f:mat:fluidvel} \footnotesize
Worldline $\mathcal{L}$ of a fluid element crossing the spacetime foliation
$(\Sigma_t)_{t\in\R}$. $\w{u}$ is the fluid 4-velocity and $\w{U}=d\w{\ell}/d\tau$
the relative velocity of the fluid with respect to the Eulerian observer, whose
4-velocity is $\w{n}$. 
$\w{U}$ is tangent to $\Sigma_t$ and enters in the orthogonal decomposition of $\w{u}$
with respect to $\Sigma_t$, via 
$\w{u} = \Gamma ( \w{n} +  \w{U} )$.
\emph{NB:} contrary to what the figure might suggest, $d\tau > d\tau_0$
(conflict between the figure's underlying Euclidean geometry and the actual 
Lorentzian geometry of spacetime). 
}
\end{figure}

Let us consider a fluid element at point $p\in\Sigma_t$ (cf. Fig.~\ref{f:mat:fluidvel}).
Let $\tau$ be the Eulerian observer's proper time at $p$. 
At the coordinate time $t+dt$, the fluid element has moved to the point $q\in\Sigma_{t+dt}$.
The date $\tau+d\tau$ attributed to the event $q$ 
by the Eulerian observer moving through
$p$ is given by the orthogonal projection $q'$ of $q$ onto the wordline of
that observer. Indeed, let us recall that the space of simultaneous events
(local rest frame) for the Eulerian
observer is the space orthogonal to his 4-velocity $\w{u}$, i.e. locally $\Sigma_t$
(cf. Sec.~\ref{s:fol:Eulerian}). Let $d\w{\ell}$ be the infinitesimal vector
connecting $q'$ to $q$. Let $d\tau_0$ be the increment of the fluid proper time
between the events $p$ and $q$. 
The \defin{Lorentz factor} of the fluid with respect to the Eulerian observer is
defined as being the proportionality factor $\Gamma$ between the proper times
$d\tau_0$ and $d\tau$:
\be \label{e:mat:def_Gamma}
	\encadre{ d\tau =: \Gamma d\tau_0 } .
\ee
One has the triangle identity (cf. Fig.~\ref{f:mat:fluidvel}):
\be \label{e:mat:fluid_triangle}
	d\tau_0 \, \w{u} = d\tau \, \w{n} + d\w{\ell} . 
\ee
Taking the scalar product with $\w{n}$ yields
\be
	d\tau_0 \, \w{n}\cdot\w{u} 
	= d\tau \, \underbrace{\w{n}\cdot\w{n}}_{=-1}
	+ \underbrace{\w{n}\cdot d\w{\ell}}_{=0} ,
\ee
hence, using relation (\ref{e:mat:def_Gamma}),
\be \label{e:mat:Gamma_scal}
	\encadre{ \Gamma = - \w{n}\cdot\w{u}} .
\ee
From a pure geometrical point of view, the Lorentz factor is thus nothing
but minus the scalar product of the two 4-velocities, the fluid's one and the
Eulerian observer's one. 
\begin{remark}
Whereas $\Gamma$ has been defined in an asymmetric way as the ``Lorentz factor
of the fluid observer \emph{with respect to} the Eulerian observer'', the above
formula shows that the Lorentz factor is actually a symmetric quantity in
terms of the two observers. 
\end{remark}
Using the components $n_\alpha$ of 
$\uu{n}$ given by Eq.~(\ref{e:dec:uun_comp}), Eq.~(\ref{e:mat:Gamma_scal}) gives
an expression of the Lorentz factor in terms of the component $u^0$ of $\w{u}$
with respect to the coordinates $(t,x^i)$:
\be
	\Gamma = N u^0 . 
\ee

The  fluid \defin{velocity relative to the Eulerian observer} is defined as the
quotient of the displacement $d\w{\ell}$ by the proper time $d\tau$, both quantities
being relative to the Eulerian observer (cf. Fig.~\ref{f:mat:fluidvel}):
\be \label{e:mat:def_U}
	\encadre{ \w{U} := \frac{d\w{\ell}}{d\tau} } . 
\ee
Notice that by construction, $\w{U}$ is tangent to $\Sigma_t$.
Dividing the identity (\ref{e:mat:fluid_triangle}) by $d\tau$ and making use
of Eq.~(\ref{e:mat:def_Gamma}) results in
\be \label{e:mat:u_U}
	\encadre{ \w{u} = \Gamma(\w{n} + \w{U}) } .
\ee
Since $\w{n}\cdot\w{U}=0$, the above writting constitutes the orthogonal 
3+1 decomposition of the fluid 4-velocity $\w{u}$.
The normalization relation of the fluid 4-velocity, i.e. $\w{u}\cdot\w{u}=-1$,
combined with Eq.~(\ref{e:mat:u_U}), results in
\be
	- 1 = \Gamma^2 (\underbrace{\w{n}\cdot\w{n}}_{=-1}
	+ 2\underbrace{\w{n}\cdot\w{U}}_{=0}
	+ \w{U}\cdot\w{U}) , 
\ee
hence
\be \label{e:mat:Gamma_U2}
	\encadre{\Gamma = \left( 1 - \w{U}\cdot\w{U} \right) ^{-1/2} } . 
\ee
Thus, in terms of the velocity $\w{U}$, the Lorentz factor is expressed by
a formula identical of that of special relativity, except of course
that the scalar product in Eq.~(\ref{e:mat:Gamma_U2}) is to be taken with 
the (curved) metric $\w{\gam}$, whereas in special relativity it is 
taken with a flat metric. 

\begin{figure}
\centerline{\includegraphics[width=0.8\textwidth]{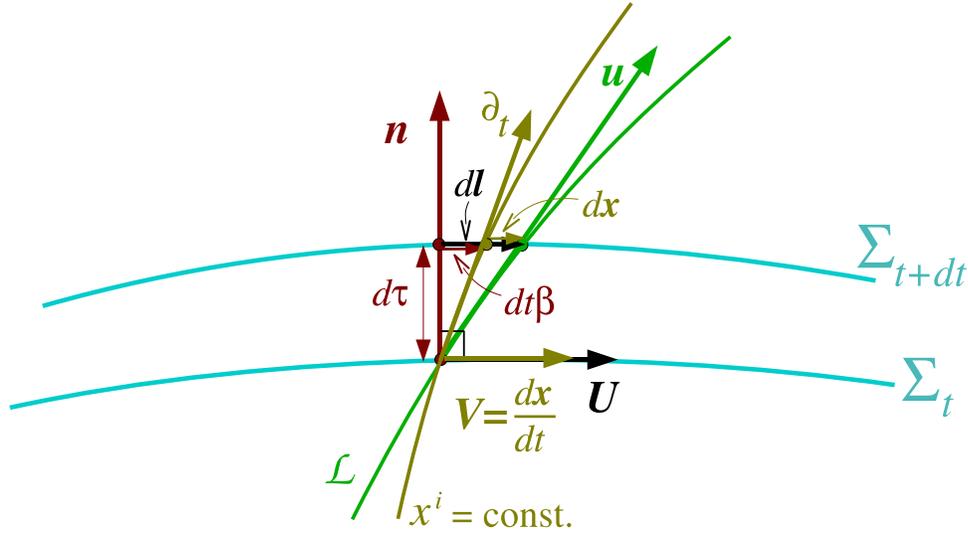}}
\caption[]{\label{f:mat:coordvel} \footnotesize
Coordinate velocity $\w{V}$ of the fluid defined as the ratio of the 
fluid displacement with respect to the line of constant spatial coordinates
to the coordinate time increment $dt$.
}
\end{figure}

It is worth to introduce another type of fluid velocity, 
namely the \defin{fluid coordinate
velocity} defined by
\be \label{e:mat:def_V}
	\encadre{ \w{V} := \frac{d\w{x}}{dt} } , 
\ee
where $d\w{x}$ is the displacement of the fluid worldline with respect to
the line of constant spatial coordinates (cf. Fig.~\ref{f:mat:coordvel}).
More precisely, if the fluid moves from the point $p$ of coordinates $(t,x^i)$ to the
point $q$ of coordinates $(t+dt,x^i+dx^i)$, the fluid coordinate velocity is
defined as the vector tangent to $\Sigma_t$, the components of which are
\be
	V^i = \frac{dx^i}{dt} . 
\ee
Noticing that the components of the fluid 4-velocity are $u^\alpha = dx^\alpha/d\tau_0$,
the above formula can be written
\be
	V^i = \frac{u^i}{u^0} . 
\ee
From the very definition of the shift vector (cf. Sec.~\ref{s:dec:shift}), 
the drift of the coordinate line
$x^i={\rm const}$ from the Eulerian observer worldline between $t$ and $t+dt$
is the vector $dt\, \w{\beta}$. Hence we have (cf. Fig.~\ref{f:mat:coordvel})
\be
	d\w{\ell} = dt\, \w{\beta} + d\w{x} . 
\ee
Dividing this relation by $d\tau$, using Eqs.~(\ref{e:mat:def_U}), (\ref{e:fol:dtau_Ndt})
and (\ref{e:mat:def_V}) yields
\be \label{e:mat:U_V}
	\encadre{ \w{U} = \frac{1}{N} \left( \w{V} + \w{\beta} \right) } . 
\ee
On this expression, it is clear that at the Newtonian limit 
as given by (\ref{e:mat:Newt_limit}), $\w{U}=\w{V}$.

\subsection{Baryon number conservation}

In addition to $\wnab\cdot\w{T}=0$, the perfect fluid must obey to the fundamental
law of baryon number conservation:
\be \label{e:mat:baryon_cons}
	\encadre{\wnab\cdot \w{j}_{\rm B} = 0} ,
\ee
where $\w{j}_{\rm B}$ is the \defin{baryon number 4-current}, expressible in terms
of the fluid 4-velocity and the fluid \defin{proper baryon number density}
$n_{\rm B}$ as 
\be \label{e:mat:jB_def}
	\encadre{ \w{j}_{\rm B} = n_{\rm B} \w{u} }.
\ee 
The \defin{baryon number density measured by the Eulerian observer} is
\be
	\mathcal{N}_{\rm B} := - \w{j}_{\rm B}\cdot \w{n} .
\ee 
Combining Eqs.~(\ref{e:mat:Gamma_scal}) and (\ref{e:mat:jB_def}), we get
\be
	\encadre{ \mathcal{N}_{\rm B} = \Gamma n_{\rm B} }.
\ee
This relation is easily interpretable by remembering that $\mathcal{N}_{\rm B}$
and $n_{\rm B}$ are volume densities and invoking the Lorentz-FitzGerald ``length
contraction'' in the direction of motion.

The \defin{baryon number current measured by the Eulerian observer} is given by
the orthogonal projection of $\w{j}_{\rm B}$ onto $\Sigma_t$:
\be
	\w{J}_{\rm B} := \vg(\w{j}_{\rm B}) .  
\ee
Taking into account that $\vg(\w{u}) = \Gamma \w{U}$ [Eq.~(\ref{e:mat:u_U})],
we get the simple relation
\be
	\encadre{ \w{J}_{\rm B} = \mathcal{N}_{\rm B} \w{U} } . 
\ee

Using the above formul\ae, as well as the orthogonal decomposition (\ref{e:mat:u_U})
of $\w{u}$, the baryon number conservation law (\ref{e:mat:baryon_cons})
can be written
\bea
	&& \wnab\cdot(n_{\rm B} \w{u}) = 0 \nonumber \\
	&\Rightarrow& \wnab\cdot[n_{\rm B} \Gamma(\w{n}+\w{U})] = 0 \nonumber \\
	&\Rightarrow& 
	\wnab\cdot[\mathcal{N}_{\rm B} \w{n} +  \mathcal{N}_{\rm B} \w{U}] = 0 	
		\nonumber \\
	&\Rightarrow& \w{n}\cdot\wnab\mathcal{N}_{\rm B}  +
	\mathcal{N}_{\rm B} \underbrace{\wnab\cdot\w{n}}_{=-K}
	+ \wnab\cdot(\mathcal{N}_{\rm B} \w{U}) = 0
\eea
Since $\mathcal{N}_{\rm B} \w{U}\in\T(\Sigma_t)$, we may use the divergence
formula (\ref{e:mat:divergence}) and obtain 
\be
	\Lie{\w{n}} \mathcal{N}_{\rm B} - K \mathcal{N}_{\rm B}
	+ \w{D}\cdot(\mathcal{N}_{\rm B} \w{U}) + \mathcal{N}_{\rm B} \w{U}
	\cdot \w{D}\ln N = 0 ,
\ee
where we have written $\w{n}\cdot\wnab\mathcal{N}_{\rm B} = 
\Lie{\w{n}} \mathcal{N}_{\rm B}$. Since
$\w{n} = N^{-1} (\wpar_t - \w{\beta})$ [Eq.~(\ref{e:dec:t_Nn_b})], 
we may rewrite the above equation as
\be \label{e:mat:baryon_cons_3p1}
	\encadre{ \left( \der{}{t} - \Lie{\beta} \right) \mathcal{N}_{\rm B}
	+ \w{D}\cdot(N \mathcal{N}_{\rm B} \w{U})
	- N K  \mathcal{N}_{\rm B} = 0 } . 
\ee
Using Eq.~(\ref{e:mat:U_V}), we can put this equation in an alternative
form
\be
	\der{}{t} \mathcal{N}_{\rm B} + \w{D}\cdot(\mathcal{N}_{\rm B} \w{V})
	+ \mathcal{N}_{\rm B} \left( \w{D}\cdot\w{\beta} - NK \right) = 0 . 
\ee

\subsection{Dynamical quantities}

The fluid energy density as measured by the Eulerian observer is given by formula
(\ref{e:dec:E_def}): $E = \w{T}(\w{n},\w{n})$, with the stress-energy tensor
(\ref{e:mat:T_fluid_parfait}). Hence
$E=(\rho + P) (\w{u}\cdot\w{n})^2 + P \w{g}(\w{n},\w{n})$.
Since $\w{u}\cdot\w{n}=-\Gamma$ [Eq.~(\ref{e:mat:Gamma_scal})]
and $\w{g}(\w{n},\w{n}) = -1$, we get
\be \label{e:mat:eps_Gamma2}
	\encadre{ E = \Gamma^2 (\rho  + P) - P }.
\ee
\begin{remark}
For pressureless matter (dust), the above formula reduces to $E=\Gamma^2 \rho$. 
The reader familiar with the formula $E=\Gamma m c^2$ may then be puzzled by the 
$\Gamma^2$ factor in (\ref{e:mat:eps_Gamma2}). However he should remind that 
$E$ is not an energy, but an energy per unit volume: the extra $\Gamma$
factor arises from  ``length contraction'' in the direction of motion.
\end{remark}

Introducing the proper baryon density $n_{\rm B}$, one may decompose the 
proper energy density $\rho$ in terms of a \defin{proper rest-mass energy density} $\rho_0$
and an \defin{proper internal energy} $\varepsilon_{\rm int}$ as 
\be
	\rho = \rho_0 + \varepsilon_{\rm int},\qquad \mbox{with}
	\quad \rho_0 := m_{\rm B} n_{\rm B} , 
\ee
$m_{\rm B}$ being a constant, namely the mean baryon rest mass
($m_{\rm B}\simeq1.66\times 10^{-27}{\ \rm kg})$.
Inserting the above relation into Eq.~(\ref{e:mat:eps_Gamma2}) and
writting $\Gamma^2\rho = \Gamma\rho + (\Gamma-1)\Gamma\rho$ leads 
to the following decomposition of $E$:
\be
	E = E_0 + E_{\rm kin} + E_{\rm int} , 
\ee
with the rest-mass energy density
\be
	E_0 := m_{\rm B} \mathcal{N}_{\rm B},
\ee
the kinetic energy density
\be
	E_{\rm kin} := (\Gamma-1) E_0 = (\Gamma-1) m_{\rm B} \mathcal{N}_{\rm B},
\ee
the internal energy density
\be
	E_{\rm int} := \Gamma^2 (\varepsilon_{\rm int}+P) - P . 
\ee
The three quantities $E_0$, $E_{\rm kin}$ and $E_{\rm int}$ are relative to the
Eulerian observer. 

At the Newtonian limit, we shall suppose that the fluid is not relativistic
[cf. (\ref{e:mat:SllE_Newt})]:
\be
	P\ll\rho_0, \quad |\epsilon_{\rm int}| \ll \rho_0,\quad
	 U^2:=\w{U}\cdot\w{U} \ll 1 .
\ee
Then we get 
\be \label{e:mat:Newt_fluid}
	\mbox{Newtonian limit:}\quad 
	\Gamma \simeq 1+\frac{U^2}{2}, \quad
	E \simeq E+P \simeq E_0 \simeq \rho_0, \quad
	E-E_0 \simeq \frac{1}{2} \rho_0 U^2 + \varepsilon_{\rm int} .
\ee

The fluid momentum density as measured by the Eulerian observer is obtained
by applying formula (\ref{e:dec:p_def}): 
\bea
	\w{p} & = & - \w{T}(\w{n}, \vg( . ) ) 
	= - (\rho+P) \underbrace{\langle \uu{u},\w{n} \rangle}_{=-\Gamma}
	 \underbrace{\langle \uu{u},\vg(.) \rangle}_{=\Gamma \uu{U}}
	- P \underbrace{\w{g}(\w{n},\vg(.))}_{=0} 
		\nonumber\\
	 & = & \Gamma^2 (\rho + P) \uu{U} , 
\eea
where Eqs.~(\ref{e:mat:Gamma_scal}) and (\ref{e:mat:u_U}) have been used to 
get the second line.
Taking into account Eq.~(\ref{e:mat:eps_Gamma2}), the above relation becomes
\be \label{e:mat:p_fluid}
	\encadre{ \w{p} = (E+P) \uu{U} } . 
\ee

Finally, by applying formula (\ref{e:dec:S_def}), we get the fluid stress tensor
with respect to the Eulerian observer:
\bea
	\w{S} & = & \vgs \w{T} = 
	(\rho+P) \underbrace{\vgs\uu{u}}_{=\Gamma\uu{U}} \otimes
	 \underbrace{\vgs\uu{u}}_{=\Gamma \uu{U}}
	+ P \underbrace{\vgs \w{g}}_{=\w{\gam}} \nonumber \\ 
	& =& P \, \w{\gam} + \Gamma^2 (\rho+P) \uu{U} \otimes \uu{U} , 
\eea
or, taking into account Eq.~(\ref{e:mat:eps_Gamma2}),
\be \label{e:mat:S_fluid}
	\encadre{ \w{S} = P \, \w{\gam} + (E+P) \uu{U} \otimes \uu{U} } . 
\ee

\subsection{Energy conservation law}

By means of Eqs.~(\ref{e:mat:p_fluid}) and (\ref{e:mat:S_fluid}), the energy conservation law (\ref{e:mat:ener_cons}) becomes
\be \label{e:mat:ener_cons_fluid}
	 \left( \der{}{t} - \Lie{\beta} \right) E 
	+ N \left\{ \w{D}\cdot\left[ (E+P) \w{U} \right] 
	- (E+P)(K+ K_{ij} U^i U^j) \right\} + 2 (E+P) \w{U}\cdot \w{D} N 
	= 0 
\ee
To take the Newtonian limit, we may combine the Newtonian limit of the baryon number 
conservation law (\ref{e:mat:baryon_cons_3p1}) with 
Eq.~(\ref{e:mat:ener_cons_Newt}) to get
\be
	\der{E'}{t} + \w{\Df} \cdot [(E'+P)\w{U}] = - \w{U} \cdot (\rho_0 \w{\Df}\Phi) ,
\ee
where $E':=E-E_0 = E_{\rm kin} + E_{\rm int}$ and we clearly recognize in the
right-hand side the power provided to a unit volume fluid element by the gravitational 
force.

\subsection{Relativistic Euler equation}

Injecting the expressions (\ref{e:mat:p_fluid}) and (\ref{e:mat:S_fluid}) into 
the momentum conservation law (\ref{e:mat:mom_cons}), we get 
\bea
	& & \left( \der{}{t} - \Liec{\beta} \right) \left[ (E+P) U_i \right]
	+ N D_j \left[ P \delta^j_{\ \, i} + (E+P) U^j U_i \right]
	+ \left[ P \gam_{ij} + (E+P) U_i U_j \right] D^j N \nonumber \\
& & \qquad \qquad - NK(E+P) U_i + E D_i N = 0 . 
\eea
Expanding and making use of Eq.~(\ref{e:mat:ener_cons_fluid}) yields
\bea
	& & \left( \der{}{t} - \Liec{\beta} \right) U_i + N U^j D_j U_i
	- U^j D_j N \, U_i + D_i N + N K_{kl} U^k U^l U_i \nonumber \\
	& & \qquad \qquad + \frac{1}{E+P} \left[ N D_i P
	+ U_i \left( \der{}{t} - \Lie{\beta} \right) P \right] = 0 .
\eea
Now, from Eq.~(\ref{e:mat:U_V}), $N U^j D_j U_i = V^j D_j U_i + \beta^j D_j U_i$,
so that $- \Liec{\beta}  U_i + N U^j D_j U_i = V^j D_j U_i - U_j D_i \beta^j$
[cf. Eq.~(\ref{e:Lie_der_1form})]. Hence the above equation can be written
\be
	\encadre{
	\begin{array}{rcl}
	\displaystyle \der{U_i}{t} + V^j D_j U_i + NK_{kl} U^k U^l U_i 
	-  U_j D_i \beta^j & = & \displaystyle - \frac{1}{E+P} \left[ N D_i P + U_i
	\left( \der{P}{t} - \beta^j \der{P}{x^j} \right) \right] \\
	& & \displaystyle - D_i N + U_i U^j D_j N .
	\end{array}
	}
\ee
The Newtonian limit of this equation is [cf. Eqs.~(\ref{e:mat:Newt_limit}) and (\ref{e:mat:Newt_fluid})]
\be
	\der{U_i}{t} + U^j \Df_j U_i = -\frac{1}{\rho_0} \Df_i P - \Df_i \Phi , 
\ee
i.e. the standard Euler equation in presence of a gravitational field of potential
$\Phi$.

\subsection{Further developments}

For further developments in 3+1 relativistic hydrodynamics, we refer to the 
review article by Font \cite{Font03}. 
Let us also point out that the 3+1 decomposition presented above
is not very convenient for discussing conservation laws, 
such as the relativistic generalizations of
Bernoulli's theorem or Kelvin's circulation theorem. For this purpose
the Carter-Lichnerowicz approach, which is based on exterior calculus, 
is much more powerfull, as discussed in Ref.~\cite{Gourg06}.

\section{Electromagnetic field} \label{s:mat:em}

not written up yet; see Ref.~\cite{ThornM82}.

\section{3+1 magnetohydrodynamics}

not written up yet; see Refs.~\cite{BaumgS03b,ShibaS05b,AntonZMMIFP06}.

%  
%    Chapitre : Conformal decomposition
%
% $Date: 2007-03-06 11:59:03 +0100 (mar, 06 mar 2007) $
% $Rev: 183 $
% $Author: e_gourgoulhon $
%%%%%%%%%%%%%%%%%%%%%%%%%%%%%

\chapter{Conformal decomposition} \label{s:cfd}

%\verb$Date: 2007-03-06 11:59:03 +0100 (mar, 06 mar 2007) $

\minitoc
\vspace{1cm}

%%%%%%%%%%%%%%%%%%%%%%%%%%%%%%%%%%%%%%%%%%%%%%%%%%%%%%%%%%%%%%%%%%%%%%%%%%%%

\section{Introduction} \label{s:cfd:intro}

Historically, conformal decompositions in 3+1 general relativity have been 
introduced in two contexts.
First of all, Lichnerowicz \cite{Lichn44} \footnote{see also Ref.~\cite{Lichn52} which 
is freely accessible on the web}
has introduced in 1944 a decomposition 
of the induced metric  $\wgm$ of the hypersurfaces $\Sigma_t$ of the type
\be \label{e:cfd:gam_gamtilde}
	\wgm = \Psi^4 \wtgm, 
\ee
where $\Psi$ is some strictly positive scalar field and $\wtgm$ an auxiliary 
metric on $\Sigma_t$, which is necessarily Riemannian (i.e. positive definite),
as $\wgm$ is. The relation (\ref{e:cfd:gam_gamtilde})
is called a \defin{conformal transformation}
and $\wtgm$ will be called hereafter the \defin{conformal metric}.
Lichnerowicz has shown that the conformal decomposition of $\wgm$, along with
some specific conformal decomposition of the extrinsic curvature provides a fruitful tool 
for the resolution of the constraint equations to get valid
initial data for the Cauchy problem.
This will be discussed in Chap.~\ref{s:ini}.

Then, in 1971-72, York \cite{York71,York72a} has shown that conformal decompositions
are also important for the time evolution problem, by demonstrating that
the two degrees of freedom of the 
gravitational field are carried by the conformal equivalence classes of 
3-metrics. A \defin{conformal equivalence class} is defined as the set of all metrics that can be related to a given metric $\wgm$ by a transform like (\ref{e:cfd:gam_gamtilde}). The argument of York is based 
on the \defin{Cotton tensor} \cite{Cotto1899}, which is a rank-3 covariant tensor
defined from the covariant derivative of the Ricci tensor $\w{R}$ of $\wgm$
by 
\be
	\mathcal{C}_{ijk} := D_k \left( R_{ij} -  \frac{1}{4} R \gam_{ij} \right)
	- D_j  \left( R_{ik} -  \frac{1}{4} R 
	\gam_{ik} \right) . 
\ee
The Cotton tensor is conformally invariant and shows the same property with respect to 
3-dimensional metric manifolds than the 
Weyl tensor [cf. Eq.~(\ref{e:hyp:Weyl})] for metric manifolds
of dimension strictly greater than 3, namely its vanishing is a necessary
and sufficient condition for the metric to be \defin{conformally flat}, 
i.e. to be expressible as $\wgm = \Psi^4 \w{f}$, where $\Psi$ is some scalar
field and $\w{f}$ a flat metric. 
Let us recall that in dimension 3, the Weyl tensor vanishes identically. 
More precisely, York \cite{York71} constructed from the Cotton tensor 
the following rank-2 tensor
\be
	C^{ij} := -\frac{1}{2} \epsilon^{ikl} \mathcal{C}_{mkl} \gam^{mj} 
	= \epsilon^{ikl} D_k \left( R^j_{\ \, l} -\frac{1}{4} R \delta^j_{\ \, l} \right) ,
\ee
where $\w{\epsilon}$ is the Levi-Civita alternating tensor associated with the 
metric $\wgm$.
This tensor is called the \defin{Cotton-York tensor} and exhibits the following
properties:
\begin{itemize}
\item symmetric: $C^{ji} = C^{ij}$
\item traceless: $\gam_{ij} C^{ij} = 0$
\item divergence-free (one says also \emph{transverse}):
$D_j C^{ij} = 0$
\end{itemize}
Moreover, if one consider, instead of $\w{C}$, the following tensor density
of weight $5/3$,
\be
	C_*^{ij} := \gam^{5/6} C^{ij},
\ee
where $\gam:=\det(\gam_{ij})$,
then one gets a conformally invariant quantity. Indeed, under a conformal transformation
of the type (\ref{e:cfd:gam_gamtilde}), 
$\epsilon^{ikl}=\Psi^{-6}\tilde\epsilon^{ikl}$,
$\mathcal{C}_{mkl}={\tilde\mathcal{C}}_{mkl}$ (conformal invariance
of the Cotton tensor),
$\gam^{ml}=\Psi^{-4}{\tilde\gam}^{ml}$
and $\gam^{5/6}=\Psi^{10}{\tilde\gam}^{5/6}$,
so that $C_*^{ij}={\tilde C}_*^{ij}$.
The traceless and transverse (TT) properties being characteristic of the pure
spin 2 representations of the gravitational field 
(cf. T. Damour's lectures \cite{Damou06}), the conformal invariance of
$C_*^{ij}$ shows that the true degrees of freedom of the gravitational
field are carried by the conformal equivalence class.
\begin{remark}
The remarkable feature of the Cotton-York tensor is to be a TT object constructed 
from the physical metric $\wgm$ alone, without the need of some extra-structure
on the manifold $\Sigma_t$. Usually, TT objects are defined with respect to 
some extra-structure, such as privileged Cartesian coordinates or a flat background
metric, as in the post-Newtonian approach to general relativity (see L. Blanchet's
lectures  \cite{Blanc06b}).
\end{remark}

\begin{remark}
The Cotton and Cotton-York tensors involve third derivatives of the metric tensor.
\end{remark}

%%%%%%%%%%%%%%%%%%%%%%%%%%%%%%%%%%%%%%%%%%%%%%%%%%%%%%%%%%%%%%%%%%%%%%%%%%%%%%%%%%%%%

\section{Conformal decomposition of the 3-metric}

\subsection{Unit-determinant conformal ``metric''}

A somewhat natural representative of a 
conformal equivalence class is the unit-determinant conformal ``metric''
\be \label{e:cfd:def_hatg}
	\w{\hat\gam} := \gam^{-1/3} \wgm ,
\ee
where $\gam := \det(\gam_{ij})$. 
This would correspond to the choice $\Psi = \gam^{1/12}$ in Eq.~(\ref{e:cfd:gam_gamtilde}).
All the metrics $\wgm$ in the same conformal equivalence class lead to the same
value of $\w{\hat\gam}$.
However, since the determinant $\gam$ depends upon the choice of coordinates
to express the components $\gam_{ij}$, $\Psi= \gam^{1/12}$ would not be a scalar field.
Actually, the quantity $\w{\hat\gam}$ is not a tensor
field, but a tensor density, of weight $-2/3$.

Let us recall that a \defin{tensor density of weight} $n\in\mathbb{Q}$ is a quantity
$\w{\tau}$ such that
\be
	\w{\tau} = \gam^{n/2} \, \w{T},
\ee
where $\w{T}$ is a tensor field. 

\begin{remark}
The conformal ``metric'' (\ref{e:cfd:def_hatg}) has been used notably 
in the BSSN formulation \cite{ShibaN95,BaumgS99} for the time evolution of
3+1 Einstein system, to be discussed in Chap.~\ref{s:evo}. An  
``associated'' connection $\w{\hat D}$ has been introduced, such that
$\w{\hat D} \w{\hat\gam}=0$.
However, since $\w{\hat\gam}$ is a tensor density and not a tensor field, there
is not a unique connection associated with it (Levi-Civita connection). In particular
one has $\w{D}\w{\hat\gam}=0$, so that the connection
$\w{D}$ associated with the metric $\wgm$
is ``associated'' with $\w{\hat\gam}$, in addition to $\w{\hat D}$.
As a consequence, some of the formul\ae\ presented in the original references
\cite{ShibaN95,BaumgS99} for the BSSN formalism have a meaning only for 
Cartesian coordinates. 
\end{remark}

\subsection{Background metric} \label{s:cfd:background_metric}

To clarify the meaning of $\w{\hat D}$ (i.e. to avoid to work with tensor 
densities) and to allow for the use of 
spherical coordinates, we introduce an extra structure on the
hypersurfaces $\Sigma_t$, namely a \defin{background metric} $\w{f}$ \cite{BonazGGN04}.
It is asked that the signature of $\w{f}$ is $(+,+,+)$, i.e. that $\w{f}$
is a Riemannian metric, as $\wgm$. Moreover, we tight $\w{f}$ to the coordinates
$(x^i)$ by demanding that the components $f_{ij}$ of $\w{f}$ with respect
to $(x^i)$ obey to 
\be \label{e:cfd:f_const}
	\der{f_{ij}}{t}  = 0 . 
\ee
An equivalent writing of this is
\be
	\w{\mathcal{L}}_{\wpar_t} \w{f} = 0 ,
\ee
i.e. the metric $\w{f}$ is Lie-dragged along the coordinate time evolution vector
$\wpar_t$. 

If the topology of $\Sigma_t$ enables it, it is quite natural to choose 
$\w{f}$ to be flat, i.e. such that its Riemann tensor vanishes. 
However, in this chapter, we shall not make such hypothesis, except in 
Sec.~\ref{s:cfd:IWM}. 

As an example of background metric, let us consider a coordinate
system $(x^i)=(x,y,z)$ on $\Sigma_t$ and define the metric $\w{f}$ as the
bilinear form whose components with respect to that coordinate system
are $f_{ij} =\mathrm{diag}(1,1,1)$ (in this example, $\w{f}$ is flat).

The inverse metric is denoted by $f^{ij}$:
\be
	f^{ik} f_{kj} = \delta^i_{\ \, j} . 
\ee
In particular note that, except for the very special case
$\gam_{ij}=f_{ij}$, one has
\be
	f^{ij} \not= \gam^{ik} \gam^{jl}\, f_{kl} .
\ee

We denote by $\w{\Df}$ the Levi-Civita connection associated
with $\w{f}$:
\be \label{e:cfd:Df_zero}
	\Df_k f_{ij} = 0 , 
\ee
and define
\be \label{e:cfd:def_upcD}
	\Df^i = f^{ij} \Df_j . 
\ee
The Christoffel symbols of the connection $\w{\Df}$
with respect to the coordinates $(x^i)$ are denoted by 
$\bar\Gamma^k_{\ \, ij}$; they are given by the
standard expression:
\be \label{e:cfd:christo_f}
	\bar\Gamma^k_{\ \, ij} = {1\over 2} f^{kl} 
		\left( \der{f_{lj}}{x^i} +\der{f_{il}}{x^j}
			- \der{f_{ij}}{x^l} \right) .
\ee

\subsection{Conformal metric} \label{s:cfd:conf_metric}

Thanks to $\w{f}$, we define 
\be \label{e:cfd:def_tildegam}
	\encadre{ \w{\tgm} := \Psi^{-4} \wgm },
\ee
where
\be \label{e:cfd:def_Psi}
	\encadre{ \Psi := \left( \frac{\gam}{f} \right)^{1/12} },\quad 
	\gam := \det(\gam_{ij}),\quad 
	f := \det(f_{ij}) .
\ee
The key point is that, contrary to $\gam$, $\Psi$ is a tensor field on $\Sigma_t$.
Indeed a change of coordinates $(x^i) \mapsto (x^{i'})$ induces the following
changes in the determinants:
\bea
	\gam'  & = & (\det J)^2 \gam \label{e:cfd:change_g} \\
	f' & = & (\det J)^2 f ,\label{e:cfd:change_f}
\eea
where $J$ denotes the Jacobian matrix
\be
	J^i_{\ \, i'} := \der{x^i}{x^{i'}} .
\ee
From Eqs.~(\ref{e:cfd:change_g})-(\ref{e:cfd:change_f}) it is obvious
that $\gam'/f' = \gam /f$, which shows that $\gam/f$, and hence $\Psi$,
is a scalar field. Of course, this scalar field depends upon the choice of
the background metric $\w{f}$. 
$\Psi$ being a scalar field, the quantity $\wtgm$ defined by
(\ref{e:cfd:def_tildegam}) is a tensor field on $\Sigma_t$. Moreover, it
is a Riemannian metric on $\Sigma_t$. 
We shall call it the \defin{conformal metric}. 
By construction, it satisfies
\be \label{e:cfd:dettgm_f}
	\encadre{ \det (\tgm_{ij}) = f }.
\ee
This is the ``unit-determinant'' condition fulfilled by $\wtgm$. Indeed, if
one uses for $(x^i)$ Cartesian-type coordinates, then $f=1$. But the condition 
(\ref{e:cfd:dettgm_f}) is more flexible and allows for the use of e.g.
spherical type coordinates $(x^i)=(r,\theta,\varphi)$, for which 
$f=r^4\sin^2\theta$.

We define the \defin{inverse conformal metric} $\tgm^{ij}$ by
the requirement
\be
	\tgm_{ik} \, \tgm^{kj} = \delta_i^{\ \, j} ,
\ee
which is equivalent to
\be
	\tgm^{ij}  = \Psi^4 \, \gam^{ij} .
\ee
Hence, combining with Eq.~(\ref{e:cfd:def_tildegam}),  
\be \label{e:cfd:gmij_up_down}
	\encadre{ \gm_{ij} = \Psi^4 \tgm_{ij} } \qquad \mbox{and} \qquad
	\encadre{ \gam^{ij}  = \Psi^{-4} \, \tgm^{ij} } .
\ee
Note also that although we are using the same notation $\tgm$
for both $\tgm_{ij}$ and $\tgm^{ij}$, one has
\be
	\tgm^{ij} \not= \gam^{ik} \gam^{jl}\, \tgm_{kl} ,
\ee
except in the special case $\Psi=1$.
\begin{example} 
A simple example of a conformal decomposition is provided by the Schwarzschild
spacetime described with \emph{isotropic coordinates} $(x^\alpha)=(t,r,\theta,\varphi)$;
the latter are related to the standard \emph{Schwarzschild coordinates}
$(t,R,\theta,\varphi)$ by $R = r \left( 1 + \frac{m}{2r} \right) ^2$.
The components of the spacetime metric tensor in the isotropic coordinates
are given by (see e.g. 
\be \label{e:cfd:Schwarz_isotropic}
    g_{\mu\nu} dx^\mu dx^\nu  = - \left( 
    \frac{1 - \frac{m}{2r}}{1 + \frac{m}{2r}} \right) ^2
         dt^2 
    + \left( 1 + \frac{m}{2r} \right) ^4 \left[ d{r}^2 
    + {r}^2 (d\theta^2 + \sin^2\theta d\varphi^2) \right] , 
\ee
where the constant $m$ is the mass of the Schwarzschild solution. 
If we define the background metric to be $f_{ij} = \mathrm{diag}(1,r^2,r^2\sin^2\theta)$,
we read on this line element that $\wgm = \Psi^4 \wtgm$ with 
\be \label{e:cfd:Psi_Schwarz}
	\Psi = 1 + \frac{m}{2r}
\ee
and $\wtgm = \w{f}$. Notice that in this example, the background metric $\w{f}$
is flat and that the conformal metric coincides with the background metric. 
\end{example}

\begin{example} 
Another example is provided by the weak field metric introduced in 
Sec.~\ref{s:mat:Newt_lim} to take Newtonian limits. We read on the line element
(\ref{e:mat:gab_weak}) that the conformal metric is $\wtgm = \w{f}$ and
that the conformal factor is
\be \label{e:cfd:Psi_Newt_lim}
	\Psi = \left( 1 - 2\Phi \right) ^{1/4} \simeq 1 - \frac{1}{2} \Phi, 
\ee
where $|\Phi| \ll 1$ and $\Phi$ reduces to the gravitational potential at the Newtonian limit.
As a side remark, notice that if we identify expressions (\ref{e:cfd:Psi_Schwarz})
and (\ref{e:cfd:Psi_Newt_lim}), we recover the standard expression 
$\Phi = - m /r$ (remember $G=1$ !) for the Newtonian gravitational potential 
outside a spherical distribution of mass. 
\end{example}

\subsection{Conformal connection}

$\wtgm$ being a well defined metric on $\Sigma_t$, let $\wtD$ be the
Levi-Civita connection associated to it:
\be	
	\wtD \wtgm = 0 . 
\ee
Let us denote by $\tilde\Gamma^k_{\ \, ij}$ the Christoffel symbols of $\wtD$ with
respect to the coordinates $(x^i)$:
\be \label{e:cfd:christo_tgm}
	\tilde\Gamma^k_{\ \, ij} = {1\over 2} \tgm^{kl} 
		\left( \der{\tgm_{lj}}{x^i} +\der{\tgm_{il}}{x^j}
			- \der{\tgm_{ij}}{x^l} \right) .
\ee

Given a tensor field $\w{T}$ of type $\left( {p\atop q} \right)$ on $\Sigma_t$,
the covariant derivatives $\wtD\w{T}$ and $\wD \w{T}$ are related by the formula
\be \label{e:cfd:DT_tDT}
	D_k T^{i_1\ldots i_p}_{\quad \quad j_1\ldots j_q}
	= \tD_k T^{i_1\ldots i_p}_{\quad \quad j_1\ldots j_q}
	+ \sum_{r=1}^p C^{i_r}_{\ \, kl} \, 
		T^{i_1\ldots l \ldots i_p}_{\quad\quad \quad j_1\ldots j_q}
	- \sum_{r=1}^q C^l_{\ \, k j_r} \, 
		T^{i_1\ldots i_p}_{\quad \quad j_1\ldots l \ldots j_q} ,
\ee  
where\footnote{The $C^k_{\ \, ij}$ are not to be confused with the components
of the Cotton tensor discussed in Sec.~\ref{s:cfd:intro}. 
Since we shall no longer make use of the latter, no
confusion may arise.}
\be \label{e:cfd:Ckij_diffGam}
	C^k_{\ \, ij} := \Gamma^k_{\ \, ij}-\tilde\Gamma^k_{\ \, ij} ,
\ee
$\Gamma^k_{\ \, ij}$ being the Christoffel symbols of the connection $\wD$. 
The formula (\ref{e:cfd:DT_tDT}) follows immediately from the expressions
of $\wD\w{T}$ and $\wtD\w{T}$ in terms of respectively the Christoffel symbols
$\Gamma^k_{\ \, ij}$ and $\tilde\Gamma^k_{\ \, ij}$.
Since
$D_k T^{i_1\ldots i_p}_{\quad \quad j_1\ldots j_q}-\tD_k T^{i_1\ldots i_p}_{\quad \quad j_1\ldots j_q}$ are the components of a tensor field, namely $\wD\w{T}-\wtD\w{T}$,
it follows from Eq.~(\ref{e:cfd:DT_tDT}) that the $C^k_{\ \, ij}$ are also the components
of a tensor field. Hence we recover a well known property: although the Christoffel 
symbols are not the components of any tensor field, the difference between two sets of
them represents the components of a tensor field.
We may express the tensor $C^k_{\ \, ij}$ in terms of the $\wtD$-derivatives
of the metric $\wgm$, by the same formula than the one
for the Christoffel symbols $\Gamma^k_{\ \, ij}$, except that the partial derivatives
are replaced by $\wtD$-derivatives:
\be \label{e:cfd:Ckij_tDgam}
	C^k_{\ \, ij} = \frac{1}{2} \gm^{kl} \left(
	\tD_i \gm_{lj} + \tD_j \gm_{il} - \tD_l \gm_{ij} \right) . 
\ee
It is easy to establish this relation by evaluating the right-hand side, expressing
the $\wtD$-derivatives of $\wgm$ in terms of the Christoffel symbols 
$\tilde\Gamma^k_{\ \, ij}$:
\bea
	\frac{1}{2} \gm^{kl} \left(
	\tD_i \gm_{lj} + \tD_j \gm_{il} - \tD_l \gm_{ij} \right) & = &
	\frac{1}{2} \gm^{kl} \bigg( 
	\der{\gm_{lj}}{x^i} - \tilde\Gamma^m_{\ \, il} \gm_{mj}
	- \tilde\Gamma^m_{\ \, ij} \gm_{lm}
	+ \der{\gm_{il}}{x^j} - \tilde\Gamma^m_{\ \, ji} \gm_{ml}
	- \tilde\Gamma^m_{\ \, jl} \gm_{im} \nonumber \\
	& & 
	- \der{\gm_{ij}}{x^l} + \tilde\Gamma^m_{\ \, li} \gm_{mj}
	+ \tilde\Gamma^m_{\ \, lj} \gm_{im} \bigg) \nonumber \\
	& = & \Gamma^k_{\ \, ij} + \frac{1}{2} \gm^{kl} (-2) 
	\tilde\Gamma^m_{\ \, ij} \gm_{lm} \nonumber \\
	& = & \Gamma^k_{\ \, ij} - \delta^k_{\ \, m} \tilde\Gamma^m_{\ \, ij} \nonumber \\
	& = & C^k_{\ \, ij} ,
\eea
where we have used the symmetry with respect to $(i,j)$ of the Christoffel symbols 
$\tilde\Gamma^k_{\ \, ij}$ to get the second line.

Let us replace $\gm_{ij}$ and $\gm^{ij}$ in Eq.~(\ref{e:cfd:Ckij_tDgam}) by their expressions (\ref{e:cfd:gmij_up_down}) in terms of $\tgm_{ij}$, $\tgm^{ij}$ 
and $\Psi$:
\bea
	 C^k_{\ \, ij} & = & \frac{1}{2} \Psi^{-4} \tgm^{kl} \left[
	\tD_i (\Psi^4 \tgm_{lj}) + \tD_j (\Psi^4\gm_{il}) 
	- \tD_l (\Psi^4\tgm_{ij}) \right] \nonumber \\ 
 & = & \frac{1}{2} \Psi^{-4} \tgm^{kl} \left( \tgm_{lj} \tD_i \Psi^4
	+ \tgm_{il} \tD_j \Psi^4 - \tgm_{ij} \tD_l \Psi^4 \right) \nonumber\\
  &= & \frac{1}{2} \Psi^{-4} \left( \delta^k_{\ \, j} \tD_i \Psi^4
	+ \delta^k_{\ \, i} \tD_j \Psi^4 - \tgm_{ij} \tD^k \Psi^4 \right) \nonumber
\eea
Hence
\be \label{e:cfd:Ckij_derPsi}
	\encadre{ C^k_{\ \, ij} = 2 \left( 
	\delta^k_{\ \, i} \tD_j \ln\Psi + \delta^k_{\ \, j} \tD_i \ln\Psi
	- \tD^k \ln\Psi \, \tgm_{ij} \right) } . 
\ee

A usefull application of this formula is to derive the relation between
the two covariant derivatives $\w{D}\w{v}$ and $\wtD \w{v}$ of a vector
field $\w{v}\in\T(\Sigma_t)$. From Eq.~(\ref{e:cfd:DT_tDT}), we have
\be
	D_j v^i = \tD_j v^i + C^i_{\ \, jk} v^k , 
\ee
so that expression (\ref{e:cfd:Ckij_derPsi}) yields
\be \label{e:cfd:der_vector}
	D_j v^i = \tD_j v^i + 2 \left( v^k \tD_k \ln\Psi \, \delta^i_{\ \, j}
		+ v^i \tD_j\ln\Psi - \tD^i\ln\Psi\, \tgm_{jk} v^k \right) .
\ee
Taking the trace, we get a relation between the two divergences:
\be \label{e:cfd:divergence_conf0}
	 D_i v^i = \tD_i v^i + 6 v^i \tD_i \ln \Psi , 
\ee
or equivalently, 
\be \label{e:cfd:divergence_conf}
	\encadre{ D_i v^i = \Psi^{-6} \tD_i \left( \Psi^6 v^i \right) } . 
\ee
\begin{remark}
The above formula could have been obtained directly from the standard
expression of the divergence of a vector field in terms of partial derivatives
and the determinant $\gm$ of $\wgm$, both with respect to some coordinate
system $(x^i)$:
\be
	D_i v^i = \frac{1}{\sqrt{\gm}} \der{}{x^i} \left( \sqrt{\gm} v^i \right) . 
\ee
Noticing that $\gm_{ij} = \Psi^4 \tgm_{ij}$ implies $\sqrt{\gm} = \Psi^6 \sqrt{\tgm}$,
we get immediately Eq.~(\ref{e:cfd:divergence_conf}). 
\end{remark}

%%%%%%%%%%%%%%%%%%%%%%%%%%%%%%%%%%%%%%%%%%%%%%%%%%%%%%%%%%%%%%%%%%%%%%%%%%%%%%%%%%%%%

\section{Expression of the Ricci tensor}

In this section, we express the Ricci tensor $\w{R}$ which appears in the 3+1 Einstein system
(\ref{e:dec:Einstein_PDE1})-(\ref{e:dec:Einstein_PDE4}), in terms of the Ricci
tensor $\w{\tilde R}$ associated with the metric $\wtgm$ and derivatives of the
conformal factor $\Psi$.

\subsection{General formula relating the two Ricci tensors} \label{s:cfd:2Ricci}

The starting point of the calculation is the Ricci
identity (\ref{e:hyp:Ricci_ident_3D}) applied to a generic vector field
$\w{v}\in\T(\Sigma_t)$:
\be
	(D_i D_j - D_j D_i) v^k = R^k_{\ \, lij} \, v^l  .
\ee
Contracting this relation on the indices $i$ and $k$ (and relabelling 
$i\leftrightarrow j$) let appear the Ricci tensor:
\be \label{e:cfd:contracted_Ricci_ident}
	R_{ij} v^j = D_j D_i v^j - D_i D_j v^j . 
\ee
Expressing the $\wD$-derivatives in term of the $\wtD$-derivatives via formula
(\ref{e:cfd:DT_tDT}), we get
\bea
	R_{ij} v^j & = & \tD_j (D_i v^j) - C^k_{\ \, ji} D_k v^j
	+ C^j_{\ \, jk} D_i v^k - \tD_i (D_j v^j) \nonumber \\
	& = & \tD_j (\tD_i v^j + C^j_{\ \, ik} v^k)
	- C^k_{\ \, ji} (\tD_k v^j + C^j_{\ \, kl} v^l)
	+ C^j_{\ \, jk} (\tD_i v^k + C^k_{\ \, il} v^l)
	- \tD_i( \tD_j v^j + C^j_{\ \, jk} v^k ) \nonumber \\
	& = & \tD_j \tD_i v^j + \tD_j C^j_{\ \, ik} \, v^k
	+ C^j_{\ \, ik} \tD_j v^k
	- C^k_{\ \, ji} \tD_k v^j - C^k_{\ \, ji} C^j_{\ \, kl} v^l
	+ C^j_{\ \, jk} \tD_i v^k + C^j_{\ \, jk} C^k_{\ \, il} v^l
	\nonumber \\
	& & - \tD_i \tD_j v^j - \tD_i C^j_{\ \, jk} \, v^k
	- C^j_{\ \, jk} \tD_i v^k \nonumber \\
	& = & \tD_j \tD_i v^j - \tD_i \tD_j v^j 
	+ \tD_j C^j_{\ \, ik} \, v^k - C^k_{\ \, ji} C^j_{\ \, kl} v^l
	+ C^j_{\ \, jk} C^k_{\ \, il} v^l  - \tD_i C^j_{\ \, jk} \, v^k . 
		\label{e:cfd:Ricci_prov}
\eea
We can replace the first two terms in the right-hand side via the contracted
Ricci identity similar to Eq.~(\ref{e:cfd:contracted_Ricci_ident}) but regarding
the connection $\wtD$:
\be \label{e:cfd:contr_Ricci_tD}
	\tD_j \tD_i v^j - \tD_i \tD_j v^j = \tilde R_{ij} v^j
\ee
Then, after some relabelling $j\leftrightarrow k$ or $j\leftrightarrow l$
of dumb indices, Eq.~(\ref{e:cfd:Ricci_prov}) becomes 
\be
	R_{ij} v^j = \tilde R_{ij} v^j + \tD_k C^k_{\ \, ij} \, v^j
	- \tD_i C^k_{\ \, jk} \, v^j
	+  C^l_{\ \, lk} C^k_{\ \, ij} v^j
	-  C^k_{\ \, li} C^l_{\ \, kj} v^j . 
\ee
This relation being valid for any vector field $\w{v}$, we conclude that
\be
	\encadre{ 
	R_{ij} = \tilde R_{ij}  + \tD_k C^k_{\ \, ij} 
	- \tD_i C^k_{\ \, kj} 
	+   C^k_{\ \, ij} C^l_{\ \, lk}
	-  C^k_{\ \, il} C^l_{\ \, kj} } , \label{e:cfd:Ricci1}
\ee
where we have used the symmetry of $C^k_{\ \, ij}$ in its two last indices. 
\begin{remark}
Eq.~(\ref{e:cfd:Ricci1}) is the general formula relating the Ricci tensors
of two connections, with the $C^k_{\ \, ij}$'s being the differences
of their Christoffel symbols [Eq.~(\ref{e:cfd:Ckij_diffGam})]. 
This formula does not rely on the fact that the metrics $\wgm$ and $\wtgm$
associated with the two connections are conformally related.
\end{remark}

\subsection{Expression in terms of the conformal factor}

Let now replace $C^k_{\ \, ij}$ in Eq.~(\ref{e:cfd:Ricci1}) by its expression 
in terms of the derivatives of $\Psi$, i.e. Eq.~(\ref{e:cfd:Ckij_derPsi}).
First of all, by contracting Eq.~(\ref{e:cfd:Ckij_derPsi}) on the indices $j$
and $k$, we have
\be
	C^k_{\ \, ki} = 2 \left( 
	\tD_i \ln\Psi + 3 \tD_i \ln\Psi
	- \tD_i \ln\Psi \right) ,
\ee
i.e. 
\be \label{e:cfd:Ckki}
	C^k_{\ \, ki} = 6 \tD_i \ln\Psi, 
\ee
whence $\tD_i C^k_{\ \, kj} = 6 \tD_i \tD_j \ln\Psi$. 
Besides,
\bea
	\tD_k C^k_{\ \, ij} & = & 2 \left( \tD_i \tD_j \ln \Psi
	+ \tD_j \tD_i \ln\Psi - \tD_k \tD^k \ln \Psi \, \tgm_{ij} \right) \nonumber \\ 
	& = & 4  \tD_i \tD_j \ln \Psi - 2 \tD_k \tD^k \ln \Psi \, \tgm_{ij} .
\eea
Consequently, Eq.~(\ref{e:cfd:Ricci1}) becomes
\bea
	R_{ij} & = & \tilde R_{ij} + 4  \tD_i \tD_j \ln \Psi 
		- 2 \tD_k \tD^k \ln \Psi \, \tgm_{ij} - 6 \tD_i \tD_j \ln\Psi \nonumber \\
	& & + 2 \left( 
	\delta^k_{\ \, i} \tD_j \ln\Psi + \delta^k_{\ \, j} \tD_i \ln\Psi
	- \tD^k \ln\Psi \, \tgm_{ij} \right) \times 6 \tD_k \ln \Psi \nonumber \\
	& & - 4  \left( 
	\delta^k_{\ \, i} \tD_l \ln\Psi + \delta^k_{\ \, l} \tD_i \ln\Psi
	- \tD^k \ln\Psi \, \tgm_{il} \right)
	\left( 
	\delta^l_{\ \, k} \tD_j \ln\Psi + \delta^l_{\ \, j} \tD_k \ln\Psi
	- \tD^l \ln\Psi \, \tgm_{kj} \right) . \nonumber 
\eea
Expanding and simplifying, we get
\be
	\encadre{R_{ij} = \tilde R_{ij} - 2 \tD_i \tD_j \ln \Psi
	- 2 \tD_k \tD^k \ln \Psi \, \tgm_{ij}
	+ 4 \tD_i \ln \Psi \, \tD_j \ln \Psi 
	- 4 \tD_k \ln \Psi \, \tD^k \ln \Psi \, \tgm_{ij} } . \label{e:cfd:Ricci_Psi}
\ee

\subsection{Formula for the scalar curvature}

The relation between the scalar curvatures is obtained by taking the trace
of Eq.~(\ref{e:cfd:Ricci_Psi}) with respect to $\wgm$:
\bea
	R & = & \gm^{ij} R_{ij} = \Psi^{-4} \tgm^{ij} R_{ij} \nonumber \\
	  & = & \Psi^{-4} \left( \tgm^{ij} \tilde R_{ij} -2 \tD_i \tD^i \ln \Psi
	- 2 \tD_k \tD^k \ln \Psi \times 3
	+ 4 \tD_i \ln \Psi \, \tD^i \ln \Psi 
	- 4 \tD_k \ln \Psi \, \tD^k \ln \Psi \times 3 \right) \nonumber \\
	R & = & \Psi^{-4} \left[ \tilde R - 8 \left(
	 \tD_i \tD^i \ln \Psi + \tD_i \ln\Psi \, \tD^i \ln \Psi\right) \right] , 
			\label{e:cfd:Ricci_scal0}
\eea
where 
\be
	\encadre{\tilde R := \tgm^{ij} \tilde R_{ij} }
\ee
is the scalar curvature associated with the conformal metric. 
Noticing that 
\be
	\tD_i \tD^i \ln \Psi = \Psi^{-1} \tD_i \tD^i \Psi 
	- \tD_i \ln\Psi \, \tD^i \ln \Psi , 
\ee
we can rewrite the above formula as
\be \label{e:cfd:Ricci_scal}
	\encadre{ R = \Psi^{-4} \tilde R - 8 \Psi^{-5} \tD_i \tD^i \Psi  } . 
\ee

%%%%%%%%%%%%%%%%%%%%%%%%%%%%%%%%%%%%%%%%%%%%%%%%%%%%%%%%%%%%%%%%%%%%%%%%%%%%%%%%%%%%%

\section{Conformal decomposition of the extrinsic curvature}

\subsection{Traceless decomposition}

The first step is to decompose the extrinsic curvature $\w{K}$ of the hypersurface
$\Sigma_t$ into a trace part and a traceless one, the trace being taken with 
the metric $\wgm$, i.e. we define
\be \label{e:cfd:A_def}
	\w{A} := \w{K} - \frac{1}{3} K \wgm ,  
\ee
where $K:=\mathrm{tr}_{\wgm}\,  \w{K} = K^i_{\ \, i}=\gm^{ij} K_{ij}$ is the trace of $\w{K}$ with respect 
to $\wgm$, i.e. (minus three times) the mean curvature of $\Sigma_t$ embedded
in $(\M,\w{g})$ (cf. Sec.~\ref{s:hyp:extr_curv}).
The bilinear form $\w{A}$ is by construction traceless:
\be \label{e:cfd:A_traceless}
	\mathrm{tr}_{\wgm}\,  \w{A} = \gm^{ij} A_{ij} = 0 .
\ee

In what follows, we shall work occasionally with the twice contravariant 
version of $\w{K}$, i.e. the tensor $\stackrel{\twoheadrightarrow}{\w{K}}$, the components of which 
are\footnote{The double arrow is extension of the single
arrow notation introduced in Sec.~\ref{s:hyp:metric_dual} 
[cf. Eq.~(\ref{e:hyp:arrow_endo})].}
\be
	K^{ij} = \gm^{ik} \gm^{jl} K_{kl} . 
\ee
Similarly, we define $\stackrel{\twoheadrightarrow}{\w{A}}$ as the twice contravariant
tensor, the components of which are
\be
	A^{ij} = \gm^{ik} \gm^{jl} A_{kl} .
\ee
Hence the traceless decomposition of $\w{K}$ and $\stackrel{\twoheadrightarrow}{\w{K}}$:
\be \label{e:cfd:K_A}
	\encadre{K_{ij} = A_{ij} + \frac{1}{3} K \gm_{ij}}
	\qquad \mbox{and} \qquad
	\encadre{K^{ij} = A^{ij} + \frac{1}{3} K \gm^{ij}} .
\ee

\subsection{Conformal decomposition of the traceless part} \label{s:cfd:conf_traceless}

Let us now perform the conformal decomposition of the traceless part of
$\w{K}$, namely, let us write 
\be \label{e:cfd:A_scale}
	A^{ij} = \Psi^\alpha {\tilde A}^{ij}
\ee 
for some power $\alpha$ to be determined. Actually there are two natural choices:
$\alpha=-4$ and $\alpha=-10$, as we discuss hereafter:

\subsubsection{1) ``Time-evolution'' scaling: $\alpha=-4$} 

Let us consider Eq.~(\ref{e:fol:Lie_m_gam}) which express the time evolution of
the $\wgm$ in terms of $\w{K}$: 
\be \label{e:cfd:Liem_gam}
	\encadre{ \Liec{m} \gm_{ij} = - 2 N K_{ij} } . 
\ee
By means of Eqs.~(\ref{e:cfd:gmij_up_down}) and (\ref{e:cfd:K_A}), this 
equation becomes
\be
	\Liec{m} \left( \Psi^4 \tgm_{ij} \right)
	= - 2N  A_{ij} - \frac{2}{3} NK \gam_{ij} ,
\ee
i.e. 
\be \label{e:cfd:Liem_tgmup0}
	\Liec{m} \tgm_{ij} = -2N \Psi^{-4} A_{ij} - \frac{2}{3} \left( 
	NK + 6 \Liec{m} \ln \Psi \right) \tgm_{ij} . 
\ee
The trace of this relation with respect to $\wtgm$ is, since $A_{ij}$ is
traceless,
\be \label{e:cfd:trace_Lmtgup}
	\tgm^{ij} \Liec{m} \tgm_{ij} = - 2 (NK + 6 \Liec{m}\ln\Psi) . 
\ee
Now 
\be \label{e:cfd:tgLtg_det}
	\tgm^{ij} \Liec{m} \tgm_{ij}  =  \Liec{m} \ln \det(\tgm_{ij}) . 
\ee
This follows from the general law of variation of the determinant of any
invertible matrix $A$:
\be \label{e:cfd:variation_det}
	\encadre{ \delta(\ln \det A) = \mathrm{tr} (A^{-1} \times \delta A) } , 
\ee
where $\delta$ denotes any variation (derivative) that fulfills the Leibniz rule,
$\mathrm{tr}$ stands for the trace and $\times$ for the matrix product. 
Applying Eq.~(\ref{e:cfd:variation_det}) to $A=(\tgm_{ij})$ and $\delta=\Liec{m}$
gives Eq.~(\ref{e:cfd:tgLtg_det}). 
By construction, $ \det (\tgm_{ij}) = f $ [Eq.~(\ref{e:cfd:dettgm_f})], so 
that, replacing $\w{m}$ by $\wpar_t - \w{\beta}$, we get 
\be
	\Liec{m} \ln \det(\tgm_{ij}) = \left( \der{}{t} - \Liec{\beta} \right)  \ln f
\ee
But, as a consequence of Eq.~(\ref{e:cfd:f_const}),
$\dert{f}{t}=0$, so that
\be
	\Liec{m} \ln \det(\tgm_{ij}) = - \Liec{\beta} \ln f =
	- \Liec{\beta}\ln\det(\tgm_{ij}).
\ee
Applying again formula (\ref{e:cfd:variation_det}) to $A=(\tgm_{ij})$ and 
$\delta=\Liec{\beta}$, we get
\bea
	\Liec{m} \ln \det(\tgm_{ij}) & = & 
	- \tgm^{ij} \Liec{\beta} \tgm_{ij} \nonumber \\
	& = &  - \tgm^{ij} \Big( \beta^k \underbrace{\tD_k\tgm_{ij}}_{=0}
	+ \tgm_{kj} \tD_i \beta^k + \tgm_{ik} \tD_j \beta^k \Big) \nonumber \\
	& = & - \delta^i_{\ \, k} \tD_i \beta^k - \delta^j_{\ \, k} \tD_j \beta^k
	\nonumber \\
	& = & - 2 \tD_i \beta^i . 
\eea
Hence Eq.~(\ref{e:cfd:tgLtg_det}) becomes
\be
	\tgm^{ij} \Liec{m} \tgm_{ij}  = - 2 \tD_i \beta^i,
\ee
so that, after substitution into Eq.~(\ref{e:cfd:trace_Lmtgup}), we get
\be \label{e:cfd:NK_LiePsi}
	NK + 6 \Liec{m}\ln\Psi = \tD_i \beta^i ,  
\ee
i.e. the following evolution equation for the conformal factor:
\be \label{e:cfd:Liem_lnPsi}
	\encadre{ \left( \der{}{t} - \Liec{\beta} \right) \ln \Psi =
	\frac{1}{6} \left( \tD_i \beta^i  - NK \right) } . 
\ee
Finally, substituting Eq.~(\ref{e:cfd:NK_LiePsi}) into Eq.~(\ref{e:cfd:Liem_tgmup0})
yields an evolution equation for the conformal metric:
\be
	 \left( \der{}{t} - \Liec{\beta} \right) \tgm_{ij}
	= -2N \Psi^{-4} A_{ij} - \frac{2}{3} \tD_k \beta^k \, \tgm_{ij} . 
\ee
This suggests to introduce the quantity
\be \label{e:cfd:tA_down}
	\encadre{\tA_{ij} := \Psi^{-4} A_{ij}}
\ee
to write
\be \label{e:cfd:evol_tgm}
	\encadre{ \left( \der{}{t} - \Liec{\beta} \right) \tgm_{ij}
	= -2N  \tA_{ij} - \frac{2}{3} \tD_k \beta^k \, \tgm_{ij} } . 
\ee
Notice that, as an immediate consequence of Eq.~(\ref{e:cfd:A_traceless}),
$\tA_{ij}$ is traceless:
\be \label{e:cfd:tA_traceless}
	\encadre{\tgm^{ij} \tA_{ij} = 0 } . 
\ee
Let us rise the indices of $\tA_{ij}$ with the conformal metric,
defining 
\be
  \tA^{ij} := \tgm^{ik} \tgm^{jl} \tA_{kl}.
\ee
Since $\tgm^{ij} = \Psi^4 \gm^{ij}$, we get 
\be \label{e:cfd:def_tA}
	\encadre{ \tA^{ij} = \Psi^4 A^{ij} } . 
\ee
This corresponds to the scaling factor
$\alpha=-4$ in Eq.~(\ref{e:cfd:A_scale}). 
This choice of scaling has been first considered by Nakamura in 1994 \cite{Nakam94}. 

We can deduce from Eq.~(\ref{e:cfd:evol_tgm}) an evolution equation for
the inverse conformal metric $\tgm^{ij}$. Indeed, raising the indices of 
Eq.~(\ref{e:cfd:evol_tgm}) with $\wtgm$, we get
\bea
	\tgm^{ik} \tgm^{jl} \Liec{m} \tgm_{kl}
	& = & - 2N \tA^{ij} - \frac{2}{3} \tD_k \beta^k \, \tgm^{ij} \nonumber \\
	\tgm^{ik} \big[ \Liec{m} (  
	\underbrace{\tgm^{jl}\tgm_{kl}}_{=\delta^j_{\ \, k}})
	- \tgm_{kl} \Liec{m} \tgm^{jl} \big]
	& = & - 2N \tA^{ij} - \frac{2}{3} \tD_k \beta^k \, \tgm^{ij} \nonumber \\
	- \underbrace{ \tgm^{ik} \tgm_{kl} }_{=\delta^i_{\ \, l}}
	\Liec{m} \tgm^{jl} 
	& = & - 2N \tA^{ij} - \frac{2}{3} \tD_k \beta^k \, \tgm^{ij} , 
\eea
hence
\be \label{e:cfd:evol_tgmup}
	\encadre{ \left( \der{}{t} - \Liec{\beta} \right) \tgm^{ij}
	= 2N  \tA^{ij} + \frac{2}{3} \tD_k \beta^k \, \tgm^{ij} } . 
\ee

\subsubsection{2) ``Momentum-constraint'' scaling: $\alpha=-10$}

Whereas the scaling $\alpha=-4$ was suggested by the evolution equation (\ref{e:cfd:Liem_gam}) (or equivalently Eq.~(\ref{e:dec:Einstein_PDE1})
of the 3+1 Einstein system), another scaling arises when contemplating the
momentum constraint equation (\ref{e:dec:Einstein_PDE4}). In this equation
appears the divergence
of the extrinsic curvature, that we can write using the twice contravariant version
of $\w{K}$ and Eq.~(\ref{e:cfd:K_A}):
\be \label{e:cfd:divK_A_K}
	D_j K^{ij} = D_j A^{ij} + \frac{1}{3} D^i K .
\ee
Now, from Eqs.~(\ref{e:cfd:DT_tDT}), (\ref{e:cfd:Ckij_derPsi}) and (\ref{e:cfd:Ckki}), 
\bea
	D_j A^{ij} & = & \tD_j A^{ij} + C^i_{\ \, jk} A^{kj}
	+ C^j_{\ \, jk} A^{ik} \nonumber \\
	& = & \tD_j A^{ij} + 2 \left( 
	\delta^i_{\ \, j} \tD_k \ln\Psi + \delta^i_{\ \, k} \tD_j \ln\Psi
	- \tD^i \ln\Psi \, \tgm_{jk} \right) A^{kj}
	+ 6 \tD_k \ln\Psi \, A^{ik} \nonumber \\
	& = & \tD_j A^{ij} + 10 A^{ij} \tD_j \ln\Psi 
	- 2 \tD^i \ln \Psi \, \tgm_{jk} A^{jk} . 
\eea
Since $\w{A}$ is traceless, $\tgm_{jk} A^{jk} = \Psi^{-4} \gm_{jk} A^{jk} = 0$.
Then the above equation reduces to $D_j A^{ij} = \tD_j A^{ij} + 10 A^{ij} \tD_j \ln\Psi $,
which can be rewritten as
\be \label{e:cfd:divA_Psi10}
	D_j A^{ij} = \Psi^{-10} \tD_j \left( \Psi^{10} A^{ij} \right) . 
\ee
Notice that this identity is valid only because $A^{ij}$ is symmetric and traceless.

Equation~(\ref{e:cfd:divA_Psi10}) suggests to introduce the quantity\footnote{notice that we have used a hat, instead of a tilde, to distinguish this quantity from that 
defined by (\ref{e:cfd:def_tA})}
\be \label{e:cfd:def_hA}
	\encadre{\hA^{ij} := \Psi^{10} A^{ij} } .
\ee
This corresponds to the scaling factor $\alpha=-10$ in Eq.~(\ref{e:cfd:A_scale}). 
It has been first introduced by Lichnerowicz in 1944 \cite{Lichn44}. 
Thanks to it and Eq.~(\ref{e:cfd:divK_A_K}), the momentum constraint equation
(\ref{e:dec:Einstein_PDE4}) can be rewritten as 
\be \label{e:cfd:mom_constr_hA1}
	\encadre{ \tD_j \hA^{ij} - \frac{2}{3} \Psi^6 \tD^i K = 8\pi \Psi^{10} p^i } . 
\ee

As for $\tA_{ij}$, we define $\hA_{ij}$ as the tensor field deduced from $\hA^{ij}$
by lowering the indices with the conformal metric:
\be
	\hA_{ij} := \tgm_{ik} \tgm_{jl} \hA^{kl} 
\ee
Taking into account Eq.~(\ref{e:cfd:def_hA}) and $\tgm_{ij} = \Psi^{-4} \gm_{ij}$,
we get
\be \label{e:cfd:hA_down}
	\encadre{\hA_{ij} = \Psi^2 A_{ij} } . 
\ee

%%%%%%%%%%%%%%%%%%%%%%%%%%%%%%%%%%%%%%%%%%%%%%%%%%%%%%%%%%%%%%%%%%%%%%%%%%%%%%%%%%%%%

\section{Conformal form of the 3+1 Einstein system}

Having performed a conformal decomposition of $\wgm$ and of the traceless
part of $\w{K}$, we are now in position to rewrite the 3+1 Einstein system
(\ref{e:dec:Einstein_PDE1})-(\ref{e:dec:Einstein_PDE4}) in terms
of conformal quantities. 

\subsection{Dynamical part of Einstein equation} \label{s:cfd:dyn_part_Einstein}

Let us consider Eq.~(\ref{e:dec:Einstein_PDE2}), i.e. the so-called dynamical
equation in the 3+1 Einstein system:
\be \label{e:cfd:Liem_Kij}
	\Liec{m} K_{ij} 
	= - D_i D_j N + N\left\{
	R_{ij} + K K_{ij} -2 K_{ik} K^k_{\ \, j} 
	+ 4\pi \left[ (S-E) \gam_{ij} - 2 S_{ij} \right] \right\} .
\ee
Let us substitute $A_{ij} + (K/3) \gm_{ij}$ for $K_{ij}$ [Eq.~(\ref{e:cfd:K_A})].
The left-hand side of the above equation becomes
\be \label{e:cfd:Liem_K_A}
	\Liec{m} K_{ij} = \Liec{m} A_{ij} + \frac{1}{3} \Liec{m} K \, \gam_{ij}
	+ \frac{1}{3} K \underbrace{\Liec{m} \gm_{ij}}_{=-2N K_{ij}} .
\ee
In this equation appears $\Liec{m} K$. We may express it by taking the trace
of Eq.~(\ref{e:cfd:Liem_Kij}) and making use of Eq.~(\ref{e:fol:trLiemK}):
\be
	\Liec{m} K  = \gm^{ij} \Liec{m} K_{ij} + 2N K_{ij} K^{ij} , 
\ee
hence
\be \label{e:cfd:evol_K0}
	\Liec{m} K = - D_i D^i N + N \left[ R + K^2 + 4\pi (S - 3E) \right] .
\ee
Let use the Hamiltonian constraint (\ref{e:dec:Einstein_PDE3}) to replace
$R+K^2$ by $16\pi E + K_{ij} K^{ij}$. Then, writing $\Liec{m} K = (\dert{}{t}
- \Liec{\beta}) K$, 
\be \label{e:cfd:evol_K}
	\encadre{ \left(\der{}{t} - \Liec{\beta} \right) K 
	= - D_i D^i N + N \left[ 4\pi (E+S) 
	+ K_{ij} K^{ij} \right] } . 
\ee
\begin{remark}
At the Newtonian limit, as defined by Eqs.~(\ref{e:mat:Newt_limit}), 
(\ref{e:mat:SllE_Newt}) and (\ref{e:mat:Newt_fluid}), Eq.~(\ref{e:cfd:evol_K})
reduces to the Poisson equation for the gravitational potential $\Phi$:
\be
	\Df_i \Df^i \Phi = 4\pi \rho_0 . 
\ee 
\end{remark}

Substituting Eq.~(\ref{e:cfd:evol_K0}) for $\Liec{m} K$ and Eq.~(\ref{e:cfd:Liem_Kij})
for $\Liec{m} K_{ij}$ into Eq.~(\ref{e:cfd:Liem_K_A}) yields
\bea
	\Liec{m} A_{ij} & = & - D_i D_j N + N\left[
	R_{ij} + \frac{5}{3} K K_{ij} - 2 K_{ik} K^k_{\ \, j} 
	- 8\pi \left( S_{ij} - \frac{1}{3}S  \gm_{ij} \right) \right] \nonumber \\
	& &+ \frac{1}{3} \left[ D_k D^k N - N (R+K^2) \right] \gm_{ij} . 
		\label{e:cfd:Liem_Aij0}	
\eea
Let us replace $K_{ij}$ by its expression in terms of $A_{ij}$ and $K$ 
[Eq.~(\ref{e:cfd:K_A})]: the terms in the right-hand side of the above equation
which involve $\w{K}$ are then written
\bea
	\frac{5K}{3} K^{ij} - 2 K_{ik} K^k_{\ \, j} - \frac{K^2}{3} \gm_{ij}
	& = & \frac{5K}{3} \left( A_{ij} + \frac{K}{3}  \gm_{ij} \right)
	- 2 \left( A_{ik} + \frac{K}{3}  \gm_{ik} \right)
	\left( A^k_{\ \, j} + \frac{K}{3}  \delta^k_{\ \, j}  \right) 
	- \frac{K^2}{3} \gm^{ij} \nonumber \\
	& = & \frac{5K}{3}  A_{ij} + \frac{5K^2}{9} \gm_{ij}
	- 2 \left( A_{ik} A^k_{\ \, j} + \frac{2K}{3} A_{ij} + \frac{K^2}{9} \gm_{ij}
	\right)
	- \frac{K^2}{3} \gm_{ij} \nonumber \\
	& = & \frac{1}{3} K A_{ij} - 2 A_{ik} A^k_{\ \, j} . 
\eea
Accordingly Eq.~(\ref{e:cfd:Liem_Aij0}) becomes
\bea
	\Liec{m} A_{ij} & = & - D_i D_j N + N\left[
	R_{ij} + \frac{1}{3} K A_{ij} - 2 A_{ik} A^k_{\ \, j} 
	- 8\pi \left( S_{ij} - \frac{1}{3}S  \gm_{ij} \right) \right] \nonumber \\
	& &+ \frac{1}{3} \left( D_k D^k N - N R \right) \gm_{ij} . 
		\label{e:cfd:Liem_Aij}	
\eea
\begin{remark}
Regarding the matter terms, this equation involves only the stress tensor $\w{S}$
(more precisely its traceless part) and not the energy density $E$, contrary
to the evolution equation (\ref{e:cfd:Liem_Kij}) for $K^{ij}$, which involves both. 
\end{remark}

At this stage, we may say that we have split the dynamical Einstein equation
(\ref{e:cfd:Liem_Kij}) in two parts: a trace part: Eq.~(\ref{e:cfd:evol_K})
and a traceless part: Eq.~(\ref{e:cfd:Liem_Aij}). 
Let us now perform the conformal decomposition of these relations, by introducing
$\tA_{ij}$. We consider $\tA_{ij}$ and not $\hA_{ij}$, i.e. the scaling
$\alpha=-4$ and not $\alpha=-10$, since we are discussing time evolution equations.

Let us first transform Eq.~(\ref{e:cfd:evol_K}).
We can express the Laplacian of the lapse
by applying the divergence relation (\ref{e:cfd:divergence_conf})
to the vector $v^i = D^i N = \gm^{ij} D_j N = \Psi^{-4} \tgm^{ij} \tD_j N
= \Psi^{-4} \tD^i N$
\bea
	D_i D^i N & = & \Psi^{-6} \tD_i \left( \Psi^6 D^i N \right) =
	 \Psi^{-6} \tD_i \left( \Psi^2 \tD^i N \right) \nonumber \\
	& = & \Psi^{-4} \left( \tD_i \tD^i N + 2 \tD_i \ln \Psi \, \tD^i N \right) .
	\label{e:cfd:laplacien_N}
\eea
Besides, from Eqs.~(\ref{e:cfd:K_A}), (\ref{e:cfd:tA_down}) and
(\ref{e:cfd:def_tA}),
\be \label{e:cfd:KijKij}
	K_{ij} K^{ij} = \left( A_{ij} + \frac{K}{3} \gm_{ij} \right)
	\left( A^{ij} + \frac{K}{3} \gm^{ij} \right)
	= A_{ij} A^{ij} + \frac{K^2}{3} = \tA_{ij} \tA^{ij} + \frac{K^2}{3} .
\ee
In view of Eqs.~(\ref{e:cfd:laplacien_N}) and (\ref{e:cfd:KijKij}), 
Eq.~(\ref{e:cfd:evol_K}) becomes
\be \label{e:cfd:evol_K_conf}
	\encadre{ \left(\der{}{t} - \Liec{\beta} \right) K 
	= - \Psi^{-4} \left( \tD_i \tD^i N + 2 \tD_i \ln \Psi \, \tD^i N \right) 
	 + N \left[ 4\pi (E+S) 
	+  \tA_{ij} \tA^{ij} + \frac{K^2}{3}\right] } . 
\ee

Let us now consider the traceless part, Eq.~(\ref{e:cfd:Liem_Aij}). 
We have, writing $A_{ij} = \Psi^4 \tA_{ij}$ and using Eq.~(\ref{e:cfd:Liem_lnPsi}),
\be \label{e:cfd:Liem_Aij_Psi}
	\Liec{m} A_{ij} = \Psi^4 \Liec{m} \tA_{ij}
	+ 4 \Psi^3 \Liec{m} \Psi \; \tA_{ij}
	= \Psi^4 \left[ \Liec{m} \tA_{ij} + \frac{2}{3} \left( 
	\tD_k \beta^k - NK \right) \tA_{ij} \right] . 
\ee
Besides, from formul\ae\ (\ref{e:cfd:DT_tDT}) and (\ref{e:cfd:Ckij_derPsi}),
\bea
	D_i D_j N & = & D_i \tD_j N 
	=  \tD_i \tD_j N - C^k_{\ \, ij} \tD_k N
		\nonumber \\
	& = & \tD_i \tD_j N - 
	2 \left( 
	\delta^k_{\ \, i} \tD_j \ln\Psi + \delta^k_{\ \, j} \tD_i \ln\Psi
	- \tD^k \ln\Psi \, \tgm_{ij} \right) \tD_k N  \nonumber \\
	& = &  \tD_i \tD_j N - 2 \left(
	\tD_i\ln\Psi \, \tD_j N + \tD_j\ln\Psi \, \tD_i N 
	- \tD^k\ln\Psi\, \tD_k N \, \tgm_{ij} \right)  . \label{e:cfd:DiDjN}
\eea
In Eq.~(\ref{e:cfd:Liem_Aij}), we can now substitute expression (\ref{e:cfd:Liem_Aij_Psi})
for $\Liec{m} A_{ij}$, (\ref{e:cfd:DiDjN}) for $D_iD_j N$, (\ref{e:cfd:Ricci_Psi})
for $R_{ij}$,
(\ref{e:cfd:laplacien_N}) for $D_k D^k N$ and (\ref{e:cfd:Ricci_scal0}) for $R$.
After some slight rearrangements, we get
\be \label{e:cfd:evol_tA}
	\encadre{
	\begin{array}{lcl}
	\displaystyle \left(\der{}{t} - \Liec{\beta} \right) \tA_{ij} & = & 
	\displaystyle - \frac{2}{3} \tD_k \beta^k\,  \tA_{ij} +
	N \left[ K\tA_{ij} - 2 \tgm^{kl} \tA_{ik} \tA_{jl}
	- 8\pi \left(\Psi^{-4} S_{ij} - \frac{1}{3} S \tgm_{ij} \right) \right] \\
	& & \displaystyle + \Psi^{-4} \bigg\{ - \tD_i \tD_j N 
		+ 2 \tD_i \ln\Psi\,  \tD_j N + 2 \tD_j \ln\Psi\, \tD_i N \\
	& & \displaystyle \qquad  \quad  
		+ \frac{1}{3}\left( \tD_k \tD^k N - 4 \tD_k\ln\Psi\, 
	\tD^k N \right) \tgm_{ij} \\
	 & & \displaystyle\qquad \quad
		 + N \bigg[ {\tilde R}_{ij} - \frac{1}{3} {\tilde R} \tgm_{ij} 
	- 2\tD_i\tD_j \ln\Psi + 4\tD_i \ln\Psi\, \tD_j\ln\Psi \\
	& & \displaystyle \qquad \qquad \quad + \frac{2}{3}
	\left( \tD_k \tD^k \ln\Psi - 2\tD_k\ln\Psi \, \tD^k \ln\Psi \right)
	\tgm_{ij} \bigg] \bigg\} .	
	\end{array}
	} 
\ee

\subsection{Hamiltonian constraint}

Substituting Eq.~(\ref{e:cfd:Ricci_scal}) for $R$ and Eq.~(\ref{e:cfd:KijKij})
into the Hamiltonian constraint equation (\ref{e:dec:Einstein_PDE3})
yields
\be \label{e:cfd:Ham_constr_comp1}
	\encadre{ \tD_i \tD^i \Psi -\frac{1}{8} {\tilde R} \Psi
	+ \left( \frac{1}{8} \tA_{ij} \tA^{ij}
	- \frac{1}{12} K^2 + 2\pi E \right) \Psi^5 = 0 } . 
\ee
Let us consider the alternative scaling $\alpha=-10$ to re-express the
term $\tA_{ij} \tA^{ij}$. By combining Eqs.~(\ref{e:cfd:def_tA}),
(\ref{e:cfd:tA_down}), (\ref{e:cfd:def_hA}) and (\ref{e:cfd:hA_down}),
we get the following relations
\be \label{e:cfd:hA_tA}
	\encadre{\hA^{ij} = \Psi^6 \tA^{ij}} \qquad \mbox{and} \qquad
	\encadre{\hA_{ij} = \Psi^6 \tA_{ij}} .
\ee
Hence $\tA_{ij} \tA^{ij} = \Psi^{-12} \hA_{ij} \hA^{ij}$ and 
Eq.~(\ref{e:cfd:Ham_constr_comp1}) becomes
\be \label{e:cfd:Lichne}
	\encadre{ 
	\tD_i \tD^i \Psi -\frac{1}{8} {\tilde R} \Psi
	+ \frac{1}{8} \hA_{ij} \hA^{ij} \, \Psi^{-7}
	+ \left( 2\pi E - \frac{1}{12} K^2 \right) \Psi^5 = 0 } . 
\ee
This is the \defin{Lichnerowicz equation}. It has been obtained by 
Lichnerowicz in 1944 \cite{Lichn44} in the special case $K=0$
(maximal hypersurface)  (cf. also Eq. (11.7) in Ref.~\cite{Lichn52}).
\begin{remark}
If one regards Eqs.~(\ref{e:cfd:Ham_constr_comp1}) and (\ref{e:cfd:Lichne})
as non-linear elliptic equations for $\Psi$, the negative power ($-7$)
of $\Psi$ in the $\hA_{ij} \hA^{ij}$ term in Eq.~(\ref{e:cfd:Lichne}), as compared 
to the positive power ($+5$) in Eq.~(\ref{e:cfd:Ham_constr_comp1}), makes a big 
difference about the mathematical properties of these two equations. 
This will be discussed in detail in Chap.~\ref{s:ini}. 
\end{remark}

\subsection{Momentum constraint}

The momentum constraint has been already written in terms of $\hA^{ij}$: it is
Eq.~(\ref{e:cfd:mom_constr_hA1}). Taking into account relation (\ref{e:cfd:hA_tA}),
we can easily rewrite it in terms of $\tA^{ij}$:
\be \label{e:cfd:mom_constr_tA}
	\encadre{ \tD_j \tA^{ij} + 6 \tA^{ij} \tD_j \ln\Psi - \frac{2}{3}
	\tD^i K = 8\pi \Psi^4 p^i }.
\ee

\subsection{Summary: conformal 3+1 Einstein system}

Let us gather Eqs.~(\ref{e:cfd:Liem_lnPsi}), (\ref{e:cfd:evol_tgm}),
(\ref{e:cfd:evol_K_conf}),  (\ref{e:cfd:evol_tA}),
(\ref{e:cfd:Ham_constr_comp1}) and (\ref{e:cfd:mom_constr_tA}):
\bea
  & &	\encadre{ \left( \der{}{t} - \Liec{\beta} \right) \Psi =
	\frac{\Psi}{6} \left( \tD_i \beta^i  - NK \right) } \label{e:cfd:Einstein1} \\
 & & \encadre{ \left( \der{}{t} - \Liec{\beta} \right) \tgm_{ij}
	= - 2N  \tA_{ij} - \frac{2}{3} \tD_k \beta^k \, \tgm_{ij} } 
	\label{e:cfd:Einstein2} \\
 &  & \encadre{ \left(\der{}{t} - \Liec{\beta} \right) K 
	= - \Psi^{-4} \left( \tD_i \tD^i N + 2 \tD_i \ln \Psi \, \tD^i N \right) 
	 + N \left[ 4\pi (E+S) 
	+  \tA_{ij} \tA^{ij} + \frac{K^2}{3}\right] } \nonumber \\
	& & \label{e:cfd:Einstein3} \\
  & & 		\encadre{
	\begin{array}{lcl}
	\displaystyle \left(\der{}{t} - \Liec{\beta} \right) \tA_{ij} & = & 
	\displaystyle - \frac{2}{3} \tD_k \beta^k\,  \tA_{ij} +
	N \left[ K\tA_{ij} - 2 \tgm^{kl} \tA_{ik} \tA_{jl}
	- 8\pi \left(\Psi^{-4} S_{ij} - \frac{1}{3} S \tgm_{ij} \right) \right] \\
	& & \displaystyle + \Psi^{-4} \bigg\{ - \tD_i \tD_j N 
		+ 2 \tD_i \ln\Psi\,  \tD_j N + 2 \tD_j \ln\Psi\, \tD_i N \\
	& & \displaystyle \qquad  \quad  
		+ \frac{1}{3}\left( \tD_k \tD^k N - 4 \tD_k\ln\Psi\, 
	\tD^k N \right) \tgm_{ij} \\
	 & & \displaystyle\qquad \quad
		 + N \bigg[ {\tilde R}_{ij} - \frac{1}{3} {\tilde R} \tgm_{ij} 
	- 2\tD_i\tD_j \ln\Psi + 4\tD_i \ln\Psi\, \tD_j\ln\Psi \\
	& & \displaystyle \qquad \qquad \quad + \frac{2}{3}
	\left( \tD_k \tD^k \ln\Psi - 2\tD_k\ln\Psi \, \tD^k \ln\Psi \right)
	\tgm_{ij} \bigg] \bigg\} .	
	\end{array}
	}   \nonumber \\
& &  \label{e:cfd:Einstein4} \\
   & & \encadre{ \tD_i \tD^i \Psi -\frac{1}{8} {\tilde R} \Psi
	+ \left( \frac{1}{8} \tA_{ij} \tA^{ij}
	- \frac{1}{12} K^2 + 2\pi E \right) \Psi^5 = 0 }  \label{e:cfd:Einstein5} \\
    & & \encadre{ \tD_j \tA^{ij} + 6 \tA^{ij} \tD_j \ln\Psi - \frac{2}{3}
	\tD^i K = 8\pi \Psi^4 p^i } . \label{e:cfd:Einstein6} 
\eea
For the last two equations, which are the constraints, we have the
alternative forms (\ref{e:cfd:Lichne}) and (\ref{e:cfd:Ham_constr_comp1})
in terms of $\hA^{ij}$ (instead of $\tA^{ij}$):
\bea
	& & 	\encadre{ 
	\tD_i \tD^i \Psi -\frac{1}{8} {\tilde R} \Psi
	+ \frac{1}{8} \hA_{ij} \hA^{ij} \, \Psi^{-7}
	+ \left( 2\pi E - \frac{1}{12} K^2 \right) \Psi^5 = 0 } , \qquad
	\qquad \qquad \qquad \qquad \ \ \label{e:cfd:Einstein5_hA} \\
	& & 
	\encadre{ \tD_j \hA^{ij} - \frac{2}{3} \Psi^6 \tD^i K = 8\pi \Psi^{10} p^i } .
		\label{e:cfd:Einstein6_hA} 
\eea

Equations~(\ref{e:cfd:Einstein1})-(\ref{e:cfd:Einstein6}) constitute the conformal
3+1 Einstein system. An alternative form is constituted by
Eqs.~(\ref{e:cfd:Einstein1})-(\ref{e:cfd:Einstein4}) and 
(\ref{e:cfd:Einstein5_hA})-(\ref{e:cfd:Einstein6_hA}). 
In terms of the original 3+1 Einstein system 
(\ref{e:dec:Einstein_PDE1})-(\ref{e:dec:Einstein_PDE4}), 
Eq.~(\ref{e:cfd:Einstein1}) corresponds to the trace of the kinematical
equation (\ref{e:dec:Einstein_PDE1}) and Eq.~(\ref{e:cfd:Einstein2}) to its
traceless part, Eq.~(\ref{e:cfd:Einstein3}) corresponds to the trace of 
the dynamical Einstein equation (\ref{e:dec:Einstein_PDE2}) and
Eq.~(\ref{e:cfd:Einstein4}) to its traceless part, Eq.~(\ref{e:cfd:Einstein5})
or Eq.~(\ref{e:cfd:Einstein5_hA})
is the Hamiltonian constraint (\ref{e:dec:Einstein_PDE3}), whereas 
Eq.~(\ref{e:cfd:Einstein6}) or Eq.~(\ref{e:cfd:Einstein6_hA}) is the momentum
constraint. 

If the system (\ref{e:cfd:Einstein1})-(\ref{e:cfd:Einstein6})
is solved in terms of $\tgm_{ij}$, $\tA_{ij}$ (or $\hA_{ij})$, $\Psi$
and $K$, then the physical metric $\wgm$ and the extrinsic curvature $\w{K}$
are recovered by 
\bea
	& & \gm_{ij} = \Psi^4 \tgm_{ij} \\
	& & K_{ij} = \Psi^4 \left( \tA_{ij} + \frac{1}{3} K \tgm_{ij} \right)  
	= \Psi^{-2} \hA_{ij} + \frac{1}{3} K \Psi^4 \tgm_{ij} . 
\eea

\section{Isenberg-Wilson-Mathews approximation to General Relativity} \label{s:cfd:IWM}

In 1978, J. Isenberg \cite{Isenb78} was looking for some approximation to
general relativity without any gravitational wave, beyond the Newtonian theory. 
The simplest of the approximations that he found amounts to impose that the 
3-metric $\wgm$ is conformally flat. In the framework of the discussion 
of Sec.~\ref{s:cfd:intro}, this is very natural since this means that $\wgm$
belongs to the conformal equivalence class of a flat metric and there
are no gravitational waves in a flat spacetime. 
This approximation has been reintroduced by Wilson and Mathews in 1989 \cite{WilsoM89}, 
who were not aware of Isenberg's work \cite{Isenb78}
(unpublished, except for the proceeding \cite{IsenbN80}).
It is now designed as the \defin{Isenberg-Wilson-Mathews approximation} 
(\defin{IWM}) to General Relativity, or sometimes the \defin{conformal flatness approximation}. 

In our notations, the IWM approximation amounts to set
\be \label{e:cfd:IWM}
	\wtgm = \w{f} 
\ee
and to demand that the background metric $\w{f}$ is flat. 
Moreover the foliation $(\Sigma_t)_{t\in\R}$ must be
chosen so that 
\be \label{e:cfd:IWM_K0}
	K = 0 , 
\ee
i.e. the hypersurfaces $\Sigma_t$ have a vanishing mean curvature.
Equivalently $\Sigma_t$ is a hypersurface of maximal volume, as it will 
be explained in Chap.~\ref{s:evo}. For this reason, foliations with $K=0$
are called \defin{maximal slicings}.

Notice that while the condition (\ref{e:cfd:IWM_K0}) can always be satisfied
by selecting a maximal slicing for the foliation $(\Sigma_t)_{t\in\R}$, 
the requirement (\ref{e:cfd:IWM})
is possible only if the Cotton tensor
of $(\Sigma_t,\wgm)$ vanishes identically, as we have seen 
in Sec.~\ref{s:cfd:intro}. Otherwise, one deviates from general
relativity. 

Immediate consequences of (\ref{e:cfd:IWM}) are
that the connection $\wtD$ is simply $\wDf$ and that the Ricci tensor $\w{\tilde R}$
vanishes identically, since $\w{f}$ is flat. 
The conformal 3+1 Einstein system (\ref{e:cfd:Einstein1})-(\ref{e:cfd:Einstein6})
then reduces to 
\bea
  & &	 \left( \der{}{t} - \Liec{\beta} \right) \Psi =
	\frac{\Psi}{6} \Df_i \beta^i  \label{e:cfd:IWM1} \\
 & &  \left( \der{}{t} - \Liec{\beta} \right) f_{ij}
	= - 2N  \tA_{ij} - \frac{2}{3} \Df_k \beta^k \, f_{ij} 
	\label{e:cfd:IWM2} \\
 &  & 0 = - \Psi^{-4} \left( \Df_i \Df^i N + 2 \Df_i \ln \Psi \, \Df^i N \right) 
	 + N \left[ 4\pi (E+S) 
	+  \tA_{ij} \tA^{ij} \right]  \label{e:cfd:IWM3} \\
  & & 	\begin{array}{lcl}
	\displaystyle \left(\der{}{t} - \Liec{\beta} \right) \tA_{ij} & = & 
	\displaystyle - \frac{2}{3} \Df_k \beta^k\,  \tA_{ij} +
	N \left[  -2 f^{kl} \tA_{ik} \tA_{jl}
	- 8\pi \left(\Psi^{-4} S_{ij} - \frac{1}{3} S f_{ij} \right) \right] \\
	& & \displaystyle + \Psi^{-4} \bigg\{ - \Df_i \Df_j N 
		+ 2 \Df_i \ln\Psi\,  \Df_j N + 2 \Df_j \ln\Psi\, \Df_i N \\
	& & \displaystyle \qquad  \quad  
		+ \frac{1}{3}\left( \Df_k \Df^k N - 4 \Df_k\ln\Psi\, 
	\Df^k N \right) f_{ij} \\
	 & & \displaystyle\qquad \quad
		 + N \bigg[ 
	- 2\Df_i\Df_j \ln\Psi + 4\Df_i \ln\Psi\, \Df_j\ln\Psi \\
	& & \displaystyle \qquad \qquad \quad + \frac{2}{3}
	\left( \Df_k \Df^k \ln\Psi - 2\Df_k\ln\Psi \, \Df^k \ln\Psi \right)
	f_{ij} \bigg] \bigg\} 	
	\end{array}
	  \label{e:cfd:IWM4} \\
   & &  \Df_i \Df^i \Psi 
	+ \left( \frac{1}{8} \tA_{ij} \tA^{ij}
	+ 2\pi E \right) \Psi^5 = 0  \label{e:cfd:IWM5} \\
    & &  \Df_j \tA^{ij} + 6 \tA^{ij} \Df_j \ln\Psi  = 
	8\pi \Psi^4 p^i  . \label{e:cfd:IWM6} 
\eea
Let us consider Eq.~(\ref{e:cfd:IWM2}). By hypothesis $\dert{f_{ij}}{t}=0$
[Eq.~(\ref{e:cfd:f_const})]. Moreover,
\be
	\Liec{\beta} f_{ij} = \beta^k \underbrace{\Df_k f_{ij}}_{=0}
	+ f_{kj} \Df_i \beta^k + f_{ik} \Df_j \beta^k
	= f_{kj} \Df_i \beta^k + f_{ik} \Df_j \beta^k , 
\ee
so that Eq.~(\ref{e:cfd:IWM2}) can be rewritten as
\be
	2N \tA_{ij} = f_{kj} \Df_i \beta^k + f_{ik} \Df_j \beta^k
	- \frac{2}{3} \Df_k \beta^k\, f_{ij} . 
\ee
Using $\tA^{ij} = f^{ik} f^{jl} \tA_{kl}$, we may rewrite this equation as
\be \label{e:cdf:IWM_tA_confKill}
	\tA^{ij} = \frac{1}{2N} (L \beta)^{ij} , 
\ee
where
\be \label{e:cfd:conf_Killing_f}
	(L \beta)^{ij} := \Df^i\beta^j + \Df^j \beta^i 
			- \frac{2}{3} \Df_k \beta^k \, f^{ij}
\ee
is the \defin{conformal Killing operator} associated with the metric $\w{f}$
(cf. Appendix~\ref{s:cko}).
Consequently, the term $\Df_j \tA^{ij}$ which appears in 
Eq.~(\ref{e:cfd:IWM6}) is expressible in terms of $\w{\beta}$ as
\bea
	\Df_j \tA^{ij} & = & \Df_j \left[ \frac{1}{2N} (L \beta)^{ij} \right]
	= \frac{1}{2N} \Df_j \left( \Df^i\beta^j + \Df^j \beta^i 
			- \frac{2}{3} \Df_k \beta^k \, f^{ij} \right)
	- \frac{1}{2N^2} (L \beta)^{ij} \Df_j N  \nonumber \\
	& = & \frac{1}{2N} \left(
	\Df_j \Df^j \beta^i + \frac{1}{3} \Df^i \Df_j \beta^j 
	- 2 \tA^{ij} \Df_j N \right) ,  	\label{e:cfd:divA_IWM}
\eea
where we have used $\Df_j \Df^i\beta^j = \Df^i\Df_j \beta^j$ since $\w{f}$ is flat. 
Inserting Eq.~(\ref{e:cfd:divA_IWM}) into Eq.~(\ref{e:cfd:IWM6}) yields
\be \label{e:cdf:IWM_mom1}
	\Df_j \Df^j \beta^i + \frac{1}{3} \Df^i \Df_j \beta^j 
	+ 2 \tA^{ij} \left( 6 N \Df_j \ln\Psi - \Df_j N \right)
	= 16 \pi N \Psi^4 p^i . 
\ee

The IWM system is formed by Eqs.~(\ref{e:cfd:IWM3}), (\ref{e:cfd:IWM5})
and (\ref{e:cdf:IWM_mom1}), which we rewrite as
\bea
  & & \encadre{\Delta N + 2\Df_i\ln\Psi \Df^i N =
  		N \left[ 4\pi (E+S) + \tA_{ij} \tA^{ij} \right] } \label{e:cfd:IWM_PDE1} \\
  & & \encadre{ \Delta \Psi 
	+ \left( \frac{1}{8} \tA_{ij} \tA^{ij}
	+ 2\pi E \right) \Psi^5 = 0  } \label{e:cfd:IWM_PDE2} \\
  & & \encadre{ \Delta \beta^i + \frac{1}{3} \Df^i \Df_j \beta^j 
	+ 2 \tA^{ij} \left( 6 N \Df_j \ln\Psi - \Df_j N \right)
	= 16 \pi N \Psi^4 p^i } , \label{e:cfd:IWM_PDE3}
\eea
where
\be
	\Delta := \Df_i \Df^i 
\ee
is the flat-space Laplacian. 
In the above equations, $\tA^{ij}$ is to be understood, not as an independent variable,
but as the function of $N$ and $\beta^i$ defined by 
Eq.~(\ref{e:cdf:IWM_tA_confKill}). 

The IWM system (\ref{e:cfd:IWM_PDE1})-(\ref{e:cfd:IWM_PDE3}) is a system
of three elliptic equations (two scalar equations and one vector equation) 
for the three unknowns $N$, $\Psi$ and $\beta^i$.
The physical 3-metric is fully determined by $\Psi$
\be \label{e:cfd:gmij_conf_flat}
	\gm_{ij} = \Psi^4 f_{ij}, 
\ee
so that, once the IWM system is solved, the full spacetime metric $\w{g}$ can 
be reconstructed via Eq.~(\ref{e:dec:g_cov}). 
\begin{remark}
In the original article \cite{Isenb78}, \label{p:cfd:Isenberg}
Isenberg has derived the system (\ref{e:cfd:IWM_PDE1})-(\ref{e:cfd:IWM_PDE3})
from a variational principle based on the Hilbert action (\ref{e:dec:S_Hilbert}),
by restricting $\gm_{ij}$ to take the form (\ref{e:cfd:gmij_conf_flat})
and requiring that the momentum conjugate to $\Psi$ vanishes.
\end{remark}
That the IWM scheme constitutes some approximation to general relativity is
clear because  the solutions $(N,\Psi,\beta^i)$ to the IWM system
(\ref{e:cfd:IWM_PDE1})-(\ref{e:cfd:IWM_PDE3}) do not in general satisfy the remaining
equations of the full conformal 3+1 Einstein system, i.e. Eqs.~(\ref{e:cfd:IWM1}) and (\ref{e:cfd:IWM4}).
However, the IWM approximation 
\begin{itemize}
\item is exact for spherically symmetric spacetimes (the Cotton tensor 
vanishes for any spherically symmetric $(\Sigma_t,\wgm)$), as shown for the
Schwarzschild spacetime in the example given in Sec.~\ref{s:cfd:conf_metric}; 
\item is very accurate for axisymmetric rotating neutron stars;
\cite{CookST96} 
\item is correct at the 1-PN order in the post-Newtonian 
expansion of general relativity.
\end{itemize}
The IWM approximation has been widely used in relativistic astrophysics, 
to compute binary neutron star mergers \cite{MatheW00,FaberGR04,OechsUPT04}
gravitational collapses of stellar cores \cite{DimmeFM02a,DimmeFM02b,DimmeNFIM05,Saijo04,Saijo05}, 
as well as quasi-equilibrium 
configurations of binary neutron stars or binary black holes 
(cf. Sec.~\ref{s:ini:binary}).

%  
%    Chapitre : Global quantities
%
% $Date: 2007-03-05 22:39:07 +0100 (lun, 05 mar 2007) $
% $Rev: 182 $
% $Author: e_gourgoulhon $
%%%%%%%%%%%%%%%%%%%%%%%%%%%%%

\chapter{Asymptotic flatness and global quantities} \label{s:glo}

%\verb$Date: 2007-03-05 22:39:07 +0100 (lun, 05 mar 2007) $

\minitoc
\vspace{1cm}

%%%%%%%%%%%%%%%%%%%%%%%%%%%%%%%%%%%%%%%%%%%%%%%%%%%%%%%%%%%%%%%%%%%%%%%%%%%%

\section{Introduction}

In this Chapter, we review the global quantities that one may associate
to the spacetime $(\M,\w{g})$ or to each slice $\Sigma_t$ of the
3+1 foliation. This encompasses various notions of mass,
linear momentum and angular momentum. 
In the absence of any symmetry, 
all these global quantities are defined only for asymptotically flat spacetimes. 
So we shall start by defining the notion of \emph{asymptotic flatness}.

%%%%%%%%%%%%%%%%%%%%%%%%%%%%%%%%%%%%%%%%%%%%%%%%%%%%%%%%%%%%%%%%%%%%%%%%%%%%

\section{Asymptotic flatness} \label{s:glo:asymp_flat}

The concept of asymptotic flatness applies to stellar type objects, modeled as
if they were alone in an otherwise empty universe (the so-called \defin{isolated bodies}). Of course, most cosmological spacetimes are not asymptotically flat.  

\subsection{Definition}

We consider a globally hyperbolic spacetime $(\M,\w{g})$ foliated by a family
$(\Sigma_t)_{t\in\R}$ of spacelike hypersurfaces. Let $\w{\gm}$ and $\w{K}$
be respectively the induced metric and extrinsic curvature of the hypersurfaces
$\Sigma_t$. One says that the spacetime is \defin{asymptotically flat} iff
there exists, on each slice $\Sigma_t$, a Riemannian ``background'' metric $\w{f}$
such that \cite{York79,York80,Strau04}
\begin{itemize}
\item $\w{f}$ is flat ($\mathbf{Riem}(\w{f})=0$), except possibly on a compact 
domain $\mathcal{B}$ of $\Sigma_t$ (the ``strong field region'');
\item there exists a coordinate system $(x^i)=(x,y,z)$ on $\Sigma_t$ such that
outside $\mathcal{B}$, the components of $\w{f}$ are $f_{ij} = \mathrm{diag}(1,1,1)$
(``Cartesian-type coordinates'') and the variable $r:=\sqrt{x^2+y^2+z^2}$ can
take arbitrarily large values on $\Sigma_t$;
\item when $r\rightarrow +\infty$, the components of $\w{\gm}$ with respect to the coordinates $(x^i)$ satisfy
\bea
	& & \gm_{ij} = f_{ij} + O(r^{-1}), \label{e:glob:aflat1}\\
	& & \der{\gm_{ij}}{x^k} = O(r^{-2});	\label{e:glob:aflat2}
\eea
\item when $r\rightarrow +\infty$, the components of $\w{K}$ with respect to the coordinates $(x^i)$ satisfy
\bea	
	& & K_{ij} = O(r^{-2}), \label{e:glob:aflat3}\\
	& & \der{K_{ij}}{x^k} = O(r^{-3}).\label{e:glob:aflat4}
\eea
\end{itemize}
The ``region'' $r\rightarrow+\infty$ is called \defin{spatial infinity}
and is denoted $i^0$.  

\begin{remark}
There exist other definitions of \emph{asymptotic flatness} which are not based
on any coordinate system nor background flat metric (see e.g. Ref.~\cite{Ashte80}
or Chap.~11 in Wald's textbook \cite{Wald84}). In particular, the spatial infinity
$i^0$ can be rigorously defined as a single point in some 
``extended'' spacetime $(\hat\M,\w{\hat g})$ in which $(\M,\w{g})$ can be
embedded with $\w{g}$  conformal to $\w{\hat g}$.  
However the present definition is perfectly adequate for our purposes.
\end{remark}

\begin{remark}
The requirement (\ref{e:glob:aflat2}) excludes the presence of gravitational waves
at spatial infinity. Indeed for gravitational waves propagating in the radial
direction:
\be
	\gm_{ij} = f_{ij} + \frac{F_{ij}(t-r)}{r} + O(r^{-2}) .
\ee
This fulfills condition (\ref{e:glob:aflat1}) but
\be
	\der{\gm_{ij}}{x^k} = - \frac{{F'}_{ij}(t-r)}{r} \frac{x^k}{r}
		- \frac{F_{ij}(t-r)}{r^2}  \frac{x^k}{r} + O(r^{-2}) 
\ee
is $O(r^{-1})$ since ${F'}_{ij} \not =0$ (otherwise  $F_{ij}$ would be a constant function
and there would be no radiation). This violates
condition (\ref{e:glob:aflat2}). 
Notice that the absence of gravitational waves at spatial infinity is 
not a serious physical restriction, since one may consider that any isolated system
has started to emit gravitational waves at a finite time ``in the past'' and that
these waves have not reached the spatial infinity yet. 
\end{remark}

\subsection{Asymptotic coordinate freedom} \label{s:glo:Spi}

Obviously the above definition of asymptotic flatness depends both 
on the foliation $(\Sigma_t)_{t\in\R}$ and on the coordinates $(x^i)$
chosen on each leaf $\Sigma_t$. It is of course important to assess
whether this dependence is strong or not. 
In other words, we would like to determine the class of 
coordinate changes
$(x^\alpha)=(t,x^i)\rightarrow({x'}^\alpha)=(t',{x'}^i)$
which preserve the asymptotic properties 
(\ref{e:glob:aflat1})-(\ref{e:glob:aflat4}). The answer is that
the coordinates $({x'}^\alpha)$ must be related to the coordinates
$({x}^\alpha)$ by \cite{Henne06}
\be \label{e:glo:spi_transforms}
	{x'}^\alpha = \Lambda^\alpha_{\ \, \mu} x^\mu 
	+ c^\alpha(\theta,\varphi) + O(r^{-1})
\ee
where $\Lambda^\alpha_{\ \, \beta}$ is a Lorentz matrix and the $c^\alpha$'s
are four functions of the angles $(\theta,\varphi)$ related to the coordinates
$(x^i)=(x,y,z)$ by the standard formul\ae:
\be
	x = r \sin\theta\cos\varphi, \quad 
	y = r \sin\theta\sin\varphi, \quad 
	z = r \cos\theta . 
\ee
The group of transformations generated by (\ref{e:glo:spi_transforms})
is related to the \defin{Spi group} (for \emph{Spatial infinity}) introduced
by Ashtekar and Hansen \cite{AshteH78,Ashte80}. However the precise relation 
is not clear because the definition of asymptotic flatness used by
these authors is not expressed as decay conditions for $\gm_{ij}$
and $K_{ij}$, as in Eqs.~(\ref{e:glob:aflat1})-(\ref{e:glob:aflat4}).

Notice that \defin{Poincar\'e transformations} are contained in transformation
group defined by (\ref{e:glo:spi_transforms}):
they simply correspond to the case $c^\alpha(\theta,\varphi)={\rm const}$. 
The transformations with $c^\alpha(\theta,\varphi)\not={\rm const}$
and $\Lambda^\alpha_{\ \, \beta} = \delta^\alpha_{\ \, \beta}$ constitute
``angle-dependent translations'' and are called \defin{supertranslations}. 

Note that if the Lorentz matrix $\Lambda^\alpha_{\ \, \beta}$
involves a boost, the transformation (\ref{e:glo:spi_transforms})
implies a change of the 3+1 foliation $(\Sigma_t)_{t\in\R}$, 
whereas if $\Lambda^\alpha_{\ \, \beta}$
corresponds only to some spatial rotation and the $c^\alpha$'s are constant,
the transformation (\ref{e:glo:spi_transforms}) describes some change 
of Cartesian-type coordinates $(x^i)$ (rotation + translation)
within the same hypersurface $\Sigma_t$.

%%%%%%%%%%%%%%%%%%%%%%%%%%%%%%%%%%%%%%%%%%%%%%%%%%%%%%%%%%%%%%%%%%%%%%%%%%%%

\section{ADM mass}

\subsection{Definition from the Hamiltonian formulation of GR} \label{s:glo:ADM_mass}

In the short introduction to the Hamiltonian formulation of general relativity
given in Sec.~\ref{s:dec:ADM}, we have for simplicity discarded any
boundary term in the action. However, because the gravitational Lagrangian 
density (the scalar curvature ${}^4\!R$) contains second order derivatives of
the metric tensor (and not only first order ones, which is a particularity of general relativity with respect to other field theories), the precise action should be
\cite{ReggeT74,Poiss04,Wald84,Henne06}
\be \label{e:glo:S_Hilbert_bound}
	S =  \int_{\mathcal{V}} {}^4\!R \sqrt{-g} \, d^4 x
	+ 2 \oint_{\partial\mathcal{V}} (Y- Y_0) \sqrt{h} \, d^3 y ,
\ee
where $\partial\mathcal{V}$ is the boundary of the domain $\mathcal{V}$
($\partial\mathcal{V}$ is assumed to be a timelike hypersurface), $Y$ the trace of the extrinsic curvature (i.e. three times the mean curvature) 
of $\partial\mathcal{V}$ embedded in $(\M,\w{g})$ and
$Y_0$ the trace of the extrinsic curvature of $\partial\mathcal{V}$ 
embedded in $(\M,\w{\eta})$, where $\w{\eta}$ is a Lorentzian metric on $\M$ which 
is flat in the region of $\partial\mathcal{V}$.  
Finally $\sqrt{h} \, d^3 y$ is the volume element induced by $\w{g}$ on
the hypersurface $\partial\mathcal{V}$, $\w{h}$ being the induced metric 
on $\partial\mathcal{V}$ and $h$ its determinant with respect to
the coordinates $(y^i)$ used on $\partial\mathcal{V}$.
The boundary term in (\ref{e:glo:S_Hilbert_bound})
guarantees that the variation of $S$ with the values
of $\w{g}$ (and not its derivatives) held fixed at $\partial\mathcal{V}$ leads to
the Einstein equation. Otherwise, from the volume term alone (Hilbert action),
one has to held fixed $\w{g}$ \emph{and} all its derivatives at $\partial\mathcal{V}$.

Let 
\be
	\Sp_t := \partial V \cap \Sigma_t .
\ee
We assume that $\Sp_t$ has the topology of a sphere. 
The gravitational Hamiltonian which can be derived from the action (\ref{e:glo:S_Hilbert_bound}) (see \cite{Poiss04} for details) contains an
additional boundary term with respect to the Hamiltonian (\ref{e:dec:Hamilt})
obtained in Sec.~\ref{s:dec:ADM} :
\be \label{e:glo:Ham}
H = - \int_{\Sigma^{\rm int}_t} \left( N C_0 - 2 \beta^i C_i \right) \sqrt{\gm} d^3 x 
	- 2 \oint_{\Sp_t} \left[ N(\kappa-\kappa_0) + \beta_i (K_{ij} - K \gm_{ij})
	s^j \right] \sqrt{q} \, d^2 y , 
\ee
where $\Sigma^{\rm int}_t$ is the part of $\Sigma_t$ bounded by $\Sp_t$,
 $\kappa$ is the trace of the extrinsic curvature of $\Sp_t$ embedded in 
$(\Sigma_t,\w{\gm})$, and $\kappa_0$ the trace of the extrinsic curvature of $\Sp_t$ embedded in $(\Sigma_t,\w{f})$ ($\w{f}$ being the metric introduced in 
Sec.~\ref{s:glo:asymp_flat}), $\w{s}$ is the unit normal to $\Sp_t$ in $\Sigma_t$,
oriented towards the asymptotic region, and $\sqrt{q} \, d^2 y$ denotes the surface
element induced by the spacetime metric on $\Sp_t$, $\w{q}$ being the induced
metric, $y^a=(y^1,y^2)$ some coordinates on $\Sp_t$ [for instance 
$y^a=(\theta,\varphi)$] and $q:=\det(q_{ab})$.

For solutions of Einstein equation, the constraints are satisfied:
$C_0=0$ and $C_i=0$, so that the value of the Hamiltonian reduces to 
\be \label{e:glo:Hsol}
	H_{\rm solution} = - 2 \oint_{\Sp_t} \left[ N(\kappa-\kappa_0) 
	+ \beta^i (K_{ij} - K \gm_{ij})
	s^j \right] \sqrt{q} \, d^2 y . 
\ee
The total energy contained in the $\Sigma_t$ is then defined as the numerical
value of the Hamiltonian for solutions, taken on a surface $\Sp_t$ at spatial infinity
(i.e. for $r\rightarrow+\infty$) and for coordinates $(t,x^i)$ that could be associated
with some asymptotically inertial observer, i.e. such that 
$N=1$ and $\w{\beta}=0$. From Eq.~(\ref{e:glo:Hsol}), we get (after restoration
of some $(16\pi)^{-1}$ factor)
\be \label{e:glo:M_ADM_def}
	\encadre{ M_{\rm ADM} := - \frac{1}{8\pi}\lim_{\Sp_t\rightarrow\infty}
	\oint_{\Sp_t}(\kappa-\kappa_0) \sqrt{q} \, d^2 y } .
\ee
This energy is called the \defin{ADM mass} of the slice $\Sigma_t$.
By evaluating the extrinsic curvature traces $\kappa$ and $\kappa_0$, it
can be shown that Eq.~(\ref{e:glo:M_ADM_def}) can be written
\be \label{e:glo:M_ADM_cov}
	\encadre{ M_{\rm ADM} = \frac{1}{16\pi}
	\lim_{\Sp_t\rightarrow\infty}
	\oint_{\Sp_t} \left[ \Df^j \gm_{ij} - \Df_i (f^{kl} \gm_{kl}) \right]
	s^i  \sqrt{q}\, d^2 y }, 
\ee
where $\w{\Df}$ stands for the connection associated with the metric $\w{f}$
and, as above, $s^i$ stands for the components of unit normal to $\Sp_t$ 
within $\Sigma_t$ and oriented towards the exterior of $\Sp_t$.
In particular, if one uses the Cartesian-type coordinates $(x^i)$ involved
in the definition of asymptotic flatness (Sec.~\ref{s:glo:asymp_flat}), 
then $\Df_i = \dert{}{x^i}$ and $f^{kl}=\delta^{kl}$ 
and the above formula becomes
\be \label{e:glo:M_ADM_cart}
	M_{\rm ADM} = \frac{1}{16\pi}
	\lim_{\Sp_t\rightarrow\infty}
	\oint_{\Sp_t} \left( \der{\gm_{ij}}{x^j} - \der{\gm_{jj}}{x^i} \right)
	s^i  \sqrt{q}\, d^2 y  . 
\ee
Notice that thanks to the asymptotic flatness requirement (\ref{e:glob:aflat2}),
this integral takes a finite value: the $O(r^2)$ part of $\sqrt{q}\, d^2 y$ 
is compensated by the $O(r^{-2})$ parts of $\dert{\gm_{ij}}{x^j}$ and  
$\dert{\gm_{jj}}{x^i}$. 

\begin{example}
Let us consider Schwarzschild spacetime and use the standard Schwarzschild coordinates
$(x^\alpha) = (t,r,\theta,\phi)$:
\be \label{e:glo:Schwarz_coord}
    g_{\mu\nu} dx^\mu dx^\nu  =  - \left( 1 - \frac{2m}{r} \right) dt^2 
    + \left( 1 - \frac{2m}{r} \right) ^{-1} dr^2 
    + r^2 (d\theta^2 + \sin^2\theta d\varphi^2) .
\ee
Let us take for $\Sigma_t$ the hypersurface of constant Schwarzschild coordinate time $t$.
Then we read on (\ref{e:glo:Schwarz_coord}) the components of the induced metric
in the coordinates $(x^i)=(r,\theta,\varphi)$:
\be \label{e:glo:gm_Schwarz}
	\gm_{ij} = \mathrm{diag}\left[ \left( 1 - \frac{2m}{r} \right) ^{-1},\ 
		r^2,\ r^2 \sin^2 \theta \right] . 
\ee
On the other side, the components of the flat metric in the same coordinates
are
\be
	f_{ij} = \mathrm{diag}\left( 1, r^2,r^2 \sin^2 \theta \right) 
	\qquad\mbox{and}\qquad 
	f^{ij} = \mathrm{diag}\left( 1, r^{-2},r^{-2} \sin^{-2} \theta \right) .
\ee
Let us now evaluate $M_{\rm ADM}$ by means of the integral (\ref{e:glo:M_ADM_cov})
(we cannot use formula (\ref{e:glo:M_ADM_cart}) because the coordinates
$(x^i)$ are not Cartesian-like). It is quite natural to take for $\Sp_t$
the sphere $r={\rm const}$ in the hypersurface $\Sigma_t$. 
Then $y^a=(\theta,\varphi)$, $\sqrt{q}=r^2\sin\theta$ and, at spatial infinity, 
$s^i  \sqrt{q}\, d^2 y = r^2 \sin\theta \, d\theta \, d\varphi\, (\wpar_r)^i $, where 
$\wpar_r$ is the natural basis
vector associated the coordinate $r$: $(\wpar_r)^i = (1,0,0)$.
Consequently, Eq.~(\ref{e:glo:M_ADM_cov}) becomes
\be \label{e:glo:M_schwarz_integ}
M_{\rm ADM} = \frac{1}{16\pi}
	\lim_{r \rightarrow\infty}
	\oint_{r={\rm const}} 
	 \left[ \Df^j \gm_{rj} - \Df_r (f^{kl} \gm_{kl}) \right]
	r^2 \sin\theta \, d\theta \, d\varphi , 
\ee
with
\be
	f^{kl} \gm_{kl} = \gm_{rr} + \frac{1}{r^2} \gm_{\theta\theta}
		+ \frac{1}{r^2\sin^2\theta} \gm_{\varphi\varphi}
			= \left( 1 - \frac{2m}{r} \right) ^{-1} + 2 ,
\ee
and since $f^{kl} \gm_{kl}$ is a scalar field, 
\be \label{e:glo:M_schwarz1}
	\Df_r (f^{kl} \gm_{kl}) = \der{}{r} (f^{kl} \gm_{kl})
		= - \left( 1 - \frac{2m}{r} \right) ^{-2} \frac{2m}{r^2} . 
\ee
There remains to evaluate $\Df^j \gm_{rj}$. One has
\be
	\Df^j \gm_{rj} = f^{jk} \Df_k \gm_{rj}
		= \Df_r \gm_{rr} + \frac{1}{r^2} \Df_\theta \gm_{r\theta}
			+ \frac{1}{r^2\sin^2\theta} \Df_\varphi \gm_{r\varphi} ,  
\ee
with the covariant derivatives given by (taking into account the form
(\ref{e:glo:gm_Schwarz}) of $\gm_{ij}$)
\bea
	& & \Df_r \gm_{rr} = \der{\gm_{rr}}{r} - 2\bar\Gamma^i_{\ \, rr} \gm_{ir} 
		= \der{\gm_{rr}}{r} - 2\bar\Gamma^r_{\ \, rr} \gm_{rr}  \\
	& & \Df_\theta \gm_{r\theta} = \der{\gm_{r\theta}}{\theta}
		- \bar\Gamma^i_{\ \, \theta r} \gm_{i\theta}
		- \bar\Gamma^i_{\ \, \theta\theta} \gm_{ri} 
		= - \bar\Gamma^\theta_{\ \, \theta r} \gm_{\theta\theta}
		- \bar\Gamma^r_{\ \, \theta\theta} \gm_{rr} 	\\	
	& & \Df_\varphi \gm_{r\varphi} = \der{\gm_{r\varphi}}{\varphi}
		- \bar\Gamma^i_{\ \, \varphi r} \gm_{i\varphi}
		- \bar\Gamma^i_{\ \, \varphi\varphi} \gm_{ri} 
		= - \bar\Gamma^\varphi_{\ \, \varphi r} \gm_{\varphi\varphi}
		- \bar\Gamma^r_{\ \, \varphi\varphi} \gm_{rr} 	,	
\eea
where the $\bar\Gamma^k_{\ \, ij}$'s are the Christoffel symbols of the connection
$\wDf$ with respect to the coordinates $(x^i)$. The non-vanishing ones are
\bea
	\bar\Gamma^r_{\ \, \theta\theta} =  -r \quad & \mbox{and} & \quad
		\bar\Gamma^r_{\ \, \varphi\varphi} = - r\sin^2\theta  \\
	\bar\Gamma^\theta_{\ \, r\theta} = \bar\Gamma^\theta_{\ \, \theta r} ={1\over r} 
		\quad & \mbox{and} & \quad
		\bar\Gamma^\theta_{\ \, \varphi\varphi} = -\cos\theta \sin\theta \\
	\bar\Gamma^\varphi_{\ \, r\varphi} = \bar\Gamma^\varphi_{\ \, \varphi r} = {1\over r}  
		\quad & \mbox{and} & \quad 
	\bar\Gamma^\varphi_{\ \, \theta\varphi} = \bar\Gamma^\varphi_{\ \, \varphi\theta} = {1\over \tan\theta}  	.
\eea
Hence 
\bea
	\Df^j \gm_{rj} & = &  \der{}{r} \left[ \left( 1 - \frac{2m}{r} \right) ^{-1}
	\right]
	+ \frac{1}{r^2} 
	\left[ - \frac{1}{r} \times r^2 + r \times \left( 1 - \frac{2m}{r} \right) ^{-1}
	\right] \nonumber\\
	& & + \frac{1}{r^2\sin^2\theta} \left[ - \frac{1}{r}\times r^2 \sin^2\theta
	+ r\sin^2\theta \times \left( 1 - \frac{2m}{r} \right) ^{-1} \right] 
	\nonumber \\
	\Df^j \gm_{rj} & = & \frac{2m}{r^2} \left( 1 - \frac{2m}{r} \right) ^{-2}
	\left( 1 - \frac{4m}{r} \right) .  \label{e:glo:M_schwarz2}
\eea
Combining Eqs.~(\ref{e:glo:M_schwarz1}) and (\ref{e:glo:M_schwarz2}), we get
\bea
	\Df^j \gm_{rj} - \Df_r (f^{kl} \gm_{kl}) & = & 
		\frac{2m}{r^2} \left( 1 - \frac{2m}{r} \right) ^{-2}
	\left( 1 - \frac{4m}{r} + 1 \right) 
	= \frac{4m}{r^2} \left( 1 - \frac{2m}{r} \right) ^{-1} \nonumber \\
	& \sim & \frac{4m}{r^2} \quad \mbox{when} \ r\rightarrow\infty , 
\eea
so that the integral (\ref{e:glo:M_schwarz_integ}) results in
\be  \label{e:glo:M_ADM_Schwarz_m}
	M_{\rm ADM} = m . 
\ee
We conclude that the ADM mass of any hypersurface $t={\rm const}$ of
Schwarzschild spacetime is nothing but the mass parameter $m$ of the Schwarzschild 
solution.
\end{example}

\subsection{Expression in terms of the conformal decomposition}

Let us introduce the conformal metric $\wtgm$ and conformal factor $\Psi$
associated to $\wgm$ according to the prescription given in Sec.~\ref{s:cfd:conf_metric},
taking for the background metric $\w{f}$ the \emph{same} metric as that 
involved in the definition of asymptotic flatness and ADM mass:
\be
	\wgm = \Psi^4 \wtgm, \label{e:glo:conf_met}
\ee 
with, in the Cartesian-type coordinates $(x^i)=(x,y,z)$ 
introduced in Sec.~\ref{s:glo:asymp_flat}:
\be \label{e:glo:det_tgm_1}
	\det(\tgm_{ij}) = 1 .
\ee
This is the property (\ref{e:cfd:dettgm_f}) since $f=\det(f_{ij})=1$ ($f_{ij}=\mathrm{diag}(1,1,1)$).
The asymptotic flatness conditions (\ref{e:glob:aflat1})-(\ref{e:glob:aflat2}) impose
\be \label{e:glo:Psi_asymp}
	\Psi = 1 + O(r^{-1}) \qquad \mbox{and} \qquad \der{\Psi}{x^k} = O(r^{-2}) 
\ee
and
\be \label{e:glo:tgm_asymp}
	\tgm_{ij} = f_{ij} + O(r^{-1}) 
	\qquad \mbox{and} \qquad \der{\tgm_{ij}}{x^k} = O(r^{-2}) .
\ee

Thanks to the decomposition (\ref{e:glo:conf_met}), the integrand of 
the ADM mass formula (\ref{e:glo:M_ADM_cov}) is
\be
	\Df^j \gm_{ij} - \Df_i (f^{kl} \gm_{kl}) =
	4 \underbrace{\Psi^3}_{\sim 1} \Df^j \Psi \, \underbrace{\tgm_{ij}}_{\sim f_{ij}}
	+ \underbrace{\Psi^4}_{\sim 1} \Df^j \tgm_{ij}
	- 4 \underbrace{\Psi^3}_{\sim 1} \Df_i \Psi \, 
	\underbrace{f^{kl}\tgm_{kl}}_{\sim 3}
	- \underbrace{\Psi^4}_{\sim 1} \Df_i (f^{kl}\tgm_{kl}) ,
\ee
where the $\sim$'s denote values when $r\rightarrow\infty$, taking into 
account (\ref{e:glo:Psi_asymp}) and (\ref{e:glo:tgm_asymp}).
Thus we have
\be \label{e:glo:integrand_conf}
	\Df^j \gm_{ij} - \Df_i (f^{kl} \gm_{kl}) \sim 
	- 8 \Df_i \Psi + \Df^j\tgm_{ij} - \Df_i (f^{kl}\tgm_{kl}) .
\ee
From (\ref{e:glo:Psi_asymp}) and (\ref{e:glo:tgm_asymp}), 
$\Df_i \Psi = O(r^{-2})$ and $\Df^j\tgm_{ij} = O(r^{-2})$.
Let us show that the unit determinant condition (\ref{e:glo:det_tgm_1})
implies $ \Df_i (f^{kl}\tgm_{kl}) = O(r^{-3})$ so that this term
actually does not contribute to the ADM mass integral.
Let us write 
\be
	\tgm_{ij} =: f_{ij} + \varepsilon_{ij},
\ee
with according to Eq.~(\ref{e:glo:tgm_asymp}), $\varepsilon_{ij}=O(r^{-1})$.
Then 
\be
	f^{kl}\tgm_{kl} = 3 + \varepsilon_{xx} + \varepsilon_{yy} 
	+ \varepsilon_{zz} 
\ee
and
\be \label{e:glo:Diftgm_eps}
	\Df_i (f^{kl}\tgm_{kl}) = \der{}{x^i} (f^{kl}\tgm_{kl})
	= \der{}{x^i} \left( \varepsilon_{xx} + \varepsilon_{yy} 
	+ \varepsilon_{zz} \right) . 
\ee
Now the determinant of $\tgm_{ij}$ is
\bea
	\det(\tgm_{ij}) & = &  \det \left(
	\begin{array}{ccc}
	1+\varepsilon_{xx} & \varepsilon_{xy} & \varepsilon_{xz} \\
	\varepsilon_{xy} & 1+ \varepsilon_{yy} & \varepsilon_{yz} \\
	\varepsilon_{xz} & \varepsilon_{yz} & 1 + \varepsilon_{zz} 
	\end{array}
	\right) \nonumber \\
	& = & 1 + \varepsilon_{xx} + \varepsilon_{yy} 
	+ \varepsilon_{zz} + \varepsilon_{xx} \varepsilon_{yy}
	+ \varepsilon_{xx} \varepsilon_{zz} + \varepsilon_{yy} \varepsilon_{zz}
	- \varepsilon_{xy}^2 - \varepsilon_{xz}^2 - \varepsilon_{yz}^2 \nonumber \\
	& & + \varepsilon_{xx} \varepsilon_{yy} \varepsilon_{zz}
	+ 2 \varepsilon_{xy} \varepsilon_{xz} \varepsilon_{yz}
	- \varepsilon_{xx} \varepsilon_{yz}^2 
	- \varepsilon_{yy} \varepsilon_{xz}^2 
	- \varepsilon_{zz} \varepsilon_{xy}^2 .
\eea
Requiring $\det(\tgm_{ij}) = 1$ implies then
\bea
	\varepsilon_{xx} + \varepsilon_{yy} + \varepsilon_{zz}  & = &
	- \varepsilon_{xx} \varepsilon_{yy}
	- \varepsilon_{xx} \varepsilon_{zz} - \varepsilon_{yy} \varepsilon_{zz}
	+ \varepsilon_{xy}^2 + \varepsilon_{xz}^2 + \varepsilon_{yz}^2 \nonumber \\
	& & - \varepsilon_{xx} \varepsilon_{yy} \varepsilon_{zz}
	- 2 \varepsilon_{xy} \varepsilon_{xz} \varepsilon_{yz}
	+ \varepsilon_{xx} \varepsilon_{yz}^2 
	+ \varepsilon_{yy} \varepsilon_{xz}^2 
	+ \varepsilon_{zz} \varepsilon_{xy}^2 .
\eea
Since according to (\ref{e:glo:tgm_asymp}),  
$\varepsilon_{ij}=O(r^{-1})$ and $\dert{\varepsilon_{ij}}{x^k}=O(r^{-2})$, 
we conclude that
\be
	\der{}{x^i}\left(
	\varepsilon_{xx} + \varepsilon_{yy} + \varepsilon_{zz} \right)
	= O(r^{-3}) ,
\ee
i.e. in view of (\ref{e:glo:Diftgm_eps}), 
\be
	\Df_i (f^{kl}\tgm_{kl}) = O(r^{-3}) . 
\ee
Thus in Eq.~(\ref{e:glo:integrand_conf}), only the first two terms in the
right-hand side contribute to the ADM mass integral, so that formula 
(\ref{e:glo:M_ADM_cov}) becomes
\be \label{e:glo:M_ADM_Psi}
	\encadre{ M_{\rm ADM} = - \frac{1}{2\pi}
	\lim_{\Sp_t\rightarrow\infty}
	\oint_{\Sp_t} s^i \left( \Df_i \Psi - \frac{1}{8} \Df^j\tgm_{ij} \right)
	 \sqrt{q}\, d^2 y } .  	
\ee
\begin{example}
Let us return to the example considered in Sec.~\ref{s:cfd:conf_metric},
namely Schwarzschild spacetime in isotropic coordinates $(t,r,\theta,\varphi)$
\footnote{although we use the same symbol, the $r$ used here is different from
the Schwarzschild coordinate $r$ of the example in Sec.~\ref{s:glo:ADM_mass}.}.
The conformal factor was found to be $\Psi = 1+m/(2r)$ [Eq.~(\ref{e:cfd:Psi_Schwarz})]
and the conformal metric to be $\wtgm = \w{f}$. Then $\Df^j\tgm_{ij}=0$ and only
the first term remains in the integral (\ref{e:glo:M_ADM_Psi}):
\be
	M_{\rm ADM} = - \frac{1}{2\pi}
	\lim_{r \rightarrow\infty}
	\oint_{r={\rm const}} 
	 \der{\Psi}{r} r^2 \sin\theta \, d\theta \, d\varphi , 
\ee
with 
\be
	\der{\Psi}{r} = \der{}{r} \left( 1+\frac{m}{2r} \right) = - \frac{m}{2r^2},
\ee
so that we get
\be
	M_{\rm ADM} = m ,
\ee
i.e. we recover the result (\ref{e:glo:M_ADM_Schwarz_m}), which was obtained by
means of different coordinates (Schwarzschild coordinates). 
\end{example}

\subsection{Newtonian limit}

To check that at the Newtonian limit, the ADM mass reduces to the usual definition
of mass, let us consider the weak field metric given by Eq.~(\ref{e:mat:gab_weak}).
We have found in Sec.~\ref{s:cfd:conf_metric} that the corresponding
conformal metric is $\wtgm = \w{f}$ and the conformal factor 
$\Psi = 1 - \Phi/2$ [Eq.~(\ref{e:cfd:Psi_Newt_lim})], where $\Phi$ reduces to
the gravitational potential at the Newtonian limit. 
Accordingly, $\Df^j\tgm_{ij} = 0$ and $\Df_i \Psi = - \frac{1}{2} \Df_i \Phi$, 
so that Eq.~(\ref{e:glo:M_ADM_Psi}) becomes
\be 
	 M_{\rm ADM} =  \frac{1}{4\pi}
	\lim_{\Sp_t\rightarrow\infty}
 	\oint_{\Sp_t} s^i \Df_i \Phi \, \sqrt{q}\, d^2 y  .  	
\ee
To take Newtonian limit, we may assume that $\Sigma_t$ has the topology of $\mathbb{R}^3$
and transform the above surface integral to a volume one by means of the
Gauss-Ostrogradsky theorem:
\be \label{e:glo:M_ADM_Newt0}
	M_{\rm ADM} =  \frac{1}{4\pi}  \int_{\Sigma_t}
		 \Df_i \Df^i \Phi \, \sqrt{f} \, d^3 x  . 
\ee
Now, at the Newtonian limit, $\Phi$ is a solution of the Poisson equation
\be
	\Df_i \Df^i \Phi = 4\pi \rho , 
\ee
where $\rho$ is the mass density (remember we are using units in which Newton's
gravitational constant $G$ is unity). 
Hence Eq.~(\ref{e:glo:M_ADM_Newt0}) becomes
\be \label{e:glo:M_ADM_Newt}
	M_{\rm ADM} =  \int_{\Sigma_t}
		 \rho \, \sqrt{f} \, d^3 x  , 
\ee
which shows that at the Newtonian limit, the ADM mass is nothing but the
total mass of the considered system. 

\subsection{Positive energy theorem} \label{s:glo:positive_ener}

Since the ADM mass represents the total energy of a gravitational system,
it is important to show that it is always positive, 
at least for ``reasonable'' models of matter (take $\rho < 0$ in
Eq.~(\ref{e:glo:M_ADM_Newt}) and you will get $M_{\rm ADM} < 0$ ...). 
If negative values of the energy would be possible, then a gravitational system could
decay to lower and lower values and thereby emit an unbounded energy 
via gravitational radiation. 

The positivity of the ADM mass has been hard to 
establish. The complete proof was eventually given in 1981 by Schoen 
and Yau \cite{SchoeY81}. A simplified proof has been found shortly
after by Witten \cite{Witte81}. 
More precisely, Schoen, Yau and Witten have shown that if the matter
content of spacetime obeys the dominant energy condition, then
$M_{\rm ADM} \geq 0$. Furthermore, $M_{\rm ADM} = 0$ if and only if
$\Sigma_t$ is a hypersurface of Minkowski spacetime.

The \defin{dominant energy condition} is the following requirement 
on the matter stress-energy tensor $\w{T}$: for any timelike and
future-directed vector $\w{v}$, the vector
$-\vec{\w{T}}(\w{v})$ defined by Eq.~(\ref{e:hyp:arrow_endo})
\footnote{in index notation, $-\vec{\w{T}}(\w{v})$ is
the vector $-T^\alpha_{\ \, \mu} v^\mu$} 
must be a future-directed timelike or null vector.
If $\w{v}$ is the 4-velocity of some observer, 
$-\vec{\w{T}}(\w{v})$ is the energy-momentum density 4-vector 
as measured by the observer and the dominant 
energy condition means that this vector must be causal. 
In particular, the dominant energy
condition implies the \defin{weak energy condition}, namely
that for any timelike and future-directed vector $\w{v}$,
$\w{T}(\w{v},\w{v}) \geq 0$. If again $\w{v}$ is the 4-velocity of some observer,
the quantity $\w{T}(\w{v},\w{v})$ is nothing but the energy density 
as measured by that observer [cf. Eq.~(\ref{e:dec:E_def})], and the
the weak energy condition simply stipulates that this energy
density must be non-negative.
In short, the dominant energy condition means that the matter energy
must be positive and that it must not travel faster than light. 

The dominant energy condition is easily expressible in terms of
the matter energy density $E$ and momentum density $\w{p}$, 
both measured by the Eulerian observer and introduced in Sec.~\ref{s:dec:T3p1}.
Indeed, from the 3+1 split (\ref{e:dec:T_3p1}) of $\w{T}$, 
the energy-momentum density 4-vector relative to the 
Eulerian observer is found to be
\be
    \w{J} := -\vec{\w{T}}(\w{n})
	= E \w{n} + \vec{\w{p}} . 
\ee
Then, since $\w{n}\cdot\vec{\w{p}}=0$,  $\w{J}\cdot\w{J} = - E^2 + \vec{\w{p}}\cdot\vec{\w{p}}$. 
Requiring that $\w{J}$ is timelike or null means
$\w{J}\cdot\w{J} \leq 0$ and that it is future-oriented amounts to $E\geq0$
(since $\w{n}$ is itself future-oriented). Hence the dominant energy
condition is equivalent to the two conditions
$E^2 \geq \vec{\w{p}}\cdot\vec{\w{p}}$ and $E\geq 0$. 
Since $\vec{\w{p}}$ is always a spacelike vector, these two conditions
are actually equivalent to the single requirement
\be \label{e:glo:dominant}
	\encadre{ E \geq \sqrt{ \vec{\w{p}}\cdot\vec{\w{p}} } } . 
\ee
This justifies the term \emph{dominant} energy condition. 

\subsection{Constancy of the ADM mass}

Since the Hamiltonian $H$ given by Eq.~(\ref{e:glo:Ham}) depends
on the configuration variables $(\gm_{ij},N,\beta^i)$ and their
conjugate momenta $(\pi^{ij},\pi^N=0,\pi^{\w{\beta}}=0)$, but not 
explicitly on the time $t$, the associated energy is a constant 
of motion:
\be
	\encadre{ \frac{d}{dt} M_{\rm ADM} = 0 } . 
\ee
Note that this property is not obvious when contemplating formula (\ref{e:glo:M_ADM_cov}), which expresses $M_{\rm ADM}$ as an integral
over $\Sp_t$. 

%%%%%%%%%%%%%%%%%%%%%%%%%%%%%%%%%%%%%%%%%%%%%%%%%%%%%%%%%%%%%%%%%%%%%%%%%%%%%%%%%%%%%%%

\section{ADM momentum}

\subsection{Definition}

As the ADM mass is associated with time translations at infinity
[taking $N=1$ and $\w{\beta}=0$ in Eq.~(\ref{e:glo:Hsol})], the 
ADM momentum is defined as the conserved quantity associated 
with the invariance of the action with respect to spatial translations. 
With respect to the Cartesian-type coordinates $(x^i)$ introduced
in Sec.~\ref{s:glo:asymp_flat}, three privileged directions for translations at spatial infinity are given by the three vectors $(\wpar_i)_{i\in\{1,2,3\}}$. 
The three conserved quantities are then obtained by setting 
$N=0$ and $\beta^i=1$ in Eq.~(\ref{e:glo:Hsol}) \cite{Henne06,ReggeT74}:
\be \label{e:glo:Pi_ADM_def}
	\encadre{ P_i := \frac{1}{8\pi} \lim_{\Sp_t\rightarrow\infty}
	\oint_{\Sp_t} \left( K_{jk} - K \gm_{jk} \right) (\wpar_i)^j \, 
	s^k \sqrt{q}\, d^2 y } , 
\qquad i\in\{1,2,3\}.
\ee
\begin{remark}
The index $i$ in the above formula is not the index of some tensor component, 
contrary to the indices $j$ and $k$. It is used to label the three vectors
$\wpar_1$, $\wpar_2$ and $\wpar_3$ and the quantities $P_1$, $P_2$ and $P_3$
corresponding to each of these vectors.
\end{remark}
Notice that the asymptotic flatness condition (\ref{e:glob:aflat3}) ensures that
$P_i$ is a finite quantity. The three numbers $(P_1,P_2,P_3)$
define the \defin{ADM momentum} of the hypersurface $\Sigma_t$. 
The values $P_i$ depend upon the choice of the coordinates $(x^i)$ but
the set $(P_1,P_2,P_3)$ transforms as the components of a linear form under
a change of Cartesian coordinates $(x^i)\rightarrow({x'}^i)$ which asymptotically
corresponds to rotation and/or a translation. Therefore 
$(P_1,P_2,P_3)$ can be regarded as a linear form which ``lives'' at the ``edge'' of
$\Sigma_t$. It can be regarded as well as a vector
since the duality vector/linear forms is trivial in the asymptotically Euclidean space.

\begin{example}
For foliations associated with the standard coordinates of Schwarzschild spacetime
(e.g. Schwarzschild coordinates (\ref{e:glo:Schwarz_coord}) or
isotropic coordinates (\ref{e:cfd:Schwarz_isotropic})),
the extrinsic curvature vanishes identically: $\w{K}=0$, so that
Eq.~(\ref{e:glo:Pi_ADM_def}) yields
\be
	P_i = 0 . 
\ee
For a non trivial example based on a ``boosted'' Schwarzschild solution,
see Ref.~\cite{York80}. 
\end{example}

\subsection{ADM 4-momentum}

Not only $(P_1,P_2,P_3)$ behaves as the components of a linear form,
but the set of four numbers
\be
	\encadre{ P_\alpha^{\rm ADM} := (-M_{\rm ADM}, P_1, P_2, P_3) }
\ee
behaves as the components of a 4-dimensional linear form any under 
coordinate change $(x^\alpha)=(t,x^i)\rightarrow({x'}^\alpha)=(t',{x'}^i)$ 
which preserves the asymptotic conditions 
(\ref{e:glob:aflat1})-(\ref{e:glob:aflat4}), i.e. any coordinate
change of the form (\ref{e:glo:spi_transforms}). 
In particular, $P_\alpha^{\rm ADM}$ is transformed 
in the proper way under the Poincar\'e group:
\be
	{P'}_\alpha^{\rm ADM} = (\Lambda^{-1})^\mu_{\ \, \alpha} \;  
	P_\mu^{\rm ADM} . 
\ee
This last property has been shown first by Arnowitt, Deser and Misner
\cite{ArnowDM62}. 
For this reason, $P_\alpha^{\rm ADM}$ is considered as a linear form
which ``lives'' at spatial infinity and is called the
\defin{ADM 4-momentum}. 

%%%%%%%%%%%%%%%%%%%%%%%%%%%%%%%%%%%%%%%%%%%%%%%%%%%%%%%%%%%%%%%%%%%%%%%%%%%%

\section{Angular momentum}

\subsection{The supertranslation ambiguity}

Generically, the angular momentum is the conserved quantity associated
with the invariance of the action with respect to rotations, in the 
same manner as the linear momentum is associated with the invariance
with respect to translations. Then 
one might naively define the total angular momentum of a given 
slice $\Sigma_t$ by an integral of the type 
(\ref{e:glo:Pi_ADM_def}) but with $\wpar_i$ being replaced by 
a rotational Killing vector $\w{\phi}$ of the 
flat metric $\w{f}$. More precisely, in terms of the Cartesian coordinates $(x^i)=(x,y,z)$ introduced in Sec.~\ref{s:glo:asymp_flat}, the
three vectors $(\w{\phi}_i)_{i\in\{1,2,3\}}$ defined by 
\bea
	& & \w{\phi}_x = - z \wpar_y + y \wpar_z  \label{e:glo:rot_flat_x} \\
	& & \w{\phi}_y = - x \wpar_z + z \wpar_x  \\
	& & \w{\phi}_z = - y \wpar_x + x \wpar_y \label{e:glo:rot_flat_z} 
\eea
are three independent Killing vectors of $\w{f}$, corresponding to a
rotation about
respectively the $x$-axis, $y$-axis and the $z$-axis. 
Then one may defined the three numbers
\be \label{e:glo:angu_mom_def}
 J_i := \frac{1}{8\pi} \lim_{\Sp_t\rightarrow\infty}
	\oint_{\Sp_t} \left( K_{jk} - K \gm_{jk} \right) (\w{\phi}_i)^j \,
	s^k \sqrt{q}\, d^2 y, 
\qquad i\in\{1,2,3\} .
\ee
The problem is that the quantities $J_i$ hence defined depend upon
the choice of the coordinates and, contrary to $P_\alpha^{\rm ADM}$, 
 do not transform as a the components of
a vector under a change 
$(x^\alpha)=(t,x^i)\rightarrow({x'}^\alpha)=(t',{x'}^i)$ 
that preserves the asymptotic properties (\ref{e:glob:aflat1})-(\ref{e:glob:aflat4}),
i.e. a transformation of the type (\ref{e:glo:spi_transforms}). 
As discussed by York \cite{York79,York80}, the problem arises because
of the existence of the supertranslations (cf. Sec.~\ref{s:glo:Spi}) in the
permissible coordinate changes (\ref{e:glo:spi_transforms}).

\begin{remark}
Independently of the above coordinate ambiguity, one may notice that the
asymptotic flatness conditions (\ref{e:glob:aflat1})-(\ref{e:glob:aflat4})
are not sufficient, \emph{by themselves}, to guarantee that the integral
(\ref{e:glo:angu_mom_def}) takes a finite value when $\Sp_t\rightarrow\infty$, 
i.e. when $r\rightarrow\infty$. Indeed, 
Eqs.~(\ref{e:glo:rot_flat_x})-(\ref{e:glo:rot_flat_z}) show that the
Cartesian components of the rotational vectors behave like
$(\w{\phi}_i)^j \sim O(r)$, so that Eq.~(\ref{e:glob:aflat3}) implies only
$\left( K_{jk} - K \gm_{jk} \right) (\w{\phi}_i)^j = O(r^{-1})$.
It is the contraction with the unit normal vector $s^k$ which 
ensures $\left( K_{jk} - K \gm_{jk} \right) (\w{\phi}_i)^j s^k = O(r^{-2})$
and hence that $J_i$ is finite. This is clear for the 
$K \gm_{jk} (\w{\phi}_i)^j s^k$ part because the vectors $\w{\phi}_i$
given by Eqs.~(\ref{e:glo:rot_flat_x})-(\ref{e:glo:rot_flat_z}) are all
orthogonal to $\w{s} \sim x/r \, \wpar_x + y/r \, \wpar_y + z/r \, \wpar_z$.
For the $K_{jk}(\w{\phi}_i)^j s^k$ part, this turns out to be true in practice,
as we shall see on the specific example of Kerr spacetime in 
Sec.~\ref{s:glo:Komar_J}.
\end{remark}

\subsection{The ``cure''} \label{s:glo:cure}

In view of the above coordinate dependence problem, 
one may define the angular momentum as a quantity which remains invariant only with
respect to a subclass of the coordinate changes (\ref{e:glo:spi_transforms}).
This is made by imposing decay conditions stronger than
(\ref{e:glob:aflat1})-(\ref{e:glob:aflat4}). For instance, York \cite{York79}
has proposed the following conditions\footnote{Actually the first condition
proposed by York, Eq.~(90) of Ref.~\cite{York79}, is 
not exactly (\ref{e:glo:QIgauge}) but can be 
shown to be equivalent to it; see also Sec.~V of Ref.~\cite{SmarrY78a}.}
on the flat divergence of the conformal metric and 
 the trace of the extrinsic curvature:
\bea 
	& & \der{\tgm_{ij}}{x^j} = O(r^{-3}), \label{e:glo:QIgauge} \\
	& & K = O(r^{-3}) . \label{e:glo:K0gauge}
\eea
Clearly these conditions are stronger than respectively 
(\ref{e:glo:tgm_asymp}) and (\ref{e:glob:aflat3}). 
Actually they are so severe that they
exclude some well known coordinates that one would like to use
to describe asymptotically flat spacetimes, for instance the 
standard Schwarzschild coordinates (\ref{e:glo:Schwarz_coord}) 
for the Schwarzschild solution.
For this reason, conditions (\ref{e:glo:QIgauge}) and
(\ref{e:glo:K0gauge}) are considered as asymptotic \emph{gauge conditions},
i.e. conditions restricting the choice of coordinates, rather than
conditions on the nature of spacetime at spatial infinity.
Condition (\ref{e:glo:QIgauge}) is called the 
\defin{quasi-isotropic gauge}. The isotropic coordinates (\ref{e:cfd:Schwarz_isotropic}) of the Schwarzschild solution trivially belong to this gauge (since
$\tgm_{ij} = f_{ij}$ for them). 
Condition (\ref{e:glo:K0gauge}) is called the \defin{asymptotically maximal
gauge}, since for maximal hypersurfaces $K$ vanishes identically.
York has shown that in the gauge (\ref{e:glo:QIgauge})-(\ref{e:glo:K0gauge}),
the angular momentum as defined by the integral 
(\ref{e:glo:angu_mom_def}) is carried by the $O(r^{-3})$ piece of $\w{K}$
(the $O(r^{-2})$ piece carrying the linear momentum $P_i$) and is invariant
(i.e. behaves as a vector) for any coordinate change within this gauge. 

Alternative decay requirements have been proposed by other authors to
fix the ambiguities in the angular momentum definition (see e.g. 
\cite{Chrus87} and references therein). For instance, Regge and Teitelboim
\cite{ReggeT74} impose a specific form and 
some parity conditions on the coefficient 
of the $O(r^{-1})$ term in Eq.~(\ref{e:glob:aflat1}) and on the coefficient
of the $O(r^{-2})$ term in Eq.~(\ref{e:glob:aflat3})
(cf. also M. Henneaux' lecture \cite{Henne06}).

As we shall see in Sec.~\ref{s:glo:Komar_J}, in the particular case of
an axisymmetric spacetime, there exists a unique definition of the
angular momentum, which is independent of any coordinate system. 

\begin{remark}
In the literature, there is often mention 
of the \emph{``ADM angular momentum''}, on the same footing as 
the ADM mass and ADM linear momentum. But as discussed above, there
is no such thing as the ``ADM angular momentum''. One has to specify
a gauge first and define the angular momentum within that gauge. 
In particular, there is no mention whatsoever of angular momentum
in the original ADM article \cite{ArnowDM62}.  
\end{remark}

\subsection{ADM mass in the quasi-isotropic gauge} \label{s:glo:M_ADM_QI}

In the quasi-isotropic gauge, the ADM mass can be expressed entirely in
terms of the flux at infinity of the gradient of the conformal factor
$\Psi$. Indeed, thanks to (\ref{e:glo:QIgauge}), the term $\Df^j\tgm_{ij}$
Eq.~(\ref{e:glo:M_ADM_Psi}) does not contribute to the integral and
we get
\be \label{e:glo:M_ADM_QI}
 \encadre{ M_{\rm ADM} = - \frac{1}{2\pi}
	\lim_{\Sp_t\rightarrow\infty}
	\oint_{\Sp_t} s^i \Df_i \Psi \, \sqrt{q}\, d^2 y } \qquad 
	\mbox{(quasi-isotropic gauge)}. 
\ee

\begin{figure}
\centerline{\includegraphics[width=0.7\textwidth]{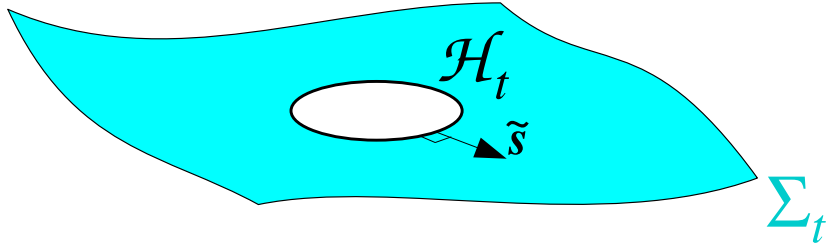}}
\caption[]{\label{f:glo:hole} \footnotesize
Hypersurface $\Sigma_t$ with a hole defining an inner boundary $\mathcal{H}_t$.}
\end{figure}

Thanks to the Gauss-Ostrogradsky theorem,
 we may transform this
formula into a volume integral. More precisely, let us assume that 
$\Sigma_t$ is diffeomorphic to either $\R^3$ or $\R^3$ minus
a ball. In the latter case, $\Sigma_t$ has an inner boundary, that we 
may call a \defin{hole} and denote by $\mathcal{H}_t$ (cf. Fig.~\ref{f:glo:hole}). 
We assume that $\mathcal{H}_t$ has the topology of a sphere. 
Actually this case is relevant for black hole spacetimes when 
black holes are treated via the so-called excision technique. %s:bho
The Gauss-Ostrogradsky formula enables to transform 
expression (\ref{e:glo:M_ADM_QI}) into
\be \label{e:glo:M_ADM_MH}
	M_{\rm ADM} = - \frac{1}{2\pi}
	\int_{\Sigma_t} \tD_i \tD^i \Psi \, \sqrt{\tgm} \, d^3 x
	+ M_{\mathcal{H}} , 
\ee
where $M_{\mathcal{H}}$ is defined by
\be
	M_{\mathcal{H}} := - \frac{1}{2\pi} \oint_{\mathcal{H}_t}
		{\tilde s}^i \tD_i \Psi \, \sqrt{\tilde q}\, d^2 y . 	
\ee
In this last equation, $\tilde q:=\det({\tilde q}_{ab})$, 
$\w{\tilde q}$ being the metric induced on $\mathcal{H}_t$ by $\wtgm$, 
and $\w{\tilde s}$ is the unit vector with respect to $\wtgm$
($\wtgm(\w{\tilde s},\w{\tilde s})=1$) tangent to $\Sigma_t$, normal to $\mathcal{H}_t$
and oriented towards the exterior of the hole (cf. Fig.~\ref{f:glo:hole}).
If $\Sigma_t$ is diffeomorphic to $\R^3$, we use formula 
(\ref{e:glo:M_ADM_MH}) with $M_{\mathcal{H}}$. 

Let now use the Lichnerowicz equation (\ref{e:cfd:Lichne}) to express
$\tD_i \tD^i \Psi$ in Eq.~(\ref{e:glo:M_ADM_MH}). We get
\be \label{e:glo:M_ADM_QI_vol}
	\encadre{ M_{\rm ADM} = \int_{\Sigma_t}  \left[ \Psi^5 E 
	+ \frac{1}{16\pi} \left( \hA_{ij} \hA^{ij} \, \Psi^{-7}
	- {\tilde R} \Psi - \frac{2}{3} K^2 \Psi^5 \right) \right]
	\sqrt{\tgm} \, d^3 x
	+ M_{\mathcal{H}} } \qquad \mbox{(QI gauge)}.
\ee
For the computation of the ADM mass in a numerical code, this formula
may be result in a greater precision that the surface integral 
at infinity (\ref{e:glo:M_ADM_QI}).
\begin{remark}
On the formula (\ref{e:glo:M_ADM_QI_vol}), 
we get immediately the Newtonian limit (\ref{e:glo:M_ADM_Newt})
by making $\Psi\rightarrow 1$, $E\rightarrow\rho$, $\hA^{ij}\rightarrow 0$,
${\tilde R} \rightarrow 0$, $K\rightarrow 0$, $\tgm\rightarrow f$ and 
$M_{\mathcal{H}}=0$.
\end{remark}

For the IWM approximation of general relativity considered in Sec.~\ref{s:cfd:IWM}, 
the coordinates belong to the quasi-isotropic gauge (since $\wtgm=\w{f}$), so
we may apply (\ref{e:glo:M_ADM_QI_vol}). Moreover, as a consequence of $\wtgm=\w{f}$,
$\tilde R = 0$ and in the IWM approximation, $K=0$. Therefore Eq.~(\ref{e:glo:M_ADM_QI_vol})
simplifies to 
\be
	M_{\rm ADM} = \int_{\Sigma_t}  \left( \Psi^5 E 
	+ \frac{1}{16\pi} \hA_{ij} \hA^{ij} \, \Psi^{-7} \right) 
	\sqrt{\tgm} \, d^3 x
	+ M_{\mathcal{H}} .
\ee
Within the framework of exact general relativity, the above formula is valid for any maximal slice $\Sigma_t$ with a conformally flat metric. 

%%%%%%%%%%%%%%%%%%%%%%%%%%%%%%%%%%%%%%%%%%%%%%%%%%%%%%%%%%%%%%%%%%%%%%%%%%%%%%%%%%%%%%%%%

\section{Komar mass and angular momentum}

In the case where the spacetime $(\M,\w{g})$ has some symmetries, one may 
define global quantities in a coordinate-independent way by means of a general technique 
introduced by Komar (1959) \cite{Komar59}. It consists in taking
flux integrals of the derivative of the Killing vector associated with the
symmetry over closed 2-surfaces surrounding the matter sources.
The quantities thus obtained are conserved in the sense that they do not depend
upon the choice of the integration 2-surface, as long as the latter stays
outside the matter. 
We discuss here two important cases: the \emph{Komar mass} resulting from time symmetry
(stationarity) and the \emph{Komar angular momentum} resulting from axisymmetry. 

\subsection{Komar mass}

Let us assume that the spacetime $(\M,\w{g})$ is \defin{stationary}. This means 
that the metric tensor $\w{g}$ is invariant by Lie transport along the field lines
of a timelike vector field $\w{k}$. The latter is called a \defin{Killing vector}. 
Provided that it is normalized so that $\w{k}\cdot\w{k}=-1$ at spatial infinity,
it is then unique. Given a 3+1 foliation $(\Sigma_t)_{t\in\R}$ of $\M$, 
and a closed 2-surface $\Sp_t$ in $\Sigma_t$, with the topology of a sphere, 
the \defin{Komar mass} is defined by
\be \label{e:glo:M_Komar_def}
	\encadre{ 
	M_{\rm K} := - \frac{1}{8\pi} \oint_{\Sp_t} \nabla^\mu k^\nu \, dS_{\mu\nu}
	} , 
\ee
with the 2-surface element
\be \label{e:glo:dS_Sp}
	dS_{\mu\nu} = (s_\mu n_\nu - n_\mu s_\nu) \sqrt{q} \, d^2 y , 
\ee
where $\w{n}$ is the unit timelike normal to $\Sigma_t$, $\w{s}$ is the unit
normal to $\Sp_t$ within $\Sigma_t$ oriented towards the exterior of $\Sp_t$,
$(y^a)=(y^1,y^2)$ are coordinates spanning $\Sp_t$, and
$q:=\det(q_{ab})$, the $q_{ab}$'s being  the components with respect to 
$(y^a)$ of the metric $\w{q}$ induced by $\wgm$
(or equivalently by $\w{g}$) on $\Sp_t$.
Actually the Komar mass can be defined over any closed 2-surface, 
but in the present context
it is quite natural to consider only 2-surfaces lying in the hypersurfaces of the 3+1
foliation.  

\begin{figure}
\centerline{\includegraphics[width=0.8\textwidth]{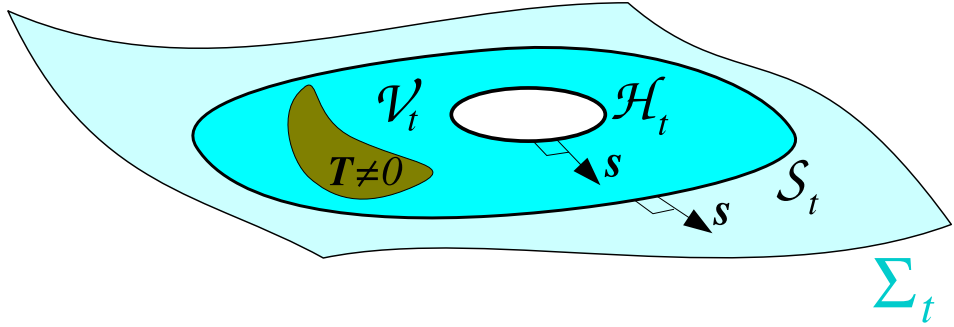}}
\caption[]{\label{f:glo:komar} \footnotesize
Integration surface $\Sp_t$ for the computation of Komar mass.
$\Sp_t$ is the external boundary of a part $\mathcal{V}_t$ of $\Sigma_t$
which contains all the matter sources ($\w{T}\not=0$).
$\mathcal{V}_t$ has possibly some inner boundary, in the form of one (or more)
hole $\mathcal{H}_t$.}
\end{figure}

A priori the quantity $M_{\rm K}$ as defined by (\ref{e:glo:M_Komar_def})
should depend on the choice of the 2-surface $\Sp_t$. However, thanks
to the fact that $\w{k}$ is a Killing vector, this is not the case, 
as long as  $\Sp_t$ is located outside any matter content of spacetime. 
In order to show this, let us transform the surface integral (\ref{e:glo:M_Komar_def})
into a volume integral. As in Sec.~\ref{s:glo:M_ADM_QI}, we suppose that 
$\Sigma_t$ is diffeomorphic to either $\R^3$ or $\R^3$ minus one hole, the results being 
easily generalized to an arbitrary number of holes (see Fig.~\ref{f:glo:komar}). 
The hole, the surface of
which is denoted by $\mathcal{H}_t$ as in Sec.~\ref{s:glo:M_ADM_QI}, must be totally
enclosed within the surface $\Sp_t$. Let us then denote by $\mathcal{V}_t$
the part of $\Sigma_t$ delimited by $\mathcal{H}_t$ and $\Sp_t$. 

The starting point is to notice that since $\w{k}$ is a Killing vector
the $\nabla^\mu k^\nu$'s in the integrand of Eq.~(\ref{e:glo:M_Komar_def})
are the components of an antisymmetric tensor. Indeed, $\w{k}$
obeys to Killing's equation\footnote{Killing's equation follows immediately
from the requirement of invariance of the metric along the field lines of
$\w{k}$, i.e. $\Lie{k}\w{g}=0$, along with the use of Eq.~(\ref{e:lie:der_comp_nab})
to express $\Lie{k}\w{g}$.}: 
\be \label{e:glo:Killing}
 	\nabla_\alpha k_\beta + \nabla_\beta k_\alpha = 0 . 
\ee
Now for any antisymmetric tensor $\w{A}$ 
of type $\left({2\atop 0}\right)$, the following
identity holds:
\be \label{e:glo:int_antisym}
	2 \int_{\mathcal{V}_t} \nabla_\nu A^{\mu\nu} \, dV_\mu = 
	\oint_{\Sp_t} A^{\mu\nu} \, dS_{\mu\nu} +
	\oint_{\mathcal{H}_t} A^{\mu\nu} \, dS^{\mathcal{H}}_{\mu\nu} , 
\ee
with $dV_\mu$ is the volume element on $\Sigma_t$:
\be \label{e:glo:dV_mu}
	dV_\mu = - n_\mu \sqrt{\gm} \, d^3 x 
\ee
and $dS^{\mathcal{H}}_{\mu\nu}$ is the surface element on $\mathcal{H}_t$ and is given
by a formula similar to Eq.~(\ref{e:glo:dS_Sp}), using the same notation for
the coordinates and the induced metric on $\mathcal{H}_t$:
\be \label{e:glo:dSprime}
	dS^{\mathcal{H}}_{\mu\nu} = (n_\mu s_\nu - s_\mu n_\nu) \sqrt{q} \, d^2 y . 
\ee
The change of sign with respect to Eq.~(\ref{e:glo:dS_Sp}) arises because we
choose the unit vector $\w{s}$ normal to $\mathcal{H}_t$ to be oriented towards
the interior of $\mathcal{V}_t$ (cf. Fig.~\ref{f:glo:komar}). 
Let us establish Eq.~(\ref{e:glo:int_antisym}). 
It is well known that for the divergence of an \emph{antisymmetric} tensor
is given by
\be
	\nabla_\nu A^{\mu\nu} = \frac{1}{\sqrt{-g}} \der{}{x^\nu}
		\left( \sqrt{-g} A ^{\mu\nu} \right) .
\ee
Using this property, as well as expression (\ref{e:glo:dV_mu}) of $dV_\mu$
with the components $n_\mu = (-N,0,0,0)$ given by Eq.~(\ref{e:dec:uun_comp}),
we get 
\be
	\int_{\mathcal{V}_t} \nabla_\nu A^{\mu\nu} \, dV_\mu
		= - \int_{\mathcal{V}_t} 
			\der{}{x^\nu}
		\left( \sqrt{-g} A^{\mu\nu} \right)
		\, n_\mu \, 
			\frac{\sqrt{\gm}}{\sqrt{-g}} \, d^3 x
		= \int_{\mathcal{V}_t} \der{}{x^\nu} 
			\left( \sqrt{\gm} N A^{0\nu} \right) \, d^3 x , 
\ee
where we have also invoked the relation (\ref{e:fol:detg_Ngetgam})
between the determinants of $\w{g}$ and $\wgm$: $\sqrt{-g} = N \sqrt{\gm}$. 
Now, since $A^{\alpha\beta}$ is antisymmetric, $A^{00}=0$ and we can write
$\dert{}{x^\nu} \left( \sqrt{\gm} N A^{0\nu} \right) =
\dert{}{x^i} \left( \sqrt{\gm} \, V^i \right)$ where $V^i = N A^{0i}$ are the components
of the vector $\w{V}\in\T(\Sigma_t)$ 
defined by $\w{V} := - \vec{\wgm}(\w{n}\cdot \w{A})$. 
The above integral then becomes
\be
	\int_{\mathcal{V}_t} \nabla_\nu A^{\mu\nu} \, dV_\mu
		= \int_{\mathcal{V}_t} \frac{1}{\sqrt{\gm}}
			\der{}{x^i} \left( \sqrt{\gm} V^i \right) \, 
				\sqrt{\gm} \, d^3 x
			= \int_{\mathcal{V}_t} D_i V^i \sqrt{\gm} \, d^3 x .
\ee
We can now use the Gauss-Ostrogradsky theorem to get 
\be
	\int_{\mathcal{V}_t} \nabla_\nu A^{\mu\nu} \, dV_\mu
		= \oint_{\partial \mathcal{V}_t}
			V^i s_i \, \sqrt{q} \, d^2 y .
\ee
Noticing that $\partial \mathcal{V}_t = \mathcal{H}_t \cup \Sp_t$ (cf. Fig.~\ref{f:glo:komar}) and
(from the antisymmetry of $A^{\mu\nu}$)
\be
	V^i s_i = V^\nu s_\nu = - n_\mu A^{\mu\nu} s_\nu = 
		\frac{1}{2} A^{\mu\nu} ( s_\mu n_\nu - n_\mu s_\nu ) , 
\ee
we get the identity~(\ref{e:glo:int_antisym}).
\begin{remark}
Equation (\ref{e:glo:int_antisym}) can also be derived by applying Stokes' theorem
to the 2-form ${}^4\!\epsilon_{\alpha\beta\mu\nu} A^{\mu\nu}$, where 
${}^4\!\epsilon_{\alpha\beta\mu\nu}$ is the Levi-Civita alternating tensor
(volume element) associated with the spacetime metric $\w{g}$ (see e.g. derivation
of Eq.~(11.2.10) in Wald's book \cite{Wald84}). 
\end{remark}

Applying formula~(\ref{e:glo:int_antisym}) to $A^{\mu\nu} = \nabla^\mu k^\nu$
we get, in view of the definition (\ref{e:glo:M_Komar_def}),
\be \label{e:glo:MK_interm}
	M_{\rm K} = -\frac{1}{4\pi} \int_{\mathcal{V}_t} 
		\nabla_\nu \nabla^\mu k^\nu \, dV_\mu
			+ M_{\rm K}^{\mathcal{H}} , 
\ee
where 
\be \label{e:glo:MKH_def}
	 M_{\rm K}^{\mathcal{H}} := 
	 	\frac{1}{8\pi} \oint_{\mathcal{H}_t} \nabla^\mu k^\nu \, dS^{\mathcal{H}}_{\mu\nu}
\ee
will be called the \defin{Komar mass of the hole}.  
Now, from the Ricci identity
\be
	\nabla_\nu \nabla^\mu k^\nu - \nabla^\mu 
		\underbrace{\nabla^\nu k^\nu}_{=0}	
		= {}^4\!R^\mu_{\ \, \nu} k^\nu , 
\ee
where the ``$=0$'' is a consequence of Killing's equation (\ref{e:glo:Killing}). 
Equation (\ref{e:glo:MK_interm}) becomes then
\be
	M_{\rm K} = -\frac{1}{4\pi} \int_{\mathcal{V}_t} 
		 {}^4\!R^\mu_{\ \, \nu} k^\nu \, dV_\mu
			+ M_{\rm K}^{\mathcal{H}} 
		= \frac{1}{4\pi} \int_{\mathcal{V}_t} 
		 {}^4\!R_{\mu\nu} k^\nu n^\mu \, \sqrt{\gm}\, d^3 x
		 	+ M_{\rm K}^{\mathcal{H}} . 
\ee
At this point, we can use Einstein equation in the form (\ref{e:dec:Einstein2})
to express the Ricci tensor ${}^4\!\w{R}$ in terms of the matter
stress-energy tensor $\w{T}$. We obtain
\be \label{e:glo:MK_Tolman0}
	M_{\rm K} = 2 \int_{\mathcal{V}_t} 
		\left( T_{\mu\nu} - \frac{1}{2} T g_{\mu\nu} \right)
			n^\mu k^\nu \sqrt{\gm}\, d^3 x
		 	+ M_{\rm K}^{\mathcal{H}} .
\ee
The support of the integral over $\mathcal{V}_t$ is reduced to the location 
of matter, i.e. the domain where $\w{T}\not =0$. It is then clear on formula
(\ref{e:glo:MK_Tolman0}) that $M_{\rm K}$ is independent of the choice of
the 2-surface $\Sp_t$, provided all the matter is contained in $\Sp_t$. 
In particular, we may extend the integration to all $\Sigma_t$ and write
formula (\ref{e:glo:MK_Tolman0}) as 
\be \label{e:glo:MK_Tolman}
	\encadre{
		M_{\rm K} = 2 \int_{\Sigma_t} 
		\left[ \w{T}(\w{n},\w{k}) - \frac{1}{2} T \, \w{n}\cdot\w{k} \right]
			\sqrt{\gm}\, d^3 x
		 	+ M_{\rm K}^{\mathcal{H}}
		} . 
\ee
The Komar mass then appears as a global quantity defined for stationary spacetimes. 
\begin{remark}
One may have $M_{\rm K}^{\mathcal{H}}<0$, with $M_{\rm K}>0$, provided that the
matter integral in Eq.~(\ref{e:glo:MK_Tolman}) compensates for the negative value
of $M_{\rm K}^{\mathcal{H}}$. Such spacetimes exist, as recently demonstrated
by Ansorg and Petroff \cite{AnsorP06}: these authors have numerically constructed
spacetimes containing a black hole with $M_{\rm K}^{\mathcal{H}}<0$
surrounded by a ring of matter (incompressible perfect fluid) such that the
total Komar mass is positive. 
\end{remark}

\subsection{3+1 expression of the Komar mass and link with the ADM mass}

In stationary spacetimes, it is natural to use coordinates adapted to the symmetry,
i.e. coordinates $(t,x^i)$ such that
\be
	\encadre{\wpar_t = \w{k} } .
\ee
Then we have the following 3+1 decomposition of the Killing vector in 
terms of the lapse and shift [cf. Eq.~(\ref{e:dec:t_Nn_b})]:
\be \label{e:glo:Killing_3p1}
	 \w{k} = N \w{n} + \w{\beta} . 
\ee
Let us inject this relation in the integrand of the definition (\ref{e:glo:M_Komar_def}) 
of the Komar mass :
\bea
	\nabla^\mu k^\nu \, dS_{\mu\nu} & = & 
		 \nabla_\mu k_\nu (s^\mu n^\nu - n^\mu s^\nu) \sqrt{q} 
		\, d^2 y\nonumber \\
		& = & 2 \nabla_\mu k_\nu \, s^\mu n^\nu \sqrt{q} 
		\, d^2 y\nonumber \\
		& = & 2 \left( \nabla_\mu N \, n_\nu + N \nabla_\mu n_\nu
			+ \nabla_\mu \beta_\nu \right) 
			s^\mu n^\nu \sqrt{q} 
		\, d^2 y\nonumber \\
		& = & 2 \left( - s^\mu \nabla_\mu N + 0 - s^\mu \beta_\nu
			\nabla_\mu n^\nu \right)  \sqrt{q} 
		\, d^2 y\nonumber \\
		& = & - 2 \left( s^i D_i N - K_{ij} s^i \beta^j \right)  \sqrt{q} 
		\, d^2 y , 	\label{e:glo:nabkdS_3p1}
\eea
where we have used Killing's equation (\ref{e:glo:Killing}) to get the second line,
the orthogonality of $\w{n}$ and $\w{\beta}$ to get the fourth one and 
expression (\ref{e:fol:nab_m_comp}) for $\nabla_\mu n^\nu$ to get the last line.
Inserting Eq.~(\ref{e:glo:nabkdS_3p1}) into Eq.~(\ref{e:glo:M_Komar_def})
yields the 3+1 expression of the Komar mass: 
\be \label{e:glo:MK_3p1}
	\encadre{ 
	M_{\rm K} =  \frac{1}{4\pi} \oint_{\Sp_t} 
	\left( s^i D_i N - K_{ij} s^i \beta^j \right)  \sqrt{q} 
		\, d^2 y
	} . 
\ee
\begin{example}
A simple prototype of a stationary spacetime is of course the Schwarzschild spacetime.
Let us compute its Komar mass by means of the above formula and the
foliation $(\Sigma_t)_{t\in\R}$ defined by the standard Schwarzschild coordinates
(\ref{e:glo:Schwarz_coord}). For this foliation, $K_{ij}=0$, which reduces
Eq.~(\ref{e:glo:MK_3p1}) to the flux of the lapse's gradient across $\Sp_t$.
Taking advantage of the spherical symmetry, we choose
$\Sp_t$ to be a surface $r={\rm const}$. Then $y^a=(\theta,\varphi)$. 
The unit normal $\w{s}$ is read from 
the line element (\ref{e:glo:Schwarz_coord}); its components with respect 
to the Schwarzschild coordinates $(r,\theta,\varphi)$ are
\be
	s^i = \left( \left(1-\frac{2m}{r}\right) ^{1/2}, 0, 0 \right) . 
\ee
$N$ and $\sqrt{q}$ are also read on the line 
element (\ref{e:glo:Schwarz_coord}): $N=(1-2m/r)^{1/2}$ and $\sqrt{q}=r^2 \sin\theta$,
so that Eq.~(\ref{e:glo:MK_3p1}) results in
\bea	
	M_{\rm K} =  \frac{1}{4\pi} \oint_{r={\rm const}} 
		\left(1-\frac{2m}{r}\right) ^{1/2} \der{}{r} \left[
			\left(1-\frac{2m}{r}\right) ^{1/2} \right]
			\, r^2 \sin\theta d\theta d\varphi . 
\eea
All the terms containing $r$ simplify and we get
\be \label{e:glo:MK_Schwarz}
	M_{\rm K} = m . 
\ee
On this particular example, we have verified that the value of $M_{\rm K}$ does
not depend upon the choice of $\Sp_t$. 
\end{example}

Let us now turn to the volume expression (\ref{e:glo:MK_Tolman}) of the Komar mass. 
By using the 3+1 decomposition (\ref{e:dec:T_3p1}) and
(\ref{e:dec:trT_SmE}) of respectively $\w{T}$ and $T$, we get
\bea
	\w{T}(\w{n},\w{k}) - \frac{1}{2} T \, \w{n}\cdot\w{k} 
	& = & - \langle \w{p}, \w{k} \rangle 
		- E \langle \uu{n}, \w{k} \rangle - \frac{1}{2} (S-E)\, \w{n}\cdot\w{k} 
			\nonumber \\
	& = & - \langle \w{p}, \w{\beta} \rangle
		+ EN + \frac{1}{2} (S-E)N 
	 =  \frac{1}{2} N(E+S) - \langle \w{p}, \w{\beta} \rangle .
\eea
Hence formula (\ref{e:glo:MK_Tolman}) becomes
\be \label{e:glo:MK_vol_3p1}
	\encadre{
		M_{\rm K} = \int_{\Sigma_t} 
		\left[ N(E+S) - 2 \langle \w{p}, \w{\beta} \rangle \right]
			\sqrt{\gm}\, d^3 x
		 	+ M_{\rm K}^{\mathcal{H}}
		} , 
\ee
with the Komar mass of the hole given by an expression identical to 
Eq.~(\ref{e:glo:MK_3p1}), except for $\Sp_t$ replaced by $\mathcal{H}_t$ 
[notice the double change of sign: first in
Eq.~(\ref{e:glo:MKH_def}) and secondly in Eq.~(\ref{e:glo:dSprime}), so
that at the end we have an expression identical to Eq.~(\ref{e:glo:MK_3p1})]:
\be
	\encadre{ 
	M_{\rm K}^{\mathcal{H}} =  \frac{1}{4\pi} \oint_{\mathcal{H}_t} 
	\left( s^i D_i N - K_{ij} s^i \beta^j \right)  \sqrt{q} 
		\, d^2 y
	} . 	
\ee

It is easy to take the Newtonian limit Eq.~(\ref{e:glo:MK_vol_3p1}), 
by making $N\rightarrow 1$, $E\rightarrow\rho$, $S\ll E$ [Eq.~(\ref{e:mat:SllE_Newt})],
$\w{\beta} \rightarrow 0$, $\gm\rightarrow f$ and 
$M_{\rm K}^{\mathcal{H}}=0$. We get 
\be
	M_{\rm K} = \int_{\Sigma_t} \rho \, \sqrt{f}\, d^3 x . 
\ee
Hence at the Newtonian limit, the Komar mass reduces to the standard total mass.
This, along with the result (\ref{e:glo:MK_Schwarz}) for Schwarzschild
spacetime, justifies the name Komar \emph{mass}. 

A natural question which arises then is how does the Komar mass relate to the
ADM mass of $\Sigma_t$ ? 
The answer is not obvious if one compares the 
defining formul\ae\  (\ref{e:glo:M_ADM_def}) and (\ref{e:glo:M_Komar_def}).
It is even not obvious if one compares the 3+1 expressions (\ref{e:glo:M_ADM_Psi})
and (\ref{e:glo:MK_3p1}): Eq.~(\ref{e:glo:M_ADM_Psi}) involves the flux of
the gradient of the conformal factor $\Psi$ of the 3-metric, whereas 
Eq.~(\ref{e:glo:MK_3p1}) involves the flux of the gradient of the lapse function $N$. 
Moreover, in Eq.~(\ref{e:glo:M_ADM_Psi}) the integral must be evaluated at
spatial infinity, whereas in Eq.~(\ref{e:glo:M_ADM_Psi}) it can be evaluated at
any finite distance (outside the matter sources). 
The answer has been obtained in 1978 by Beig \cite{Beig78}, 
as well as by Ashtekar and Magnon-Ashtekar the year after \cite{AshteM79}:
for any foliation $({\Sigma_t})_{t\in\R}$ whose unit normal
vector $\w{n}$ coincides with the timelike Killing vector $\w{k}$ at spatial
infinity [i.e. $N\rightarrow 1$ and $\w{\beta}\rightarrow 0$ in 
Eq.~(\ref{e:glo:Killing_3p1})], 
\be
	\encadre{M_{\rm K} = M_{\rm ADM}} . 
\ee

\begin{remark}
In the quasi-isotropic gauge, we have obtained a volume expression of the ADM mass,
Eq.~(\ref{e:glo:M_ADM_QI_vol}), that we may compare to the volume expression 
(\ref{e:glo:MK_vol_3p1}) of the Komar mass. Even when there is no hole, the
two expressions are pretty different. In particular, the Komar mass integral
has a compact support (the matter domain), whereas the ADM mass integral has
not. 
\end{remark}
\subsection{Komar angular momentum} \label{s:glo:Komar_J}

If the spacetime $(\M,\w{g})$ is axisymmetric, its 
\defin{Komar angular momentum} is defined by a surface integral similar
that of the Komar mass, Eq.~(\ref{e:glo:M_Komar_def}), but with the Killing
vector $\w{k}$ replaced by the Killing vector $\w{\phi}$ associated with
the axisymmetry:
\be  \label{e:glo:JK_def}
	\encadre{ 
	J_{\rm K} :=  \frac{1}{16\pi} \oint_{\Sp_t} \nabla^\mu \phi^\nu \, dS_{\mu\nu}
	} . 
\ee
Notice a factor $-2$ of difference with respect to formula (\ref{e:glo:M_Komar_def})
(the so-called \emph{Komar's anomalous factor} \cite{Katz85}). 

For the same reason as for $M_{\rm K}$, $J_{\rm K}$ is actually independent of 
the surface $\Sp_t$ as long as the latter is outside all the possible matter sources
and $J_{\rm K}$ can be expressed by a volume integral over the matter
by a formula similar to (\ref{e:glo:MK_Tolman}) (except for the factor $-2$):
\be \label{e:glo:JK_Tolman}
		J_{\rm K} = - \int_{\Sigma_t} 
		\left[ \w{T}(\w{n},\w{\phi}) - \frac{1}{2} T \, \w{n}\cdot\w{\phi} \right]
			\sqrt{\gm}\, d^3 x
		 	+ J_{\rm K}^{\mathcal{H}} , 
\ee
with 
\be
	 J_{\rm K}^{\mathcal{H}} := 
	 	- \frac{1}{16\pi} \oint_{\mathcal{H}_t} \nabla^\mu \phi^\nu 
		\, dS^{\mathcal{H}}_{\mu\nu} . 
\ee

Let us now establish the 3+1 expression of the Komar angular momentum. 
It is natural to choose a foliation adapted to the axisymmetry in the sense
that the Killing vector $\w{\phi}$ is tangent to the hypersurfaces $\Sigma_t$. 
Then $\w{n}\cdot\w{\phi}=0$ and the integrand in the definition (\ref{e:glo:JK_def}) is
\bea
	\nabla^\mu \phi^\nu \, dS_{\mu\nu} & = & 
		 \nabla_\mu \phi_\nu (s^\mu n^\nu - n^\mu s^\nu) \sqrt{q} 
		\, d^2 y\nonumber \\
		& = & 2 \nabla_\mu \phi_\nu \, s^\mu n^\nu \sqrt{q} 
		\, d^2 y\nonumber \\
		& = &  -2 s^\mu \phi_\nu \nabla_\mu  n^\nu \sqrt{q} 
		\, d^2 y\nonumber \\
		& = & 2 K_{ij} s^i \phi^j \sqrt{q} \, d^2 y . 
\eea
Accordingly Eq.~(\ref{e:glo:JK_def}) becomes
\be \label{e:glol:JK_3p1}
	\encadre{ 
	J_{\rm K} =  \frac{1}{8\pi} \oint_{\Sp_t} 
		K_{ij} s^i \phi^j \sqrt{q} \, d^2 y
	} . 
\ee
\begin{remark}
Contrary to the 3+1 expression of the Komar mass which turned out to be very
different from the expression of the ADM mass, the 3+1 expression of the Komar
angular momentum as given by Eq.~(\ref{e:glol:JK_3p1}) is very similar to
the expression of the angular momentum deduced from the Hamiltonian formalism,
i.e. Eq.~(\ref{e:glo:angu_mom_def}). The only differences are that it is no longer
necessary to take the limit $\Sp_t\rightarrow\infty$ and that there is no
trace term $K\gm_{ij} s^i \phi^j$ in Eq.~(\ref{e:glol:JK_3p1}). Moreover, 
if one evaluates the Hamiltonian expression
in the asymptotically maximal gauge (\ref{e:glo:K0gauge}) then $K=O(r^{-3})$
and thanks to the asymptotic orthogonality of $\w{s}$ and $\w{\phi}$, 
$\gm_{ij} s^i \phi^j=O(1)$, so that $K\gm_{ij} s^i \phi^j$ does not contribute to
the integral and expressions (\ref{e:glol:JK_3p1}) and (\ref{e:glo:angu_mom_def})
are then identical. 
\end{remark}

\begin{example}
A trivial example is provided by Schwarzschild spacetime, which among other
things is axisymmetric. For the 3+1 foliation associated with the Schwarzschild coordinates
(\ref{e:glo:Schwarz_coord}), the extrinsic curvature tensor $\w{K}$ vanishes
identically, so that Eq.~(\ref{e:glol:JK_3p1}) yields immediately $J_{\rm K}=0$.
For other foliations, like that associated with Eddington-Finkelstein coordinates,
$\w{K}$ is no longer zero but is such that $K_{ij} s^i \phi^j=0$, yielding
again $J_{\rm K}=0$ (as it should be since the Komar angular momentum is
independent of the foliation). 
Explicitely for Eddington-Finkelstein coordinates,
\be
 K_{ij}s^i = \left( - \frac{2m}{r^2} \frac{1+\frac{m}{r}}{1+\frac{2m}{r}},\ 0, \ 0
	\right) , 
\ee
(see e.g. Eq.~(D.25) in Ref.~\cite{GourgJ06a})
and $\phi^j=(0,0,1)$, so that obviously $K_{ij} s^i \phi^j=0$.
\end{example}

\begin{example}
The most natural non trivial example is certainly that of Kerr spacetime.
Let us use the 3+1 foliation associated with the standard Boyer-Lindquist coordinates
$(t,r,\theta,\varphi)$ and evaluate the integral (\ref{e:glol:JK_3p1}) by choosing
for $\Sp_t$ a sphere $r={\rm const}$. Then $y^a=(\theta,\varphi)$. 
The Boyer-Lindquist components of $\w{\phi}$ are $\phi^i=(0,0,1)$ and
those of $\w{s}$ are $s^i=(s^r,0,0)$ since $\gm_{ij}$ is diagonal is these
coordinates. 
The formula (\ref{e:glol:JK_3p1}) then reduces to 
\be
	J_{\rm K} =  \frac{1}{8\pi} \oint_{r={\rm const}} 
		K_{r\varphi} s^r \sqrt{q} \, d\theta\, d\varphi . 
\ee
The extrinsic curvature component $K_{r\varphi}$ can be evaluated via formula (\ref{e:dec:Einstein_PDE1}),
which reduces to $2N K_{ij} = \Liec{\beta} \gm_{ij}$ since $\dert{\gm_{ij}}{t}=0$.
From the Boyer-Lindquist line element
(see e.g. Eq.~(5.29) in Ref.~\cite{HawkiE73}), we read the components of the shift:
\be \label{e:glo:beta_Kerr_BL}
	(\beta^r,\beta^\theta,\beta^\varphi) = 
	\left(0,\ 0,\ - \frac{2 a m r}{(r^2+a^2)(r^2+a^2\cos^2\theta)
	+ 2 a^2 m r \sin^2\theta} \right) , 
\ee
where $m$ and $a$ are the two parameters of the Kerr solution. 
Then, using Eq.~(\ref{e:Lie_der_comp}),
\be
    K_{r\varphi} = \frac{1}{2N}  \Liec{\beta} \gm_{r\varphi}
	= \frac{1}{2N}  \bigg( \beta^\varphi \underbrace{\der{\gm_{r\varphi}}{\varphi}}_{=0}
	+ \gm_{\varphi\varphi} \der{\beta^\varphi}{r}
	+ \gm_{r\varphi}
	\underbrace{\der{\beta^\varphi}{\varphi}}_{=0} \bigg) 
	= \frac{1}{2N}  \gm_{\varphi\varphi} \der{\beta^\varphi}{r} . 
\ee
Hence 
\be
 	J_{\rm K} =  \frac{1}{16\pi} \oint_{r={\rm const}} 
	\frac{s^r}{N}  \gm_{\varphi\varphi} \der{\beta^\varphi}{r} 
	\sqrt{q} \, d\theta\, d\varphi .
\ee
The values of $s^r$, $N$, $\gm_{\varphi\varphi}$ and $\sqrt{q}$ can all be read
on the Boyer-Lindquist line element. However this is a bit tedious. 
To simplify things, let us evaluate $J_{\rm K}$ only in the limit
$r\rightarrow\infty$. Then $s^r \sim 1$, $N\sim 1$, $\gm_{\varphi\varphi}\sim r^2\sin^2\theta$,
$\sqrt{q}\sim r^2\sin\theta$ and, from Eq.~(\ref{e:glo:beta_Kerr_BL}),
$\beta^\varphi \sim -2 a m /r^3$, so that
\be
	J_{\rm K} =  \frac{1}{16\pi} \oint_{r={\rm const}}
	r^2\sin^2 \frac{6 a m }{r^4} r^2 \sin\theta \, d\theta\, d\varphi
	= \frac{3 am}{8\pi} \times 2\pi \times \int_0^\pi \sin^3 \theta\, d\theta .
\ee
Hence, as expected, 
\be
	J_{\rm K} = a m . 
\ee
\end{example}

Let us now find the 3+1 expression of the volume version (\ref{e:glo:JK_Tolman})
of the Komar angular momentum.  
We have $\w{n}\cdot\w{\phi} = 0$ and, from the 
3+1 decomposition  (\ref{e:dec:T_3p1}) of $\w{T}$:
\be
	\w{T}(\w{n},\w{\phi}) = - \langle \w{p} , \w{\phi} \rangle . 
\ee
Hence formula (\ref{e:glo:JK_Tolman}) becomes
\be
	\encadre{ 
		J_{\rm K} = \int_{\Sigma_t} 
		\langle \w{p} , \w{\phi} \rangle
			\sqrt{\gm}\, d^3 x
		 	+ J_{\rm K}^{\mathcal{H}}
	 } , 
\ee
with 
\be
	\encadre{
	J_{\rm K}^{\mathcal{H}} =  \frac{1}{8\pi} \oint_{\mathcal{H}_t} 
		K_{ij} s^i \phi^j \sqrt{q} \, d^2 y } . 
\ee
\begin{example}
Let us consider a perfect fluid. Then $\w{p} = (E+P) \uu{U}$ [Eq.~(\ref{e:mat:p_fluid})],
so that
\be
			J_{\rm K} = \int_{\Sigma_t} 
		(E+P) \, \w{U}\cdot\w{\phi} \, 
			\sqrt{\gm}\, d^3 x
		 	+ J_{\rm K}^{\mathcal{H}} . 
\ee
Taking $\w{\phi} = - y\wpar_x + x\wpar y$ (symmetry axis = $z$-axis), 
the Newtonian limit of this expression is then
\be
	J_{\rm K} = \int_{\Sigma_t} \rho (-y U^x + x U^y)  \, dx\, dy\, dz , 
\ee
i.e. we recognize the standard expression for the angular momentum around 
the $z$-axis. 
\end{example}

%  
%    Chapitre : Initial data
%
% $Date: 2007-03-06 11:59:03 +0100 (mar, 06 mar 2007) $
% $Rev: 183 $
% $Author: e_gourgoulhon $
%%%%%%%%%%%%%%%%%%%%%%%%%%%%%

\chapter{The initial data problem} \label{s:ini}

%\verb$Date: 2007-03-06 11:59:03 +0100 (mar, 06 mar 2007) $

\minitoc
\vspace{1cm}

%%%%%%%%%%%%%%%%%%%%%%%%%%%%%%%%%%%%%%%%%%%%%%%%%%%%%%%%%%%%%%%%%%%%%%%%%%%%

\section{Introduction}

\subsection{The initial data problem} \label{s:ini:idp}

We have seen in Chap.~\ref{s:dec} that thanks to the 3+1 decomposition,
the resolution of Einstein equation amounts to solving a Cauchy
problem, namely to evolve ``forward in time'' some initial data.
However this is a Cauchy problem with constraints. This makes the set up
of initial data a non trivial task, because these data must obey
the constraints. Actually one may distinguish two problems:
\begin{itemize}
\item \emph{The mathematical problem:} given some hypersurface $\Sigma_0$, 
find a Riemannian metric $\wgm$, a symmetric bilinear form $\w{K}$
and some matter distribution $(E,\w{p})$ on $\Sigma_0$
such that the Hamiltonian constraint
(\ref{e:dec:Einstein_PDE3}) and the momentum constraint (\ref{e:dec:Einstein_PDE4})
are satisfied:
\bea
 & & \encadre{ R + K^2 - K_{ij} K^{ij} = 16\pi E } \label{e:ini:Ham_constr}\\
 & & \encadre{ D_j K^j_{\ \, i} - D_i K = 8\pi  p_i } . \label{e:ini:mom_constr}
\eea
In addition, the matter distribution $(E,\w{p})$ may have some constraints
from its own. We shall not discuss them here.
\item \emph{The astrophysical problem:} make sure that the solution
to the constraint equations has something to do with the physical system 
that one wish to study. 
\end{itemize}
Notice that Eqs.~(\ref{e:ini:Ham_constr})-(\ref{e:ini:mom_constr}) involve 
a single hypersurface $\Sigma_0$, not a foliation $\left(\Sigma_t\right)_{t\in \R}$. 
In particular, neither the lapse function nor the shift vector appear in these equations. 
Facing them, 
a naive way to proceed would be to choose freely the metric $\wgm$, thereby
fixing the connection $\w{D}$ and the scalar curvature $R$, and to solve 
Eqs.~(\ref{e:ini:Ham_constr})-(\ref{e:ini:mom_constr}) for $\w{K}$.
Indeed, for fixed $\wgm$, $E$, and $\w{p}$, Eqs.~(\ref{e:ini:Ham_constr})-(\ref{e:ini:mom_constr}) form a quasi-linear system 
of first order for the components $K_{ij}$. 
However, as discussed by Choquet-Bruhat \cite{Foure56}, this 
approach is not satisfactory
because we have only four equations for six unknowns $K_{ij}$ and there is
no natural prescription for choosing arbitrarily two among the six components
$K_{ij}$. 

Lichnerowicz (1944) \cite{Lichn44} has shown that a
much more satisfactory split of the initial data $(\wgm,\w{K})$
between freely choosable parts and parts obtained by solving Eqs.~(\ref{e:ini:Ham_constr})-(\ref{e:ini:mom_constr})
is provided by the conformal decomposition introduced in Chap.~\ref{s:cfd}. 
Lichnerowicz method has been extended by Choquet-Bruhat (1956, 1971) 
\cite{Foure56,Choqu71},
by York and \'O Murchadha (1972, 1974, 1979)
\cite{York72b,York73,OMurcY74,York79} and more recently by York and Pfeiffer
(1999, 2003) \cite{York99,PfeifY03}. 
Actually, conformal decompositions are by far the 
most widely spread techniques to get initial data for the 3+1
Cauchy problem. 
Alternative methods exist, such as the quasi-spherical ansatz introduced by
Bartnik in 1993 \cite{Bartn93} or a procedure developed by Corvino (2000)
\cite{Corvi00} and by Isenberg,
Mazzeo and Pollack (2002) \cite{IsenbMP02} for gluing together known solutions of
the constraints, thereby producing new ones. 
Here we shall limit ourselves to the conformal methods. 
Standard reviews on this subject are the articles by
York (1979) \cite{York79} and Choquet-Bruhat and York (1980) 
\cite{ChoquY80}. Recent reviews are the articles by
Cook (2000) \cite{Cook00}, Pfeiffer (2004) \cite{Pfeif04}
and Bartnik and Isenberg (2004) \cite{BartnI04}.

\subsection{Conformal decomposition of the constraints}

The conformal form of the constraint equations has been derived
in Chap.~\ref{s:cfd}. We have introduced there the conformal metric $\wtgm$ and the conformal factor $\Psi$
such that the metric $\wgm$ induced by the spacetime metric on some hypersurface
$\Sigma_0$ is
[cf. Eq.~(\ref{e:cfd:gmij_up_down})]
\be \label{e:ini:gm_Psi_tgm}
	\gm_{ij} = \Psi^4 \tgm_{ij} , 
\ee
and have decomposed the traceless part $A^{ij}$ of the extrinsic curvature $K^{ij}$
according to [cf. Eq.~(\ref{e:cfd:def_hA})]
\be \label{e:ini:A_Psi_hA}
	A^{ij} = \Psi^{-10} \hA^{ij} .
\ee
We consider here the decomposition involving $\hA^{ij}$ [$\alpha=-10$ in 
Eq.~(\ref{e:cfd:A_scale})] and not the alternative one, which uses
$\tA^{ij}$ ($\alpha=-4$), because we have seen in Sec.~\ref{s:cfd:conf_traceless}
that the former is well adapted to the momentum constraint. 
Using the decompositions (\ref{e:ini:gm_Psi_tgm}) and (\ref{e:ini:A_Psi_hA}),
we have rewritten the Hamiltonian constraint (\ref{e:ini:Ham_constr}) and
the momentum constraint (\ref{e:ini:mom_constr}) 
as respectively the Lichnerowicz equation 
[Eq.~(\ref{e:cfd:Einstein5_hA})] and an
equation involving the divergence of $\hA^{ij}$ with respect to the
conformal metric [Eq.~(\ref{e:cfd:Einstein6_hA})] :
\bea
	& & 	\encadre{ 
	\tD_i \tD^i \Psi -\frac{1}{8} {\tilde R} \Psi
	+ \frac{1}{8} \hA_{ij} \hA^{ij} \, \Psi^{-7}
	+ 2\pi {\tilde E} \Psi^{-3} - \frac{1}{12} K^2 \Psi^5 = 0 } , \qquad
	\qquad \qquad \qquad \qquad \ \ \label{e:ini:Ham_conf} \\
	& & 
	\encadre{ \tD_j \hA^{ij} - \frac{2}{3} \Psi^6 \tD^i K = 8\pi {\tilde p}^i } , 
		\label{e:ini:mom_conf} 
\eea
where we have introduce the rescaled matter quantities
\be \label{e:ini:tE_def}
	{\tilde E} := \Psi^8 E
\ee
and
\be \label{e:ini:tp_def}
	{\tilde p}^i := \Psi^{10} p^i .
\ee
The definition of ${\tilde p}^i$ is clearly motivated by Eq.~(\ref{e:cfd:Einstein6_hA}).
On the contrary the power $8$ in the definition of $\tilde E$ is not the 
only possible choice. As we shall see in \S~\ref{s:ini:Lichne}, 
it is chosen (i) to guarantee a negative power of $\Psi$ in the $\tilde E$ term in Eq.~(\ref{e:ini:Ham_conf}), resulting in some uniqueness property of the solution
and (ii) to allow for an easy implementation of the dominant energy condition. 

\section{Conformal transverse-traceless method} \label{s:ini:CTT}

\subsection{Longitudinal/transverse decomposition of $\hA^{ij}$} \label{s:ini:long_trans}

In order to solve the system (\ref{e:ini:Ham_conf})-(\ref{e:ini:mom_conf}),
York (1973,1979) \cite{York73,York74,York79} has decomposed
$\hA^{ij}$ into a longitudinal part and a transverse one, by setting
\be \label{e:ini:decomp_hA}
	\encadre{ \hA^{ij} = (\tilde L X)^{ij} + \hA^{ij}_{\rm TT} } , 
\ee
where $\hA^{ij}_{\rm TT}$ is both traceless and transverse (i.e. divergence-free)
with respect to the metric $\wtgm$:
\be \label{e:ini:hA_TT}
	\tgm_{ij} \hA^{ij}_{\rm TT} = 0
	\qquad \mbox{and} \qquad
	\tD_j \hA^{ij}_{\rm TT} = 0 ,
\ee
and $(\tilde L X)^{ij}$ is the \defin{conformal Killing operator} associated
with the metric $\wtgm$ and acting on the vector field $\w{X}$:
\be \label{e:ini:conf_Killing_def}
	\encadre{ (\tilde L X)^{ij} := \tD^i X^j + \tD^j X^i 
			- \frac{2}{3} \tD_k X^k \, \tgm^{ij} } . 
\ee
The properties of this linear differential operator are detailed in Appendix~\ref{s:cko}.
Let us retain here that $(\tilde L X)^{ij}$ is by construction traceless:
\be
	\tgm_{ij} (\tilde L X)^{ij} = 0  
\ee
(it must be so
because in Eq.~(\ref{e:ini:decomp_hA}) both $\hA^{ij}$ and $\hA^{ij}_{\rm TT}$
are traceless)
and the kernel of $\w{\tilde L}$ is made of the \defin{conformal Killing vectors}
of the metric $\wtgm$, i.e. the generators of the conformal 
isometries (cf. Sec.~\ref{s:cko:ckv}). The symmetric tensor $(\tilde L X)^{ij}$ 
is called the \defin{longitudinal part} of $\hA^{ij}$, whereas $\hA^{ij}_{\rm TT}$
is called the \defin{transverse part}. 

Given $\hA^{ij}$, the vector $\w{X}$ is determined by taking the divergence 
of Eq.~(\ref{e:ini:decomp_hA}): taking into account property (\ref{e:ini:hA_TT}), 
we get
\be \label{e:ini:tDLX_divA}
	\tD_j (\tilde L X)^{ij} = \tD_j \hA^{ij} .
\ee
The second order operator $\tD_j (\tilde L X)^{ij}$ acting on the vector $\w{X}$
is the \defin{conformal vector Laplacian} $\w{\tilde\Delta}_L$:
\be
  \encadre{ \tilde\Delta_L \, X^i := \tD_j (\tilde L X)^{ij}
	= \tD_j \tD^j X^i + \frac{1}{3} \tD^i \tD_j X^j
	+ {\tilde R}^i_{\ \, j} X^j } , 
\ee
where the second equality follows from Eq.~(\ref{e:cko:DeltaL_DD}).
The basic properties of $\w{\tilde\Delta}_L$ are investigated in Appendix~\ref{s:cko},
where it is shown that this operator is elliptic and that its kernel is, 
in practice, reduced to the conformal Killing vectors of $\wtgm$, if any. We rewrite Eq.~(\ref{e:ini:tDLX_divA}) as
\be \label{e:ini:DeltaLX_divA}
	\tilde\Delta_L \, X^i = \tD_j \hA^{ij} .
\ee
The existence and uniqueness of the longitudinal/transverse decomposition
(\ref{e:ini:decomp_hA}) depend on the existence and uniqueness of solutions
$\w{X}$ to Eq.~(\ref{e:ini:DeltaLX_divA}).
We shall consider two cases:  
\begin{itemize}
\item $\Sigma_0$ is a \emph{closed manifold}, i.e. is compact without boundary;
\item $(\Sigma_0,\wgm)$ is an \emph{asymptotically flat manifold}, in the sense
made precise in Sec.~\ref{s:glo:asymp_flat}. 
\end{itemize}
In the first case, it is shown in Appendix~\ref{s:cko} that solutions
to Eq.~(\ref{e:ini:DeltaLX_divA}) exist provided that the source $\tD_j \hA^{ij}$
is orthogonal to all conformal Killing vectors of $\wtgm$, in the sense
that [cf. Eq.~(\ref{e:cko:intCS})]:
\be 
	\forall \w{C}\in\mathrm{ker}\, \w{\tilde L},\quad 
	 \int_{\Sigma}  \tgm_{ij}  C^i \tD_k \hA^{jk} \sqrt{\tgm} \, d^3 x = 0  . 
\ee
But this is easy to verify:
using the fact that the source is a pure divergence and that $\Sigma_0$
is closed, we may integrate by parts and get, for any vector field $\w{C}$,
\be
	\int_{\Sigma_0}  \tgm_{ij}  C^i \, \tD_k \hA^{jk} \sqrt{\tgm} \, d^3 x = 
	- \frac{1}{2}
	\int_{\Sigma_0}  \tgm_{ij}  \tgm_{kl}  (\tilde L C)^{ik} \hA^{jl}
		\sqrt{\tgm} \, d^3 x .
\ee
Then, obviously, when $\w{C}$ is a conformal Killing vector, the right-hand
side of the above equation vanishes. 
So there exists a solution to Eq.~(\ref{e:ini:DeltaLX_divA}) and
this solution is unique up to the addition of a conformal Killing vector. 
However, given a solution $\w{X}$, for any conformal Killing vector $\w{C}$,
the solution $\w{X}+\w{C}$ yields to the same value of $\w{\tilde L}\w{X}$,
since $\w{C}$ is by definition in the kernel of $\w{\tilde L}$. 
Therefore we conclude that the decomposition (\ref{e:ini:decomp_hA}) of $\hA^{ij}$
is unique, although the vector $\w{X}$ may not be if $(\Sigma_0,\wtgm)$
admits some conformal isometries. 

In the case of an asymptotically flat manifold, the existence and uniqueness
is guaranteed by the Cantor 
theorem mentioned in Sec.~\ref{s:cko:Poisson}. We shall then require the decay
condition
\be
	\dderp{\tgm_{ij}}{x^k}{x^l} = O(r^{-3})	\label{e:ini:aflat_extra}
\ee
in addition to the asymptotic flatness conditions (\ref{e:glo:tgm_asymp})
introduced in Chap.~\ref{s:glo}. This guarantees that [cf. Eq.~(\ref{e:cko:decay_Ricci})]
\be
	{\tilde R}_{ij} = O(r^{-3}) . 
\ee
In addition, we notice that
$\hA^{ij}$ obeys the decay condition $\hA^{ij}=O(r^{-2})$
which is inherited from the asymptotic flatness condition (\ref{e:glob:aflat3}).
Then  $\tD_j \hA^{ij}=O(r^{-3})$ so that condition (\ref{e:cko:decay_source_Poisson})
is satisfied. Then all conditions are fulfilled to conclude that 
Eq.~(\ref{e:ini:DeltaLX_divA}) admits a unique solution $\w{X}$ 
which vanishes at infinity. 

To summarize, for all considered cases (asymptotic flatness and
closed manifold), any symmetric and traceless tensor $\hA^{ij}$
(decaying as $O(r^{-2})$ in the asymptotically flat case) 
admits a unique longitudinal/transverse decomposition of
the form (\ref{e:ini:decomp_hA}). 

\subsection{Conformal transverse-traceless form of the constraints}

Inserting the longitudinal/transverse decomposition (\ref{e:ini:decomp_hA})
into the constraint equations (\ref{e:ini:Ham_conf}) and (\ref{e:ini:mom_conf})
and making use of Eq.~(\ref{e:ini:DeltaLX_divA}) yields to the system
\bea
	& & 	\encadre{ 
	\tD_i \tD^i \Psi -\frac{1}{8} {\tilde R} \Psi
	+ \frac{1}{8} \left[(\tilde L X)_{ij} +  \hA_{ij}^{\rm TT}\right]
	 \left[(\tilde L X)^{ij} +  \hA^{ij}_{\rm TT}\right]\, \Psi^{-7}
	+ 2\pi {\tilde E} \Psi^{-3} - \frac{1}{12} K^2 \Psi^5 = 0 } , 	\nonumber \\ 
	& & \label{e:ini:Ham_CTT} \\
	& & 
	\encadre{ \tilde\Delta_L \, X^i - \frac{2}{3} \Psi^6 \tD^i K 
	= 8\pi {\tilde p}^i } ,	\label{e:ini:mom_CTT}
\eea
where
\bea
	& & (\tilde L X)_{ij} := \tgm_{ik} \tgm_{jl} (\tilde L X)^{kl} \\
	& & \hA_{ij}^{\rm TT} := \tgm_{ik} \tgm_{jl} \hA^{kl}_{\rm TT} .
\eea

With the constraint equations written as (\ref{e:ini:Ham_CTT}) and (\ref{e:ini:mom_CTT}),
we see clearly which part of the initial data on $\Sigma_0$ can be freely chosen
and which part is ``constrained'': 
\begin{itemize}
\item free data: 
\begin{itemize}
\item conformal metric $\wtgm$;
\item symmetric traceless and transverse tensor $\hA^{ij}_{\rm TT}$ (traceless and transverse are meant with respect to $\wtgm$: $\tgm_{ij} \hA^{ij}_{\rm TT} = 0$
and $\tD_j \hA^{ij}_{\rm TT} = 0$);
\item scalar field $K$; 
\item conformal matter variables: $(\tilde E,{\tilde p}^i)$;
\end{itemize}
\item constrained data (or ``determined data''): 
\begin{itemize}
\item conformal factor $\Psi$, obeying the non-linear elliptic equation
(\ref{e:ini:Ham_CTT}) (Lichnerowicz equation)
\item vector $\w{X}$, obeying the linear elliptic equation
(\ref{e:ini:mom_CTT}) .
\end{itemize}
\end{itemize}
Accordingly the general strategy to get valid initial data for the Cauchy problem 
is to choose $(\tgm_{ij},\hA^{ij}_{\rm TT},K,\tilde E,{\tilde p}^i)$ on $\Sigma_0$
and solve the system (\ref{e:ini:Ham_CTT})-(\ref{e:ini:mom_CTT}) to get
$\Psi$ and $X^i$. Then one constructs
\bea
	\gm_{ij} & = & \Psi^4 \tgm_{ij} \label{e:ini:recons_gm}\\
	K^{ij} & = & \Psi^{-10} \left( (\tilde L X)^{ij} + \hA^{ij}_{\rm TT} \right)
	+ \frac{1}{3} \Psi^{-4} K \tgm^{ij}  \label{e:ini:recons_K} \\
	E & = & \Psi^{-8} {\tilde E} \\
	p^i & = & \Psi^{-10} {\tilde p}^i 
\eea
and obtains a set $(\wgm,\w{K},E,\w{p})$ which satisfies the constraint
equations (\ref{e:ini:Ham_constr})-(\ref{e:ini:mom_constr}).
This method has been proposed by York (1979) \cite{York79} and is naturally
called the \defin{conformal transverse traceless} (\defin{CTT}) method. 

\subsection{Decoupling on hypersurfaces of constant mean curvature}

Equations (\ref{e:ini:Ham_CTT}) and (\ref{e:ini:mom_CTT}) are coupled, but we
notice that if, among the free data, we choose $K$ to be a constant field
on $\Sigma_0$,
\be \label{e:ini:CMC}
	K = {\rm const},
\ee 
then they decouple partially : condition (\ref{e:ini:CMC}) implies $\tD^i K = 0$,
so that the momentum constraint (\ref{e:ini:mom_constr}) becomes independent 
of $\Psi$:
\be \label{e:ini:mom_CMC}
 \tilde\Delta_L \, X^i  = 8\pi {\tilde p}^i \qquad (K={\rm const}) . 
\ee
The condition (\ref{e:ini:CMC}) on the extrinsic curvature of $\Sigma_0$
defines what is called a \defin{constant mean curvature} (\defin{CMC}) hypersurface.
Indeed let us recall that $K$ is nothing but minus three times the
mean curvature of $(\Sigma_0,\wgm)$ embedded in $(\M,\w{g})$
[cf. Eq.~(\ref{e:hyp:K_mean_curvature})]. 
A maximal hypersurface, having $K=0$, is of course a special case of a CMC hypersurface.
On a CMC hypersurface, the task of obtaining initial data is greatly simplified: 
one has first 
to solve the linear elliptic equation (\ref{e:ini:mom_CMC}) to get $\w{X}$
and plug the solution in Eq.~(\ref{e:ini:Ham_CTT}) to form an equation for $\Psi$.
Equation~(\ref{e:ini:mom_CMC}) is the conformal vector Poisson equation studied
in Appendix~\ref{s:cko}. It is shown in Sec.~\ref{s:cko:Poisson} that it always
solvable for the two cases of interest mentioned in 
Sec.~\ref{s:ini:long_trans}: closed or asymptotically flat manifold.
Moreover, the solutions $\w{X}$ are such that the value of $\w{\tilde L} \w{X}$
is unique.

\subsection{Lichnerowicz equation} \label{s:ini:Lichne}

Taking into account the CMC decoupling, 
the difficult problem is to solve Eq.~(\ref{e:ini:Ham_CTT}) for $\Psi$.
This equation is elliptic and highly non-linear\footnote{although it is 
\emph{quasi-linear} in the technical sense, i.e. linear with respect 
to the highest-order derivatives}. 
It has been first studied by Lichnerowicz \cite{Lichn44,Lichn52}
in the case $K=0$ ($\Sigma_0$ maximal) and $\tilde E=0$ (vacuum). 
Lichnerowicz has shown that given 
the value of $\Psi$ at the boundary of a bounded domain of $\Sigma_0$
(Dirichlet problem), there exists at most one solution to Eq.~(\ref{e:ini:Ham_CTT}).
Besides, he showed the existence of a solution provided that $\hA_{ij} \hA^{ij}$
is not too large. 
These early results have been much improved since then. 
In particular  Cantor \cite{Canto77} 
has shown that in the asymptotically flat case, still with
$K=0$ and $\tilde E=0$, Eq.~(\ref{e:ini:Ham_CTT}) is solvable if and only 
if the metric $\wtgm$ is conformal to a metric with vanishing scalar curvature
(one says then that $\wtgm$ belongs to the \defin{positive Yamabe class})
(see also Ref.~\cite{Maxwe04b}).
In the case of closed manifolds, the complete analysis of the CMC case
has been achieved by Isenberg (1995) \cite{Isenb95}.

For more details and further references, we recommend the review articles
by Choquet-Bruhat and York \cite{ChoquY80} and Bartnik and Isenberg \cite{BartnI04}. 
Here we shall simply repeat the argument of York \cite{York99} to 
justify the rescaling (\ref{e:ini:tE_def}) of $E$. This rescaling is indeed 
related to the uniqueness of solutions to the Lichnerowicz equation.
Consider a solution $\Psi_0$ to Eq.~(\ref{e:ini:Ham_CTT}) in the
case $K=0$, to which we restrict ourselves. Another solution close to 
$\Psi_0$ can be written $\Psi=\Psi_0 + \epsilon$, with $|\epsilon| \ll \Psi_0$:
\be
\tD_i \tD^i (\Psi_0+\epsilon) -\frac{1}{8} {\tilde R} (\Psi_0+\epsilon)
	+ \frac{1}{8} \hA_{ij} \hA^{ij} \, (\Psi_0+\epsilon)^{-7}
	+ 2\pi {\tilde E} (\Psi_0+\epsilon)^{-3}  = 0 . 
\ee
Expanding to the first order in $\epsilon/\Psi_0$ leads to 
the following linear equation
for $\epsilon$:
\be \label{e:ini:eq_epsilon}
	\tD_i \tD^i \epsilon - \alpha \epsilon = 0 , 
\ee
with
\be \label{e:ini:def_alpha_Lich}
	\alpha := \frac{1}{8} {\tilde R} 
	+ \frac{7}{8} \hA_{ij} \hA^{ij} \Psi_0^{-8} + 
	6\pi {\tilde E} \Psi_0^{-4} .
\ee
Now, if $\alpha\geq 0$,  one can show, by means of the maximum principle,
that the solution of (\ref{e:ini:eq_epsilon}) which vanishes at spatial
infinity is necessarily $\epsilon=0$ (see Ref.~\cite{ChoquC81} or \S~B.1 of Ref.~\cite{ChoquIY00}). We therefore conclude that the solution $\Psi_0$
to Eq.~(\ref{e:ini:Ham_CTT}) is unique (at least locally) in this case.
On the contrary, if $\alpha<0$, non trivial oscillatory solutions of 
Eq.~(\ref{e:ini:eq_epsilon}) exist, making the solution $\Psi_0$ not unique.
The key point is that the scaling (\ref{e:ini:tE_def}) of $E$ yields the 
term $+6\pi{\tilde E} \Psi_0^{-4}$ in Eq.~(\ref{e:ini:def_alpha_Lich}),
which contributes to make $\alpha$ positive. If we had not rescaled $E$, 
i.e. had considered the original Hamiltonian constraint equation (\ref{e:cfd:Einstein5_hA}), the contribution to $\alpha$ would
have been instead $-10\pi E \Psi_0^4$, i.e. would have been negative. 
Actually, any rescaling $\tilde E = \Psi^s E$ with $s>5$ would have work
to make $\alpha$ positive. The choice $s=8$ in Eq.~(\ref{e:ini:tE_def}) 
is motivated by the fact that if the conformal data 
$(\tilde E,\tilde p^i)$ obey the ``conformal''
dominant energy condition (cf. Sec.~\ref{s:glo:positive_ener})
\be
	\tilde E \geq \sqrt{ \tgm_{ij} \tilde p^i \tilde p^j} , 
\ee
then, via the scaling (\ref{e:ini:tp_def}) of $p^i$, the reconstructed physical data
$(E,p^i)$ will automatically obey the dominant energy condition as stated by Eq.~(\ref{e:glo:dominant}):
\be
	E \geq \sqrt{ \gm_{ij} p^i p^j} .
\ee

\subsection{Conformally flat and momentarily static initial data}
\label{s:ini:cflat_static}

In this section we search for asymptotically flat initial data 
$(\Sigma_0,\wgm,\w{K})$. 
Let us then consider the simplest case one may think of, namely choose the freely
specifiable data $(\tgm_{ij},\hA^{ij}_{\rm TT},K,\tilde E,{\tilde p}^i)$ to
be a flat metric:
\be \label{e:ini:tgm_f}
	\tgm_{ij} = f_{ij} ,
\ee
a vanishing transverse-traceless part of the extrinsic curvature:
\be
	\hA^{ij}_{\rm TT} = 0 ,
\ee
a vanishing mean curvature (maximal hypersurface)
\be
	K = 0 ,
\ee 
and a vacuum spacetime:
\be
	\tilde E = 0, \qquad {\tilde p}^i = 0 .
\ee
Then $\tD_i = \Df_i$, $\tilde R = 0$, $\w{\tilde L} = \w{L}$ 
[cf. Eq.~(\ref{e:cfd:conf_Killing_f})] and the constraint equations (\ref{e:ini:Ham_CTT})-(\ref{e:ini:mom_CTT})
reduce to 
\bea
	& & \Delta \Psi 
	+ \frac{1}{8} (L X)_{ij} (L X)^{ij} \,  \Psi^{-7} = 0 \label{e:ini:Ham_ex1} \\
	& & \Delta_L X^i = 0 , \label{e:ini:mom_ex1}
\eea
where $\Delta$ and $\Delta_L$ are respectively the scalar Laplacian 
and the conformal vector Laplacian associated with the flat metric $\w{f}$:
\be \label{e:ini:def_Delta}
	\Delta := \Df_i \Df^i
\ee
and 
\be
	\Delta_L X^i := \Df_j \Df^j X^i + \frac{1}{3} \Df^i \Df_j X^j .
\ee
Equations (\ref{e:ini:Ham_ex1})-(\ref{e:ini:mom_ex1}) must be solved with the boundary conditions
\bea
	& & \Psi = 1 \qquad \mbox{when} \quad r\rightarrow \infty  \label{e:ini:BC_Psi1}\\
	& & \w{X} = 0 \qquad \mbox{when} \quad r\rightarrow \infty , \label{e:ini:BC_X1}
\eea
which follow from the asymptotic flatness requirement. 
The solution depends on the topology of $\Sigma_0$, since the latter may introduce some
inner boundary conditions in addition to (\ref{e:ini:BC_Psi1})-(\ref{e:ini:BC_X1})

Let us start with the simplest case: $\Sigma_0 = \R^3$. 
Then the solution of Eq.~(\ref{e:ini:mom_ex1}) subject to the boundary condition
(\ref{e:ini:BC_X1}) is
\be
	\w{X} = 0 
\ee
and there is no other solution (cf. Sec.~\ref{s:cko:Poisson}).
Then obviously $(L X)^{ij} = 0$, so that Eq.~(\ref{e:ini:Ham_ex1}) reduces
to Laplace equation for $\Psi$:
\be
	\Delta \Psi = 0 . 
\ee
With the boundary condition (\ref{e:ini:BC_Psi1}), there is a unique regular solution
on $\R^3$: 
\be
	\Psi = 1 . 
\ee
The initial data reconstructed from Eqs.~(\ref{e:ini:recons_gm})-(\ref{e:ini:recons_K})
is then
\bea
	& & \wgm = \w{f}  \\
	& & \w{K} = 0 .
\eea
These data correspond to a spacelike hyperplane of Minkowski spacetime. 
Geometrically the condition $\w{K}=0$ is that of a \emph{totally geodesic hypersurface}
(cf. Sec.~\ref{e:hyp:link_nab_D}). Physically data with $\w{K}=0$ 
are said to be \defin{momentarily static} or \defin{time symmetric}. 
Indeed, from Eq.~(\ref{e:fol:nab_m_comp}),
\be
	\Lie{m} \w{g} = - 2N \w{K} - 2 \wnab_{\w{n}}N \, \uu{n}\otimes\uu{n} . 
\ee
So if $\w{K}=0$ and if moreover one chooses a geodesic slicing around $\Sigma_0$
(cf. Sec.~\ref{s:dec:Gaussian_normal}), which yields $N=1$ and $\wnab_{\w{n}}N=0$, 
then 
\be
	\Lie{m} \w{g} = 0 . 
\ee
This means that, locally (i.e. on $\Sigma_0$), the normal evolution vector $\w{m}$ is  a spacetime Killing vector.
This vector being timelike, the configuration is then \defin{stationary}.
Moreover, the Killing vector $\w{m}$ being orthogonal to some hypersurface 
(i.e. $\Sigma_0$), the stationary configuration is called \defin{static}.
Of course, this staticity properties holds a priori only on $\Sigma_0$ since 
there is no guarantee that the time development of Cauchy data with $\w{K}=0$
at $t=0$ maintains $\w{K}=0$ at $t>0$. Hence the qualifier \emph{`momentarily'}
in the expression \emph{`momentarily static'} for data with $\w{K}=0$.
\begin{figure}
\centerline{\includegraphics[width=0.8\textwidth]{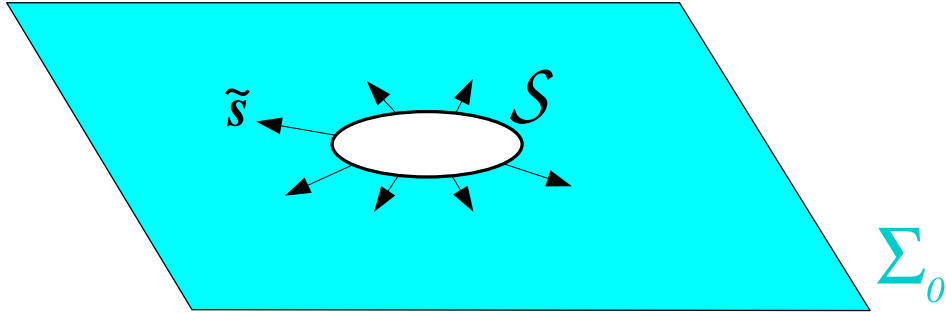}}
\caption[]{\label{f:ini:r3ball} \footnotesize
Hypersurface $\Sigma_0$ as $\R^3$ minus a ball, displayed via an embedding
diagram based on 
the metric $\wtgm$, which coincides with the Euclidean metric on $\R^3$. 
Hence $\Sigma_0$ appears to be flat. 
The unit normal of the inner boundary $\Sp$
with respect to the metric $\wtgm$ 
is $\w{\tilde s}$. Notice that $\wtD\cdot \w{\tilde s}>0$.}
\end{figure}

\begin{figure}
\centerline{\includegraphics[width=0.8\textwidth]{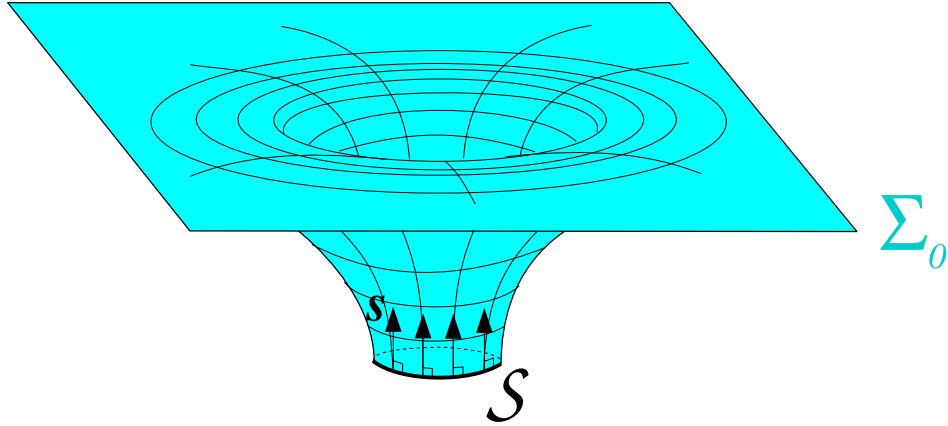}}
\caption[]{\label{f:ini:sminimal} \footnotesize
Same hypersurface $\Sigma_0$ as in Fig.~\ref{f:ini:r3ball} but
displayed via an embedding
diagram based on the metric $\wgm$ instead of $\wtgm$. The unit normal of the inner boundary $\Sp$
with respect to that metric 
is $\w{s}$. Notice that $\wD\cdot \w{s}=0$, which means that $\Sp$ is a 
minimal surface of $(\Sigma_0,\wgm)$.}
\end{figure}

To get something less trivial than a slice of Minkowski spacetime, 
let us consider a slightly more complicated
topology for $\Sigma_0$, namely $\R^3$ minus a ball (cf. Fig.~\ref{f:ini:r3ball}). 
The sphere $\Sp$ delimiting the ball is then the inner boundary of $\Sigma_0$
and we must provide boundary conditions for $\Psi$ and $\w{X}$ on $\Sp$
to solve Eqs.~(\ref{e:ini:Ham_ex1})-(\ref{e:ini:mom_ex1}).
For simplicity, let us choose
\be
	\left.\w{X}\right| _{\Sp} =  0 .
\ee
Altogether with the outer boundary condition (\ref{e:ini:BC_X1}),
this leads to $\w{X}$ being identically zero as the unique solution of Eq.~(\ref{e:ini:mom_ex1}). So, again, the Hamiltonian constraint reduces
to Laplace equation
\be \label{e:ini:Laplace_Psi}
	\Delta \Psi = 0 . 
\ee
If we choose the boundary condition $\left.\Psi\right| _{\Sp} = 1$, then 
the unique solution is $\Psi=1$ and we are back to the previous example
(slice of Minkowski spacetime). 
In order to have something non trivial, i.e. to ensure 
that the metric $\wgm$ will not be flat, let us demand that $\wgm$
admits a \emph{closed minimal surface}, that we will choose to be $\Sp$. 
This will necessarily translate as a boundary condition for $\Psi$ since
all the information on the metric is encoded in $\Psi$ (let us recall that
from the choice (\ref{e:ini:tgm_f}), $\wgm=\Psi^4\w{f}$).
$\Sp$ is a \defin{minimal surface} of $(\Sigma_0,\wgm)$
iff its mean curvature vanishes, or equivalently if its unit normal $\w{s}$
is divergence-free (cf. Fig.~\ref{f:ini:sminimal}):
\be \label{e:ini:div_s_0}
	\left. D_i s^i \right| _{\Sp}= 0 .
\ee
This is the analog of $\wnab\cdot\w{n}=0$ for maximal hypersurfaces,
the change from \emph{minimal} to \emph{maximal} being due to the change
of signature, from the Riemannian to the Lorentzian one. 
By means of Eq.~(\ref{e:cfd:divergence_conf}), condition (\ref{e:ini:div_s_0})
is equivalent to
\be \label{e:ini:div_s_0_1}
	\left. \Df_i (\Psi^6 s^i) \right| _{\Sp}= 0 , 
\ee
where we have used $\tD_i = \Df_i$, since $\wtgm=\w{f}$.
Let us rewrite this expression in terms of
the unit vector $\w{\tilde s}$ normal to $\Sp$ with respect to the metric $\wtgm$
(cf. Fig.~\ref{f:ini:r3ball}); we have
\be
	\w{\tilde s} = \Psi^{-2} \w{s} ,
\ee
since $\wtgm(\w{\tilde s},\w{\tilde s}) = \Psi^{-4} \wtgm(\w{s},\w{s})
= \wgm(\w{s},\w{s}) = 1$. Thus Eq.~(\ref{e:ini:div_s_0_1}) becomes
\be \label{e:ini:div_s_0_2}
	\left. \Df_i (\Psi^4 {\tilde s}^i) \right| _{\Sp}
 = \left. \frac{1}{\sqrt{f}} \der{}{x^i} \left( \sqrt{f} \Psi^4 {\tilde s}^i \right)
	\right| _{\Sp} = 0 .
\ee
Let us introduce on $\Sigma_0$ a coordinate system of spherical type,
$(x^i)=(r,\theta,\varphi)$, such that (i) $f_{ij} = \mathrm{diag}(1,r^2,r^2\sin^2\theta)$
and (ii) $\Sp$ is the sphere $r=a$, where
$a$ is some positive constant. Since in these coordinates $\sqrt{f} = r^2\sin\theta$ 
and ${\tilde s}^i=(1,0,0)$, the minimal surface condition (\ref{e:ini:div_s_0_2})
is written as
\be
	\left. \frac{1}{r^2} \der{}{r} \left( \Psi^4 r^2 \right) 
		\right| _{r=a}= 0  , 
\ee
i.e. 
\be \label{e:ini:bc_Psi_S}
	\left. \left( \der{\Psi}{r} + \frac{\Psi}{2r} \right) \right| _{r=a} = 0 
\ee
This is a boundary condition of mixed Newmann/Dirichlet type for $\Psi$. 
The unique solution of the Laplace equation (\ref{e:ini:Laplace_Psi}) 
which satisfies boundary conditions
(\ref{e:ini:BC_Psi1}) and (\ref{e:ini:bc_Psi_S}) is
\be
	 \Psi = 1 + \frac{a}{r} . 
\ee
The parameter $a$ is then easily related to the ADM 
mass $m$ of the hypersurface $\Sigma_0$. Indeed using formula
(\ref{e:glo:M_ADM_QI}), $m$ is evaluated as
\be \label{e:ini:m_2a}
	m = -\frac{1}{2\pi} \lim_{r \rightarrow\infty}
	\oint_{r={\rm const}} 
	 \der{\Psi}{r} r^2 \sin\theta \, d\theta \, d\varphi 
	 = -\frac{1}{2\pi} \lim_{r \rightarrow\infty}
	 	4\pi r^2 \der{}{r} \left( 1 + \frac{a}{r} \right) 
	= 2 a. 
\ee
Hence $a=m/2$ and we may write
\be
	\encadre{ \Psi = 1 + \frac{m}{2r} }. 
\ee
Therefore, in terms of the 
coordinates $(r,\theta,\varphi)$, the obtained initial data $(\wgm,\w{K})$ are 
\bea
	& & \gm_{ij} = \left( 1 + \frac{m}{2r} \right) ^4
		\mathrm{diag} (1,r^2,r^2\sin\theta) \label{e:ini:gm_Schwarz_iso} \\
	& & K_{ij} = 0 . \label{e:ini:Kij_Schwarz}
\eea
So, as above, the initial data are momentarily static.
Actually, we recognize on (\ref{e:ini:gm_Schwarz_iso})-(\ref{e:ini:Kij_Schwarz})
a slice $t={\rm const}$ of \emph{Schwarzschild spacetime}
in isotropic coordinates [compare with Eq.~(\ref{e:cfd:Schwarz_isotropic})].
\begin{figure}
\centerline{\includegraphics[width=0.6\textwidth]{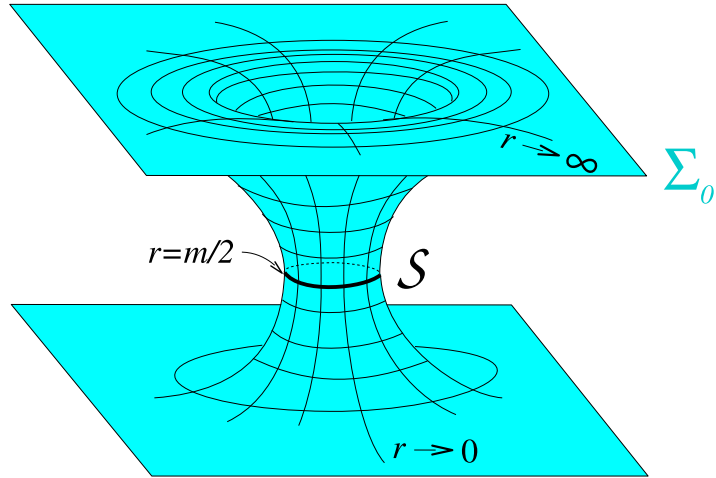}}
\caption[]{\label{f:ini:einst_rosen} \footnotesize
Extended hypersurface $\Sigma'_0$ obtained by gluing a copy of $\Sigma_0$
at the minimal surface $\Sp$ and defining an Einstein-Rosen bridge between
two asymptotically flat regions.}
\end{figure}

\begin{figure}
\centerline{\includegraphics[width=0.6\textwidth]{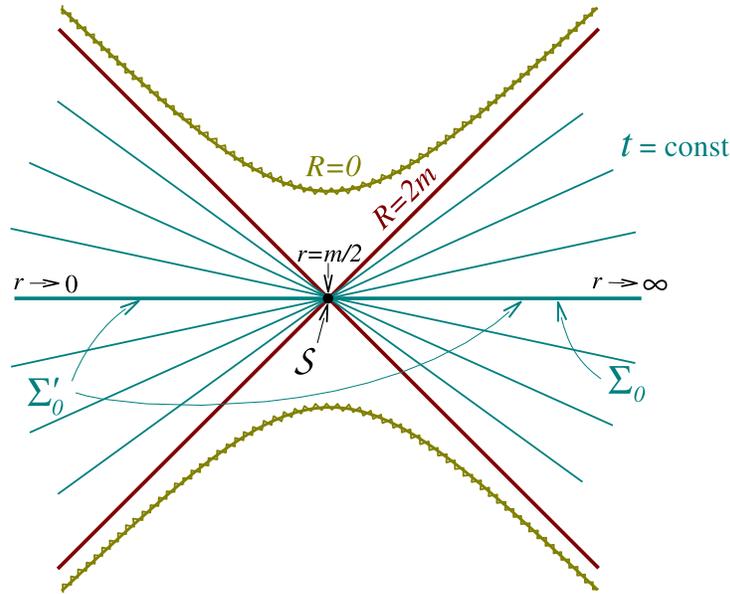}}
\caption[]{\label{f:ini:kruskal} \footnotesize
Extended hypersurface $\Sigma'_0$ depicted in the
Kruskal-Szekeres representation of Schwarzschild spacetime. 
$R$ stands for Schwarzschild radial coordinate and $r$ for the isotropic radial coordinate.
$R=0$ is the singularity and $R=2m$ the event horizon.
$\Sigma'_0$ is nothing but a hypersurface $t={\rm const}$, where $t$ is the
Schwarzschild time coordinate. 
In this diagram, these hypersurfaces are straight lines and the Einstein-Rosen bridge $\Sp$ is reduced to a point.}
\end{figure}

The isotropic coordinates $(r,\theta,\varphi)$ covering the manifold $\Sigma_0$
are such that the range of $r$ is $[m/2,+\infty)$. But thanks to the 
minimal character of the inner boundary $\Sp$, we can extend $(\Sigma_0,\wgm)$ to a
larger Riemannian manifold $(\Sigma'_0,\wgm')$ with
$\left.\wgm'\right| _{\Sigma_0} = \wgm$ and $\wgm'$ smooth at $\Sp$.
This is made possible by gluing a copy of $\Sigma_0$ at $\Sp$
(cf. Fig.~\ref{f:ini:einst_rosen}).
The topology of $\Sigma'_0$ is $\mathbb{S}^2\times \R$ and the
range of $r$ in $\Sigma'_0$ is $(0,+\infty)$. 
The extended metric $\wgm'$ keeps exactly the same form
as (\ref{e:ini:gm_Schwarz_iso}):
\be
	\gm'_{ij}\, dx^i\, dx^j  = \left( 1 + \frac{m}{2r} \right) ^4
	\left( dr^2 + r^2 d\theta^2 + r^2\sin^2\theta d\varphi^2 \right) .
\ee
By the change of variable
\be \label{e:ini:inversion}
	r \mapsto r' = \frac{m^2}{4r}
\ee
it is easily shown that the region $r\rightarrow 0$ does not correspond to some
``center'' but is actually a second asymptotically flat region (the lower one in Fig.~\ref{f:ini:einst_rosen}). Moreover the transformation (\ref{e:ini:inversion}),
with $\theta$ and $\varphi$ kept fixed, is an isometry of $\wgm'$. It maps a
point $p$ of $\Sigma_0$ to the point located at the vertical of $p$ in 
Fig.~\ref{f:ini:einst_rosen}. The minimal sphere $\Sp$ is invariant under
this isometry. The region around $\Sp$
is called an \defin{Einstein-Rosen bridge}. 
$(\Sigma'_0,\wgm')$ is still a slice of Schwarzschild spacetime. 
It connects two asymptotically flat regions without
entering below the event horizon, as shown in the Kruskal-Szekeres diagram 
of Fig.~\ref{f:ini:kruskal}.

\subsection{Bowen-York initial data} \label{s:ini:Bowen_York}

Let us select the same simple free data as above, namely
\be \label{e:ini:BY:free_data}
	\tgm_{ij} = f_{ij}, \quad \hA^{ij}_{\rm TT} = 0, \quad K=0,
	\quad \tilde E = 0
	\quad \mbox{and}\quad {\tilde p}^i = 0 .
\ee
For the hypersurface $\Sigma_0$, instead of $\R^3$ minus a ball, 
we choose $\R^3$ minus a point:
\be
\Sigma_0 = \mathbb{R}^3\backslash\{O\}	.
\ee
The removed point $O$ is called a \defin{puncture} \cite{BrandB97}.
The topology of $\Sigma_0$ is $\mathbb{S}^2\times \R$;
it differs from the topology considered in Sec.~\ref{s:ini:cflat_static}
($\R^3$ minus a ball); actually it is the
same topology as that of the extended manifold $\Sigma'_0$
(cf. Fig.~\ref{f:ini:einst_rosen}).

Thanks to the choice (\ref{e:ini:BY:free_data}), the system to be solved is still (\ref{e:ini:Ham_ex1})-(\ref{e:ini:mom_ex1}). If we choose the trivial solution
$\w{X}=0$ for Eq.~(\ref{e:ini:mom_ex1}), we are back
to the slice of Schwarzschild spacetime considered in Sec.~\ref{s:ini:cflat_static},
except that now $\Sigma_0$ is the extended manifold previously denoted
$\Sigma'_0$. 

Bowen and York \cite{BowenY80} have obtained a simple non-trivial solution of
Eq.~(\ref{e:ini:mom_ex1}) (see also Ref.~\cite{BeigK04}).
Given a Cartesian coordinate system $(x^i)=(x,y,z)$ on $\Sigma_0$ 
(i.e. a coordinate system such that $f_{ij}=\mathrm{diag}(1,1,1)$) with respect
to which the coordinates of the puncture $O$ are $(0,0,0)$, this solution 
writes 
\be \label{e:ini:BY_X}
X^i = - \frac{1}{4 r} \left( 7 f^{ij} P_j +  \frac{P_j x^j x^i }{r^2}
	\right) - \frac{1}{r^3} \epsilon^{ij}_{\ \ k} S_j x^k ,
\ee
where $r:=\sqrt{x^2+y^2+z^2}$, $\epsilon^{ij}_{\ \ k}$ is the Levi-Civita alternating
tensor associated with the flat metric $\w{f}$ and 
$(P_i,S_j)=(P_1,P_2,P_3,S_1,S_2,S_3)$
are six real numbers, which constitute the six parameters of the Bowen-York solution.
Notice that since $r\not=0$ on $\Sigma_0$, the Bowen-York solution is a regular 
and smooth solution on the entire $\Sigma_0$.
\begin{example}
Choosing $P_i=(0,P,0)$ and $S_i=(0,0,S)$, where $P$ and $S$ are two real numbers,
leads to the following expression of the Bowen-York solution:
\be \label{e:ini:BY_exaX}
	\left\{ \begin{array}{lcl}
	X^x & = & \displaystyle - \frac{P}{ 4} \frac{x y}{ r^3} 
	+ S \frac{y}{ r^3}   \\[2ex]
	X^y & = & \displaystyle - \frac{P}{ 4r } \left( 7 + \frac{y^2}{ r^2} \right)
						 - S \frac{x}{ r^3}   \\[2ex]
	X^z & = & \displaystyle - \frac{P}{ 4} \frac{x z}{ r^3} 
	\end{array} \right. 
\ee
\end{example} 
The conformal traceless extrinsic curvature corresponding to the solution
(\ref{e:ini:BY_X}) is deduced from formula (\ref{e:ini:decomp_hA}), which in 
the present case reduces to $\hA^{ij} = (L X)^{ij}$; one gets
\be \label{e:ini:BY_hA}
\encadre{ \hA^{ij} =
		\frac{3}{2 r^3} \left[ x^i P^j + x^j P^i  - \left(
			f^{ij} - \frac{x^i x^j}{r^2} \right)
			P_k x^k \right]
			+ \frac{3}{r^5} \left( \epsilon^{ik}_{\ \  l} S_k x^l x^j
			+ \epsilon^{jk}_{\ \  l} S_k x^l x^i \right)
},
\ee
where $P^i := f^{ij} P_j$.
The tensor $\hA^{ij}$ given by Eq.~(\ref{e:ini:BY_hA}) is called the 
\defin{Bowen-York extrinsic curvature}. Notice that 
the $P_i$ part of $\hA^{ij}$ decays asymptotically as $O(r^{-2})$, whereas the
$S_i$ part decays as $O(r^{-3})$.
\begin{remark}
Actually the expression of $\hA^{ij}$ given in the original Bowen-York article 
\cite{BowenY80} contains an additional term with respect to Eq.~(\ref{e:ini:BY_hA}),
but the role of this extra term is only to ensure that the solution is isometric
through an inversion across some sphere. We are not interested by such a property
here, so we have dropped this term. Therefore, strictly speaking, we should
name expression (\ref{e:ini:BY_hA}) the \emph{simplified} Bowen-York extrinsic curvature.
\end{remark}

\begin{example}
Choosing $P_i=(0,P,0)$ and $S_i=(0,0,S)$ as in the previous example
[Eq.~(\ref{e:ini:BY_exaX})], we get
\bea
	\hA^{xx} & = & - \frac{3P}{2r^3} y \left(1-\frac{x^2}{r^2} \right)
	- \frac{6S}{r^5} xy  \\
	\hA^{xy} & = & \frac{3P}{2r^3} x \left(1+\frac{y^2}{r^2} \right)
	+ \frac{3S}{r^5} (x^2-y^2) \\
	\hA^{xz} & = & \frac{3P}{2r^5} xyz - \frac{3S}{r^5} yz \\
	\hA^{yy} & = &  \frac{3P}{2r^3} y \left(1+\frac{y^2}{r^2} \right)
	+ \frac{6S}{r^5} xy \\
	\hA^{yz} & = &  \frac{3P}{2r^3} z  \left(1+\frac{y^2}{r^2} \right)
	  + \frac{3S}{r^5} xz \\
	\hA^{zz} & = & - \frac{3P}{2r^3} y  \left(1-\frac{z^2}{r^2} \right) . 
\eea 
In particular we verify that $\hA^{ij}$ is traceless:
$\tgm_{ij} \hA^{ij} = f_{ij} \hA^{ij} = \hA^{xx}+\hA^{yy}+\hA^{zz} = 0 $. 
\end{example}

The Bowen-York extrinsic curvature provides an analytical solution of the
momentum constraint (\ref{e:ini:mom_ex1}) but there remains to solve the
Hamiltonian constraint (\ref{e:ini:Ham_ex1}) for $\Psi$, with
the asymptotic flatness boundary condition $\Psi=1$ when $r\rightarrow \infty$. 
Since $\w{X}\not=0$, Eq.~(\ref{e:ini:Ham_ex1}) is no
longer a simple Laplace equation, as in Sec.~\ref{s:ini:cflat_static}, but a
 non-linear elliptic equation. There is no hope to get any analytical
solution and one must solve Eq.~(\ref{e:ini:Ham_ex1}) numerically to get $\Psi$
and reconstruct the full initial data $(\wgm,\w{K})$ via 
Eqs.~(\ref{e:ini:recons_gm})-(\ref{e:ini:recons_K}).

Let us now discuss the physical significance of the parameters $(P_i,S_i)$
of the Bowen-York solution. First of all, 
the ADM momentum of the initial data $(\Sigma_0,\wgm,\w{K})$ is computed 
via formula (\ref{e:glo:Pi_ADM_def}). Taking into account that $\Psi$
is asymptotically one and $K$ vanishes, we can write
\be
	 P_i^{\rm ADM} = \frac{1}{8\pi} \lim_{r\rightarrow\infty}
	\oint_{r={\rm const}} \hA_{ik} \, x^k r \sin\theta \, d\theta\, d\varphi  , 
\qquad i\in\{1,2,3\}, 
\ee
where we have used the fact that, within the Cartesian coordinates
$(x^i)= (x,y,z)$, $(\wpar_i)^j = \delta^j_{\ \, i}$ and $s^k = x^k/r$.
If we insert expression (\ref{e:ini:BY_hA}) for $\hA_{jk}$ in this formula,
we notice that the $S_i$ part decays too fast to contribute to the integral; 
there remains only
\bea
	P_i^{\rm ADM} & = & \frac{1}{8\pi} \lim_{r\rightarrow\infty}
	\oint_{r={\rm const}} \frac{3}{2r^2}
	\bigg[ x_i P_j x^j + r^2 P_i - 
	\underbrace{ \left(x_i - \frac{x^i r^2}{r^2}\right)}_{=0} 
	P_k x^k \bigg] \sin\theta \, d\theta\, d\varphi \nonumber \\
	& = & \frac{3}{16\pi}  
	\bigg( P_j \oint_{r={\rm const}} \frac{x^i x^j}{r^2} \sin\theta \, 
	d\theta\, d\varphi
 	+ P_i 
	\underbrace{\oint_{r={\rm const}} \sin\theta \, d\theta\, d\varphi}_{=4\pi} 
	 \bigg)  . \label{e:ini:BY_PADM1}
\eea
Now 
\be
	\oint_{r={\rm const}} \frac{x^i x^j}{r^2} \sin\theta \, 
	d\theta\, d\varphi = \delta^{ij} \oint_{r={\rm const}} \frac{(x^j)^2}{r^2} \sin\theta \, 
	d\theta\, d\varphi = 
	\delta^{ij} \frac{1}{3} \oint_{r={\rm const}} \frac{r^2}{r^2} \sin\theta \, 
	d\theta\, d\varphi 
	 = \frac{4\pi}{3} \delta^{ij} , 
\ee
so that Eq.~(\ref{e:ini:BY_PADM1}) becomes
\be
	P_i^{\rm ADM} = \frac{3}{16\pi} \left( \frac{4\pi}{3} + 4\pi \right) P_i , 
\ee
i.e. 
\be
	\encadre{ P_i^{\rm ADM} = P_i }.
\ee
Hence the parameters $P_i$ of the Bowen-York solution are nothing but the
three components of the ADM linear momentum of the hypersurface $\Sigma_0$. 

Regarding the angular momentum, we notice that since $\tgm_{ij}=f_{ij}$
in the present case, the Cartesian coordinates $(x^i)=(x,y,z)$ belong to the quasi-isotropic gauge introduced in Sec.~\ref{s:glo:cure} 
(condition (\ref{e:glo:QIgauge}) is trivially fulfilled).
We may then use formula (\ref{e:glo:angu_mom_def}) to 
define the angular momentum of Bowen-York initial.  
Again, since $\Psi\rightarrow 1$ at spatial infinity and $K=0$, we can write
\be
 J_i = \frac{1}{8\pi} \lim_{r\rightarrow\infty}
	\oint_{r={\rm const}} \hA_{jk} (\w{\phi}_i)^j \,
	x^k r  \sin\theta \, d\theta\, d\varphi, 
\qquad i\in\{1,2,3\} .
\ee
Substituting expression (\ref{e:ini:BY_hA}) for $\hA_{jk}$ as well as
expressions (\ref{e:glo:rot_flat_x})-(\ref{e:glo:rot_flat_z}) 
for $(\w{\phi}_i)^j$, we get
that only the $S_i$ part contribute to this integral. After some computation, 
we find
\be
	\encadre{ J_i = S_i } . 	
\ee
Hence the parameters $S_i$ of the Bowen-York solution are nothing but the
three components of the angular momentum of the hypersurface $\Sigma_0$. 

\begin{remark}
The Bowen-York solution with $P^i=0$ and $S^i=0$ reduces to the
momentarily static  solution found in Sec.~\ref{s:ini:cflat_static}, i.e. 
is a slice $t={\rm const}$ of the Schwarzschild spacetime ($t$ being the
Schwarzschild time coordinate). 
However Bowen-York initial data with $P^i=0$ and $S^i\not=0$ do not 
constitute a slice of Kerr spacetime. Indeed, it has been
shown \cite{GaratP00} that there does not exist any foliation of Kerr spacetime by hypersurfaces
which (i) are
axisymmetric, (ii) smoothly reduce in the non-rotating limit to the hypersurfaces
of constant Schwarzschild time and (iii) are conformally flat, i.e. 
have induced metric $\wtgm=\w{f}$, as the Bowen-York hypersurfaces have. 
This means that a Bowen-York solution with $S^i\not=0$ does represent 
initial data for a rotating black hole, but this black hole is not stationary:
it is ``surrounded'' by gravitational radiation, as demonstrated by the
time development of these initial data \cite{BrandS95b,GleisNPP98}.
\end{remark}

\section{Conformal thin sandwich method}

\subsection{The original conformal thin sandwich method} \label{s:ini:CTS_ori}

An alternative to the conformal transverse-traceless method for computing
initial data has been introduced by York in 1999 \cite{York99}. 
It is motivated by expression (\ref{e:cfd:evol_tgmup}) for the 
traceless part of the extrinsic curvature scaled with $\alpha=-4$:
\be \label{e:ini:tA_Lmg}
	\tA^{ij} = \frac{1}{2N} \left[ \left( \der{}{t} - \Liec{\beta} \right)
	 \tgm^{ij}  - \frac{2}{3} \tD_k \beta^k \, \tgm^{ij}  \right] .
\ee
Noticing that [cf. Eq.~(\ref{e:ini:conf_Killing_def})]
\be
- \Liec{\beta} \tgm^{ij} = (\tilde L \beta)^{ij}+ \frac{2}{3} \tD_k \beta^k , 
\ee
and introducing the short-hand notation
\be
	\dot\tgm^{ij} := \der{}{t}\tgm^{ij} , 
\ee
we can rewrite Eq.~(\ref{e:ini:tA_Lmg}) as
\be  \label{e:ini:tA_Lbeta}
	\tA^{ij} = \frac{1}{2N} \left[ \dot\tgm^{ij} + (\tilde L \beta)^{ij} \right] .
\ee
The relation between $\tA^{ij}$ and $\hA^{ij}$ is [cf. Eq.~(\ref{e:cfd:hA_tA})]
\be
	\hA^{ij} = \Psi^6 \tA^{ij} . 
\ee
Accordingly, Eq.~(\ref{e:ini:tA_Lbeta}) yields
\be \label{e:ini:hA_dg_beta}
  \encadre{ \hA^{ij} = \frac{1}{2\tilde N} \left[ \dot\tgm^{ij} + 
	(\tilde L \beta)^{ij} \right] } , 
\ee
where we have introduced the \defin{conformal lapse}
\be \label{e:ini:def_tN}
	\encadre{ \tilde N := \Psi^{-6} N } . 
\ee
Equation~(\ref{e:ini:hA_dg_beta}) constitutes a decomposition of $\hA^{ij}$
alternative to the longitudinal/transverse decomposition (\ref{e:ini:decomp_hA}). 
Instead of expressing $\hA^{ij}$ in terms of a vector $\w{X}$ and
a TT tensor $\hA^{ij}_{\rm TT}$, it expresses it in terms of the shift vector
$\w{\beta}$, the time derivative of the conformal metric, $\dot\tgm^{ij}$, 
and the conformal lapse $\tilde N$. 

The Hamiltonian constraint, written as the Lichnerowicz equation
(\ref{e:ini:Ham_conf}), takes the same form as before:
\be \label{e:ini:Ham_CTS}
\encadre{\tD_i \tD^i \Psi -\frac{{\tilde R}}{8}  \Psi
	+ \frac{1}{8} \hA_{ij} \hA^{ij} \, \Psi^{-7}
	+ 2\pi {\tilde E} \Psi^{-3} - \frac{K^2}{12}  \Psi^5 = 0 },	
\ee
except that now $\hA^{ij}$ is to be understood as the
combination (\ref{e:ini:hA_dg_beta}) of $\beta^i$, $\dot\tgm^{ij}$ and
$\tilde N$. 
On the other side, the momentum constraint (\ref{e:ini:mom_conf}) becomes, 
once expression (\ref{e:ini:hA_dg_beta}) is substituted for $\hA^{ij}$, 
\be \label{e:ini:mom_CTS}
	\encadre{ \tD_j \left( \frac{1}{\tilde N} (\tilde L \beta)^{ij} \right)
 + \tD_j \left( \frac{1}{\tilde N} \dot\tgm^{ij} \right)
  - \frac{4}{3} \Psi^6 \tD^i K = 16\pi {\tilde p}^i }. 
\ee
In view of the system (\ref{e:ini:Ham_CTS})-(\ref{e:ini:mom_CTS}), the 
method to compute initial data consists in choosing freely 
$\tgm_{ij}$,  $\dot\tgm^{ij}$, $K$, ${\tilde N}$, ${\tilde E}$
and ${\tilde p}^i$ on $\Sigma_0$ and solving (\ref{e:ini:Ham_CTS})-(\ref{e:ini:mom_CTS}) to get $\Psi$ and $\beta^i$. 
This method is called \defin{conformal thin sandwich} (\defin{CTS}), 
because one input
is the time derivative $\dot\tgm^{ij}$, which can be obtained from 
the value of the conformal metric on two neighbouring hypersurfaces
$\Sigma_t$ and $\Sigma_{t+\delta t}$ (``thin sandwich'' view point). 

\begin{remark}
The term ``thin sandwich'' originates from a previous method 
devised in the early sixties by Wheeler and his collaborators 
\cite{BaierSW62,Wheel64}. Contrary to the methods exposed here, the
thin sandwich method
was not based on a conformal decomposition: it considered the constraint
equations (\ref{e:ini:Ham_constr})-(\ref{e:ini:mom_constr}) as a system
to be solved for the lapse $N$ and the shift vector $\w{\beta}$, given
the metric $\wgm$ and its time derivative. The extrinsic curvature 
which appears in (\ref{e:ini:Ham_constr})-(\ref{e:ini:mom_constr}) was
then considered as the function of $\wgm$, $\dert{\wgm}{t}$, $N$ and $\w{\beta}$
given by Eq.~(\ref{e:dec:Einstein_PDE1}). However, this method does
not work in general \cite{BartnF93}. On the contrary the \emph{conformal} 
thin sandwich method introduced by York \cite{York99} and exposed
above was shown to work
\cite{ChoquIY00}. 
\end{remark}

As for the conformal transverse-traceless method treated in Sec.~\ref{s:ini:CTT}, 
on CMC hypersurfaces, Eq.~(\ref{e:ini:mom_CTS})
decouples from Eq.~(\ref{e:ini:Ham_CTS}) and becomes an elliptic linear equation 
for $\w{\beta}$. 

\subsection{Extended conformal thin sandwich method}

An input of the above method is the conformal lapse $\tilde N$.
Considering the astrophysical problem stated in Sec.~\ref{s:ini:idp},
it is not clear how to pick a relevant value for $\tilde N$.
Instead of choosing an arbitrary value, 
Pfeiffer and York \cite{PfeifY03} have suggested
to compute $\tilde N$ from the Einstein equation
giving the time derivative of the trace $K$ of the extrinsic curvature,
i.e. Eq.~(\ref{e:cfd:Einstein3}):
\be \label{e:ini:evol_K}
\left(\der{}{t} - \Liec{\beta} \right) K 
	= - \Psi^{-4} \left( \tD_i \tD^i N + 2 \tD_i \ln \Psi \, \tD^i N \right) 
	 + N \left[ 4\pi (E+S) 
	+  \tA_{ij} \tA^{ij} + \frac{K^2}{3}\right] .
\ee
This amounts to add this equation to the initial data system. 
More precisely, Pfeiffer and York \cite{PfeifY03} suggested to combine 
Eq.~(\ref{e:ini:evol_K}) with the Hamiltonian constraint to get an equation
involving the quantity $N\Psi = \tilde N \Psi^7$ and containing no
scalar products of gradients as the $\tD_i \ln\Psi \tD^i N$ term in 
Eq.~(\ref{e:ini:evol_K}), thanks to the identity
\be
	\tD_i \tD^i N + 2 \tD_i \ln \Psi \, \tD^i N 
	= \Psi^{-1} \left[ \tD_i \tD^i (N\Psi) + N \tD_i \tD^i \Psi \right] . 
\ee
Expressing the left-hand side of the above equation in terms of Eq.~(\ref{e:ini:evol_K})
and substituting $\tD_i \tD^i \Psi$ in the right-hand side by its expression
deduced from Eq.~(\ref{e:ini:Ham_CTS}), we get 
\be \label{e:ini:eq_NPsi7}
\tD_i \tD^i (\tilde N \Psi^7) - (\tilde N \Psi^7)
	\left[ \frac{1}{8} \tilde R + \frac{5}{12} K^2 \Psi^4
	+ \frac{7}{8} \hA_{ij} \hA^{ij} \Psi^{-8} 
	+ 2 \pi (\tilde E+2\tilde S) \Psi^{-4} \right] 
	+ \left( \dot K - \beta^i \tD_i K \right) \Psi^5 = 0 ,
\ee
where we have used the short-hand notation
\be
	\dot K := \der{K}{t} 
\ee
and have set
\be
	\tilde S := \Psi^8 S . 
\ee
Adding Eq.~(\ref{e:ini:eq_NPsi7}) to Eqs.~(\ref{e:ini:Ham_CTS}) and
(\ref{e:ini:mom_CTS}), the initial data system becomes
\bea
 & & \encadre{\tD_i \tD^i \Psi -\frac{{\tilde R}}{8}  \Psi
	+ \frac{1}{8} \hA_{ij} \hA^{ij} \, \Psi^{-7}
	+ 2\pi {\tilde E} \Psi^{-3} - \frac{K^2}{12}  \Psi^5 = 0 } \label{e:ini:XCTS1} \\
 & & \encadre{ \tD_j \left( \frac{1}{\tilde N} (\tilde L \beta)^{ij} \right)
 + \tD_j \left( \frac{1}{\tilde N} \dot\tgm^{ij} \right)
  - \frac{4}{3} \Psi^6 \tD^i K = 16\pi {\tilde p}^i } \label{e:ini:XCTS2} \\
 & & \encadre{
	\begin{array}{ll}
	\displaystyle \tD_i \tD^i (\tilde N \Psi^7) - (\tilde N \Psi^7)
	\bigg[ \frac{\tilde R}{8} + \frac{5}{12} K^2 \Psi^4
	& \displaystyle\!\!\!\! + \frac{7}{8} \hA_{ij} \hA^{ij} \Psi^{-8} 
	+ 2 \pi (\tilde E+2\tilde S) \Psi^{-4} \bigg] \\ 
	& \displaystyle \qquad + \left( \dot K - \beta^i \tD_i K \right) \Psi^5 = 0  
	\end{array} } , \label{e:ini:XCTS3}
\eea
where $\hA^{ij}$ is the function of $\tilde N$, $\beta^i$, $\tgm_{ij}$ and
$\dot\tgm^{ij}$ defined by Eq.~(\ref{e:ini:hA_dg_beta}).
Equations (\ref{e:ini:XCTS1})-(\ref{e:ini:XCTS3}) constitute the
\defin{extended conformal thin sandwich} (\defin{XCTS}) 
system for the initial data problem.
The free data are the conformal metric $\wtgm$, its coordinate time derivative
$\w{\dot\tgm}$, the extrinsic curvature trace $K$, its coordinate time derivative
$\dot K$, and the rescaled matter variables $\tilde E$, $\tilde S$ and 
$\tilde p^i$. The constrained data are the conformal factor $\Psi$, the 
conformal lapse $\tilde N$ and the shift vector $\w{\beta}$. 
\begin{remark}
The XCTS system (\ref{e:ini:XCTS1})-(\ref{e:ini:XCTS3}) is a coupled system.
Contrary to the CTT system (\ref{e:ini:Ham_CTT})-(\ref{e:ini:mom_CTT}), the
assumption of constant mean curvature, and in particular of maximal slicing,
does not allow to decouple it.
\end{remark}

\subsection{XCTS at work: static black hole example} \label{s:ini:XCTS_work}

Let us illustrate the extended conformal thin sandwich method on a simple
example. Take for the hypersurface $\Sigma_0$ the punctured manifold considered
in Sec.~\ref{s:ini:Bowen_York}, namely
\be
\Sigma_0 = \R^3\backslash\{O\}	.
\ee
For the free data, let us perform the simplest choice:
\be \label{e:ini:XCTS_ex_free}
	\tgm_{ij} = f_{ij}, 
	\quad \dot\tgm^{ij} = 0, \quad K=0, \quad \dot K =0,
	\quad \tilde E = 0,
	\quad \tilde S = 0,
	\quad \mbox{and}\quad {\tilde p}^i = 0 ,
\ee
i.e. we are searching for vacuum initial data on a maximal and conformally flat 
hypersurface with all the freely specifiable time derivatives set to zero.
Thanks to (\ref{e:ini:XCTS_ex_free}), the XCTS system (\ref{e:ini:XCTS1})-(\ref{e:ini:XCTS3}) reduces to
\bea
 & & \Delta \Psi 
	+ \frac{1}{8} \hA_{ij} \hA^{ij} \, \Psi^{-7} = 0 \label{e:ini:XCTS_ex1} \\
 & & \Df_j \left( \frac{1}{\tilde N} (L \beta)^{ij} \right) = 0 \label{e:ini:XCTS_ex2} \\
& & \Delta (\tilde N \Psi^7) - 
	\frac{7}{8} \hA_{ij} \hA^{ij} \Psi^{-1} \tilde N 
		= 0 \label{e:ini:XCTS_ex3} .
\eea
Aiming at finding the simplest solution, we notice that
\be \label{e:ini:XCTS_ex_beta}
	\w{\beta} = 0
\ee
is a solution of Eq.~(\ref{e:ini:XCTS_ex2}). Together with $\dot\tgm^{ij} = 0$, it 
leads to [cf. Eq.~(\ref{e:ini:hA_dg_beta})]
\be
	\hA^{ij} = 0 .
\ee
The system (\ref{e:ini:XCTS_ex1})-(\ref{e:ini:XCTS_ex3}) reduces then further:
\bea
  	& & \Delta \Psi = 0 \label{e:ini:XCTS_ex1-2} \\
	& & \Delta (\tilde N \Psi^7) = 0 . \label{e:ini:XCTS_ex3-2}
\eea
Hence we have only two Laplace equations to solve. Moreover Eq.~(\ref{e:ini:XCTS_ex1-2})
decouples from Eq.~(\ref{e:ini:XCTS_ex3-2}). For simplicity, let us assume spherical
symmetry around the puncture $O$. 
We introduce an adapted spherical coordinate
system $(x^i) = (r,\theta,\varphi)$ on $\Sigma_0$.
The puncture $O$ is then at $r=0$. The
simplest non-trivial solution of (\ref{e:ini:XCTS_ex1-2}) which obeys the asymptotic
flatness condition $\Psi\rightarrow 1$ as $r\rightarrow+\infty$ is 
\be \label{e:ini:XCTS_ex_Psi}
	\Psi = 1 + \frac{m}{2r} , 
\ee
where as in Sec.~\ref{s:ini:cflat_static}, the constant $m$ is the ADM mass of $\Sigma_0$
[cf. Eq.~(\ref{e:ini:m_2a})]. Notice that since $r=0$ is excluded from $\Sigma_0$,
$\Psi$ is a perfectly regular solution on the entire manifold $\Sigma_0$.
Let us recall that the Riemannian manifold $(\Sigma_0,\wgm)$ corresponding to this
value of $\Psi$ via $\wgm=\Psi^4 \w{f}$ is the Riemannian manifold denoted $(\Sigma'_0,\wgm)$ in Sec.~\ref{s:ini:cflat_static} and depicted in 
Fig.~\ref{f:ini:einst_rosen}. In particular it has two asymptotically flat ends:
$r\rightarrow+\infty$ and $r\rightarrow 0$ (the puncture). 

As for Eq.~(\ref{e:ini:XCTS_ex1-2}), 
the simplest solution of Eq.~(\ref{e:ini:XCTS_ex3-2}) obeying the
asymptotic flatness requirement $\tilde N\Psi^7 \rightarrow 1$ as 
$r\rightarrow+\infty$ is
\be \label{e:ini:ex_tN_Psi7}
	\tilde N \Psi^7 = 1 + \frac{a}{r} , 
\ee
where $a$ is some constant.
Let us determine $a$ from the value of the lapse function at the second
asymptotically flat end $r\rightarrow 0$. The lapse being related to $\tilde N$
via Eq.~(\ref{e:ini:def_tN}), Eq.~(\ref{e:ini:ex_tN_Psi7}) is equivalent
to
\be \label{e:ini:XCTS_ex_N}
	N = \left( 1 + \frac{a}{r} \right) \Psi^{-1} 
	=  \left(1 + \frac{a}{r} \right) \left( 1 + \frac{m}{2r} \right) ^{-1} 
	= \frac{r+a}{r+m/2} . 
\ee
Hence
\be  \label{e:ini:XCTS_ex_limN_a}
	\lim_{r\rightarrow 0} N = \frac{2a}{m} . 
\ee
There are two natural choices for $\lim_{r\rightarrow 0} N$. The first one
is 
\be \label{e:ini:XCTS_ex_limN_1}
  \lim_{r\rightarrow 0} N = 1,
\ee
 yielding $a=m/2$. Then, from Eq.~(\ref{e:ini:XCTS_ex_N})
$N=1$ everywhere on $\Sigma_0$. This value of $N$ corresponds to a geodesic
slicing (cf. Sec.~\ref{s:dec:Gaussian_normal}). The second choice is
\be \label{e:ini:XCTS_ex_limN_m1}
	\lim_{r\rightarrow 0} N = -1. 
\ee
This choice is compatible with asymptotic flatness: it simply means that
the coordinate time $t$ is running ``backward'' near the asymptotic flat end
$r\rightarrow 0$. This contradicts the assumption $N>0$ in 
the definition of the lapse function given in Sec.~\ref{s:fol:def_lapse}. 
However, we shall generalize here the definition of the lapse to allow for
negative values: whereas the unit vector $\w{n}$ is always future-oriented, the
scalar field $t$ is allowed to decrease towards the future. Such a situation 
has already been encountered for the part of the slices $t={\rm const}$ located on the
left side of Fig.~\ref{f:ini:kruskal}. Once reported 
into Eq.~(\ref{e:ini:XCTS_ex_limN_a}),
the choice (\ref{e:ini:XCTS_ex_limN_m1}) yields $a=-m/2$, so that
\be \label{e:ini:XCTS_ex_N-2}
	N = \left(1 - \frac{m}{2r} \right) \left( 1 + \frac{m}{2r} \right) ^{-1} .
\ee
Gathering relations (\ref{e:ini:XCTS_ex_beta}), (\ref{e:ini:XCTS_ex_Psi}) and
(\ref{e:ini:XCTS_ex_N-2}), we arrive at the following expression of the spacetime metric
components:
\be \label{e:ini:XCTS_ex_gab}
	   g_{\mu\nu} dx^\mu dx^\nu  = - \left( 
    \frac{1 - \frac{m}{2r}}{ 1 + \frac{m}{2r}} \right) ^2
         dt^2 
    + \left( 1 + \frac{m}{2r} \right) ^4 \left[ d{r}^2 
    + {r}^2 (d\theta^2 + \sin^2\theta d\varphi^2) \right]  .
\ee
We recognize the line element of Schwarzschild spacetime in isotropic coordinates
[cf. Eq.~(\ref{e:cfd:Schwarz_isotropic})]. Hence we recover the same initial
data as in Sec.~\ref{s:ini:cflat_static} and depicted in Figs.~\ref{f:ini:einst_rosen}
and \ref{f:ini:kruskal}. The bonus is that we have the complete expression
of the metric $\w{g}$ on $\Sigma_0$, and not only the induced metric $\wgm$.
\begin{remark}
The choices (\ref{e:ini:XCTS_ex_limN_1}) and (\ref{e:ini:XCTS_ex_limN_m1})
for the asymptotic value of the lapse both lead to a momentarily static
initial slice in Schwarzschild spacetime. The difference is that the
time development corresponding to choice (\ref{e:ini:XCTS_ex_limN_1}) 
(geodesic slicing) will depend on $t$, whereas the time development corresponding to choice (\ref{e:ini:XCTS_ex_limN_m1}) will not, since in the latter case $t$
coincides with the standard Schwarzschild time coordinate, 
which makes $\wpar_t$ a Killing vector.
\end{remark}

\subsection{Uniqueness of solutions}

Recently, Pfeiffer and York \cite{PfeifY05} have exhibited a choice of vacuum free
data $(\tgm_{ij},\dot\tgm^{ij},K,\dot K)$ for which the solution 
$(\Psi,\tilde N,\beta^i)$ to the XCTS system (\ref{e:ini:XCTS1})-(\ref{e:ini:XCTS3}) 
is not unique (actually two solutions are found). 
The conformal metric $\wtgm$ is the flat metric plus 
a linearized quadrupolar gravitational wave, as obtained by Teukolsky 
\cite{Teuko82}, with a tunable amplitude. $\dot\tgm^{ij}$ corresponds to the
time derivative of this wave, and both $K$ and $\dot K$ are chosen to zero.
On the contrary, for the same free data, with $\dot K=0$ substituted by 
$\tilde N=1$, Pfeiffer and York have shown that the original conformal
thin sandwich method as described in Sec.~\ref{s:ini:CTS_ori} leads to a unique
solution (or no solution at all if the amplitude of the wave is two large). 

Baumgarte, \'O Murchadha and Pfeiffer \cite{BaumgOP06} have argued that the lack
of uniqueness for the XCTS system may be due to the term
\be
	-(\tilde N \Psi^7) \frac{7}{8} \hA_{ij} \hA^{ij} \Psi^{-8} = 
	 - \frac{7}{32} \Psi^6 \tgm_{ik} \tgm_{jl}
	\left[ \dot\tgm^{ij} + (\tilde L \beta)^{ij} \right]
	\left[ \dot\tgm^{kl} + (\tilde L \beta)^{kl} \right]
	\, (\tilde N \Psi^7)^{-1}
\ee
in Eq.~(\ref{e:ini:XCTS3}). Indeed, if we proceed as for the analysis of
Lichnerowicz equation in Sec.~\ref{s:ini:Lichne}, we notice that this
term, with the minus sign and the
negative power of $(\tilde N \Psi^7)^{-1}$, makes the linearization of Eq.~(\ref{e:ini:XCTS3}) of the type $\tD_i \tD^i \epsilon + \alpha\epsilon=\sigma$,
with $\alpha>0$. This ``wrong'' sign of $\alpha$ prevents the application of the
maximum principle to guarantee the uniqueness of the solution. 

The non-uniqueness of solution of the XCTS system for certain choice of 
free data has been confirmed by Walsh \cite{Walsh06} by means of bifurcation
theory.

\subsection{Comparing CTT, CTS and XCTS}

The conformal transverse traceless (CTT) method exposed in Sec.~\ref{s:ini:CTT}
and the (extended) conformal thin sandwich (XCTS) method considered here
differ by the choice of free data:
whereas both methods use the conformal metric
$\wtgm$ and the trace of the extrinsic curvature $K$ as free data,
CTT employs in addition $\hA^{ij}_{\rm TT}$, whereas for CTS (resp. XCTS) 
the additional
free data is $\dot\tgm^{ij}$, as well as $\tilde N$ (resp. $\dot K$).
Since $\hA^{ij}_{\rm TT}$ is directly related to the extrinsic curvature
and the latter is linked to the canonical momentum of the gravitational field
in the Hamiltonian formulation of general relativity (cf. Sec.~\ref{s:dec:ADM}), 
the CTT method can be considered as the approach to the initial data problem
in the \emph{Hamiltonian representation}. On the other side, $\dot\tgm^{ij}$ being
the ``velocity'' of $\tgm^{ij}$, the (X)CTS method constitutes the approach
in the \emph{Lagrangian representation} \cite{York04}.

\begin{remark}
The (X)CTS method assumes that the conformal metric
is unimodular: $\det (\tgm_{ij}) = f$ [Eq.~(\ref{e:cfd:dettgm_f})] 
(since Eq.~(\ref{e:ini:hA_dg_beta}) follows from this assumption),
whereas the CTT method can be applied with any
conformal metric. 
\end{remark}

The advantage of CTT is that its mathematical theory is well developed, 
yielding existence and uniqueness theorems, at least for constant mean curvature
(CMC) slices. The mathematical theory of CTS is very close to CTT. In particular, the
momentum constraint decouples from the Hamiltonian constraint on CMC slices. 
On the contrary, XCTS has a much more involved mathematical structure. In
particular the CMC condition does not yield to any decoupling. 
The advantage of XCTS is then to be better suited to the description of
quasi-stationary spacetimes, since 
$\dot\tgm^{ij}=0$ and $\dot K=0$ are necessary conditions for 
$\wpar_t$ to be a  Killing vector. This makes XCTS the method to be used in 
order to prepare initial data in quasi-equilibrium.
For instance, it has been shown \cite{GrandGB02,DamouGG02}
that XCTS yields orbiting binary black hole configurations 
in much better agreement with post-Newtonian computations
than the CTT treatment based on a superposition of two Bowen-York solutions. 

A detailed comparison of CTT and XCTS for a single spinning or boosted 
black hole has been performed by Laguna \cite{Lagun04}.

%%%%%%%%%%%%%%%%%%%%%%%%%%%%%%%%%%%%%%%%%%%%%%%%%%%%%%%%%%%%%%%%%%%%%%%%%%%%%%%%%%%%%%

\section{Initial data for binary systems} \label{s:ini:binary}

A major topic of contemporary numerical relativity is the computation of the 
merger of a binary system of black holes or neutron stars, for such systems
are among the most promising sources of gravitational radiation for the
interferometric detectors either groundbased (LIGO, VIRGO, GEO600, TAMA) or 
in space (LISA). 
The problem of preparing initial data for these systems has therefore  
received a lot of attention in the past decade. 

\begin{figure}
\centerline{\includegraphics[width=0.6\textwidth]{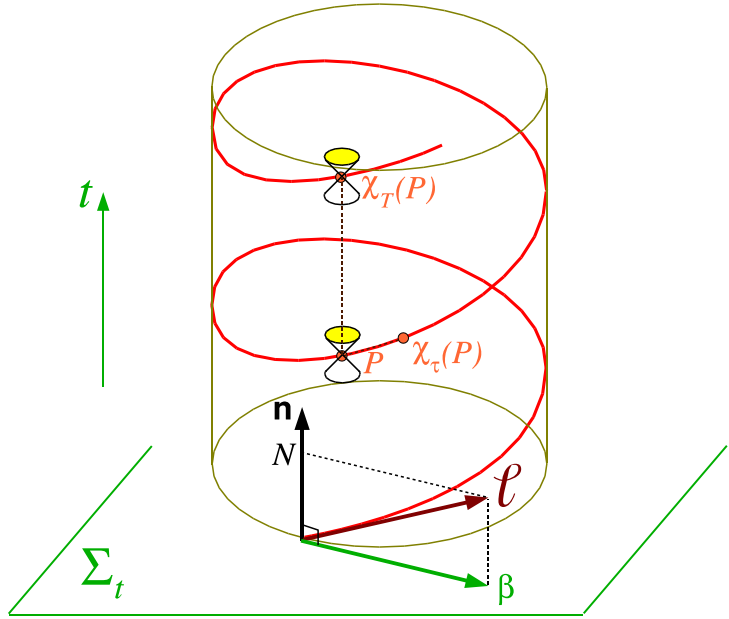}}
\caption[]{\label{f:ini:helical} \footnotesize
Action of the helical symmetry group, with Killing vector $\w{\ell}$.
$\chi_\tau(P)$ is the displacement of the point $P$ 
by the member of the symmetry group of parameter $\tau$.
$N$ and $\w{\beta}$ are respectively the lapse function and the shift vector
associated with coordinates adapted to the symmetry, i.e. coordinates $(t,x^i)$
such that $\wpar_t = \w{\ell}$.}
\end{figure}

\subsection{Helical symmetry}

Due to the gravitational-radiation reaction, a relativistic binary system 
has an inspiral
motion, leading to the merger of the two components. However, when the 
two bodies are are sufficiently far apart, one may approximate the spiraling
orbits by closed ones. Moreover, it is well known that gravitational radiation circularizes the orbits very efficiently, at least for comparable mass systems
\cite{Blanc06a}. 
We may then consider that the motion is described by a
sequence of \emph{closed circular orbits}. 

The geometrical translation of this physical assumption is that the
spacetime $(\M,\w{g})$ is endowed with some symmetry, 
called \defin{helical symmetry}.
Indeed exactly circular orbits imply the existence of a one-parameter symmetry
group such that the associated Killing vector $\w{\ell}$ obeys 
the following properties \cite{FriedUS02}:
(i) $\w{\ell}$ is timelike near the system, (ii) far from it, $\w{\ell}$ is spacelike
but there exists a smaller number $T>0$ such that the separation between any point 
$P$ and its image $\chi_T(P)$ under the symmetry group is timelike
(cf. Fig.~\ref{f:ini:helical}). $\w{\ell}$ is called a 
\defin{helical Killing vector}, its field lines in a spacetime 
diagram being helices (cf. Fig.~\ref{f:ini:helical}).

Helical symmetry is exact in theories of gravity where gravitational radiation
does not exist, namely: 
\begin{itemize}
\item in Newtonian gravity,
\item in post-Newtonian gravity, up to the second order,
\item in the Isenberg-Wilson-Mathews approximation to general relativity 
discussed in Sec.~\ref{s:cfd:IWM}.
\end{itemize}
Moreover helical symmetry can be exact in full general relativity for a 
non-axisymmetric system (such
as a binary) with standing gravitational waves \cite{Detwe94}.
But notice that a spacetime with helical symmetry and standing gravitational
waves cannot be asymptotically flat \cite{GibboS83}. 

To treat helically symmetric spacetimes, it is natural to choose coordinates
$(t,x^i)$ that are adapted to the symmetry, i.e. such that 
\be
	\wpar_t = \w{\ell} . 
\ee
Then all the fields are independent of the coordinate $t$. In particular,
\be \label{e:ini:dot_helic}
	\dot\tgm^{ij} = 0 \qquad \mbox{and} \qquad \dot K = 0 . 
\ee
If we employ the XCTS formalism to compute initial data, we therefore get some definite prescription for the free data $\dot\tgm^{ij}$ and $\dot K$. 
On the contrary, the requirements (\ref{e:ini:dot_helic}) do not have any immediate
translation in the CTT formalism. 

\begin{remark}
Helical symmetry can also be usefull to treat binary black holes outside
the scope of the 3+1 formalism, as shown by Klein \cite{Klein04}, who
developed a quotient space formalism to reduce the problem to a 
three dimensional $\mathrm{SL}(2,\R)/\mathrm{SO}(1,1)$ sigma model.
\end{remark}

Taking into account (\ref{e:ini:dot_helic}) and choosing maximal slicing 
($K=0$), the XCTS system (\ref{e:ini:XCTS1})-(\ref{e:ini:XCTS3}) becomes
\bea
 & & \tD_i \tD^i \Psi -\frac{{\tilde R}}{8}  \Psi
	+ \frac{1}{8} \hA_{ij} \hA^{ij} \, \Psi^{-7}
	+ 2\pi {\tilde E} \Psi^{-3}  = 0  \label{e:ini:XCTS-heli1} \\
 & &  \tD_j \left( \frac{1}{\tilde N} (\tilde L \beta)^{ij} \right)
 - 16\pi {\tilde p}^i  = 0 \label{e:ini:XCTS-heli2} \\
 & & \tD_i \tD^i (\tilde N \Psi^7) - (\tilde N \Psi^7)
	\left[ \frac{\tilde R}{8} 
	 + \frac{7}{8} \hA_{ij} \hA^{ij} \Psi^{-8} 
	+ 2 \pi (\tilde E+2\tilde S) \Psi^{-4} \right]  = 0  ,\label{e:ini:XCTS-heli3}
\eea
where [cf. Eq.~(\ref{e:ini:hA_dg_beta})]
\be 
   \hA^{ij} = \frac{1}{2\tilde N} 
	(\tilde L \beta)^{ij} . 
\ee

\subsection{Helical symmetry and IWM approximation}

If we choose, as part of the free data, the conformal metric to be flat,
\be \label{e:ini:helic_tgm_f}
	\tgm_{ij} = f_{ij} , 
\ee
then the helically symmetric XCTS system  
(\ref{e:ini:XCTS-heli1})-(\ref{e:ini:XCTS-heli3}) reduces to 
\bea
 & & \Delta \Psi 
	+ \frac{1}{8} \hA_{ij} \hA^{ij} \, \Psi^{-7}
	+ 2\pi {\tilde E} \Psi^{-3} = 0 \label{e:ini:XCTS-heli-flat1} \\
 & & \Delta \beta^i + \frac{1}{3} \Df^i \Df_j \beta^j
	- (L\beta)^{ij} \Df_j \ln \tilde N
  =  16\pi {\tilde N} {\tilde p}^i  \label{e:ini:XCTS-heli-flat2} \\
& & \Delta(\tilde N \Psi^7) - (\tilde N \Psi^7)
	\left[  \frac{7}{8} \hA_{ij} \hA^{ij} \Psi^{-8} 
	+ 2 \pi (\tilde E+2\tilde S) \Psi^{-4} \right]= 0 , 
					\label{e:ini:XCTS-heli-flat3}
\eea
where 
\be 
   \hA^{ij} = \frac{1}{2\tilde N}  (L \beta)^{ij}  
\ee
and $\wDf$ is the connection associated with the flat metric $\w{f}$, 
$\Delta :=  \Df_i \Df^i$ is the flat Laplacian [Eq.~(\ref{e:ini:def_Delta})], and 
$(L\beta)^{ij} := \Df^i \beta^j + \Df^j \beta^i - \frac{2}{3} \Df_k\beta^k \, f^{ij}$ 
[Eq.~(\ref{e:cfd:conf_Killing_f})]. 

We remark that the system (\ref{e:ini:XCTS-heli-flat1})-(\ref{e:ini:XCTS-heli-flat3})
is identical to the Isenberg-Wilson-Mathews (IWM) system 
(\ref{e:cfd:IWM_PDE1})-(\ref{e:cfd:IWM_PDE3})
presented in Sec.~\ref{s:cfd:IWM}: given that $\tilde E = \Psi^8 E$, 
$\tilde p^i = \Psi^{10} p^i$, $\tilde N = \Psi^{-6} N$, $\hA^{ij} = \Psi^6 \tA^{ij}$
and $\hA_{ij} \hA^{ij} = \Psi^{12} \tA_{ij} \tA^{ij}$, Eq.~(\ref{e:ini:XCTS-heli-flat1})
coincides with Eq.~(\ref{e:cfd:IWM_PDE2}),  Eq.~(\ref{e:ini:XCTS-heli-flat2})
coincides with Eq.~(\ref{e:cfd:IWM_PDE3}) and Eq.~(\ref{e:ini:XCTS-heli-flat3})
is a combination of Eqs.~(\ref{e:cfd:IWM_PDE1}) and (\ref{e:cfd:IWM_PDE2}).
Hence, within helical symmetry, the XCTS system with the choice $K=0$
and $\wtgm = \w{f}$ is equivalent to the IWM system. 
\begin{remark}
Contrary to IWM, XCTS is not some approximation to general relativity: it 
provides exact initial data. The only thing that may be questioned is
the astrophysical relevance of the XCTS
data with $\wtgm = \w{f}$. 
\end{remark}

\subsection{Initial data for orbiting binary black holes}

The concept of helical symmetry for generating orbiting binary black hole
initial data has been introduced in 2002 by Gourgoulhon, Grandcl\'ement and
Bonazzola \cite{GourgGB02,GrandGB02}. The system of equations that 
these authors have derived is equivalent to the XCTS system with $\wtgm = \w{f}$, 
their work being previous to the formulation of the XCTS method by Pfeiffer 
and York (2003) \cite{PfeifY03}. Since then other groups have combined 
XCTS with helical symmetry to compute binary black hole
initial data \cite{CookP04,Ansor05,Ansor07,CaudiCGP06}. Since all these studies
are using a flat conformal metric [choice (\ref{e:ini:helic_tgm_f})], 
the PDE system to be solved is
(\ref{e:ini:XCTS-heli-flat1})-(\ref{e:ini:XCTS-heli-flat3}), with 
the additional simplification $\tilde E = 0$ and ${\tilde p}^i=0$
(vacuum). The initial data manifold $\Sigma_0$ is chosen to be 
$\R^3$ minus two balls:
\be \label{e:ini:S0_R3_balls}
	\Sigma_0 = \R^3 \backslash (\mathcal{B}_1 \cup \mathcal{B}_2 ) .
\ee
In addition to the asymptotic flatness conditions, some boundary conditions
must be provided on the surfaces $\Sp_1$ and $\Sp_2$ of 
$\mathcal{B}_1$ and $\mathcal{B}_2$. 
One choose boundary conditions corresponding to a \emph{non-expanding horizon},
\index{non-expanding horizon} since this concept
characterizes black holes in equilibrium. We shall not detail these boundary
conditions here; they can be found in Refs.~\cite{CookP04,GourgJ06a}. 
The condition of non-expanding horizon provides 3 among the 5 required boundary conditions [for the 5 components $(\Psi,\tilde N,\beta^i)$]. The two remaining 
boundary conditions are given by (i) the choice of the foliation (choice of the
value of $N$ at $\Sp_1$ and $\Sp_2$) and (ii) the choice of the rotation state 
of each black hole (``individual spin''), as explained in Ref.~\cite{CaudiCGP06}. 

Numerical codes for solving the above system have been constructed by
\begin{itemize}
\item  Grandcl\'ement, Gourgoulhon and Bonazzola (2002) \cite{GrandGB02}
for corotating binary black holes;
\item Cook, Pfeiffer, Caudill and Grigsby (2004, 2006) \cite{CookP04,CaudiCGP06}
for corotating and irrotational binary black holes;
\item  Ansorg (2005, 2007) \cite{Ansor05,Ansor07}
for corotating binary black holes.
\end{itemize}
Detailed comparisons with post-Newtonian initial data (either from the standard 
post-Newtonian formalism \cite{Blanc02} or from the Effective One-Body
approach \cite{BuonaD99,Damou01}) have revealed a very good agreement,
as shown in Refs.~\cite{DamouGG02,CaudiCGP06}. 

An alternative to (\ref{e:ini:S0_R3_balls}) for the initial data manifold
would be to consider the twice-punctured $\R^3$:
\be \label{e:ini:S0_R3_points}
	\Sigma_0 = \R^3 \backslash \{O_1,O_2\} ,
\ee
where $O_1$ and $O_2$ are two points of $\R^3$. 
This would constitute some extension to the two bodies case of the punctured initial
data discussed in Sec.~\ref{s:ini:XCTS_work}. 
However, as shown by Hannam, Evans, Cook and Baumgarte in 2003 \cite{HannaECB03},
it is not possible to find a solution of the helically symmetric XCTS system
with a regular lapse in this case\footnote{see however Ref.~\cite{Hanna05} for some
attempt to circumvent this}. 
For this reason, initial data based on the puncture manifold (\ref{e:ini:S0_R3_points})
are computed within the CTT framework discussed in Sec.~\ref{s:ini:CTT}.
As already mentioned, there is no natural way to implement helical symmetry in
this framework. One instead selects the free data $\hA^{ij}_{\rm TT}$ to vanish
identically, as in the single black hole case treated in Secs.~\ref{s:ini:cflat_static}
and \ref{s:ini:Bowen_York}. Then
\be
	\hA^{ij} = (\tilde L X)^{ij} .
\ee
The vector $\w{X}$ must obey Eq.~(\ref{e:ini:mom_ex1}), which arises from the
momentum constraint. Since this equation is linear, 
one may choose for $\w{X}$ a linear superposition of two Bowen-York
solutions (Sec.~\ref{s:ini:Bowen_York}):
\be
	\w{X} = \w{X}_{(\w{P}^{(1)},\w{S}^{(1)})}
	+ \w{X}_{(\w{P}^{(2)},\w{S}^{(2)})} ,
\ee
where $\w{X}_{(\w{P}^{(a)},\w{S}^{(a)})}$ ($a=1,2$) is the Bowen-York
solution (\ref{e:ini:BY_X}) centered on $O_a$.
This method has been first implemented by Baumgarte in 2000 \cite{Baumg00}. 
It has been since then used by Baker, Campanelli, Lousto and Takashi
(2002) \cite{BakerCLT02} and Ansorg, Br\"ugmann and Tichy (2004) \cite{AnsorBT04}.
The initial data hence obtained are closed from helically symmetric XCTS initial 
data at large separation but deviate significantly from them, as well as
from post-Newtonian initial data, when the two black holes are very close.
This means that the Bowen-York extrinsic curvature is bad for close binary systems
in quasi-equilibrium
(see discussion in Ref.~\cite{DamouGG02}).
\begin{remark}
Despite of this, CTT Bowen-York configurations have been used as initial data
for the recent binary black hole inspiral and merger computations 
by Baker et al. \cite{BakerCCKV06a,BakerCCKV06b,VanMeBKC06} and Campanelli et al. 
\cite{CampaLMZ06,CampaLZ06a,CampaLZ06b,CampaLZ06c}. Fortunately, these initial
data had a relative large separation, so that they differed only slightly
from the helically symmetric XCTS ones. 
\end{remark}

Instead of choosing somewhat arbitrarily the free data of the CTT and XCTS methods,
notably setting $\wtgm=\w{f}$, one may deduce them from post-Newtonian results. 
This has been done for the binary black hole problem by Tichy, Br\"ugmann,
Campanelli and Diener (2003) \cite{TichyBCD03}, who have used the CTT
method with the free data $(\tgm_{ij},\hA^{ij}_{\rm TT})$ given by the second order
post-Newtonian (2PN) metric. In the same spirit, Nissanke (2006) \cite{Nissa06}
has provided 2PN free data for both the CTT and XCTS methods. 

\subsection{Initial data for orbiting binary neutron stars}

For computing initial data corresponding to orbiting binary neutron stars,
one must solve equations for the fluid motion in addition to the Einstein constraints. 
Basically this amounts to 
solving $\vec{\wnab} \cdot \w{T} = 0$ [Eq.~(\ref{e:mat:divT})]
in the context of helical symmetry. One can then show that a first integral
of motion exists in two cases: (i) the stars are corotating, i.e. the fluid 4-velocity
is colinear to the helical Killing vector (rigid motion), 
(ii) the stars are irrotational, i.e. 
the fluid vorticity vanishes. The most straightforward way to get the first 
integral of motion
is by means of the Carter-Lichnerowicz formulation of relativistic hydrodynamics,
as shown in Sec.~7 of Ref.~\cite{Gourg06}. Other derivations have been obtained
in 1998 by Teukolsky \cite{Teuko98} and Shibata \cite{Shiba98}. 

From the astrophysical point of view, the irrotational motion is much more interesting
than the corotating one, because the viscosity of neutron star matter is far too 
low to ensure the synchronization of the stellar spins with the orbital motion. 
On the other side, the irrotational state is a very good approximation for neutron
stars that are not millisecond rotators. Indeed, for these stars 
the spin frequency is much lower than the orbital frequency at 
the late stages of the inspiral and thus can be neglected. 

The first initial data for binary neutron stars on circular orbits have been computed
by Baumgarte, Cook, Scheel, Shapiro and Teukolsky in 1997 
\cite{BaumgCSST97,BaumgCSST98}
in the corotating case, and by Bonazzola, Gourgoulhon and Marck in 1999 \cite{BonazGM99a} in the irrotational case. These results were based on a polytropic equation of state.
Since then configurations in the irrotational regime have been obtained
\begin{itemize}
\item for a polytropic equation of state 
\cite{MarroMW99,UryuE00,UryuSE00,GourgGTMB01,TanigG02b,TanigG03};
\item for nuclear matter equations of state issued from recent nuclear physics
computations \cite{BejgeGGHTZ05,OechsJM07};
\item for strange quark matter \cite{OechsUPT04,LimouGG05}.
\end{itemize}
All these computation are based on a flat conformal metric [choice (\ref{e:ini:helic_tgm_f})], by solving the helically symmetric XCTS system
(\ref{e:ini:XCTS-heli-flat1})-(\ref{e:ini:XCTS-heli-flat3}), supplemented by
an elliptic equation for the velocity potential.
Only very recently, configurations based on a non flat conformal metric
have been obtained by Uryu, Limousin, Friedman, Gourgoulhon and Shibata
\cite{UryuLFGS06}. The conformal metric is then deduced from a waveless approximation
developed by Shibata, Uryu and Friedman \cite{ShibaUF04} and 
which goes beyond the IWM approximation.

\subsection{Initial data for black hole - neutron star binaries}

Let us mention briefly that initial data for a mixed binary system, i.e. 
a system composed of a black hole and a neutron star, have been obtained
very recently by Grandcl\'ement \cite{Grand06} and Taniguchi, Baumgarte, 
Faber and Shapiro \cite{TanigBFS06,TanigBFS07}.
Codes aiming at computing such systems have also been presented by 
Ansorg \cite{Ansor07} and Tsokaros and Uryu \cite{TsokaU07}.

%  
%    Chapitre : Choice of foliation and coordinates 
%
% $Date: 2007-03-05 22:39:07 +0100 (lun, 05 mar 2007) $
% $Rev: 182 $
% $Author: e_gourgoulhon $
%%%%%%%%%%%%%%%%%%%%%%%%%%%%%

\chapter{Choice of foliation and spatial coordinates} \label{s:evo}

%\verb$Date: 2007-03-05 22:39:07 +0100 (lun, 05 mar 2007) $

\minitoc
\vspace{1cm}

%%%%%%%%%%%%%%%%%%%%%%%%%%%%%%%%%%%%%%%%%%%%%%%%%%%%%%%%%%%%%%%%%%%%%%%%%%%%

\section{Introduction}

Having investigated the initial data problem in the preceding chapter, 
the next logical step is to discuss the evolution problem, i.e. 
the development $(\Sigma_t,\wgm)$ of initial data $(\Sigma_0,\wgm,\w{K})$.
This constitutes the integration of the Cauchy problem introduced
in Sec.~\ref{s:dec:Cauchy}. As discussed in Sec.~\ref{s:dec:geometrodynamics}, 
a key feature of this problem is the freedom of choice for the lapse function $N$
and the shift vector $\w{\beta}$, reflecting respectively the choice of foliation
$(\Sigma_t)_{t\in\R}$ and the choice of coordinates $(x^i)$ on each leaf $\Sigma_t$ of the foliation. These choices are crucial because they determine the
specific form of the 3+1 Einstein system 
(\ref{e:dec:Einstein_PDE1})-(\ref{e:dec:Einstein_PDE4}) that one has actually 
to deal with. In particular, depending of the choice of $(N,\w{\beta})$, this
system can be made more hyperbolic or more elliptic. 

Extensive discussions about the various possible choices of foliations and
spatial coordinates can be found in the seminal articles by Smarr and York
\cite{SmarrY78b,York79} as well as in the review articles by 
Alcubierre \cite{Alcub05},  Baumgarte and Shapiro \cite{BaumgS03},
and Lehner \cite{Lehne01}.

%%%%%%%%%%%%%%%%%%%%%%%%%%%%%%%%%%%%%%%%%%%%%%%%%%%%%%%%%%%%%%%%%%%%%%%%%%%%%%%%%%%%%

\section{Choice of foliation}

\subsection{Geodesic slicing} \label{s:evo:geodesic}

The simplest choice of foliation one might think about is the 
\defin{geodesic slicing}, for it corresponds
to a unit lapse:
\be \label{e:evo:lapse_geod}
	\encadre{N = 1} .
\ee 
Since the 4-acceleration $\w{a}$ of the Eulerian observers is nothing but
the spatial gradient of $\ln N$ [cf. Eq.~(\ref{e:fol:a_DN})], the
choice (\ref{e:evo:lapse_geod}) implies $\w{a}=0$, i.e. the 
worldlines of the Eulerian observers are geodesics, hence the name
\emph{geodesic slicing}. Moreover the choice 
(\ref{e:evo:lapse_geod}) implies that 
the proper time along these worldlines coincides with the coordinate time $t$. 

We have already used the geodesic slicing to discuss the basics feature of the 3+1
Einstein system in Sec.~\ref{s:dec:Gaussian_normal}. We have also argued there
that, due to the tendency of timelike geodesics without vorticity (as the worldlines
of the Eulerian observers are) to focus
and eventually cross, this type of foliation can become pathological within a finite
range of $t$.

\begin{example}
A simple example of geodesic slicing is provided by the use of 
Painlev\'e-Gullstrand coordinates $(t,R,\theta,\varphi)$ in 
Schwarzschild spacetime (see e.g. Ref.~\cite{MarteP01}). 
These coordinates are defined as follows:
$R$ is nothing but the standard Schwarzschild radial 
coordinate\footnote{in this chapter, we systematically use the notation $R$
for Schwarzschild radial coordinate (areal radius), leaving the notation $r$
for other types of radial coordinates, such that the isotropic one 
[cf. Eq.~(\ref{e:cfd:Schwarz_isotropic})]}, 
whereas the Painlev\'e-Gullstrand coordinate $t$ is related to the
Schwarzschild time coordinate $t_{\rm S}$ by
\be
	t = t_{\rm S} + 4m \left( \sqrt{\frac{R}{2m}}
	+ \frac{1}{2} \ln \left| \frac{\sqrt{R/2m}-1}{\sqrt{R/2m}+1} \right| \right) . 
\ee
The metric components with respect to Painlev\'e-Gullstrand coordinates are 
extremely simple, being given by
\be
	g_{\mu\nu} dx^\mu dx^\nu = - dt^2 + \left(dR + \sqrt{\frac{2m}{R}} \, dt
	\right)^2
	+ R^2 (d\theta^2 + \sin^2\theta \, d\varphi^2) . 
\ee  
By comparing with the general line element (\ref{e:dec:g_gam_N_beta}), we read
on the above expression that $N=1$, $\beta^i=(\sqrt{2m/R},0,0)$ and
$\gam_{ij} = \textrm{diag}(1,R^2,R^2\sin^2\theta)$. Thus the hypersurfaces 
$t={\rm const}$ are geodesic slices. Notice that the induced metric $\wgm$ is flat. 
\end{example}

\begin{figure}
\centerline{\includegraphics[width=0.7\textwidth]{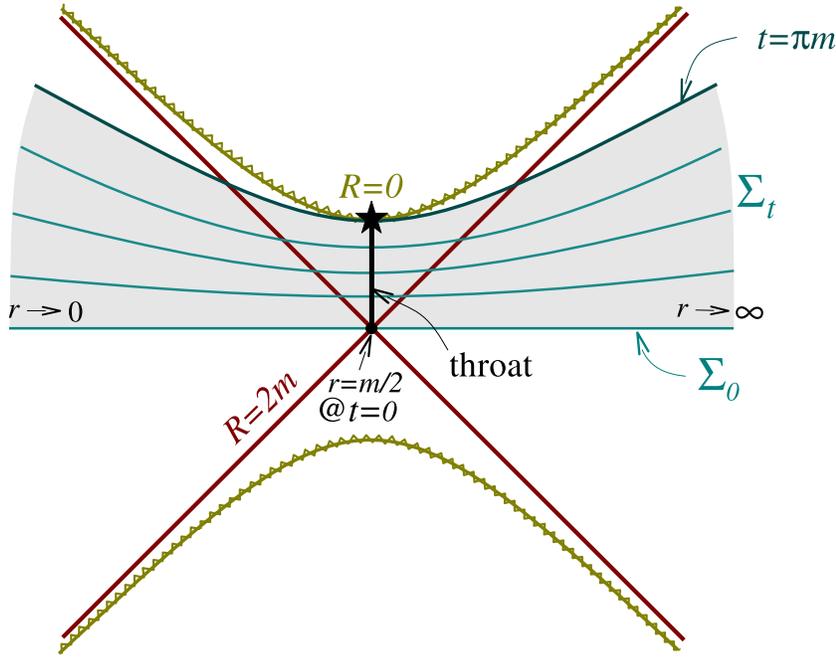}}
\caption[]{\label{f:evo:geod} \footnotesize
Geodesic-slicing evolution from the initial slice $t=t_{\rm S}=0$ 
of Schwarzschild spacetime depicted in a Kruskal-Szekeres diagram. $R$ stands for Schwarzschild radial
coordinate (areal radius), so that $R=0$ is the singularity and
$R=2m$ is the event horizon (figure adapted from Fig.~2a of 
\cite{SmarrY78b}). 
}
\end{figure}

\begin{example}
Another example of geodesic slicing, still in Schwarzschild spacetime, is provided
by the time development with $N=1$ of the initial data constructed in Secs.~\ref{s:ini:cflat_static} and \ref{s:ini:XCTS_work}, namely the momentarily 
static slice $t_{\rm S}=0$ of Schwarzschild spacetime, with topology
$\mathbb{R}\times\mathbb{S}^2$ (Einstein-Rosen bridge). 
The resulting foliation is depicted in Fig.~\ref{f:evo:geod}. It 
hits the singularity at $t=\pi m$, reflecting the bad behavior of geodesic
slicing.
\end{example}

In numerical relativity, geodesic slicings have been used by Nakamura, 
Oohara and Kojima to perform in 1987 the first 3D evolutions of vacuum spacetimes with gravitational waves \cite{NakamOK87}. However, as discussed in Ref.~\cite{ShibaN95},
the evolution was possible only for a pretty limited range of $t$, 
because of the focusing property mentioned above.

\subsection{Maximal slicing} \label{s:evo:maximal}

\begin{figure}
\centerline{\includegraphics[width=0.7\textwidth]{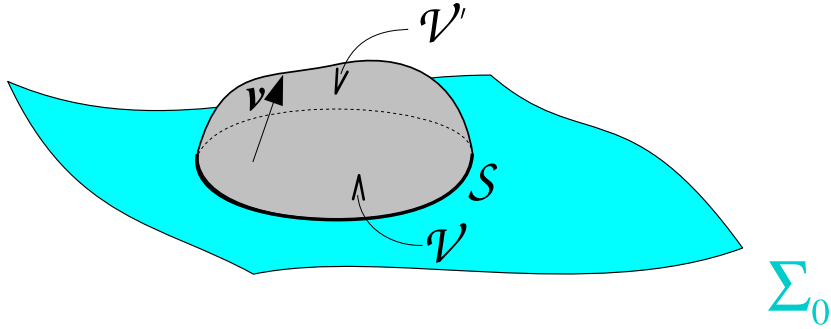}}
\caption[]{\label{f:evo:maximal} \footnotesize
Deformation of a volume $\mathcal{V}$ delimited by the surface $\Sp$
in the hypersurface $\Sigma_0$.
}
\end{figure}

A very famous type of foliation is maximal slicing, already encountered
in Sec.~\ref{s:cfd:IWM} and in Chap.~\ref{s:ini}, where it plays a great
role in decoupling the constraint equations. The \defin{maximal slicing}
corresponds to 
the vanishing of the mean curvature of the hypersurfaces $\Sigma_t$:
\be \label{e:evo:trK_zero}
	\encadre{K=0} . 
\ee
The fact that this condition leads to hypersurfaces of \emph{maximal volume} 
can be seen as follows. Consider some hypersurface $\Sigma_0$ and a closed two-dimensional
surface $\Sp$ lying in $\Sigma_0$  (cf. Fig.~\ref{f:evo:maximal}). 
The volume of the domain $\mathcal{V}$ enclosed in $\Sp$ is
\be \label{e:evo:vol_domV}
	V = \int_{\mathcal{V}} \sqrt{\gm} \, d^3 x ,
\ee
where $\gm=\det\gm_{ij}$ is the determinant of the metric $\wgm$ with respect
to some coordinates $(x^i)$ used in $\Sigma_t$.
Let us consider a small deformation $\mathcal{V}'$ of $\mathcal{V}$ that 
keeps the boundary $\Sp$ fixed. $\mathcal{V}'$ is generated by a small displacement
along a vector field $\w{v}$ of every point of $\mathcal{V}$, such that
$\left. \w{v}\right| _{\Sp} = 0$. Without any loss of generality, we may
consider that $\mathcal{V}'$ lies in a hypersurface $\Sigma_{\delta t}$
that is a member of some ``foliation'' $(\Sigma_t)_{t\in\R}$
such that $\Sigma_{t=0}=\Sigma_0$. The hypersurfaces $\Sigma_t$ intersect
each other at $\Sp$, which violates condition (\ref{e:fol:non_intersect})
in the definition of a foliation given in Sec.~\ref{s:fol:def_foliat},
hence the quotes around the word ``foliation''.
Let us consider a 3+1 coordinate system $(t,x^i)$ associated with the
``foliation'' $(\Sigma_t)_{t\in\R}$ and adapted to $\Sp$ in the sense
that the position of $\Sp$ in these coordinates does not depend upon $t$.
The vector $\wpar_t$ associated to these coordinates is then related
to the displacement vector $\w{v}$ by
\be
	\w{v} = \delta t \, \wpar_t . 
\ee
Introducing the lapse function $N$ and shift vector $\w{\beta}$ associated
with the coordinates $(t,x^i)$, the above relation becomes
[cf. Eq.~(\ref{e:dec:t_Nn_b})] $\w{v} = \delta t \, (N\w{n} + \w{\beta})$.
Accordingly, the condition $\left. \w{v}\right| _{\Sp} = 0$ implies
\be \label{e:evo:beta_S_zero}
	\left. N \right| _{\Sp} = 0 \qquad \mbox{and} \qquad
	\left. \w{\beta} \right| _{\Sp} = 0 .
\ee
Let us define $V(t)$ as the volume of the domain $\mathcal{V}_t$ delimited by $\Sp$
in $\Sigma_t$. It is given by a formula identical to Eq.~(\ref{e:evo:vol_domV}),
except of course that the integration domain has to be replaced by
$\mathcal{V}_t$. 
Moreover, the domains $\mathcal{V}_t$ lying at fixed values of the coordinates
$(x^i)$, we have
\be \label{e:evo:dVdt1}
	\frac{dV}{dt} = \int_{\mathcal{V}_t} \der{\sqrt{\gm}}{t}  \, d^3 x . 
\ee
Now, contracting Eq.~(\ref{e:dec:Einstein_PDE1}) with $\gm^{ij}$ and using
Eq.~(\ref{e:dec:Lie_beta_gam}), we get
\be \label{e:evo:tr_dt_gm}
	\gm^{ij} \der{}{t} \gm_{ij} = - 2 N K + 2 D_i \beta^i . 
\ee
From the general rule (\ref{e:cfd:variation_det}) for the variation of a
determinant, 
\be
	\gm^{ij} \der{}{t} \gm_{ij} = \der{}{t} (\ln \gm) = 
	\frac{2}{\sqrt{\gm}} \der{\sqrt{\gm}}{t} , 
\ee
so that Eq.~(\ref{e:evo:tr_dt_gm}) becomes
\be \label{e:evo:evol_ln_gm}
	\encadre{ \frac{1}{\sqrt{\gm}} \der{\sqrt{\gm}}{t} = - NK + D_i \beta^i }. 
\ee
Let us use this relation to express Eq.~(\ref{e:evo:dVdt1}) as
\be \label{e:evo:dVdt2}
	\frac{dV}{dt} = \int_{\mathcal{V}_t} \left[ - N K + D_i \beta^i \right]
	\sqrt{\gm} \, d^3 x . 
\ee
Now from the Gauss-Ostrogradsky theorem, 
\be
	\int_{\mathcal{V}_t} D_i \beta^i \sqrt{\gm} \, d^3 x  
	= \oint_{\Sp} \beta^i s_i \sqrt{q} \, d^2 y , 
\ee
where $\w{s}$ is the unit normal to $\Sp$ lying in $\Sigma_t$, $\w{q}$ is the
induced metric on $\Sp$, $(y^a)$ are coordinates on $\Sp$ and $q=\det q_{ab}$. 
Since $\w{\beta}$ vanishes on $\Sp$ [property (\ref{e:evo:beta_S_zero})], the
above integral is identically zero and Eq.~(\ref{e:evo:dVdt2}) reduces
to
\be
	\encadre{ \frac{dV}{dt} = - \int_{\mathcal{V}_t} N K \sqrt{\gm} \, d^3 x }. 
\ee
We conclude that if $K=0$ on $\Sigma_0$, the volume $V$ enclosed in $\Sp$
is extremal with respect to variations of the domain delimited by $\Sp$, provided
that the boundary of the domain remains $\Sp$.  
In the Euclidean space, such an extremum would define a \emph{minimal surface},
the corresponding variation problem being a \defin{Plateau problem} [named after
the Belgian physicist Joseph Plateau (1801-1883)]: given a closed contour
$\Sp$ (wire loop),  find the surface $\mathcal{V}$ (soap film)
of minimal area (minimal surface tension energy) bounded by $\Sp$. 
However, in the present
case of a metric of Lorentzian signature, it can be shown that the extremum is
actually a maximum, hence the name \emph{maximal slicing}. 
For the same reason, a timelike geodesic between two points in 
spacetime is the curve of \emph{maximum} length joining these two points.

Demanding that the maximal slicing condition (\ref{e:evo:trK_zero}) holds for
all hypersurfaces $\Sigma_t$, once combined with the evolution equation 
(\ref{e:cfd:evol_K}) for $K$, yields the following elliptic equation
for the lapse function:
\be \label{e:evo:eq_lapse_max}
	\encadre{ 
	D_i D^i N = N \left[ 4\pi (E+S) 
	+ K_{ij} K^{ij} \right] } . 
\ee

\begin{remark}
We have already noticed that at the Newtonian limit, Eq.~(\ref{e:evo:eq_lapse_max}) reduces to the Poisson equation for the gravitational potential $\Phi$
(cf. Sec.~\ref{s:cfd:dyn_part_Einstein}). Therefore the maximal slicing
can be considered as a natural generalization to the relativistic case
of the canonical slicing of Newtonian spacetime by hypersurfaces of constant 
absolute time. In this respect, let us notice that 
the ``beyond Newtonian'' approximation of general relativity constituted
by the Isenberg-Wilson-Mathews approach discussed in Sec.~\ref{s:cfd:IWM}
is also based on maximal slicing.  
\end{remark}

\begin{figure}
\centerline{\includegraphics[width=0.7\textwidth]{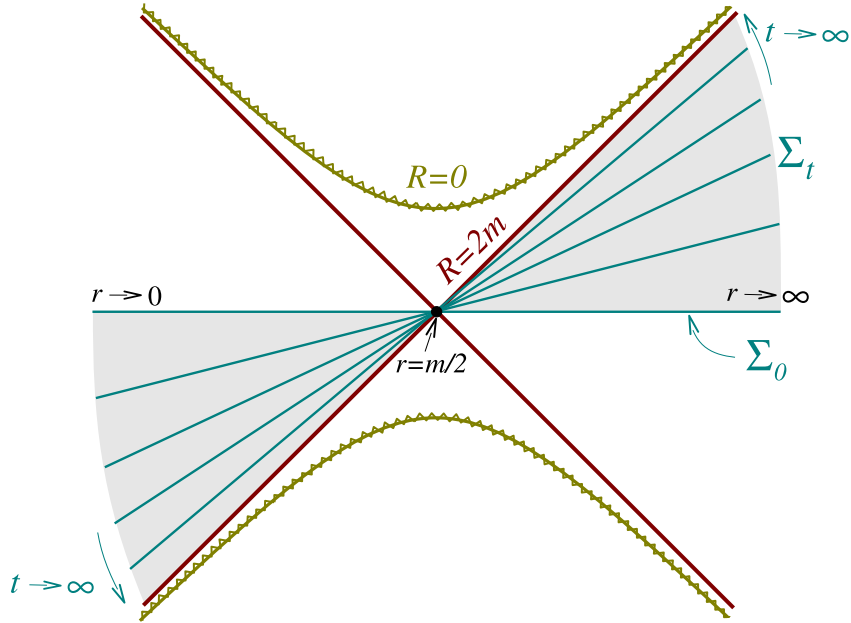}}
\caption[]{\label{f:evo:max_antisym} \footnotesize
Kruskal-Szekeres diagram showing the maximal slicing of Schwarzschild spacetime defined by the standard Schwarzschild time coordinate $t$. As for Fig.~\ref{f:evo:geod},
$R$ stands for Schwarzschild radial
coordinate (areal radius), so that $R=0$ is the singularity and
$R=2m$ is the event horizon, whereas $r$ stands for the isotropic radial
coordinate [cf. Eq.~(\ref{e:ini:XCTS_ex_gab})].}
\end{figure}

\begin{example}
In Schwarzschild spacetime, the standard Schwarzschild time coordinate 
$t$ defines maximal hypersurfaces $\Sigma_t$, 
which are spacelike for $R>2m$ ($R$ being Schwarzschild radial coordinate). Indeed these hypersurfaces are totally geodesic:
$\w{K}=0$ (cf. \S~\ref{e:hyp:link_nab_D}), 
so that, in particular, $K=\mathrm{tr}_{\gm}\w{K}=0$. 
This maximal slicing is shown in 
Fig.~\ref{f:evo:max_antisym}. The corresponding lapse function expressed in terms
of the isotropic radial coordinate $r$ is
\be \label{e:evo:ex_N_antisym}
	N = \left(1 - \frac{m}{2r} \right) \left( 1 + \frac{m}{2r} \right) ^{-1} .
\ee
As shown in Sec.~\ref{s:ini:XCTS_work}, the above expression can be 
derived by means of the XCTS formalism. 
Notice that the foliation $(\Sigma_t)_{t\in\R}$ does not penetrate under the
event horizon ($R=2m$) and that the lapse
is negative for $r<m/2$ (cf. discussion in 
Sec.~\ref{s:ini:XCTS_work} about negative lapse values).
\end{example}

Besides its nice geometrical definition, 
an interesting property of maximal slicing is the 
\defin{singularity avoidance}. 
This is related to the fact that the set of the Eulerian observers of
a maximal foliation define an \emph{incompressible flow}: 
indeed, thanks to  Eq.~(\ref{e:hyp:K_div_n}), the condition 
$K=0$ is equivalent to the incompressibility condition 
\be
 \wnab\cdot \w{n} = 0 
\ee
for the 4-velocity field $\w{n}$ of the Eulerian observers. 
If we compare with the Eulerian observers of geodesic
slicings (Sec.~\ref{s:evo:geodesic}), who have the tendency to squeeze, we may
say that maximal-slicing Eulerian observers do not converge
because they are accelerating ($\w{D} N \not =0$) in order to balance the
focusing effect of gravity. 
Loosely speaking, the incompressibility prevents the Eulerian
observers from converging towards the central singularity if the latter forms
during the time evolution. This is illustrated by the following example
in Schwarzschild spacetime.

\begin{figure}
\centerline{\includegraphics[width=0.7\textwidth]{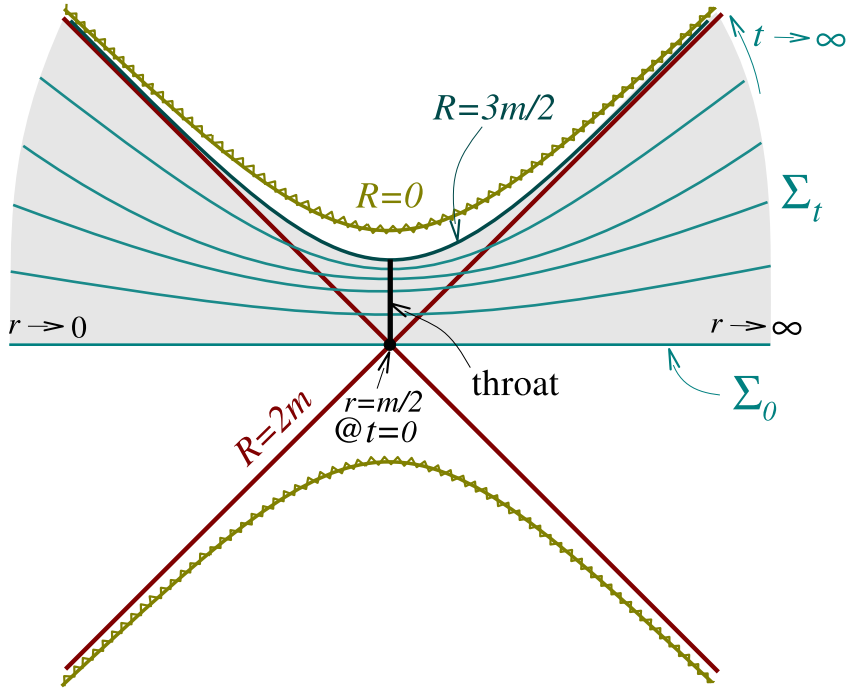}}
\caption[]{\label{f:evo:max_sym} \footnotesize
Kruskal-Szekeres diagram depicting the maximal slicing of Schwarzschild spacetime 
defined by the Reinhart/Estabrook et al. time function $t$
[cf. Eq.~(\ref{e:evo:g_comp_Estabrook})]. 
As for Figs.~\ref{f:evo:geod} and \ref{f:evo:max_antisym},
$R$ stands for Schwarzschild radial
coordinate (areal radius), so that $R=0$ is the singularity and
$R=2m$ is the event horizon, whereas $r$ stands for the isotropic radial
coordinate. At the throat (minimal surface), $R=R_C$ where $R_C$ is the function
of $t$ defined below Eq.~(\ref{e:evo:ex_t_C})
(figure adapted from Fig.~1 of Ref.~\cite{EstabWCDST73}).}
\end{figure}

\begin{example}
Let us consider the time development of the initial data constructed in Secs.~\ref{s:ini:cflat_static} and \ref{s:ini:XCTS_work}, namely the momentarily 
static slice $t_{\rm S}=0$ of Schwarzschild spacetime (with the Einstein-Rosen 
bridge). 
A first maximal slicing development
of these initial data is that based on Schwarzschild time coordinate $t_{\rm S}$
and discussed above (Fig.~\ref{f:evo:max_antisym}). 
The corresponding lapse function is given by Eq.~(\ref{e:evo:ex_N_antisym}) and
is \emph{antisymmetric} about the minimal surface $r=m/2$ (throat).
There exists a second maximal-slicing development of the same initial data
but with a lapse which is \emph{symmetric} about the throat. It has been found
in 1973 by Estabrook, Wahlquist, Christensen, DeWitt, Smarr and Tsiang 
\cite{EstabWCDST73}, as well as Reinhart \cite{Reinh73}. 
The corresponding time coordinate $t$ is different from Schwarzschild time
coordinate $t_{\rm S}$, except for $t=0$ (initial slice $t_{\rm S}=0$). 
In the coordinates $(x^\alpha)=(t,R,\theta,\varphi)$, where $R$ is Schwarzschild
radial coordinate, the metric components obtained by 
Estabrook et al. \cite{EstabWCDST73} (see also Refs.~\cite{BeigO98,Beig00,ReimaB04})
take the form
\be \label{e:evo:g_comp_Estabrook}
	g_{\mu\nu} dx^\mu dx^\nu = - N^2 dt^2 + \left( 1 -\frac{2m}{R}
	+ \frac{C(t)^2}{R^4} \right) ^{-1} \left( dR + \frac{C(t)}{R^2} N \, dt
	\right) ^2
	+ R^2 (d\theta^2 + \sin^2\theta d\varphi^2) ,
\ee
where 
\be
	N = N(R,t) = \sqrt{1 - \frac{2m}{R} + \frac{C(t)^2}{R^4}}
	\left\{ 1 + \frac{dC}{dt} \int_R^{+\infty} 
	\frac{x^4 \, dx}{\left[ x^4 - 2m x^3 + C(t)^2 \right] ^{3/2}} \right\}, 
\ee
and  $C(t)$ is the function of $t$ defined implicitly by 
\be \label{e:evo:ex_t_C}
	t = - C \int_{R_C}^{+\infty} \frac{dx}{(1-2m/x)\sqrt{x^4-2mx^3 + C^2}} ,
\ee
$R_C$ being the unique root of the polynomial $P_C(x):= x^4 - 2m x^3 + C^2$
in the interval $(3m/2,\, 2m]$. $C(t)$ varies from $0$ at $t=0$
to $C_\infty := (3\sqrt{3}/4) m^2$ as $t\rightarrow+\infty$. 
Accordingly, $R_C$ decays from $2m$ ($t=0$) to $3m/2$ ($t\rightarrow+\infty$). 
Actually, for $C=C(t)$, $R_C$  represents the smallest value of the radial
coordinate $R$ in the slice $\Sigma_t$. 
This maximal slicing of Schwarzschild spacetime is represented in 
Fig.~\ref{f:evo:max_sym}. We notice that, as $t\rightarrow+\infty$, the
slices $\Sigma_t$ accumulate on a limiting hypersurface: the
hypersurface $R=3m/2$ (let us recall that for
$R<2m$, the hypersurfaces $R={\rm const}$ are spacelike and are thus
eligible for a 3+1 foliation). Actually, it can be seen that the hypersurface
$R=3m/2$ is the only hypersurface $R={\rm const}$ which is spacelike and
maximal \cite{Beig00}. 
If we compare with Fig.~\ref{f:evo:geod}, we notice that, contrary 
to the geodesic slicing, 
the present foliation never encounters the singularity. 
\end{example}

The above example illustrates the singularity-avoidance property of 
maximal slicing: while the entire spacetime outside the event horizon is
covered by the foliation, the hypersurfaces ``pile up'' in the black hole
region so that they never reach the singularity. As a consequence, 
in that region, the proper time (of Eulerian observers)
between two neighbouring hypersurfaces tends to zero as $t$ increases. 
According to Eq.~(\ref{e:fol:dtau_Ndt}), this implies
\be
	N \rightarrow 0 \qquad \mbox{as} \quad t\rightarrow+\infty . 
\ee
This ``phenomenon'' is called \defin{collapse of the lapse}.
Beyond the Schwarzschild case discussed above, the collapse of the lapse
is a generic feature of maximal slicing of spacetimes
describing black hole formation via gravitational collapse.
For instance, it occurs in the analytic solution obtained by Petrich, Shapiro and
Teukolsky \cite{PetriST85} for the maximal slicing of the Oppenheimer-Snyder
spacetime (gravitational collapse of a spherically symmetric
homogeneous ball of pressureless matter). 

In numerical relativity, maximal slicing has been used in 
the computation of the (axisymmetric) head-on collision of two black holes by
Smarr, \v Cade\v z and Eppley in the seventies \cite{SmarrCDE76,Smarr79}, 
as well as in computations of axisymmetric 
gravitational collapse by Nakamura and Sato (1981) 
\cite{Nakam81,NakamS81}, Stark and Piran (1985) \cite{StarkP85}
and Evans (1986) \cite{Evans86}. 
Actually Stark and Piran used a mixed type of foliation introduced
by Bardeen and Piran \cite{BardeP83}: maximal slicing near the
origin ($r=0$) and polar slicing far from it. The \defin{polar slicing}
is defined in spherical-type coordinates $(x^i)=(r,\theta,\varphi)$
by 
\be
	K^\theta_{\ \, \theta} + K^\varphi_{\ \, \varphi}=0, 
\ee
instead of $K^r_{\ \, r}+K^\theta_{\ \, \theta} + K^\varphi_{\ \, \varphi}=0$
for maximal slicing.

Whereas maximal slicing is a nice choice of foliation, with a clear geometrical
meaning, a natural Newtonian limit and a singularity-avoidance feature, 
it has not been much used in 3D (no spatial symmetry) numerical relativity. 
The reason is a technical one: imposing maximal slicing requires to solve the
elliptic equation (\ref{e:evo:eq_lapse_max}) for the lapse
and elliptic equations are usually CPU-time consuming, 
except if one make uses of fast elliptic solvers \cite{GrandBGM01,BonazGGN04}.
For this reason, most of the recent computations of binary black hole inspiral and merger have been performed with the 1+log slicing, 
to be discussed in Sec.~\ref{s:evo:1plog}. 
Nevertheless, it is worth to note that maximal slicing has been used for the first grazing collisions of binary black holes, as computed 
by Br\"ugmann (1999) \cite{Brueg99}. 

To avoid the resolution of an elliptic equation while preserving most of the
good properties of maximal slicing, an \defin{approximate maximal slicing}
has been introduced in 1999 by Shibata \cite{Shiba99a}. It consists in transforming
Eq.~(\ref{e:evo:eq_lapse_max}) into a \emph{parabolic} equation by adding 
a term of the type $\dert{N}{\lambda}$ in the right-hand side and to compute
the ``$\lambda$-evolution'' for some range of the parameter $\lambda$. 
This amounts to resolve a heat like equation. Generically the solution 
converges towards a stationary one, so that $\dert{N}{\lambda}\rightarrow 0$
and the original elliptic equation (\ref{e:evo:eq_lapse_max}) is solved. 
The approximate maximal slicing has been used by Shibata, Uryu and
Taniguchi to compute the merger of binary neutron stars 
\cite{Shiba99c,ShibaU00,ShibaU02,ShibaTU03,ShibaTU05,ShibaT06}, as well as
by Shibata and Sekiguchi for 2D (axisymmetric) gravitational collapses
\cite{Shiba03a,Shiba03b,SekigS05} or 3D ones \cite{ShibaS05}. 

\subsection{Harmonic slicing} \label{s:evo:harm_slic}

Another important category of time slicing is deduced from the standard
\defin{harmonic} or \defin{De Donder} condition for the spacetime
coordinates $(x^\alpha)$:
\be \label{e:evo:harm_coord_def}
	 \square_{\w{g}} x^\alpha = 0 , 
\ee
where $\square_{\w{g}} := \nabla_\mu \nabla^\mu$ is the d'Alembertian associated
with the metric $\w{g}$ and each coordinate $x^\alpha$ is considered as a 
scalar field on $\M$. 
Harmonic coordinates have been introduced by De Donder in 1921 \cite{DeDon21}
and have played an important role in theoretical developments, notably 
in Choquet-Bruhat's demonstration (1952, \cite{Foure52}) of the well-posedness of the Cauchy problem for 3+1 Einstein equations (cf. Sec.~\ref{s:dec:existence_uniqueness}).

The \defin{harmonic slicing} is defined by requiring that the harmonic
condition holds for the $x^0=t$ coordinate, 
but not necessarily for the other coordinates, leaving the freedom to choose
any coordinate $(x^i)$ in each hypersurface $\Sigma_t$:
\be \label{e:evo:harm_slicing0}
	\encadre{ \square_{\w{g}} t = 0 } .
\ee
Using the standard expression for the d'Alembertian, we get
\be
	\frac{1}{\sqrt{-g}} \der{}{x^\mu} \bigg( \sqrt{-g} g^{\mu\nu} 
	\underbrace{\der{t}{x^\nu}}_{=\delta^0_{\ \, \nu}} \bigg) = 0 , 
\ee
i.e. 
\be
	\der{}{x^\mu} \left( \sqrt{-g} g^{\mu0} \right) = 0 .  
\ee
Thanks to the relation $\sqrt{-g} = N \sqrt{\gm}$ [Eq.~(\ref{e:fol:detg_Ngetgam})],
this equation becomes
\be
	\der{}{t} \left( N \sqrt{\gm} g^{00} \right)
	+ \der{}{x^i} \left( N \sqrt{\gm} g^{i0} \right) = 0 . 
\ee
From the expression of $g^{\alpha\beta}$ given by Eq.~(\ref{e:dec:g_con}),  
$g^{00} = - 1/N^2$ and $g^{i0} = \beta^i / N^2$. Thus
\be \label{e:evo:harm_slicing1}
	- \der{}{t} \left( \frac{\sqrt{\gm}}{N} \right)
	+ \der{}{x^i} \left( \frac{\sqrt{\gm}}{N} \beta^i \right) = 0 . 
\ee
Expanding and reordering gives
\be
	\der{N}{t} - \beta^i \der{N}{x^i} - N \bigg[
	\frac{1}{\sqrt{\gm}}\der{\sqrt{\gm}}{t}
	- 
 \underbrace{\frac{1}{\sqrt{\gm}} \der{}{x^i} \left( \sqrt{\gm} 
	\beta^i \right)}_{=D_i \beta^i}
\bigg] = 0 . 
\ee
Thanks to Eq.~(\ref{e:evo:evol_ln_gm}), the term in brackets can be replaced by
$-NK$, so that the harmonic slicing condition becomes
\be \label{e:evo:harm_slicing_N}
	\encadre{ \left( \der{}{t} - \Lie{\beta} \right) N = - K N^2 } . 
\ee
Thus we get an \emph{evolution}
equation for the lapse function. This contrasts with
Eq.~(\ref{e:evo:lapse_geod}) for geodesic slicing and
Eq.~(\ref{e:evo:eq_lapse_max}) for maximal slicing.

The harmonic slicing has been introduced by Choquet-Bruhat and Ruggeri (1983)
\cite{ChoquR83} as a way to put the 3+1 Einstein system in a hyperbolic form.
It has been considered more specifically in the context of numerical relativity
by Bona and Masso (1988) \cite{BonaM88}. For a review and more references
see Ref.~\cite{Reula98}. 
\begin{remark}
The harmonic slicing equation (\ref{e:evo:harm_slicing_N}) was already
laid out by Smarr and York in 1978 \cite{SmarrY78a}, 
as a part of the expression of
de Donder coordinate condition in terms of 3+1 variables. 
\end{remark}

\begin{example}
In Schwarzschild spacetime, the hypersurfaces of constant 
standard Schwarzschild time coordinate $t=t_{\rm S}$ and depicted in 
Fig.~\ref{f:evo:max_antisym} constitute some harmonic slicing, in addition to being maximal (cf.~Sec.~\ref{s:evo:maximal}). 
Indeed, using Schwarzschild coordinates $(t,R,\theta,\varphi)$ or
isotropic coordinates $(t,r,\theta,\varphi)$, we have 
$\dert{N}{t}=0$ and $\w{\beta}=0$. Since $K=0$ for these hypersurfaces, 
we conclude that the harmonic slicing condition (\ref{e:evo:harm_slicing_N}) is satisfied. 
\end{example}

\begin{example}
The above slicing does not penetrate under the event horizon. 
A harmonic slicing of Schwarzschild spacetime (and more generally
Kerr-Newman spacetime) which passes smoothly through 
the event horizon has been found by Bona and
Mass\'o \cite{BonaM88}, as well as Cook and Scheel \cite{CookS97}. 
It is given by a time coordinate $t$ that is related to Schwarzschild
time $t_{\rm S}$ by
\be
	t = t_{\rm S} + 2m\ln\left| 1 - \frac{2m}{R} \right| , 
\ee
where $R$ is Schwarzschild radial coordinate (areal radius).
The corresponding
expression of Schwarzschild metric is \cite{CookS97}
\be
g_{\mu\nu} dx^\mu dx^\nu = - N^2 dt^2 + \frac{1}{N^2} 
	\left( dR + \frac{4m^2}{R^2} N^2 \, dt
	\right) ^2
	+ R^2 (d\theta^2 + \sin^2\theta d\varphi^2) ,	
\ee
where 
\be
	N = 
	\left[ 
	\left( 1 + \frac{2m}{R} \right) \left( 1 + \frac{4m^2}{R^2} \right) 
	\right] ^{-1/2} . 
\ee
Notice that all metric coefficients are regular at the event horizon
($R=2m$). 
This harmonic slicing is represented in a Kruskal-Szekeres diagram 
in Fig.~1 of Ref.~\cite{CookS97}. It is clear from that figure 
that the hypersurfaces $\Sigma_t$ never hit the singularity
(contrary to those of the geodesic slicing shown in Fig.~\ref{f:evo:geod}),
but they come arbitrary close to it as $t\rightarrow+\infty$. 
\end{example}
We infer from the above example that the harmonic slicing has some
singularity avoidance feature, but weaker than that of maximal 
slicing: for the latter, the hypersurfaces $\Sigma_t$ never come
close to the singularity as $t\rightarrow+\infty$ 
(cf. Fig.~\ref{f:evo:max_sym}). This has been confirmed by means of numerical
computations by Shibata and Nakamura \cite{ShibaN95}.
\begin{remark}
If one uses normal coordinates, i.e. spatial coordinates $(x^i)$
such that $\w{\beta}=0$, then the harmonic slicing condition
in the form (\ref{e:evo:harm_slicing1}) is easily integrated to
\be \label{e:evo:harm_slic_integr}
	N = C(x^i) \sqrt{\gm} , 
\ee
where $C(x^i)$ is an arbitrary function of the spatial coordinates, 
which does not depend upon $t$. 
Equation~(\ref{e:evo:harm_slic_integr}) is as easy to implement as
the geodesic slicing condition ($N=1$). It is related to the \defin{conformal
time slicing} introduced by Shibata and Nakamura \cite{ShibaN92}. 
\end{remark}

\subsection{1+log slicing} \label{s:evo:1plog}

Bona, Mass\'o, Seidel and Stela (1995) \cite{BonaMSS95} have generalized 
the harmonic slicing condition (\ref{e:evo:harm_slicing_N}) to
\be \label{e:evo:gen_harm_slicing}
    \left( \der{}{t} - \Lie{\beta} \right) N = - K N^2 f(N)  , 
\ee
where $f$ is an arbitrary function. The harmonic slicing corresponds to
$f(N)=1$. The geodesic slicing also fulfills this relation with $f(N)=0$. 
The choice $f(N)=2/N$ leads to 
\be \label{e:evo:1plog_dNdt}
    \encadre{ \left( \der{}{t} - \Lie{\beta} \right) N = - 2 K N  }. 
\ee
Substituting Eq.~(\ref{e:evo:evol_ln_gm}) for $-KN$, we obtain
\be
	\left( \der{}{t} - \Lie{\beta} \right) N  = \der{}{t}\ln\gm
	- 2 D_i \beta^i .
\ee
If normal coordinates are used, $\w{\beta}=0$ and the above equation 
reduces to 
\be
	\der{N}{t} = \der{}{t}\ln\gm , 
\ee
a solution of which is
\be \label{e:evo:1plog_def}
	 N = 1 + \ln\gm  . 
\ee
For this reason, a foliation whose lapse function obeys Eq.~(\ref{e:evo:1plog_dNdt})
is called a \defin{1+log slicing}. The original 1+log condition 
(\ref{e:evo:1plog_def}) has been 
 introduced by Bernstein (1993)
\cite{Berns93} and Anninos et al. (1995) \cite{AnninMSST95} (see also
Ref.~\cite{BonaMSS97}). 
Notice that, even when $\w{\beta}\not=0$, we still define the 1+log slicing
by condition (\ref{e:evo:1plog_dNdt}), although the ``1+log'' relation 
(\ref{e:evo:1plog_def}) does no longer hold. 

\begin{remark}
As for the geodesic slicing [Eq.~(\ref{e:evo:lapse_geod})], 
the harmonic slicing with zero shift [Eq.~(\ref{e:evo:harm_slic_integr})], 
the original 1+log slicing with zero shift [Eq.~(\ref{e:evo:1plog_def})] 
belongs to the family
of \defin{algebraic slicings} \cite{Piran83,BaumgS03}: the determination of the lapse
function does not require to solve any equation. 
It is therefore very easy to implement. 
\end{remark}

The 1+log slicing has stronger singularity
avoidance properties than harmonic slicing: it has been found to ``mimic''
maximal slicing \cite{AnninMSST95}. 

Alcubierre has shown in 1997 \cite{Alcub97} that 
for any slicing belonging to the 
family (\ref{e:evo:gen_harm_slicing}), and in particular for the
harmonic and 1+log slicings, some smooth initial data $(\Sigma_0,\wgm)$
can be found such that the foliation $(\Sigma_t)$ become singular for a
finite value of $t$.
\begin{remark}
The above finding does not contradict the well-posedness
of the Cauchy problem established by Choquet-Bruhat in 1952 \cite{Foure52}
for generic smooth initial data by means of harmonic coordinates
(which define a harmonic slicing) (cf. Sec.~\ref{s:dec:existence_uniqueness}). Indeed it must be remembered that Choquet-Bruhat's
theorem is a \emph{local} one, whereas the pathologies found by Alcubierre
develop for a finite value of time. Moreover, these pathologies are far
from being generic, as the tremendous successes of the 1+log slicing in 
numerical relativity have shown (see below). 
\end{remark}

The 1+log slicing has been used the 3D investigations of the dynamics
of relativistic stars by Font et al. in 2002 \cite{FontGIMRSSST02}. 
It has also been used in most of the recent computations of
binary black hole inspiral and merger : Baker et al. 
\cite{BakerCCKV06a,BakerCCKV06b,VanMeBKC06}, Campanelli et al. 
\cite{CampaLMZ06,CampaLZ06a,CampaLZ06b,CampaLZ06c}, 
Sperhake \cite{Sperh07}, Diener et al. \cite{DieneHPSSTTV06}, 
Br\"ugmann et al. \cite{BruegGHHST07,MarroTBGHHS07},
and Herrmann et al. \cite{HerrmSL07,HerrmHSLM07}.
The works \cite{DieneHPSSTTV06} and \cite{HerrmSL07} and 
 The first three groups employ exactly
Eq.~(\ref{e:evo:1plog_dNdt}), whereas the last two groups are using 
a modified (``zero-shift'') version:
\be
 \der{N}{t}  = - 2 K N . 
\ee
The recent 3D gravitational collapse calculations of Baiotti et al. 
\cite{BaiotHMLRSFS05,BaiotHRS05,BaiotR06} are based on a slight modification 
of the 1+log slicing: instead of Eq.~(\ref{e:evo:1plog_dNdt}), these authors
have used 
\be 
     \left( \der{}{t} - \Lie{\beta} \right) N = - 2 N (K-K_0),   
\ee
where $K_0$ is the value of $K$ at $t=0$.

\begin{remark}
There is a basic difference between maximal slicing and 
the other types of foliations presented above (geodesic, harmonic
and 1+log slicings): the property of being maximal
is applicable to a \emph{single hypersurface} $\Sigma_0$, whereas the 
property of being geodesic, harmonic or 1+log are meaningful only
for a \emph{foliation} $(\Sigma_t)_{t\in\R}$. This is reflected in the basic definition
of these slicings: the maximal slicing is defined from the extrinsic curvature
tensor only ($K=0$), which characterizes a single hypersurface 
(cf. Chap.~\ref{s:hyp}), whereas the
definitions of geodesic, harmonic and 1+log slicings all involve the lapse
function $N$, which of course makes sense only for a foliation 
(cf. Chap.~\ref{s:fol}). 
\end{remark}

%%%%%%%%%%%%%%%%%%%%%%%%%%%%%%%%%%%%%%%%%%%%%%%%%%%%%%%%%%%%%%%%%%%%%%%%%%%%%%%%%%%%%%

\section{Evolution of spatial coordinates}  \label{s:evo:spat_coord}

Having discussed the choice of the foliation $(\Sigma_t)_{t\in\R}$, let us turn
now to the choice of the coordinates $(x^i)$ in each hypersurface $\Sigma_t$.
As discussed in Sec.~\ref{s:dec:coord_lapse_shift}, this is done via the 
shift vector $\w{\beta}$. More precisely, once some coordinates $(x^i)$ are
set in the initial slice $\Sigma_0$, the shift vector governs the propagation 
of these coordinates to all the slices $\Sigma_t$. 

\subsection{Normal coordinates}

As for the lapse choice $N=1$ (geodesic slicing, Sec.~\ref{s:evo:geodesic}), 
the simplest choice for the shift vector is to set it to zero:
\be
	\encadre{\w{\beta} = 0} . 
\ee
For this choice, the lines $x^i = {\rm const}$ are normal to the hypersurfaces
$\Sigma_t$ (cf. Fig.~\ref{f:dec:shift}), hence the name \defin{normal coordinates}. 
The alternative name is \defin{Eulerian coordinates}, defining the
so-called \defin{Eulerian gauge} \cite{Barde83}. This is of course justified 
by the fact that the lines $x^i = {\rm const}$ are then the worldlines
of the Eulerian observers introduced in Sec.~\ref{s:fol:Eulerian}. 

Besides their simplicity, an advantage of normal coordinates is to 
be as regular as the foliation itself: they cannot introduce some pathology
per themselves. On the other hand,
the major drawback of these coordinates is that they may lead to a large coordinate 
shear, resulting
in  large values of the metric coefficients $\gm_{ij}$. 
This is specially true if rotation is present. For instance, in Kerr or rotating
star spacetimes, the field lines of the stationary Killing vector $\w{\xi}$
are not orthogonal to the hypersurfaces $t={\rm const}$. Therefore, if
one wishes to have coordinates adapted to stationarity, i.e. to 
have $\wpar_t = \w{\xi}$, one must allow for $\w{\beta}\not=0$. 

Despite of the shear problem mentioned above, normal coordinates have been
used because of their simplicity in early treatments of two famous axisymmetric problems in numerical relativity: the head-on collision of black holes
by Smarr, Eppley and \v Cade\v z in 1976-77 \cite{SmarrCDE76,Smarr79}
and the gravitational collapse of a rotating star by Nakamura in 1981
\cite{Nakam81,NakamS81}. More recently, normal coordinates have also been used in 
the 3D evolution of gravitational waves performed by Shibata and Nakamura
(1995) \cite{ShibaN95} and Baumgarte and Shapiro (1999) \cite{BaumgS99}, 
as well as in the 3D grazing collisions of binary black holes 
computed by Br\"ugmann (1999) \cite{Brueg99} and 
Alcubierre et al. (2001) \cite{AlcubBBLNST01}.

\subsection{Minimal distortion} \label{s:evo:min_distort}

\begin{figure}
\centerline{\includegraphics[width=0.7\textwidth]{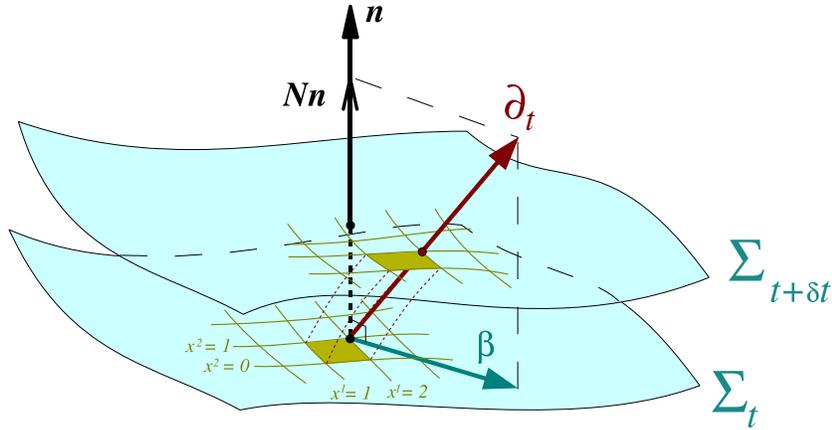}}
\caption[]{\label{f:evo:distor} \footnotesize
Distortion of a spatial domain defined by fixed values of the coordinates
$(x^i)$. 
}
\end{figure}

A very well motivated choice of spatial coordinates has been introduced in 1978 
by Smarr and York \cite{SmarrY78a,SmarrY78b} (see also Ref.~\cite{York79}).
As discussed in Sec.~\ref{s:cfd:intro}, the physical degrees of freedom of the
gravitational field are carried by the conformal 3-metric $\wtgm$.
The evolution of the latter with respect to the coordinates $(t,x^i)$ is
given by the derivative $\w{\dot{\tilde\gm}} := \bm{\mathcal{L}}_{\wpar_t} \wtgm$, the
components of which are
\be
	\dot\tgm_{ij} = \der{\tgm_{ij}}{t} . 
\ee
Given a foliation $(\Sigma_t)_{t\in\R}$, the idea of Smarr and York is to choose the coordinates $(x^i)$, and hence the vector $\wpar_t$, in order to minimize this
time derivative. There is not a unique way to minimize $\dot\tgm_{ij}$; 
this can be realized by counting the degrees of freedom: $\dot\tgm_{ij}$
has 5 independent components\footnote{as a symmetric $3\times 3$ matrix,
$\dot\tgm_{ij}$ has a priori 6 components, but one degree of freedom is lost
in the demand $\det\tgm_{ij} = \det f_{ij}$ [Eq.~(\ref{e:cfd:dettgm_f})], 
which implies $\det\dot\tgm_{ij} = 0$ via Eq.~(\ref{e:cfd:f_const}).}
and, for a given foliation, only 3 degrees of freedom can be controled via the 3 coordinates $(x^i)$. One then proceeds as follows. First one notices that
$\w{\dot{\tilde\gm}}$ is related to the 
\defin{distortion tensor} $\w{Q}$, the latter being defined as
the trace-free part of the time derivative of the physical metric $\wgm$:
\be \label{e:evo:Sigma_def}
	\w{Q} := \bm{\mathcal{L}}_{\wpar_t} \wgm - \frac{1}{3} 
	\left(\mathrm{tr}_{\gm} \bm{\mathcal{L}}_{\wpar_t} \wgm \right) \wgm ,
\ee
or in components,
\be \label{e:evo:Sigma_def_comp}
	Q_{ij} = \der{\gm_{ij}}{t} - \frac{1}{3} \gm^{kl} \der{\gm_{kl}}{t}
	\, \gm_{ij} .
\ee
$\w{Q}$ measures the change in shape from $\Sigma_t$ to $\Sigma_{t+\delta t}$ 
of any spatial domain $\mathcal{V}$ which lies at fixed values of 
the coordinates $(x^i)$ (the evolution of $\mathcal{V}$ is then along 
the vector $\wpar_t$, cf. Fig.~\ref{f:evo:distor}). 
Thanks to the trace removal, $\w{Q}$ does not take into account
the change of volume, but only the change in shape (shear). 
From the law (\ref{e:cfd:variation_det}) of variation of a determinant,
\be
	\gm^{kl} \der{\gm_{kl}}{t} = \der{}{t} \ln \gm = 12 \der{}{t} \ln\Psi 
	+ \underbrace{\der{}{t}\ln f}_{=0}
	= 12 \der{}{t} \ln\Psi,
\ee
where we have used the relation (\ref{e:cfd:def_Psi}) between the determinant
$\gm$ and the conformal factor $\Psi$, as well as the property (\ref{e:cfd:f_const}).
Thus we may rewrite Eq.~(\ref{e:evo:Sigma_def_comp}) as
\be
	Q_{ij} = \der{\gm_{ij}}{t} - 4 \der{}{t} \ln\Psi \, \gm_{ij} 
	= \der{}{t} (\Psi^4 \tgm_{ij}) - 4 \Psi^3 \der{\Psi}{t} \, \tgm_{ij}
	= \Psi^4 \der{\tgm_{ij}}{t} . 
\ee
Hence the relation between the distortion tensor and the time derivative of the 
conformal metric:
\be \label{e:evo:Sigma_Psi_dot_tgm}
	\w{Q} = \Psi^4 \w{\dot{\tilde\gm}} .
\ee
The rough idea would be to choose the coordinates $(x^i)$ in order to 
minimize $\w{Q}$. Taking into account that it is symmetric and 
traceless, $\w{Q}$ has 5 independent components. Thus it cannot
be set identically to zero since we have only 3 degrees of freedom 
in the choice of the coordinates $(x^i)$.
To select which part of $\w{Q}$ to set to zero, let us 
decompose it into a longitudinal part and a TT part, in a manner similar to 
Eq.~(\ref{e:ini:decomp_hA}):
\be
	Q_{ij} = (LX)_{ij} + Q_{ij}^{\rm TT} .
\ee
$LX$ denotes the conformal Killing operator associated with the metric
$\wgm$ and acting on some vector field $\w{X}$ 
(cf. Appendix~\ref{s:cko}) \footnote{In Sec.~\ref{s:cfd:IWM}, we have also used
the notation $L$ for the conformal Killing operator associated with the
flat metric $\w{f}$, but no confusion should arise in the present context.}:
\be
(LX)_{ij} := D_i X_j + D_j X_i - \frac{2}{3} D_k X^k \, \gm_{ij}
\ee
and $Q_{ij}^{\rm TT}$ is both traceless and transverse (i.e. divergence-free)
with respect to the metric $\wgm$: $D^j Q_{ij}^{\rm TT} = 0$.
$\w{X}$ is then related to the divergence of $\w{Q}$ by
$D^j (LX)_{ij} = D^j Q_{ij}$.
It is legitimate to relate the TT part to the dynamics of the gravitational
field and to attribute the longitudinal part to the change in $\gm_{ij}$
which arises because of the variation of coordinates
from $\Sigma_t$ to $\Sigma_{t+\delta t}$.
This longitudinal part has 3 degrees of freedom (the 3 components of the 
vector $\w{X}$) and we might set it to zero by
some judicious choice of the coordinates $(x^i)$. 
The \defin{minimal distortion} coordinates are thus defined 
by the requirement $\w{X}=0$ or 
\be
	Q_{ij} = Q_{ij}^{\rm TT} , 
\ee
i.e.
\be \label{e:evo:def_min_dist}
	\encadre{ D^j Q_{ij} = 0} .
\ee

Let us now express $\w{Q}$ in terms of the shift vector to turn the
above condition into an equation for the evolution of spatial coordinates. 
By means of Eqs.~(\ref{e:dec:Einstein_PDE1}) and (\ref{e:evo:tr_dt_gm}), 
Eq.~(\ref{e:evo:Sigma_def_comp}) becomes
\be
	Q_{ij} = - 2N K_{ij} \Liec{\beta} \gm_{ij}+ 
	- \frac{1}{3} \left( -2NK + 2D_k \beta^k \right) \, \gm_{ij} ,
\ee
i.e. (since $\Liec{\beta} \gm_{ij} = D_i \beta_j + D_j \beta_i$)
\be \label{e:evo:Sigma_A_beta}
	Q_{ij} = - 2 N A_{ij} + (L\beta)_{ij} ,
\ee
where we let appear the trace-free part $\w{A}$ of the extrinsic curvature
$\w{K}$ [Eq.~(\ref{e:cfd:A_def})].
If we insert this expression into the minimal distortion requirement 
(\ref{e:evo:def_min_dist}), we get
\be
	-2N D_j A^{ij} - 2 A^{ij} D_j N + D_j (L\beta)^{ij} = 0 . 
\ee
Let then use the momentum constraint (\ref{e:dec:Einstein_PDE4}) to express
the divergence of $\w{A}$ as
\be
	D_j A^{ij} = 8 \pi p^i + \frac{2}{3} D^i K . 
\ee 
Besides, we recognize in $D_j (L\beta)^{ij}$ the conformal vector Laplacian
associated with the metric $\wgm$, so that we can write [cf. Eq.~(\ref{e:cko:tDelta_DD_Ricci})]
\be
	D_j (L\beta)^{ij} = D_j D^j \beta^i + \frac{1}{3} D^i D_j \beta^j
	+ R^i_{\ \, j} \beta^j , 
\ee
where $\w{R}$ is the Ricci tensor associated with $\wgm$. 
Thus we arrive at 
\be \label{e:evo:min_dist_beta}
	\encadre{
	D_j D^j \beta^i + \frac{1}{3} D^i D_j \beta^j
	+ R^i_{\ \, j} \beta^j = 16\pi N p^i + \frac{4}{3} ND^i K 
	+ 2 A^{ij} D_j N } .
\ee
This is the elliptic equation on the shift vector that one has to
solve in order to enforce the minimal distortion. 

\begin{remark}
For a constant mean curvature (CMC) slicing, and in particular for
a maximal slicing, the term $D^i K$ vanishes and the above equation
is slightly simplified. Incidentally, this is the form originally derived
by Smarr and York (Eq.~(3.27) in Ref.~\cite{SmarrY78a}). 
\end{remark}

Another way to introduce minimal distortion amounts to 
minimizing the integral
\be \label{e:evo:S_min_dist}
	S = \int_{\Sigma_t} Q_{ij} Q^{ij} \sqrt{\gm} \, d^3x 
\ee
with respect to the shift vector $\w{\beta}$, keeping the slicing fixed
(i.e. fixing $\wgm$, $\w{K}$ and $N$).
Indeed, if we replace $\w{Q}$ by its expression (\ref{e:evo:Sigma_A_beta}),
we get
\be
	S = \int_{\Sigma_t} \left[ 4N^2 A_{ij} A^{ij} - 4 N A_{ij} (L\beta)^{ij}
	+  (L\beta)_{ij}  (L\beta)^{ij} \right] \sqrt{\gm} \, d^3x .
\ee
At fixed values of $\wgm$, $\w{K}$ and $N$, $\delta N=0$, $\delta A_{ij}=0$
and $\delta (L\beta)^{ij} = (L\delta\beta)^{ij}$, so that the variation of $S$
with respect to $\w{\beta}$ is
\be
	\delta S = \int_{\Sigma_t} \left[ - 4 N A_{ij} (L\delta\beta)^{ij}
	+ 2 (L\beta)_{ij} (L\delta\beta)^{ij} \right] \sqrt{\gm} d^3x 	
		= 2 \int_{\Sigma_t} Q_{ij} (L\delta\beta)^{ij} 
	\, \sqrt{\gm} \, d^3x .
\ee
Now, since $\w{Q}$ is symmetric and traceless,
$Q_{ij} (L\delta\beta)^{ij} = Q_{ij} (D^i\delta \beta^j + D^j\delta\beta^i 
- 2/3\, D_k\delta\beta^k\, \gm^{ij}) =  Q_{ij} (D^i\delta\beta^j + D^j\delta\beta^i)
= 2 Q_{ij} D^i\delta\beta^j$. Hence
\bea
  \delta S & = & 4 \int_{\Sigma_t} Q_{ij} D^i\delta\beta^j \, \sqrt{\gm} \, d^3x
	\nonumber \\
	& = & 4 \int_{\Sigma_t} 
	\left[ D^i\left( Q_{ij} \delta \beta^j \right)
	- D^i Q_{ij} \, \delta \beta^j \right] \, \sqrt{\gm}\,  d^3x \nonumber \\
	& = & 4 \oint_{\partial\Sigma_t}
	Q_{ij} \delta \beta^j s^i \, \sqrt{q} \, d^2 y
	- 4 \int_{\Sigma_t} D^i Q_{ij} \,\delta  \beta^j\, \sqrt{\gm} \,  d^3x 
\eea
Assuming that $\delta\beta^i=0$ at the boundaries of $\Sigma_t$
(for instance at spatial infinity), we deduce from the above relation
that $\delta S=0$ for any variation of the shift vector if and only if
$D^i Q_{ij} = 0$. Hence we recover condition (\ref{e:evo:def_min_dist}). 

In stationary spacetimes, an important property of the minimal distortion gauge is 
to be fulfilled by coordinates adapted to the stationarity
(i.e. such that $\wpar_t$ is a Killing vector):
it is immediate from Eq.~(\ref{e:evo:Sigma_def})
that $\w{Q} = 0$ when $\wpar_t$ is a symmetry generator, so 
that condition (\ref{e:evo:def_min_dist}) is trivially satisfied.
Another nice feature of the minimal distortion gauge is that in 
the weak field region (radiative zone), it
includes the standard TT gauge of linearized gravity \cite{SmarrY78a}. Actually
Smarr and York \cite{SmarrY78a} have advocated for maximal slicing
combined with minimal distortion as a very good coordinate choice for
radiative spacetimes, calling such choice the \defin{radiation gauge}. 

\begin{remark}
A ``new minimal distortion'' gauge has been introduced in 2006 by
Jantzen and York \cite{JantzY06}. It corrects the time
derivative of $\wtgm$ in the original minimal distortion condition 
by the lapse function $N$ [cf. relation (\ref{e:fol:dtau_Ndt}) between the coordinate
time $t$ and the Eulerian observer's proper time $\tau$], i.e. one requires
\be
	D^j \left( \frac{1}{N} Q_{ij} \right) = 0 
\ee
instead of (\ref{e:evo:def_min_dist}). This amounts to minimizing the
integral
\be
	S' = \int_{\Sigma_t} (N^{-1}Q_{ij}) (N^{-1}Q^{ij}) 
	\sqrt{-g} \, d^3x 
\ee
with respect to the shift vector. Notice the spacetime measure
$\sqrt{-g} = N\sqrt{\gm}$ instead of the spatial measure $\sqrt{\gm}$ in
Eq.~(\ref{e:evo:S_min_dist}). 
\end{remark}

The minimal distortion condition can be expressed in terms of the time
derivative of the conformal metric by combining Eqs.~(\ref{e:evo:Sigma_Psi_dot_tgm})
and (\ref{e:evo:def_min_dist}):
\be
	\encadre{ D^j (\Psi^4 \dot\tgm_{ij} ) = 0 }. 
\ee
Let us write this relation in terms of the connection $\wtD$ (associated with 
the metric $\wtgm$) instead of the connection $\wD$ (associated with 
the metric $\wgm$). 
To this purpose, let us use Eq.~(\ref{e:cfd:divA_Psi10})
which relates the $\wD$-divergence of a traceless symmetric tensor
to its $\wtD$-divergence: since $Q^{ij}$ is traceless and symmetric,
we obtain
\be
	D_j Q^{ij} = \Psi^{-10} \tD_j \left( \Psi^{10} Q^{ij} \right) .
\ee
Now $Q^{ij} = \gm^{ik} \gm^{jl} Q_{kl} 
= \Psi^{-8} \tgm^{ik} \tgm^{jl} Q_{kl} = 
\Psi^{-4} \tgm^{ik} \tgm^{jl}\dot\tgm_{kl}$; hence
\be
	D_j Q^{ij} = \Psi^{-10} \tD_j 
	\left( \Psi^6 \tgm^{ik} \tgm^{jl}\dot\tgm_{kl} \right) 
	=  \Psi^{-10} \tgm^{ik} \tD^l 
	\left( \Psi^6 \dot\tgm_{kl} \right) .
\ee
The minimal distortion condition is therefore
\be \label{e:evo:min_dist_tD}
	\encadre{ \tD^j ( \Psi^6 \dot\tgm_{ij} ) = 0 } . 
\ee

\subsection{Approximate minimal distortion} \label{s:evo:AMDist}

In view of Eq.~(\ref{e:evo:min_dist_tD}), 
it is natural to consider the simpler condition
\be \label{e:evo:altern_min_dist}
	\tD^j \dot\tgm_{ij}  = 0 ,
\ee
which of course differs from the true minimal distortion (\ref{e:evo:min_dist_tD})
by a term $6 \dot\tgm_{ij} \tD^j\ln\Psi$. 
Nakamura (1994) \cite{Nakam94,OoharNS97} has then introduced the
\defin{pseudo-minimal distortion} condition by replacing 
(\ref{e:evo:altern_min_dist}) by
\be \label{e:evo:pseudo-min_dist}
	\encadre{\Df^j \dot\tgm_{ij}  = 0} ,
\ee
where $\wDf$ is the connection associated with the flat metric $\w{f}$. 

An alternative has been introduced by Shibata (1999) \cite{Shiba99b} as follows. 
Starting from Eq.~(\ref{e:evo:altern_min_dist}), let us express $\dot\tgm_{ij}$
in terms of $\w{A}$ and $\w{\beta}$: from Eq.~(\ref{e:ini:tA_Lbeta}), we
deduce that 
\bea
	2N \tA_{ij}  & = &
	\tgm_{ik} \tgm_{jl} \left[ \dot\tgm^{kl} + (\tilde L \beta)^{kl} \right]
	= \tgm_{jl} \bigg[ \der{}{t} 
	(\underbrace{\tgm_{ik} \tgm^{kl}}_{=\delta^l_{\ \, i}}) 
	- \tgm^{kl} \der{\tgm_{ik}}{t} + \tgm_{ik}  (\tilde L \beta)^{kl} \bigg]
	\nonumber \\
	& = & - \dot\tgm_{ij} + \tgm_{ik} \tgm_{jl}  (\tilde L \beta)^{kl}  ,
\eea
where $\tA_{ij} := \tgm_{ik} \tgm_{jl} \tA^{kl} = \Psi^{-4} A_{ij}$.
Equation~(\ref{e:evo:altern_min_dist}) becomes then
\be
	\tD^j \left[ \tgm_{ik} \tgm_{jl}  (\tilde L \beta)^{kl} - 2N \tA_{ij} 
	\right] = 0 ,  
\ee
or equivalently (cf. Sec.~\ref{s:cko:def_cvlap}),
\be
	\tD_j \tD^j \beta^i + \frac{1}{3} \tD^i \tD_j \beta^j
	+ {\tilde R}^i_{\ \, j} \beta^j
	- 2 \tA^{ij} \tD_j N - 2N \tD_j \tA^{ij} = 0 . 
\ee
We can express $\tD_j \tA^{ij}$ via the momentum constraint (\ref{e:cfd:Einstein6})
and get
\be \label{e:evo:altern_min_dist_beta}
	\tD_j \tD^j \beta^i + \frac{1}{3} \tD^i \tD_j \beta^j
	+ {\tilde R}^i_{\ \, j} \beta^j
	- 2 \tA^{ij} \tD_j N + 4N 
	\left[ 3 \tA^{ij} \tD_j \ln\Psi - \frac{1}{3} \tD^i K
 	- 4\pi \Psi^4 p^i \right] = 0 . 
\ee
At this stage, Eq.~(\ref{e:evo:altern_min_dist_beta}) is nothing but a rewriting
of Eq.~(\ref{e:evo:altern_min_dist}) as an elliptic equation for the shift vector.
Shibata \cite{Shiba99b} then proposes to replace in this equation the conformal
vector Laplacian relative to $\wtgm$ and acting on $\w{\beta}$ by the conformal
vector Laplacian relative to the flat metric $\w{f}$, thereby writing
\be \label{e:evo:approx_min_dist}
	\Df_j \Df^j \beta^i + \frac{1}{3} \Df^i \Df_j \beta^j
	- 2 \tA^{ij} \tD_j N + 4N 
	\left[ 3 \tA^{ij} \tD_j \ln\Psi - \frac{1}{3} \tD^i K
 	- 4\pi \Psi^4 p^i \right] = 0 . 
\ee
The choice of coordinates defined by solving Eq.~(\ref{e:evo:approx_min_dist})
instead of (\ref{e:evo:min_dist_beta}) is called  
\defin{approximate minimal distortion}.

The approximate minimal distortion has been used by Shibata and Uryu  
\cite{ShibaU00,ShibaU02}
for their first computations of the merger of binary neutron stars, 
as well as by Shibata, Baumgarte and Shapiro 
for computing the collapse of supramassive neutron stars at the mass-shedding
limit (Keplerian angular velocity) \cite{ShibaBS00a} 
and for studying the dynamical bar-mode instability in differentially
rotating neutron stars \cite{ShibaBS00b}. 
It has also been used by Shibata \cite{Shiba03a} to devise a 2D (axisymmetric)
code to compute the long-term evolution of rotating neutron stars and 
gravitational collapse.  

\subsection{Gamma freezing} \label{s:evo:Gamma_freez}

The \defin{Gamma freezing} prescription for the evolution of spatial coordinates  
is very much related to Nakamura's pseudo-minimal distortion (\ref{e:evo:pseudo-min_dist}):
it differs from it only in the replacement of $\Df^j$ by $\Df_j$ and 
$\dot\tgm_{ij}$ by 
$\dot\tgm^{ij}:= \dert{\tgm^{ij}}{t}$:
\be \label{e:evo:Gam_freez1}
	\encadre{\Df_j \dot\tgm^{ij} = 0 } . 
\ee
The name \emph{Gamma freezing} is justified as follows: 
since $\dert{}{t}$ and $\wDf$
commute [as a consequence of (\ref{e:cfd:f_const})],
Eq.~(\ref{e:evo:Gam_freez1}) is equivalent to
\be \label{e:evo:Gam_freez2}
	\der{}{t} \left( \Df_j \tgm^{ij} \right) = 0 . 
\ee
Now, expressing the covariant derivative $\Df_j$ in terms of the Christoffel symbols
$\bar\Gamma^i_{\ \, jk}$ of the metric $\w{f}$ with respect to the
coordinates $(x^i)$, we get
\bea
	\Df_j \tgm^{ij} & = &  \der{\tgm^{ij}}{x^j}
	+ \bar\Gamma^i_{\ \, jk} \tgm^{kj}
	+ \underbrace{\bar\Gamma^j_{\ \, jk}}_{=\frac{1}{2} \der{}{x^k}\ln f}
	 \tgm^{ik} \nonumber \\
	& = &  \der{\tgm^{ij}}{x^j} + \tilde\Gamma^i_{\ \, jk} \tgm^{kj}
	+ \left(\bar\Gamma^i_{\ \, jk} - \tilde\Gamma^i_{\ \, jk} \right) \tgm^{kj}
	+ \underbrace{\frac{1}{2} \der{}{x^k}\ln\tgm}_{= \tilde\Gamma^j_{\ \, jk}}
	\; \tgm^{ik} \nonumber \\
	& = & \underbrace{\der{\tgm^{ij}}{x^j}
	+ \tilde\Gamma^i_{\ \, jk} \tgm^{kj} + \tilde\Gamma^j_{\ \, jk} \tgm^{ik}
	}_{=\tD_j \tgm^{ij} = 0}
	+ \left(\bar\Gamma^i_{\ \, jk} - \tilde\Gamma^i_{\ \, jk} \right)
	 \tgm^{kj} \nonumber \\
	& = & \tgm^{jk} \left(\bar\Gamma^i_{\ \, jk} - \tilde\Gamma^i_{\ \, jk} \right) ,
\eea
where $\tilde\Gamma^i_{\ \, jk}$ denote the Christoffel symbols
of the metric $\wtgm$ with respect to the coordinates $(x^i)$ and
we have used $\tgm=f$ [Eq.~(\ref{e:cfd:dettgm_f})] to write the second line. 
If we introduce the notation
\be \label{e:evo:tGam_diffGam}
	\encadre{ \tilde\Gamma^i := \tgm^{jk}
	\left(\tilde\Gamma^i_{\ \, jk} - \bar\Gamma^i_{\ \, jk} \right) } ,
\ee
then the above relation becomes
\be \label{e:evo:tGam_divtgm}
	\encadre{ \Df_j \tgm^{ij}  = - \tilde\Gamma^i }. 
\ee
\begin{remark}
If one uses Cartesian-type coordinates, then $\bar\Gamma^i_{\ \, jk}=0$
and the $\tilde\Gamma^i$'s reduce to the contracted Christoffel symbols introduced
by Baumgarte and Shapiro \cite{BaumgS99} [cf. their Eq.~(21)]. 
In the present case, the $\tilde\Gamma^i$'s are the components of a \emph{vector
field} $\w{\tilde\Gamma}$ on $\Sigma_t$,
as it is clear from relation (\ref{e:evo:tGam_divtgm}), or from
expression (\ref{e:evo:tGam_diffGam}) if one remembers that, although the
Christoffel symbols are not the components of any tensor field, 
the differences between two sets of them are. 
Of course the vector field  $\w{\tilde\Gamma}$ depends on the choice of the background metric $\w{f}$. 
\end{remark}
By combining Eqs.~(\ref{e:evo:tGam_divtgm}) and (\ref{e:evo:Gam_freez2}), we see that
the  Gamma freezing condition is equivalent to 
\be \label{e:evo:Gam_freez3}
	\encadre{ \der{\tilde\Gamma^i}{t} = 0 }, 
\ee
hence the name \emph{Gamma freezing}: for such a choice, the vector $\w{\tilde\Gamma}$
does not evolve, in the sense that
$\w{\mathcal{L}}_{\wpar_t}  \w{\tilde\Gamma} = 0$. 
The Gamma freezing prescription has been introduced 
by Alcubierre and Br\"ugmann  in 2001 \cite{AlcubB01},
in the form of Eq.~(\ref{e:evo:Gam_freez3}).

Let us now derive the equation that the shift vector must obey in order
to enforce the Gamma freezing condition. 
If we express the Lie derivative in the evolution equation 
(\ref{e:cfd:Einstein2}) for $\tgm^{ij}$ 
in terms of the covariant derivative $\wDf$
[cf. Eq.~(\ref{e:Lie_der_comp})], we get 
\be
	\dot\tgm^{ij} = 2 N \tA^{ij} + \beta^k \Df_k \tgm^{ij}
	- \tgm^{kj} \Df_k \beta^i - \tgm^{ik} \Df_k \beta^j
	+ \frac{2}{3} \Df_k \beta^k \, \tgm^{ij} . 
\ee
Taking the flat-divergence of this relation and 
using relation~(\ref{e:evo:tGam_divtgm}) (with the commutation property
of $\dert{}{t}$ and $\wDf$) yields
\bea
   \der{\tilde\Gamma^i}{t} = 
   & = & - 2 N \Df_j \tA^{ij} - 2 A^{ij} \Df_j N
	+ \beta^k \Df_k \tilde\Gamma^i 
	- \tilde\Gamma^k \Df_k \beta^i + \frac{2}{3} \tilde\Gamma^i \Df_k\beta^k
	\nonumber \\
	 & &  \tgm^{jk} \Df_j \Df_k \beta^i + \frac{1}{3} \tgm^{ij}
	\Df_j \Df_k \beta^k  . \label{e:evo:dGamdt0}
\eea
Now, we may use the momentum constraint (\ref{e:cfd:Einstein6}) to 
express $\Df_j \tA^{ij}$:
\be
	 \tD_j \tA^{ij}  = - 6 \tA^{ij} \tD_j \ln\Psi + \frac{2}{3}
	\tD^i K + 8\pi \Psi^4 p^i , 
\ee
with 
\be
	 \tD_j \tA^{ij} = \Df_j \tA^{ij}
	+ \left(\tilde\Gamma^i_{\ \, jk} - \bar\Gamma^i_{\ \, jk} \right)
	\tA^{kj}
	+ \underbrace{\left(\tilde\Gamma^j_{\ \, jk} - \bar\Gamma^j_{\ \, jk} \right)
	}_{=0} \tA^{ik} , 
\ee
where the ``$=0$'' results from the fact that 
$2 \tilde\Gamma^j_{\ \, jk} = \dert{\ln\tgm}{x^k}$
and $2 \bar\Gamma^j_{\ \, jk} = \dert{\ln f}{x^k}$,
with $\tgm :=\det\tgm_{ij} = \det f_{ij} =: f$ [Eq.~(\ref{e:cfd:dettgm_f})].
Thus Eq.~(\ref{e:evo:dGamdt0}) becomes
\bea
	\der{\tilde\Gamma^i}{t} 
   & = & \tgm^{jk} \Df_j \Df_k \beta^i + \frac{1}{3} \tgm^{ij}
	\Df_j \Df_k \beta^k + \frac{2}{3} \tilde\Gamma^i \Df_k\beta^k
	- \tilde\Gamma^k \Df_k \beta^i + \beta^k \Df_k \tilde\Gamma^i
	\nonumber \\
	& & - 2N \left[ 8\pi \Psi^4 p^i 
	- \tA^{jk}  \left(\tilde\Gamma^i_{\ \, jk} - \bar\Gamma^i_{\ \, jk} \right)
	- 6 \tA^{ij} \Df_j\ln \Psi 
	+ \frac{2}{3} \tgm^{ij} \Df_j K \right]
	- 2 \tA^{ij} \Df_j N . 	\nonumber\\
	& & 		\label{e:evo:dGamdt}
\eea	
We conclude that the Gamma freezing condition (\ref{e:evo:Gam_freez3})
is equivalent to 
\be \label{e:evo:Gam_freez_beta}
	\encadre{
	\begin{array}{l}
	 \displaystyle \tgm^{jk} \Df_j \Df_k \beta^i + \frac{1}{3} \tgm^{ij}
	\Df_j \Df_k \beta^k + \frac{2}{3} \tilde\Gamma^i \Df_k\beta^k
	- \tilde\Gamma^k \Df_k \beta^i + \beta^k \Df_k \tilde\Gamma^i
	= \\
	 \displaystyle  \ \qquad
	2N \left[ 8\pi \Psi^4 p^i 
	- \tA^{jk}  \left(\tilde\Gamma^i_{\ \, jk} - \bar\Gamma^i_{\ \, jk} \right)
	- 6 \tA^{ij} \Df_j\ln \Psi 
	+ \frac{2}{3} \tgm^{ij} \Df_j K \right]
	+ 2 \tA^{ij} \Df_j N .
	\end{array}
	} 
\ee
This is an elliptic equation for the shift vector, which bears some resemblance
with Shibata's approximate minimal distortion, Eq.~(\ref{e:evo:approx_min_dist}). 

\subsection{Gamma drivers} \label{s:evo:Gam_drivers}

As seen above the Gamma freezing condition (\ref{e:evo:Gam_freez3})
yields to the elliptic equation (\ref{e:evo:Gam_freez_beta}) 
for the shift vector. 
Alcubierre and Br\"ugmann \cite{AlcubB01} have proposed to turn it
into a parabolic equation by considering, instead of Eq.~(\ref{e:evo:Gam_freez3}),
the relation
\be \label{e:evo:par_Gam_driver}
	\encadre{ \der{\beta^i}{t} = k \der{\tilde\Gamma^i}{t}  },
\ee
where $k$ is a positive function.  
The resulting coordinate choice is called a \defin{parabolic Gamma driver}. 
Indeed, if we inject Eq.~(\ref{e:evo:par_Gam_driver}) into Eq.~(\ref{e:evo:dGamdt}),
we clearly get a parabolic equation for the shift vector, of the
type $\dert{\beta^i}{t} = k \left[ \tgm^{jk} \Df_j \Df_k \beta^i 
+ \frac{1}{3} \tgm^{ij} \Df_j \Df_k \beta^k + \cdots \right]$.

An alternative has been introduced in 2003 
by Alcubierre, Br\"ugmann, Diener, Koppitz,
Pollney, Seidel and Takahashi \cite{AlcubBDKPST03} (see also Refs.~\cite{LindbS03}
and \cite{BonaLP05}); 
it requires
\be \label{e:evo:hyp_Gam_driver}
	\encadre{ \dder{\beta^i}{t} = k \der{\tilde\Gamma^i}{t} 
	- \left( \eta - \der{}{t}\ln k \right) \der{\beta^i}{t} },
\ee
where $k$ and $\eta$ are two positive functions. 
The prescription (\ref{e:evo:hyp_Gam_driver}) is called a
\defin{hyperbolic Gamma driver} \cite{AlcubBDKPST03,LindbS03,BonaLP05}. 
Indeed, thanks to Eq.~(\ref{e:evo:dGamdt}), it is equivalent to 
\bea
	& & \dder{\beta^i}{t} + \left( \eta - \der{}{t}\ln k \right)
	\der{\beta^i}{t} = k\bigg\{
	\tgm^{jk} \Df_j \Df_k \beta^i + \frac{1}{3} \tgm^{ij}
	\Df_j \Df_k \beta^k + \frac{2}{3} \tilde\Gamma^i \Df_k\beta^k
	- \tilde\Gamma^k \Df_k \beta^i + \beta^k \Df_k \tilde\Gamma^i
	\nonumber \\
	& & \quad - 2N \left[ 8\pi \Psi^4 p^i 
	- \tA^{jk}  \left(\tilde\Gamma^i_{\ \, jk} - \bar\Gamma^i_{\ \, jk} \right)
	- 6 \tA^{ij} \Df_j\ln \Psi 
	+ \frac{2}{3} \tgm^{ij} \Df_j K \right]
	- 2 \tA^{ij} \Df_j N \bigg\} ,
\eea
which is a hyperbolic equation for the shift vector, of the type of the
telegrapher's equation. 
The term with the coefficient $\eta$ is a dissipation term. 
It has been found by Alcubierre et al. \cite{AlcubBDKPST03} crucial
to add it to avoid strong oscillations in the shift.

The hyperbolic Gamma driver condition (\ref{e:evo:hyp_Gam_driver})
is equivalent to the following first order system
\be \label{e:evo:Gam_driver_1order}
	\left\{ \begin{array}{lcl}
	\displaystyle \der{\beta^i}{t} & = & k B^i \\[1ex]
	\displaystyle \der{B^i}{t} & = & \displaystyle 
	\der{\tilde\Gamma^i}{t} - \eta B^i .
	\end{array} 
	\right. 
\ee
\begin{remark}
In the case where $k$ does not depend on $t$, the Gamma driver condition
(\ref{e:evo:hyp_Gam_driver}) reduces to a previous hyperbolic condition
proposed by Alcubierre, Br\"ugmann, Pollney, Seidel 
and Takahashi \cite{AlcubBPST01}, namely 
\be
\dder{\beta^i}{t} = k \der{\tilde\Gamma^i}{t} 
	- \eta  \der{\beta^i}{t} . 
\ee
\end{remark}

Hyperbolic Gamma driver conditions
have been employed in many recent numerical computations:
\begin{itemize}
\item  3D gravitational collapse calculations by Baiotti et al. (2005, 2006) \cite{BaiotHMLRSFS05,BaiotR06}, with $k=3/4$ and 
$\eta=3/M$, where $M$ is the ADM mass; 
\item the first evolution of a binary black hole system lasting for about
one orbit by Br\"ugmann, Tichy and Jansen (2004) \cite{BruegTJ04},
with $k = 3/4\, N \Psi^{-2}$ and $\eta = 2/M$;
\item binary black hole mergers by 
\begin{itemize}
\item Campanelli, Lousto, Marronetti and Zlochower
(2006) \cite{CampaLMZ06,CampaLZ06a,CampaLZ06b,CampaLZ06c}, with
$k=3/4$;
\item Baker et al. (2006)
\cite{BakerCCKV06a,BakerCCKV06b}, with 
$k=3N/4$ and a slightly modified version of Eq.~(\ref{e:evo:Gam_driver_1order}),
namely $\dert{\tilde\Gamma^i}{t}$ replaced by 
$\dert{\tilde\Gamma^i}{t} - \beta^j\dert{\tilde\Gamma^i}{x^j}$ 
in the second equation;
\item Sperhake \cite{Sperh07}, with 
$k=1$ and $\eta=1/M$.
\end{itemize} 
\end{itemize}
Recently, van Meter et al.  \cite{VanMeBKC06} and Br\"ugmann et al. \cite{BruegGHHST07}
have considered a modified version of Eq.~(\ref{e:evo:Gam_driver_1order}), by
replacing all the derivatives $\dert{}{t}$ by 
$\dert{}{t} - \beta^j \dert{}{x^j}$, i.e. writing
\be \label{e:evo:Gam_driver_advect}
	\left\{ \begin{array}{lcl}
	\displaystyle \der{\beta^i}{t} - \beta^j \der{\beta^i}{x^j} & = & k B^i \\[1ex]
	\displaystyle \der{B^i}{t} - \beta^j \der{B^i}{x^j}& = & \displaystyle 
	\der{\tilde\Gamma^i}{t} - \beta^j \der{\tilde\Gamma^i}{x^j} - \eta B^i .
	\end{array} 
	\right. 
\ee
In particular, Br\"ugmann et al. \cite{BruegGHHST07,MarroTBGHHS07} have computed
binary black hole mergers using (\ref{e:evo:Gam_driver_advect}) with $k=3/4$
and $\eta$ ranging from $0$ to $3.5/M$, whereas Herrmann et al. \cite{HerrmHSLM07}
have used (\ref{e:evo:Gam_driver_advect}) with $k=3/4$
and $\eta=2/M$.

\subsection{Other dynamical shift gauges}

Shibata (2003) \cite{Shiba03b} has introduced a spatial gauge that
is closely related to the hyperbolic Gamma driver: it is defined by the
requirement
\be \label{e:evo:dyn_shift_Shiba}
	\der{\beta^i}{t} = \tgm^{ij} \left( F_j + \delta t\, \der{F_j}{t} \right) , 
\ee
where $\delta t$ is the time step used in the numerical computation 
and\footnote{let us recall that $\Df^i := f^{ij} \Df_j$}
\be
	F_i := \Df^j \tgm_{ij} . 
\ee
From the definition of the inverse metric $\tgm^{ij}$, namely the identity
$\tgm^{ik} \tgm_{kj} = \delta^i_{\ \, j}$, and relation (\ref{e:evo:tGam_divtgm}), 
it is easy to show that $F_i$ is related to $\tilde\Gamma^i$ by 
\be \label{e:evo:F_Gam}
	F_i = \tgm_{ij} \tilde\Gamma^j - \left(\tgm^{jk} - f^{jk}\right)
	\Df_k \tgm_{ij} . 
\ee
Notice that in the weak field region, i.e. where $\tgm^{ij} = f^{ij} + h^{ij}$
with $f_{ik} f_{jl} h^{kl} h^{ij} \ll 1$, the second term in Eq.~(\ref{e:evo:F_Gam})
is of second order in $\w{h}$, so that at first order in $\w{h}$, 
Eq.~(\ref{e:evo:F_Gam}) reduces to $F_i \simeq  \tgm_{ij} \tilde\Gamma^j$.
Accordingly Shibata's prescription (\ref{e:evo:dyn_shift_Shiba}) becomes
\be
	\der{\beta^i}{t} \simeq \tilde\Gamma^i 
	+ \tgm^{ij} \delta t\, \der{F_j}{t} . 
\ee
If we disregard the $\delta t$ term in the right-hand side and 
take the time derivative of this equation, we obtain
the Gamma-driver condition (\ref{e:evo:hyp_Gam_driver}) with $k=1$
and $\eta=0$. The term in $\delta t$ has been introduced by Shibata 
\cite{Shiba03b} in order to stabilize the numerical code. 

The spatial gauge (\ref{e:evo:dyn_shift_Shiba}) has been used by 
Shibata (2003) \cite{Shiba03b} and Sekiguchi and Shibata (2005) \cite{SekigS05}
to compute axisymmetric gravitational collapse of
rapidly rotating neutron stars to black holes, as well as by Shibata 
and Sekiguchi (2005) \cite{ShibaS05} to compute 3D gravitational collapses,
allowing for the development of nonaxisymmetric instabilities. 
It has also been used by Shibata, Taniguchi and Uryu (2003-2006)
\cite{ShibaTU03,ShibaTU05,ShibaT06} to compute the merger of binary neutron stars,
while their preceding computations \cite{ShibaU00,ShibaU02} rely on the
approximate minimal distortion gauge (Sec.~\ref{s:evo:AMDist}). 

%%%%%%%%%%%%%%%%%%%%%%%%%%%%%%%%%%%%%%%%%%%%%%%%%%%%%%%%%%%%%%%%%%%%%%%%%%%%%%%%%%%%%%

\section{Full spatial coordinate-fixing choices}

The spatial coordinate choices discussed in Sec.~\ref{s:evo:spat_coord},
namely vanishing shift, minimal distortion, Gamma freezing, Gamma driver and
related prescriptions, are relative to the \emph{propagation}
of the coordinates $(x^i)$ away from the initial hypersurface $\Sigma_0$.
They do not restrict at all the choice of coordinates in $\Sigma_0$. 
Here we discuss some coordinate choices that fix completely the coordinate
freedom, including in the initial hypersurface. 

\subsection{Spatial harmonic coordinates}

The first full coordinate-fixing choice we shall discuss is that of
\defin{spatial harmonic coordinates}. They are defined by 
\be \label{e:evo:spat_harm_def}
	\encadre{ D_j D^j x^i = 0 }, 
\ee
in full analogy with the spacetime harmonic coordinates [cf. 
Eq.~(\ref{e:evo:harm_coord_def})]. 
The above condition is equivalent to 
\be
	\frac{1}{\sqrt{\gm}} \der{}{x^j} \bigg(
	\sqrt{\gm} \gm^{jk} 
	\underbrace{\der{x^i}{x^k}}_{=\delta^i_{\ \, k}} \bigg) = 0 ,  
\ee
i.e. 
\be \label{e:evo:spat_harm_def2}
	 \der{}{x^j} \left( \sqrt{\gm} \gm^{ij} \right) = 0 .  
\ee
This relation restricts the coordinates to be of Cartesian type. 
Notably, it forbids the use of spherical-type coordinates, even
in flat space, for it is violated by 
$\gm_{ij} = \mathrm{diag}(1,r^2,r^2\sin^2\theta)$. 
To allow for any type of coordinates, let us rewrite condition 
(\ref{e:evo:spat_harm_def2}) in terms 
of a background flat metric $\w{f}$ (cf. discussion in 
Sec.~\ref{s:cfd:background_metric}), as 
\be
	\encadre{ \Df_j \left[ \left( \frac{\gm}{f} \right) ^{1/2} \gm^{ij}
	\right] = 0 }, 
\ee
where $\wDf$ is the connection associated with $\w{f}$ and 
$f:=\det f_{ij}$ is the determinant of $\w{f}$ with respect
to the coordinates $(x^i)$. 

Spatial harmonic coordinates have been considered by \v Cade\v z
\cite{Cadez75} for binary black holes and by Andersson and Moncrief
\cite{AnderMo03} in order to put the 3+1 Einstein system into 
an elliptic-hyperbolic form and to show that the corresponding Cauchy problem
is well posed. 

\begin{remark}
The spatial harmonic coordinates discussed above should not be confused with \emph{spacetime} harmonic coordinates; the latter would be defined by  
$\square_{\w{g}} x^i = 0$ [spatial part
of Eq.~(\ref{e:evo:harm_coord_def})]
instead of (\ref{e:evo:spat_harm_def}). 
Spacetime harmonic coordinates, as well as some generalizations, are
considered e.g. in Ref.~\cite{AlcubCGNRS05}.
\end{remark}

\subsection{Dirac gauge} \label{s:evo:Dirac}

As a natural way to fix the coordinates in his Hamiltonian formulation 
of general relativity (cf. Sec.~\ref{s:dec:ADM}), Dirac \cite{Dirac59}
has introduced in 1959 the following condition: 
\be
	\der{}{x^j} \left( \gm^{1/3} \gm^{ij} \right) = 0 . 
\ee
It differs from the definition (\ref{e:evo:spat_harm_def2})
of spatial harmonic coordinates only by the power of the determinant $\gm$. 
Similarly, we may rewrite it more covariantly in terms of the background
flat metric $\w{f}$ as \cite{BonazGGN04}
\be
	\encadre{ \Df_j \left[ \left( \frac{\gm}{f} \right) ^{1/3} \gm^{ij}
	\right] = 0 } . 
\ee
We recognize in this equation the inverse conformal metric 
[cf. Eqs.~(\ref{e:cfd:def_Psi}) and (\ref{e:cfd:gmij_up_down})],
so that we may write:
\be \label{e:evo:Dirac_def}
	\encadre{ \Df_j \tgm^{ij} = 0 } . 
\ee
We call this condition the \defin{Dirac gauge}. 
It has been first discussed in the context of numerical relativity
in 1978 by Smarr and York \cite{SmarrY78a} but disregarded in profit of the minimal 
distortion gauge (Sec.~\ref{s:evo:min_distort}), for the latter leaves the freedom
to choose the coordinates in the initial hypersurface. 
In terms of the vector $\w{\tilde\Gamma}$ introduced in Sec.~\ref{s:evo:Gamma_freez},
the Dirac gauge has a simple expression, thanks to relation (\ref{e:evo:tGam_divtgm}):
\be	\label{e:evo:Dirac_Gam}
	\encadre{\tilde\Gamma^i = 0}. 
\ee
It is then clear that if the coordinates $(x^i)$ obey the Dirac gauge at all
times $t$, then they belong to the Gamma freezing class discussed in Sec.~\ref{s:evo:Gamma_freez}, 
for Eq.~(\ref{e:evo:Dirac_Gam}) implies Eq.~(\ref{e:evo:Gam_freez3}).
Accordingly, the shift vector of Dirac-gauge coordinates has to satisfy the 
Gamma freezing elliptic equation (\ref{e:evo:Gam_freez_beta}), with the additional simplification 
$\tilde\Gamma^i=0$:
\be 
	\encadre{
	\begin{array}{lcl}
	 \displaystyle \tgm^{jk} \Df_j \Df_k \beta^i + \frac{1}{3} \tgm^{ij}
	\Df_j \Df_k \beta^k 
	& = & 
	 \displaystyle  
	2N \left[ 8\pi \Psi^4 p^i 
	- \tA^{jk}  \left(\tilde\Gamma^i_{\ \, jk} - \bar\Gamma^i_{\ \, jk} \right)
	- 6 \tA^{ij} \Df_j\ln \Psi 
	+ \frac{2}{3} \tgm^{ij} \Df_j K \right] \\[2ex]
	& & + 2 \tA^{ij} \Df_j N .
	\end{array}
	} 
\ee

The Dirac gauge, along with maximal slicing, has been employed by 
Bonazzola, Gourgoulhon, Grandcl\'ement and Novak \cite{BonazGGN04} 
to devise a constrained scheme\footnote{the concept of \emph{constrained
scheme} will be discussed in Sec.~\ref{s:sch:constr}}
for numerical relativity, that has been applied  
to 3D evolutions of gravitational waves. 
It has also been used by Shibata, Uryu and Friedman \cite{ShibaUF04}
to formulate waveless approximations of general relativity
that go beyond the IWM approximation discussed in Sec.~\ref{s:cfd:IWM}. 
Such a formulation has been employed recently to compute 
quasi-equilibrium configurations of 
binary neutron stars \cite{UryuLFGS06}.
Since Dirac gauge is a full coordinate-fixing gauge, the initial
data must fulfill it. Recently, Lin and Novak \cite{LinN06} have
computed equilibrium configurations of rapidly rotating stars
within the Dirac gauge, which may serve as initial data for 
gravitational collapse.

%  
%    Chapitre : Evolution schemes
%
% $Date: 2007-03-05 22:39:07 +0100 (lun, 05 mar 2007) $
% $Rev: 182 $
% $Author: e_gourgoulhon $
%%%%%%%%%%%%%%%%%%%%%%%%%%%%%

\chapter{Evolution schemes} \label{s:sch}

%\verb$Date: 2007-03-05 22:39:07 +0100 (lun, 05 mar 2007) $

\minitoc
\vspace{1cm}

%%%%%%%%%%%%%%%%%%%%%%%%%%%%%%%%%%%%%%%%%%%%%%%%%%%%%%%%%%%%%%%%%%%%%%%%%%%%

\section{Introduction}

Even after having selected the foliation and the spatial coordinates propagation
(Chap.~\ref{s:evo}),
there remains various strategies to integrate the 3+1 Einstein equations, 
either in their original form (\ref{e:dec:Einstein_PDE1})-(\ref{e:dec:Einstein_PDE4}),
or in the conformal form (\ref{e:cfd:Einstein1})-(\ref{e:cfd:Einstein6}). 
In particular, the constraint equations (\ref{e:dec:Einstein_PDE3})-(\ref{e:dec:Einstein_PDE4}) or
(\ref{e:cfd:Einstein5})-(\ref{e:cfd:Einstein6}) may be solved or not
during the evolution, giving rise to respectively the
so-called \emph{free evolution schemes} and the \emph{constrained schemes}. 
We discuss here the two types of schemes (Sec.~\ref{s:sch:constr} and \ref{s:sch:free}),
and present afterwards a widely used free evolution scheme: the BSSN one 
(Sec.~\ref{s:sch:BSSN}).  

Some review articles on the subject are those by Stewart (1998) \cite{Stewa98}, 
Friedrich and Rendall (2000) \cite{FriedR00},
Lehner (2001) \cite{Lehne01}, Shinkai and Yoneda (2002,2003)
\cite{ShinkY02,Shink03}, Baumgarte and Shapiro (2003) \cite{BaumgS03},
and Lehner and Reula (2004) \cite{LehneR04}. 

%%%%%%%%%%%%%%%%%%%%%%%%%%%%%%%%%%%%%%%%%%%%%%%%%%%%%%%%%%%%%%%%%%%%%%%%%%%%%%%%%%%

\section{Constrained schemes} \label{s:sch:constr}

A \defin{constrained scheme} is a time scheme for integrating the 3+1
Einstein system in which some (\defin{partially constrained scheme}) or
all (\defin{fully constrained scheme}) of the four constraints are used
to compute some of the metric coefficients at each step of the numerical
evolution. 
 
In the eighties, partially constrained schemes, 
with only the Hamiltonian constraint enforced,
have been widely used in 2-D (axisymmetric) computations 
(e.g. Bardeen and Piran \cite{BardeP83}, 
Stark and Piran \cite{StarkP85}, Evans \cite{Evans86}).
Still in the 2-D axisymmetric case, 
fully constrained schemes have been used 
by Evans \cite{Evans89}
and Shapiro and Teukolsky \cite{ShapiT92} 
for non-rotating spacetimes, and by Abrahams, Cook, 
Shapiro and Teukolsky \cite{AbrahCST94} for rotating ones. 
More recently the (2+1)+1 axisymmetric code of Choptuik,
Hirschmann, Liebling and Pretorius (2003)
\cite{ChoptHLP03} is based on a constrained scheme too. 

Regarding 3D numerical relativity, a fully constrained scheme based on the
original 3+1 Einstein system (\ref{e:dec:Einstein_PDE1})-(\ref{e:dec:Einstein_PDE4})
has been used to evolve a single black hole by Anderson and Matzner (2005) 
\cite{AnderM05}. Another fully constrained scheme has been devised by
Bonazzola, Gourgoulhon, Grandcl\'ement and Novak (2004) \cite{BonazGGN04},
but this time for the conformal 3+1 Einstein system (\ref{e:cfd:Einstein1})-(\ref{e:cfd:Einstein6}). The latter scheme makes use
of maximal slicing and Dirac gauge (Sec.~\ref{s:evo:Dirac}). 

%%%%%%%%%%%%%%%%%%%%%%%%%%%%%%%%%%%%%%%%%%%%%%%%%%%%%%%%%%%%%%%%%%%%%%%%%%%%%%%%%%%%%

\section{Free evolution schemes} \label{s:sch:free}

\subsection{Definition and framework}

A \defin{free evolution scheme} is a time scheme for integrating the 3+1
Einstein system in which the constraint
equations are solved only to get the initial data, 
e.g. by following one of the prescriptions  
discussed in Chap.~\ref{s:ini}. 
The subsequent evolution is performed via the dynamical equations only, without
enforcing the constraints.
Actually, facing the 3+1 Einstein system 
(\ref{e:dec:Einstein_PDE1})-(\ref{e:dec:Einstein_PDE4}), we realize that 
the dynamical equation (\ref{e:dec:Einstein_PDE2}), coupled with 
the kinematical relation (\ref{e:dec:Einstein_PDE1}) and some choices for
the lapse function and shift vector (as discussed in Chap.~\ref{s:evo}), 
is sufficient to get the values of $\wgm$, $\w{K}$, $N$ and $\w{\beta}$
at all times $t$, from which we can reconstruct the full spacetime metric
$\w{g}$. 

A natural question which arises then is : to which extent
does the metric $\w{g}$ hence obtained fulfill the Einstein equation 
(\ref{e:dec:Einstein}) ?
The dynamical part, Eq.~(\ref{e:dec:Einstein_PDE2}), is fulfilled by construction,
but what about the constraints (\ref{e:dec:Einstein_PDE3}) and (\ref{e:dec:Einstein_PDE4}) ? If they were violated by the solution 
$(\wgm,\w{K})$ of the dynamical equation, then the obtained metric $\w{g}$
would not satisfy Einstein equation. The key point is that, as we shall see
in Sec.~\ref{s:sch:prop_constraints}, provided that the constraints are 
satisfied at $t=0$, the dynamical equation (\ref{e:dec:Einstein_PDE2})
ensures that they are satisfied for all $t>0$.

\subsection{Propagation of the constraints} \label{s:sch:prop_constraints}

Let us derive evolution equations for the constraints, or more 
precisely, for the constraint violations. These evolution equations
will be consequences of the 
Bianchi identities\footnote{the following computation is inspired from Frittelli's
article \cite{Fritt97}}. We denote by $\w{G}$ the \defin{Einstein tensor}:
\be \label{e:sch:EinsTens_def}
	\w{G} := {}^4\!\w{R} - \frac{1}{2} {}^4\!R \, \w{g} , 
\ee
so that the Einstein equation (\ref{e:dec:Einstein}) is written
\be \label{e:sch:Einstein_G}
	\w{G} = 8\pi \w{T} . 
\ee
The \defin{Hamiltonian constraint violation} is the scalar field defined by
\be
	\encadre{ H := \w{G}(\w{n},\w{n}) - 8\pi \w{T}(\w{n},\w{n}) },
\ee
i.e. 
\be \label{e:sch:H_Rnn}
	H = {}^4\!\w{R}(\w{n},\w{n}) + \frac{1}{2} {}^4\!R  - 8 \pi E,  
\ee
where we have used the relations $\w{g}(\w{n},\w{n})=-1$ and 
$\w{T}(\w{n},\w{n})=E$ [Eq.~(\ref{e:dec:E_def})]. 
Thanks to the scalar Gauss equation (\ref{e:hyp:Gauss_scalar})
we may write
\be 
	\encadre{ H = \frac{1}{2} \left( R + K^2 - K_{ij} K^{ij} \right) - 8\pi E} .
\ee
Similarly we define the \defin{momentum constraint violation} as the 1-form field
\be
	\encadre{\w{M} := - \w{G}(\w{n},\vg(.)) + 8\pi \w{T} (\w{n},\vg(.)) } . 
\ee
By means of the contracted Codazzi equation (\ref{e:hyp:Codazzi_contract}) and
the relation $\w{T}(\w{n}, \vg(.) )=-\w{p}$ [Eq.~(\ref{e:dec:p_def})], we
get
\be 
	\encadre{ M_i = D_j K^j_{\ \, i} - D_i K - 8\pi  p_i     } ,
\ee
From the above expressions, we see that 
the Hamiltonian constraint (\ref{e:dec:Einstein_PDE3})
and the momentum constraint (\ref{e:dec:Einstein_PDE4})
are equivalent to respectively
\bea
	H & = & 0 \\
	M_i & = & 0 . 
\eea

Finally we define the \defin{dynamical equation violation} as the spatial 
tensor field
\be 
	\encadre{ \w{F}  := \vgs\,  {}^4\!\w{R} - 8\pi \vgs \left(  \w{T} 
	- \frac{1}{2} T \, \w{g} \right) }. 
\ee
Indeed, let us recall that the dynamical part of the 3+1 Einstein system,
Eq.~(\ref{e:dec:Einstein_PDE2}) is nothing but the spatial projection of 
the Einstein equation written in terms of the Ricci tensor ${}^4\!\w{R}$, 
i.e. Eq.~(\ref{e:dec:Einstein2}), instead of the Einstein tensor, 
i.e. Eq.~(\ref{e:sch:Einstein_G}) (cf. Sec.~\ref{s:dec:project_Einstein}). 
Introducing the stress tensor $\w{S}=\vgs \w{T}$ [Eq.~(\ref{e:dec:S_def})]
and using the relations $T=S-E$ [Eq.~(\ref{e:dec:trT_SmE})]
and $\vgs\w{g}=\wgm$, we can
write $\w{F}$ as
\be \label{e:sch:F_S}
	\w{F}  = \vgs\, {}^4\!\w{R} - 8\pi \left[ \w{S}
	+ \frac{1}{2} (E-S) \wgm \right] . 
\ee
From Eq.~(\ref{e:dec:EEdyn_proj}), we see that the dynamical part of Einstein 
equation is equivalent to 
\be \label{e:sch:dyn_eq}
	\w{F} = 0 . 
\ee
This is also clear if we replace $\vgs\,  {}^4\!\w{R}$ in Eq.~(\ref{e:sch:F_S})
by the expression (\ref{e:fol:proj_Ricci}): we immediately get 
Eq.~(\ref{e:dec:Einstein_PDE2}). 

Let us express $\vgs (\w{G}-8\pi\w{T})$ in terms of $\w{F}$. 
Using Eq.~(\ref{e:sch:EinsTens_def}), we have
\be
	\vgs (\w{G}-8\pi\w{T}) = \vgs\, {}^4\!\w{R}
	- \frac{1}{2}  {}^4\!R  \wgm - 8\pi \w{S} .
\ee
Comparing with Eq.~(\ref{e:sch:F_S}), we get
\be \label{e:sch:G_proj}
	\vgs (\w{G}-8\pi\w{T}) = \w{F} - \frac{1}{2} \left[
	 {}^4\!R  + 8\pi(S-E) \right] \wgm . 
\ee
Besides, the trace of Eq.~(\ref{e:sch:F_S}) is
\bea
	F & = & \mathrm{tr}_{\wgm} \w{F} = \gam^{ij} F_{ij} 
	= \gam^{\mu\nu} F_{\mu\nu} \nonumber \\  
	& = & \underbrace{\gam^{\mu\nu} \gam^\rho_{\ \, \mu}}_{=\gam^{\rho\nu}}
	 \gam^\sigma_{\ \, \nu}
	 {}^4\!R_{\rho\sigma} - 8\pi \left[ S + \frac{1}{2} (E-S) \times 3 \right]
	\nonumber \\
	 & = & \gam^{\rho\sigma} {}^4\!R_{\rho\sigma} 
	+4\pi (S-3E) 
	= {}^4\!R + {}^4\!R_{\rho\sigma} \, n^\rho n^\sigma 
	+ 4\pi (S-3E) . 
\eea
Now, from Eq.~(\ref{e:sch:H_Rnn}), 
${}^4\!R_{\rho\sigma} \, n^\rho n^\sigma = H - {}^4\!R / 2 + 8\pi E$, 
so that the above relation becomes
\bea
	F & = &  {}^4\!R + H - \frac{1}{2} {}^4\!R + 8\pi E + 4\pi (S-3E) 
	\nonumber \\
	 & = & H + \frac{1}{2} \left[ {}^4\!R + 8\pi (S-E) \right]  . 
\eea
This enables us to write Eq.~(\ref{e:sch:G_proj}) as
\be \label{e:sch:vgs_G_F}
	\vgs (\w{G}-8\pi\w{T}) = \w{F} + (H-F) \wgm . 
\ee

Similarly to the 3+1 decomposition (\ref{e:dec:T_3p1}) of the stress-energy
tensor, the 3+1 decomposition of $\w{G}-8\pi\w{T}$ is
\be
	\w{G}-8\pi\w{T} = \vgs(\w{G}-8\pi\w{T}) 
	+ \uu{n}\otimes\w{M} + \w{M}\otimes\uu{n}
	+ H \, \uu{n}\otimes\uu{n} , 
\ee
$\vgs(\w{G}-8\pi\w{T})$ playing the role of $\w{S}$, 
$\w{M}$ that of $\w{p}$ and $H$ that of $E$. 
Thanks to Eq.~(\ref{e:sch:vgs_G_F}), we may write
\be \label{e:sch:G_T_decomp3p1}
	\encadre{ \w{G}-8\pi\w{T} = \w{F} + (H-F) \wgm 
	+ \uu{n}\otimes\w{M} + \w{M}\otimes\uu{n}
	+ H \, \uu{n}\otimes\uu{n} } , 
\ee
or, in index notation,
\be
	G_{\alpha\beta} -8\pi T_{\alpha\beta} = 
	F_{\alpha\beta} + (H-F) \gm_{\alpha\beta} 
	+ n_\alpha M_\beta + M_\alpha n_\beta + H n_\alpha n_\beta .
\ee
This identity can be viewed as the 3+1 decomposition of Einstein equation
(\ref{e:sch:Einstein_G}) in terms of the dynamical equation violation
$\w{F}$, the Hamiltonian constraint violation $H$ and the momentum constraint
violation $\w{M}$. 

The next step consists in invoking the contracted \defin{Bianchi identity}:
\be \label{e:sch:Bianchi}
	\encadre{\vec{\wnab} \cdot \w{G} = 0 } ,
\ee
i.e.
\be
	\encadre{\nabla^\mu G_{\alpha\mu} = 0 } . 
\ee
Let us recall that this identity is purely geometrical and holds independently 
of Einstein equation. 
In addition, we assume that the matter obeys the energy-momentum conservation
law (\ref{e:mat:divT}) : 
\be \label{e:sch:divT}
	\encadre{\vec{\wnab} \cdot\w{T} = 0} .
\ee
In view of the Bianchi identity (\ref{e:sch:Bianchi}), Eq.~(\ref{e:sch:divT})
is a necessary condition for the Einstein equation (\ref{e:sch:Einstein_G}) to hold. 
\begin{remark}
We assume here specifically that Eq.~(\ref{e:sch:divT}) holds, because 
in the following we do not demand that the whole Einstein equation is satisfied,
but only its dynamical part, i.e. Eq.~(\ref{e:sch:dyn_eq}).  
\end{remark}
As we have seen in Chap.~\ref{s:mat}, in order for Eq.~(\ref{e:sch:divT}) to be
satisfied, the matter energy density $E$ and momentum density $\w{p}$ 
(both relative to the Eulerian observer) must obey to the evolution equations (\ref{e:mat:ener_cons}) and (\ref{e:mat:mom_cons}). 

Thanks to the Bianchi identity (\ref{e:sch:Bianchi}) and to the energy-momentum
conservation law (\ref{e:sch:divT}), the divergence of Eq.~(\ref{e:sch:G_T_decomp3p1})
leads to, successively,
\bea
	& & \nabla_\mu \left(G^\mu_{\ \, \alpha} - 8\pi T^\mu_{\ \, \alpha}\right)
	 = 0 \nonumber \\
	& & \nabla_\mu \left[ F^\mu_{\ \, \alpha} +(H-F)\gm^\mu_{\ \, \alpha}
	+ n^\mu M_\alpha + M^\mu n_\alpha
	+ H n^\mu n_\alpha  \right] = 0,  \nonumber \\
	& & \nabla_\mu F^\mu_{\ \, \alpha} + D_\alpha(H-F) + (H-F) 
	\left( \nabla_\mu n^\mu n_\alpha + n^\mu \nabla_\mu n_\alpha \right)
	- K M_\alpha + n^\mu \nabla_\mu M_\alpha
	 \nonumber \\
	& & \qquad \qquad + \nabla_\mu M^\mu \, n_\alpha- M^\mu K_{\mu\alpha}
	+ n^\mu \nabla_\mu H \, n_\alpha
	- H K n_\alpha + H D_\alpha\ln N 
	 = 0 , \nonumber \\
	& & \nabla_\mu F^\mu_{\ \, \alpha} + D_\alpha(H-F)
	+ (2 H -F) (D_\alpha \ln N - K n_\alpha) - K M_\alpha
	+  n^\mu \nabla_\mu M_\alpha,
	 \nonumber \\
	& & \qquad \qquad + \nabla_\mu M^\mu \, n_\alpha
	- K_{\alpha\mu} M^\mu +  n^\mu \nabla_\mu H \, n_\alpha = 0 ,
		\label{e:divGmT}
\eea
where we have used Eq.~(\ref{e:fol:nab_n_K_comp}) to express the $\w{\nabla}\uu{n}$
in terms of $\w{K}$ and $\w{D}\ln N$ (in particular $\nabla_\mu n^\mu=-K$).
Let us contract Eq.~(\ref{e:divGmT}) with $\w{n}$: we get, successively,
\bea
	& & n^\nu  \nabla_\mu F^\mu_{\ \, \nu}
	+ (2H-F) K + n^\nu n^\mu\nabla_\mu M_\nu - \nabla_\mu M^\mu
	- n^\mu \nabla_\mu H = 0, \nonumber \\
	& & - F^\mu_{\ \, \nu} \nabla_\mu n^\nu
	+ (2H-F) K - M_\nu n^\mu\nabla_\mu n^\nu- \nabla_\mu M^\mu
	- n^\mu \nabla_\mu H = 0, \nonumber \\
	& & K^{\mu\nu} F_{\mu\nu} + (2H-F) K - M^\nu D_\nu \ln N 
	- \nabla_\mu M^\mu
	- n^\mu \nabla_\mu H = 0  . \label{e:sch:evol_Ham_prov1}
\eea
Now the $\wnab$-divergence of $\w{M}$ is related to the $\wD$-one by
\bea
	D_\mu M^\mu & = & \gm^\rho_{\ \, \mu} \gm^\sigma_{\ \, \nu} 
		\nabla_\rho M^\sigma
	= \gm^\rho_{\ \, \sigma} \nabla_\rho M^\sigma
	= \nabla_\rho M^\rho + n^\rho n_\sigma \nabla_\rho M^\sigma \nonumber \\
	& = & \nabla_\mu M^\mu - M^\mu D_\mu \ln N . 
\eea
Thus Eq.~(\ref{e:sch:evol_Ham_prov1}) can be written
\be
	n^\mu \nabla_\mu H = - D_\mu M^\mu - 2 M^\mu D_\mu \ln N 
	+ K(2H-F) + K^{\mu\nu} F_{\mu\nu} .
\ee
Noticing that 
\be
	n^\mu \nabla_\mu H = \frac{1}{N} m^\mu \nabla_\mu H 
	= \frac{1}{N} \Lie{m} H = 
	\frac{1}{N} \left( \der{}{t} - \Liec{\beta} \right) H,
\ee
where $\w{m}$ is the normal evolution vector (cf. Sec.~\ref{s:fol:norm_evol}), 
we get the following evolution equation for the Hamiltonian constraint violation
\be
	\encadre{ \left( \der{}{t} - \Liec{\beta} \right) H
	= - D_i (N M^i) -  M^i D_i N + N K(2H-F) + 
	N K^{ij} F_{ij} } .  \label{e:sch:evol_Ham0}
\ee

Let us now project Eq.~(\ref{e:divGmT}) onto $\Sigma_t$:
\be \gm^{\nu\alpha} \nabla_\mu F^\mu_{\ \, \nu} 
	+ D^\alpha(H-F)
	+ (2 H -F) D^\alpha \ln N  - K M^\alpha
	+  \gm^\alpha_{\ \, \nu} n^\mu \nabla_\mu M^\nu
	- K^\alpha_{\ \, \mu} M^\mu = 0 .  \label{e:sch:evol_mom_prov1}
\ee
Now the $\wnab$-divergence of $\w{F}$ is related to the $\wD$-one by
\bea
	D_\mu F^{\mu\alpha} & = & \gm^\rho_{\ \, \mu} \gm^\mu_{\ \, \sigma}
	\gm^{\nu\alpha} \nabla_\rho F^\sigma_{\ \, \nu}
 	 = \gm^\rho_{\ \, \sigma} \gm^{\nu\alpha}
	   \nabla_\rho F^\sigma_{\ \, \nu}
	= \gm^{\nu\alpha} \left( \nabla_\rho F^\rho_{\ \, \nu}
	+ n^\rho n_\sigma \nabla_\rho F^\sigma_{\ \, \nu} \right) 
	\nonumber \\
 & = & \gm^{\nu\alpha} \left( \nabla_\rho F^\rho_{\ \, \nu}
	- F^\sigma_{\ \, \nu} n^\rho \nabla_\rho n_\sigma \right) 
	\nonumber \\
& = & \gm^{\nu\alpha} \nabla_\mu F^\mu_{\ \, \nu}
	- F^{\alpha\mu} D_\mu \ln N . \label{e:sch:DF_nabF}
\eea
Besides, we have
\bea
	\gm^\alpha_{\ \, \nu} n^\mu \nabla_\mu M^\nu
	& = & \frac{1}{N} \gm^\alpha_{\ \, \nu} m^\mu \nabla_\mu M^\nu
	= \frac{1}{N} \gm^\alpha_{\ \, \nu} \left( \Liec{m} M^\nu
	+ M^\mu \nabla_\mu m^\nu \right) \nonumber \\
	& = & \frac{1}{N} \left[ \Liec{m} M^\alpha
	+ \gm^\alpha_{\ \, \nu} M^\mu (\nabla_\mu N\, n^\nu
	+ N  \nabla_\mu n^\nu) \right] \nonumber \\
	& = & \frac{1}{N} \Liec{m} M^\alpha - K^\alpha_{\ \, \mu} M^\mu ,
		\label{e:sch:ngradM}
\eea
where property (\ref{e:fol:liem_preserve}) has been used 
to write $\gm^\alpha_{\ \, \nu} \Liec{m} M^\nu = \Liec{m} M^\alpha$. 

Thanks to Eqs.~(\ref{e:sch:DF_nabF}) and (\ref{e:sch:ngradM}),
and to the relation $\Lie{m} = \dert{}{t} - \Lie{\beta}$, 
Eq.~(\ref{e:sch:evol_mom_prov1}) yields an evolution equation 
for the momentum constraint violation:
\be
	\encadre{\left( \der{}{t} - \Liec{\beta} \right) M^i 
	= - D_j(N F^{ij}) + 2N K^i_{\ \, j} M^j + NK M^i
	+ ND^i(F-H) + (F-2H) D^i N } . \label{e:sch:evol_mom0}
\ee

Let us now assume that the dynamical Einstein equation is satisfied,
then $\w{F}=0$ [Eq.~(\ref{e:sch:dyn_eq})] and Eqs.~(\ref{e:sch:evol_Ham0})
and (\ref{e:sch:evol_mom0}) reduce to 
\bea
	\left( \der{}{t} - \Liec{\beta} \right) H & = &
		 - D_i (N M^i) + 2 N K H -  M^i D_i N   \label{e:sch:evol_Ham1} \\
	\left( \der{}{t} - \Liec{\beta} \right) M^i 
	& = & - D^i(N H) + 2N K^i_{\ \, j} M^j + NK M^i
	+ H D^i N . \label{e:sch:evol_mom1}
\eea
If the constraints are satisfied at $t=0$, then $H| _{t=0}=0$ and $M^i| _{t=0}=0$. 
The above system gives then
\bea
	& & \left. \der{H}{t} \right| _{t=0} = 0 \\[2ex]
	& & \left.  \der{M^i}{t} \right |_{t=0} = 0 . 
\eea
We conclude that, at least in the case where all the fields are analytical
(in order to invoke the Cauchy-Kovalevskaya theorem), 
\be
	\forall t\geq 0,\quad H=0 \quad \mbox{and} \quad M^i=0 , 
\ee
i.e. the constraints are preserved by the dynamical evolution 
equation (\ref{e:dec:Einstein_PDE2}). 
Even if the hypothesis of analyticity is relaxed, the result 
still holds because the system 
(\ref{e:sch:evol_Ham1})-(\ref{e:sch:evol_mom1}) is symmetric
hyperbolic \cite{Fritt97}. 

\begin{remark}
The above result on the preservation of the constraints in a free evolution 
scheme holds only if the matter source obeys the 
energy-momentum conservation law (\ref{e:sch:divT}). 
\end{remark}

\subsection{Constraint-violating modes}

The constraint preservation property established in the preceding section 
adds some substantial support to the concept of free evolution scheme. However this is
a mathematical result and it does not guarantee that numerical solutions
will not violate the constraints. Indeed 
numerical codes based on free evolution schemes have been plagued 
for a long time by the so-called \defin{constraint-violating modes}. 
The latter are solutions
$(\wgm,\w{K},N,\w{\beta})$ which satisfy $\w{F}=0$ up to numerical accuracy
but with $H\not =0$ and $\w{M}\not =0$, although if initially
$H= 0$ and $\w{M}=0$ (up to numerical accuracy). The reasons for the 
appearance of these constraint-violating modes are twofold:
(i) due to numerical errors, the conditions $H=0$ and $\w{M}=0$
are slightly violated in the initial data, and the evolution equations
amplify (in most cases exponentially !) this violation 
and (ii) constraint violations
may flow into the computational domain from boundary conditions imposed at 
timelike boundaries.  
Notice that the demonstration in Sec.~\ref{s:sch:prop_constraints}
did not take into account any boundary and could not rule out (ii).  

An impressive amount of works have then been devoted to this issue
(see \cite{ShinkY02} for a review and Ref.~\cite{KiddeLSBP05,SarbaT05} for
recent solutions to problem (ii)). We mention hereafter shortly the
symmetric hyperbolic formulations, before discussing the most
successful approach to date: the BSSN scheme.

\subsection{Symmetric hyperbolic formulations}

The idea is to introduce auxiliary variables so that the dynamical equations become a first-order symmetric hyperbolic 
system, because these systems are known to be well posed 
(see e.g. \cite{Stewa98,Reula06}). 
This comprises the formulation developed in 2001 by Kidder, Scheel and
Teukolsky \cite{KiddeST01} (\defin{KST formulation}), 
which constitutes some generalization of previous
formulations developed by Frittelli and Reula (1996) \cite{FrittR96} and
by Andersson and York (1999) \cite{AnderY99}, the latter 
being known as the \defin{Einstein-Christoffel system}.

\section{BSSN scheme} \label{s:sch:BSSN}

\subsection{Introduction}

The \defin{BSSN scheme} is a free evolution scheme for the conformal 3+1 Einstein system (\ref{e:cfd:Einstein1})-(\ref{e:cfd:Einstein6}) which has been devised by
Shibata and Nakamura in 1995 \cite{ShibaN95}. It has been re-analyzed by
Baumgarte and Shapiro in 1999 \cite{BaumgS99}, with a slight modification, 
and bears since then the name
\emph{BSSN} for \emph{Baumgarte-Shapiro-Shibata-Nakamura}.  

\subsection{Expression of the Ricci tensor of the conformal metric}

The starting point of the BSSN formulation is the conformal 3+1 Einstein system (\ref{e:cfd:Einstein1})-(\ref{e:cfd:Einstein6}). 
One then proceeds by expressing the Ricci tensor $\w{\tilde R}$ of the
conformal metric $\wtgm$, which
appears in Eq.~(\ref{e:cfd:Einstein4}), in terms of the 
derivatives of $\wtgm$. 
To this aim, we consider the standard expression of the Ricci tensor
in terms of the Christoffel symbols $\tilde\Gamma^k_{\ \, ij}$
of the metric $\wtgm$ with respect to the coordinates $(x^i)$:
\be \label{e:sch:Rij1}
	\tilde R_{ij} = \der{}{x^k} \tilde\Gamma^k_{\ \, ij}
	- \der{}{x^j} \tilde\Gamma^k_{\ \, ik}
	+ \tilde\Gamma^k_{\ \, ij} \tilde\Gamma^l_{\ \, kl}
 	- \tilde\Gamma^k_{\ \, il} \tilde\Gamma^l_{\ \, kj} . 
\ee
Let us introduce the type $\left({1\atop 2}\right)$ tensor field $\w{\Delta}$ 
defined by 
\be \label{e:sch:def_Delta}
	\encadre{ \Delta^k_{\ \, ij} := \tilde\Gamma^k_{\ \, ij}
	- \bar\Gamma^k_{\ \, ij} }, 
\ee
where the $\bar\Gamma^k_{\ \, ij}$'s denote the Christoffel symbols
of the flat metric $\w{f}$ with respect to the coordinates $(x^i)$. 
As already noticed in Sec.~\ref{s:evo:Gamma_freez}, the identity
(\ref{e:sch:def_Delta}) does define a tensor field, although each set
of Christoffel symbols, $\tilde\Gamma^k_{\ \, ij}$ or $\bar\Gamma^k_{\ \, ij}$,
is by no means the set of components of any tensor field. 
Actually an alternative expression of $\Delta^k_{\ \, ij}$, which is
manifestly covariant, is 
\be \label{e:sch:Delta_Dftgm}
	\encadre{ \Delta^k_{\ \, ij} = \frac{1}{2} \tgm^{kl} \left(
	\Df_i \tgm_{lj} + \Df_j \tgm_{il} - \Df_l \tgm_{ij} \right) }, 
\ee 
where $\Df_i$ stands for the covariant derivative associated with the 
flat metric $\w{f}$. It is not difficult to establish the equivalence
of Eqs.~(\ref{e:sch:def_Delta}) and (\ref{e:sch:Delta_Dftgm}): starting from
the latter, we have
\bea
	\Delta^k_{\ \, ij} & = & \frac{1}{2} \tgm^{kl} \bigg( 
	\der{\tgm_{lj}}{x^i} - \bar\Gamma^m_{\ \, il} \tgm_{mj}
	- \bar\Gamma^m_{\ \, ij} \tgm_{lm} 
	+ \der{\tgm_{il}}{x^j} - \bar\Gamma^m_{\ \, ji} \tgm_{ml}
	- \bar\Gamma^m_{\ \, jl} \tgm_{im} \nonumber \\
	& & \qquad \quad - \der{\tgm_{ij}}{x^l}  + \bar\Gamma^m_{\ \, li} \tgm_{mj}
	+ \bar\Gamma^m_{\ \, lj} \tgm_{im} \bigg) \nonumber \\
	& = & \tilde\Gamma^k_{\ \, ij} + \frac{1}{2} \tgm^{kl} 
	\left( -2 \bar\Gamma^m_{\ \, ij} \tgm_{lm} \right)
	\ =\  \tilde\Gamma^k_{\ \, ij} 
	- \underbrace{\tgm^{kl} \tgm_{lm}}_{=\delta^k_{\ \, m}}
	\bar\Gamma^m_{\ \, ij} \nonumber \\
	& = & \tilde\Gamma^k_{\ \, ij} - \bar\Gamma^k_{\ \, ij} , 
\eea
hence we recover Eq.~(\ref{e:sch:def_Delta}). 
\begin{remark}
While it is a well defined tensor field, $\w{\Delta}$ depends upon the 
background flat metric $\w{f}$, which is not unique on the hypersurface 
$\Sigma_t$. 
\end{remark}
A useful relation is obtained by contracting
Eq.~(\ref{e:sch:def_Delta}) on the indices $k$ and $j$:
\be
	\Delta^k_{\ \, ik} = \tilde\Gamma^k_{\ \, ik}
	- \bar\Gamma^k_{\ \, ik}
	= \frac{1}{2} \der{}{x^i} \ln \tgm
	- \frac{1}{2} \der{}{x^i} \ln f , 
\ee 
where $\tgm:=\det \tgm_{ij}$ and $f:=\det f_{ij}$. 
Since by construction $\tgm=f$ [Eq.~(\ref{e:cfd:dettgm_f})], we get
\be \label{e:sch:trace_Delta}
	\encadre{ \Delta^k_{\ \, ik} = 0 } . 
\ee

\begin{remark}
If the coordinates $(x^i)$ are of Cartesian type, then 
$\bar\Gamma^k_{\ \, ij}=0$, $\Delta^k_{\ \, ij} = \tilde\Gamma^k_{\ \, ij}$
and $\Df_i = \dert{}{x^i}$. 
This is actually the case considered in the original articles of the BSSN
formalism \cite{ShibaN95,BaumgS99}. We follow here the method of 
Ref.~\cite{BonazGGN04} to allow for non Cartesian coordinates, e.g. spherical
ones. 
\end{remark}

Replacing  $\tilde\Gamma^k_{\ \, ij}$ by 
$\Delta^k_{\ \, ij} + \bar\Gamma^k_{\ \, ij}$ [Eq.~(\ref{e:sch:def_Delta})]
in the expression (\ref{e:sch:Rij1}) of the Ricci tensor yields
\bea
	\tilde R_{ij} & = & 
	\der{}{x^k} (\Delta^k_{\ \, ij} + \bar\Gamma^k_{\ \, ij})
	- \der{}{x^j} (\Delta^k_{\ \, ik} + \bar\Gamma^k_{\ \, ik})
	+ (\Delta^k_{\ \, ij} + \bar\Gamma^k_{\ \, ij})
	(\Delta^l_{\ \, kl} + \bar\Gamma^l_{\ \, kl}) \nonumber \\
	& & - (\Delta^k_{\ \, il} + \bar\Gamma^k_{\ \, il})
	(\Delta^l_{\ \, kj} + \bar\Gamma^l_{\ \, kj}) \nonumber \\
	& = &  \der{}{x^k} \Delta^k_{\ \, ij} + \der{}{x^k}  \bar\Gamma^k_{\ \, ij}
	- \der{}{x^j} \Delta^k_{\ \, ik} - \der{}{x^j} \bar\Gamma^k_{\ \, ik}
	+ \Delta^k_{\ \, ij} \Delta^l_{\ \, kl} 
	+ \bar\Gamma^l_{\ \, kl} \Delta^k_{\ \, ij} \nonumber \\
	& & + \bar\Gamma^k_{\ \, ij} \Delta^l_{\ \, kl}
 	+ \bar\Gamma^k_{\ \, ij} \bar\Gamma^l_{\ \, kl} 
 	- \Delta^k_{\ \, il} \Delta^l_{\ \, kj} 
	- \bar\Gamma^l_{\ \, kj} \Delta^k_{\ \, il}
	- \bar\Gamma^k_{\ \, il} \Delta^l_{\ \, kj}
	- \bar\Gamma^k_{\ \, il} \bar\Gamma^l_{\ \, kj} . \label{e:sch:Rij2}
\eea
Now since the metric $\w{f}$ is flat, its Ricci tensor vanishes identically, 
so that
\be
	\der{}{x^k}  \bar\Gamma^k_{\ \, ij}
	- \der{}{x^j} \bar\Gamma^k_{\ \, ik}
	+ \bar\Gamma^k_{\ \, ij} \bar\Gamma^l_{\ \, kl} 
	- \bar\Gamma^k_{\ \, il} \bar\Gamma^l_{\ \, kj} = 0 . 
\ee
Hence Eq.~(\ref{e:sch:Rij2}) reduces to 
\bea
	\tilde R_{ij} & = &  \der{}{x^k} \Delta^k_{\ \, ij} 
	- \der{}{x^j} \Delta^k_{\ \, ik}
	+ \Delta^k_{\ \, ij} \Delta^l_{\ \, kl} 
	+ \bar\Gamma^l_{\ \, kl} \Delta^k_{\ \, ij}
	+ \bar\Gamma^k_{\ \, ij} \Delta^l_{\ \, kl}
	- \Delta^k_{\ \, il} \Delta^l_{\ \, kj}  \nonumber \\
	& & - \bar\Gamma^l_{\ \, kj} \Delta^k_{\ \, il}
	- \bar\Gamma^k_{\ \, il} \Delta^l_{\ \, kj} .
\eea
Property (\ref{e:sch:trace_Delta}) enables us to simplify this
expression further:
\bea
	\tilde R_{ij} & = & \der{}{x^k} \Delta^k_{\ \, ij} 
	+ \bar\Gamma^l_{\ \, kl} \Delta^k_{\ \, ij}
- \bar\Gamma^l_{\ \, kj} \Delta^k_{\ \, il}
- \bar\Gamma^k_{\ \, il} \Delta^l_{\ \, kj}
- \Delta^k_{\ \, il} \Delta^l_{\ \, kj} \nonumber \\
	& = & \der{}{x^k} \Delta^k_{\ \, ij} 
	+ \bar\Gamma^k_{\ \, kl} \Delta^l_{\ \, ij}
- \bar\Gamma^l_{\ \, ki} \Delta^k_{\ \, lj}
- \bar\Gamma^l_{\ \, kj} \Delta^k_{\ \, il}
- \Delta^k_{\ \, il} \Delta^l_{\ \, kj} .
\eea
We recognize in the first four terms of the right-hand side
the covariant derivative $\Df_k \Delta^k_{\ \, ij}$, hence
\be \label{e:sch:Rij3}
	\tilde R_{ij} = \Df_k \Delta^k_{\ \, ij}
	- \Delta^k_{\ \, il} \Delta^l_{\ \, kj} . 
\ee
\begin{remark}
Even if $\Delta^k_{\ \, ik}$ would not vanish, we would have obtained an expression
of the Ricci tensor with exactly the same structure as Eq.~(\ref{e:sch:Rij1}), 
with the 
partial derivatives $\dert{}{x^i}$ replaced by the covariant derivatives 
$\Df_i$ and the Christoffel symbols $\tilde\Gamma^k_{\ \, ij}$ replaced by the
tensor components $\Delta^k_{\ \, ij}$. 
Indeed Eq.~(\ref{e:sch:Rij3}) can be seen as being nothing but a particular case 
of the more general formula obtained in Sec.~\ref{s:cfd:2Ricci} and relating
the Ricci tensors associated with two different metrics, namely Eq.~(\ref{e:cfd:Ricci1}). 
Performing in the latter the substitutions $\wgm\rightarrow\wtgm$, $\wtgm\rightarrow\w{f}$,
$R_{ij}\rightarrow\tilde R_{ij}$, $\tilde R_{ij}\rightarrow 0$ (for $\w{f}$ is
flat), $\tD_i\rightarrow \Df_i$ and $C^k_{\ \, ij}\rightarrow \Delta^k_{\ \, ij}$
[compare Eqs.~(\ref{e:cfd:Ckij_diffGam}) and (\ref{e:sch:def_Delta})]
and using property (\ref{e:sch:trace_Delta}), 
we get immediately Eq.~(\ref{e:sch:Rij3}). 
\end{remark}

If we substitute expression (\ref{e:sch:Delta_Dftgm}) for $\Delta^k_{\ \, ij}$
into Eq.~(\ref{e:sch:Rij3}), we get
\bea
	\tilde R_{ij} & = & \frac{1}{2} 
	\Df_k \left[ \tgm^{kl} \left(
	\Df_i \tgm_{lj} + \Df_j \tgm_{il} - \Df_l \tgm_{ij} \right) \right]
	- \Delta^k_{\ \, il} \Delta^l_{\ \, kj} \nonumber \\
	& = & \frac{1}{2} \bigg\{
	\Df_k  \bigg[ \Df_i (\underbrace{\tgm^{kl} \tgm_{lj}}_{\delta^k_{\ \, j}})
	- \tgm_{lj} \Df_i \tgm^{kl}
	+  \Df_j (\underbrace{\tgm^{kl} \tgm_{il}}_{\delta^k_{\ \, i}})
	- \tgm_{il} \Df_j \tgm^{kl} \bigg]
	- \Df_k \tgm^{kl} \, \Df_l \tgm_{ij}
	-  \tgm^{kl} \Df_k \Df_l \tgm_{ij} \bigg\} \nonumber \\
	& & - \Delta^k_{\ \, il} \Delta^l_{\ \, kj} \nonumber \\
	& = &  \frac{1}{2} \bigg(
	- \Df_k \tgm_{lj} \, \Df_i \tgm^{kl}
	- \tgm_{lj} \Df_k \Df_i \tgm^{kl}
	- \Df_k \tgm_{il} \, \Df_j \tgm^{kl}
	- \tgm_{il} \Df_k \Df_j \tgm^{kl} 
	- \Df_k \tgm^{kl} \, \Df_l \tgm_{ij}\nonumber \\
	& & \qquad -  \tgm^{kl} \Df_k \Df_l \tgm_{ij}
	\bigg) - \Delta^k_{\ \, il} \Delta^l_{\ \, kj} . 
\eea
Hence we can write, using $\Df_k \Df_i = \Df_i \Df_k$ (since $\w{f}$ is flat)
and exchanging some indices $k$ and $l$, 
\be \label{e:sch:Rij4}
	\encadre{ \tilde R_{ij} = - \frac{1}{2} 
	\left( \tgm^{kl} \Df_k \Df_l \tgm_{ij}
	+ \tgm_{ik} \Df_j \Df_l \tgm^{kl} 
	+ \tgm_{jk} \Df_i \Df_l \tgm^{kl} \right)
	+ \mathcal{Q}_{ij}(\wtgm,\wDf\wtgm) }  , 
\ee
where 
\be \label{e:sch:def_Qij}
	\mathcal{Q}_{ij}(\wtgm,\wDf\wtgm) := - \frac{1}{2} 
	\left(  
	\Df_k \tgm_{lj} \, \Df_i \tgm^{kl} 
	+ \Df_k \tgm_{il} \, \Df_j \tgm^{kl} 
	+ \Df_k \tgm^{kl} \, \Df_l \tgm_{ij}
	\right) 
	- \Delta^k_{\ \, il} \Delta^l_{\ \, kj}
\ee
is a term which does not contain any second derivative of $\wtgm$
and which is quadratic in the first derivatives. 

\subsection{Reducing the Ricci tensor to a Laplace operator}

If we consider the Ricci tensor as a differential operator acting on 
the conformal metric $\wtgm$, its principal part (or \emph{principal symbol},
cf. Sec.~\ref{s:cko:ellip_char}) is given by the three terms involving
second derivatives in the right-hand side of Eq.~(\ref{e:sch:Rij4}).
We recognize in the first term, $ \tgm^{kl} \Df_k \Df_l \tgm_{ij}$, 
a kind of Laplace operator acting on $\tgm_{ij}$. 
Actually, for a weak gravitational field, i.e. for $\tgm^{ij} = f^{ij} + h^{ij}$
with $f_{ik} f_{jl} h^{kl} h^{ij} \ll 1$, we have, at the linear order in $\w{h}$,
$ \tgm^{kl} \Df_k \Df_l \tgm_{ij} \simeq \Delta_{\w{f}} \tgm_{ij}$,
where $\Delta_{\w{f}}=f^{kl}\Df_k\Df_l$ is the Laplace operator associated with 
the metric $\w{f}$. If we combine Eqs.~(\ref{e:cfd:Einstein2}) and 
(\ref{e:cfd:Einstein4}), the Laplace operator in $\tilde R_{ij}$
gives rise to a \emph{wave operator} 
for $\tgm_{ij}$, namely
\be
	\left[ \left(\der{}{t} - \Liec{\beta} \right) ^2 
	- \frac{N^2}{\Psi^4} \tgm^{kl} \Df_k \Df_l \right]
	\tgm_{ij} = \cdots 
\ee
Unfortunately the other two terms that involve second derivatives
in Eq.~(\ref{e:sch:Rij4}), namely $\tgm_{ik} \Df_j \Df_l \tgm^{kl}$ and
$\tgm_{jk} \Df_i \Df_l \tgm^{kl}$, spoil the elliptic character of the operator
acting on $\tgm_{ij}$ in $\tilde R_{ij}$, so that the combination of 
Eqs.~(\ref{e:cfd:Einstein2}) and (\ref{e:cfd:Einstein4}) does no longer lead to
a wave operator. 

To restore the Laplace operator, Shibata and Nakamura \cite{ShibaN95} have
considered the term $\Df_l \tgm^{kl}$ which appears in the 
second and third terms of Eq.~(\ref{e:sch:Rij4}) as a variable independent
from $\tgm_{ij}$. We recognize in this term the opposite of the vector 
$\w{\tilde\Gamma}$ that has been introduced in Sec.~\ref{s:evo:Gamma_freez}
[cf. Eq.~(\ref{e:evo:tGam_divtgm})]:
\be \label{e:sch:tGam_divtgm}
	\encadre{ \tilde\Gamma^i = - \Df_j \tgm^{ij} } .
\ee
Equation~(\ref{e:sch:Rij4}) then becomes
\be \label{e:sch:Rij_tGam}
	\encadre{ \tilde R_{ij} =  \frac{1}{2} 
	\left( - \tgm^{kl} \Df_k \Df_l \tgm_{ij}
	+ \tgm_{ik} \Df_j \tilde\Gamma^k
	+ \tgm_{jk} \Df_i \tilde\Gamma^k \right)
	+ \mathcal{Q}_{ij}(\wtgm,\wDf\wtgm) }  .
\ee
\begin{remark}
Actually, Shibata and Nakamura \cite{ShibaN95} have introduced the covector
$F_i := \Df^j \tgm_{ij}$ instead of $\tilde\Gamma^i$. As 
Eq.~(\ref{e:evo:F_Gam}) shows, the two quantities are closely related. 
They are even equivalent in the linear regime. 
The quantity $\tilde\Gamma^i$ has been introduced by Baumgarte and Shapiro 
\cite{BaumgS99}. It has the advantage over $F_i$ to encompass all the 
second derivatives of $\tgm_{ij}$ that are not part of the Laplacian. 
If one use $F_i$, this is true only at the linear order (weak field region). 
Indeed, by means of Eq.~(\ref{e:evo:F_Gam}), we can write
\be
	\tilde R_{ij} = \frac{1}{2} 
	\left( - \tgm^{kl} \Df_k \Df_l \tgm_{ij}
	+ \Df_j F_i + \Df_i F_j + h^{kl} \Df_i \Df_k \tgm_{jl}
	+ h^{kl} \Df_i \Df_k \tgm_{jl} \right)
	+ {\mathcal{Q}'}_{ij}(\wtgm,\wDf\wtgm) , 
\ee
where $h^{kl} := \tgm^{kl} - f^{kl}$. When compared with
(\ref{e:sch:Rij_tGam}), the above expression contains the additional 
terms $h^{kl} \Df_i \Df_k \tgm_{jl}$ and $h^{kl} \Df_i \Df_k \tgm_{jl}$,
which are quadratic in the deviation of $\wtgm$ from the flat metric. 
\end{remark} 

The Ricci scalar $\tilde R$, which appears along $\tilde R_{ij}$
in Eq.~(\ref{e:cfd:Einstein4}), is deduced from the 
trace of Eq.~(\ref{e:sch:Rij_tGam}): 
\bea
	\tilde R & = & \tgm^{ij} \tilde R_{ij}
	\ =\   \frac{1}{2} 
	\bigg( - \tgm^{kl} \tgm^{ij}\Df_k \Df_l \tgm_{ij}
	+ \underbrace{\tgm^{ij}\tgm_{ik}}_{=\delta^j_{\ \, k}} 
		\Df_j \tilde\Gamma^k
	+ \underbrace{\tgm^{ij}\tgm_{jk}}_{=\delta^i_{\ \, k}}  
	\Df_i \tilde\Gamma^k \bigg)
	+ \tgm^{ij} \mathcal{Q}_{ij}(\wtgm,\wDf\wtgm) \nonumber \\
	& = & \frac{1}{2} \left[ 
	\tgm^{kl} \Df_k \left( \tgm^{ij} \Df_l \tgm_{ij} \right)
	+ \tgm^{kl} \Df_k \tgm^{ij} \, \Df_l \tgm_{ij} 
	+ 2 \Df_k \tilde\Gamma^k \right]
	+ \tgm^{ij} \mathcal{Q}_{ij}(\wtgm,\wDf\wtgm) . 
\eea
Now, from Eq.~(\ref{e:sch:Delta_Dftgm}), 
$\tgm^{ij} \Df_l \tgm_{ij} = 2 \Delta^k_{\ \, lk}$, and from 
Eq.~(\ref{e:sch:trace_Delta}), $\Delta^k_{\ \, lk}=0$. 
Thus the first term in the right-hand side of the above equation vanishes
and we get 
\be \label{e:sch:tR_scal}
	\encadre{\tilde R = \Df_k \tilde\Gamma^k + 
	\mathcal{Q}(\wtgm,\wDf\wtgm) },  
\ee
where
\be \label{e:sch:def_Q}
	\mathcal{Q}(\wtgm,\wDf\wtgm) := 
	\frac{1}{2} \tgm^{kl} \Df_k \tgm^{ij} \, \Df_l \tgm_{ij}
	+ \tgm^{ij} \mathcal{Q}_{ij}(\wtgm,\wDf\wtgm)
\ee
is a term that does not contain any second derivative of $\wtgm$
and is quadratic in the first derivatives. 

The idea of introducing auxiliary variables, such as $\tilde\Gamma^i$ or $F_i$, 
to reduce the Ricci tensor to a Laplace-like operator  
traces back to Nakamura, Oohara and Kojima (1987) \cite{NakamOK87}.
In that work, such a treatment was performed for the Ricci tensor $\w{R}$ of the physical 
metric $\wgm$, whereas in Shibata and Nakamura's study (1995) \cite{ShibaN95},
it was done for the Ricci tensor $\w{\tilde R}$ of the conformal metric $\wtgm$.
The same considerations had been put forward much earlier for the four-dimensional
Ricci tensor ${}^4\!\w{R}$. 
Indeed, this is the main motivation for the \emph{harmonic coordinates} 
mentioned in Sec.~\ref{s:evo:harm_slic}: de Donder \cite{DeDon21} introduced
these coordinates in 1921 in order to write the principal part of the 
Ricci tensor as a wave operator acting on the metric coefficients $g_{\alpha\beta}$:
\be \label{e:sch:4R_harm}
	{}^4\! R_{\alpha\beta} = - \frac{1}{2} g^{\mu\nu} \der{}{x^\mu} 
	\der{}{x^\nu} g_{\alpha\beta} 
	+ \mathcal{Q}_{\alpha\beta}(\w{g},\w{\partial g}) , 
\ee
where $\mathcal{Q}_{\alpha\beta}(\w{g},\w{\partial g})$ is a term which does not
contain any second derivative of $\w{g}$ and which is quadratic in the
first derivatives. 
In the current context, the analogue of harmonic coordinates would be
to set $\tilde\Gamma^i = 0$, for then Eq.~(\ref{e:sch:Rij_tGam}) would
resemble Eq.~(\ref{e:sch:4R_harm}). The choice $\tilde\Gamma^i = 0$
corresponds to the \emph{Dirac gauge} discussed in Sec.~\ref{s:evo:Dirac}. 
However the philosophy of the BSSN formulation is to leave 
free the coordinate choice, allowing for any value of 
$\tilde\Gamma^i$. In this respect, a closer 4-dimensional analogue of BSSN 
is the \emph{generalized harmonic decomposition} 
\index{harmonic decomposition!generalized} introduced
by Friedrich (1985) \cite{Fried85} and Garfinkle (2002) \cite{Garfi02}
(see also Ref.~\cite{GundlCHM05,LindbSKOR06}) and implemented by Pretorius for the binary
black hole problem \cite{Preto05a,Preto05b,Preto06}. 

The allowance for any coordinate system means that $\tilde\Gamma^i$ becomes a new variable, 
in addition to $\tgm_{ij}$, $\tA_{ij}$, $\Psi$, $K$, $N$ and $\beta^i$. 
One then needs an evolution
equation for it. But we have already derived such an equation in 
Sec.~\ref{s:evo:Gamma_freez}, namely Eq.~(\ref{e:evo:dGamdt}). 
Equation~(\ref{e:sch:tGam_divtgm}) is then a constraint on the system, 
in addition to the Hamiltonian and momentum constraints. 

\subsection{The full scheme}

By collecting together Eqs.~(\ref{e:cfd:Einstein1})-(\ref{e:cfd:Einstein4}), 
(\ref{e:sch:Rij_tGam}), (\ref{e:sch:tR_scal}) and (\ref{e:evo:dGamdt}), we can write 
the complete system of evolution equations for the BSSN scheme: 
\bea
  & &	\encadre{ \left( \der{}{t} - \Liec{\beta} \right) \Psi =
	\frac{\Psi}{6} \left( \tD_i \beta^i  - NK \right) } \label{e:sch:BSSN1} \\
 & & \encadre{ \left( \der{}{t} - \Liec{\beta} \right) \tgm_{ij}
	= - 2N  \tA_{ij} - \frac{2}{3} \tD_k \beta^k \, \tgm_{ij} } 
	\label{e:sch:BSSN2} \\
 &  & \encadre{ \left(\der{}{t} - \Liec{\beta} \right) K 
	= - \Psi^{-4} \left( \tD_i \tD^i N + 2 \tD_i \ln \Psi \, \tD^i N \right) 
	 + N \left[ 4\pi (E+S) 
	+  \tA_{ij} \tA^{ij} + \frac{K^2}{3}\right] } \nonumber \\
	& & \label{e:sch:BSSN3} \\
  & & 		\encadre{
	\begin{array}{lcl}
	\displaystyle \left(\der{}{t} - \Liec{\beta} \right) \tA_{ij} & = & 
	\displaystyle - \frac{2}{3} \tD_k \beta^k\,  \tA_{ij} +
	N \left[ K\tA_{ij} - 2 \tgm^{kl} \tA_{ik} \tA_{jl}
	- 8\pi \left(\Psi^{-4} S_{ij} - \frac{1}{3} S \tgm_{ij} \right) \right] \\
	& & \displaystyle + \Psi^{-4} \bigg\{ - \tD_i \tD_j N 
		+ 2 \tD_i \ln\Psi\,  \tD_j N + 2 \tD_j \ln\Psi\, \tD_i N \\
	& & \displaystyle \qquad  \quad  
		+ \frac{1}{3}\left( \tD_k \tD^k N - 4 \tD_k\ln\Psi\, 
	\tD^k N \right) \tgm_{ij} \\
	 & & \displaystyle\qquad \quad
		 + N \bigg[ \frac{1}{2} 
	\left( - \tgm^{kl} \Df_k \Df_l \tgm_{ij}
	+ \tgm_{ik} \Df_j \tilde\Gamma^k
	+ \tgm_{jk} \Df_i \tilde\Gamma^k \right)
	+ \mathcal{Q}_{ij}(\wtgm,\wDf\wtgm) \\
& & \displaystyle\qquad \quad - \frac{1}{3} 
\left(\Df_k \tilde\Gamma^k + 
	\mathcal{Q}(\wtgm,\wDf\wtgm) \right) \tgm_{ij} 
	- 2\tD_i\tD_j \ln\Psi + 4\tD_i \ln\Psi\, \tD_j\ln\Psi \\
	& & \displaystyle \qquad \qquad \quad + \frac{2}{3}
	\left( \tD_k \tD^k \ln\Psi - 2\tD_k\ln\Psi \, \tD^k \ln\Psi \right)
	\tgm_{ij} \bigg] \bigg\} .	
	\end{array}
	}   \nonumber \\
& &  \label{e:sch:BSSN4} \\
& & \encadre{
	\begin{array}{lcl}
	\displaystyle \left( \der{}{t} - \Liec{\beta} \right) \tilde\Gamma^i 
   & = & \displaystyle \frac{2}{3} \Df_k\beta^k \, \tilde\Gamma^i  +
	\tgm^{jk} \Df_j \Df_k \beta^i + \frac{1}{3} \tgm^{ij}
	\Df_j \Df_k \beta^k 
	- 2 \tA^{ij} \Df_j N \\
  & & \displaystyle - 2N \left[ 8\pi \Psi^4 p^i 
	- \tA^{jk}  \Delta^i_{\ \, jk}
	- 6 \tA^{ij} \Df_j\ln \Psi 
	+ \frac{2}{3} \tgm^{ij} \Df_j K \right]
	\end{array}
	} , \label{e:sch:BSSN5} 
\eea
where $\mathcal{Q}_{ij}(\wtgm,\wDf\wtgm)$ and $\mathcal{Q}(\wtgm,\wDf\wtgm)$
are defined by Eqs.~(\ref{e:sch:def_Qij}) and (\ref{e:sch:def_Q}) and we have used $\Liec{\beta}\tilde\Gamma^i = \beta^k \Df_k \tilde\Gamma^i 
- \tilde\Gamma^k \Df_k \beta^i$ to rewrite Eq.~(\ref{e:evo:dGamdt}).
These equations must be supplemented with the constraints 
(\ref{e:cfd:Einstein5}) (Hamiltonian constraint), (\ref{e:cfd:Einstein6})
(momentum constraint), (\ref{e:cfd:dettgm_f}) (``unit'' determinant of $\tgm_{ij}$),
(\ref{e:cfd:tA_traceless}) ($\w{\tA}$ traceless)
and  (\ref{e:sch:tGam_divtgm}) (definition of $\w{\tilde\Gamma}$):
\bea
   & & \encadre{ \tD_i \tD^i \Psi -\frac{1}{8} {\tilde R} \Psi
	+ \left( \frac{1}{8} \tA_{ij} \tA^{ij}
	- \frac{1}{12} K^2 + 2\pi E \right) \Psi^5 = 0 }  \label{e:sch:BSSN6} \\
    & & \encadre{ \tD^j \tA_{ij} + 6 \tA_{ij} \tD^j \ln\Psi - \frac{2}{3}
	\tD_i K = 8\pi p_i }  \label{e:sch:BSSN7} \\
    & & \encadre{ \det (\tgm_{ij}) = f } \label{e:sch:BSSN8} \\
    & & \encadre{ \tgm^{ij} \tA_{ij} = 0 } \label{e:sch:BSSN9} \\
    & & \encadre{ \tilde\Gamma^i + \Df_j \tgm^{ij} = 0 } . \label{e:sch:BSSN10}
\eea

The unknowns for the BSSN system are $\Psi$, $\tgm_{ij}$, $K$, $\tA_{ij}$ and
$\tilde\Gamma^i$. They involve $1+6+1+6+3=17$ components, which are evolved via
the 17-component equations (\ref{e:sch:BSSN1})-(\ref{e:sch:BSSN5}).
The constraints (\ref{e:sch:BSSN6})-(\ref{e:sch:BSSN10}) involve 
$1+3+1+1+3=9$ components, reducing the number of degrees of freedom to $17-9=8$. 
The coordinate choice, via the lapse function $N$ and the shift vector $\beta^i$, 
reduces this number to $8-4=4=2\times 2$, 
which corresponds to the 2 degrees of freedom
of the gravitational field expressed in terms of the couple 
$(\tgm_{ij},\tA_{ij})$.

The complete system to be solved must involve some additional equations
resulting from the choice of lapse $N$ and shift vector $\w{\beta}$, as discussed
in Chap.~\ref{s:evo}. The well-posedness of the whole system is	
discussed in Refs.~\cite{BeyerS04} and \cite{GundlM06}, for some usual 
coordinate choices, like harmonic slicing (Sec.~\ref{s:evo:harm_slic})
with hyperbolic gamma driver (Sec.~\ref{s:evo:Gam_drivers}).

\subsection{Applications}

The BSSN scheme is by far the most widely used evolution scheme in contemporary
numerical relativity. It has notably been used for computing 
gravitational collapses 
\cite{ShibaBS00a,Shiba03a,SekigS05,ShibaS05,BaiotHMLRSFS05,BaiotHRS05,BaiotR06},
mergers of binary
neutron stars \cite{Shiba99c,ShibaU00,ShibaU02,ShibaTU03,ShibaTU05,ShibaT06}
and mergers of binary black holes \cite{BakerCCKV06a,BakerCCKV06b,VanMeBKC06,CampaLMZ06,CampaLZ06a,CampaLZ06b,%
CampaLZ06c,Sperh07,DieneHPSSTTV06,BruegGHHST07,MarroTBGHHS07,HerrmSL07,HerrmHSLM07}.
In addition, most recent codes for general relativistic MHD employ the BSSN formulation
\cite{DuezLSS05,ShibaS05b,ShibaLSS06,GiacoR07}.

\appendix
%  
%    Annexe : Lie derivative
%
% $Date: 2007-03-06 11:59:03 +0100 (mar, 06 mar 2007) $
% $Rev: 183 $
% $Author: e_gourgoulhon $
%%%%%%%%%%%%%%%%%%%%%%%%%%%%%

\chapter{Lie derivative} \label{s:lie}

%\verb$Date: 2007-03-06 11:59:03 +0100 (mar, 06 mar 2007) $

\minitoc
\vspace{1cm}

%%%%%%%%%%%%%%%%%%%%%%%%%%%%%%%%%%%%%%%%%%%%%%%%%%%%%%%%%%%%%%%%%%%%%%%%%%%%

\section{Lie derivative of a vector field}

\subsection{Introduction}

Genericaly the ``derivative'' of some vector field $\w{v}$ on $\M$
is to be constructed for the variation $\delta\w{v}$
of $\w{v}$ between two neighbouring points $p$ and $q$. 
Naively, one would write $\delta\w{v} = \w{v}(q)-\w{v}(p)$.
However $\w{v}(q)$ and $\w{v}(p)$ belong to different vector spaces:
$\T_q(\M)$ and $\T_p(\M)$.
Consequently the subtraction $\w{v}(q)-\w{v}(p)$ is ill defined.
To proceed in the definition of the derivative of a vector field, one must
introduce some extra-structure on the manifold $\M$: this can be either
some \emph{connection} $\w{\nabla}$ (as the Levi-Civita connection associated with
the metric tensor $\w{g}$), leading to the \emph{covariant derivative}
$\w{\nabla}\w{v}$ or 
another vector field $\w{u}$, leading to the derivative of $\w{v}$ along $\w{u}$
which is the \emph{Lie derivative} discussed in this Appendix.  
These two types of derivative generalize straightforwardly to
any kind of tensor field. For the specific kind of tensor fields constituted by
differential forms, there exists a third type of derivative, which does not 
require any extra structure on $\M$: the \emph{exterior derivative}
(see the classical textbooks \cite{MisneTW73,Wald84,Strau04} or
Ref.~\cite{Gourg06} for an introduction). 

\subsection{Definition}

\begin{figure}
\centerline{\includegraphics[width=0.6\textwidth]{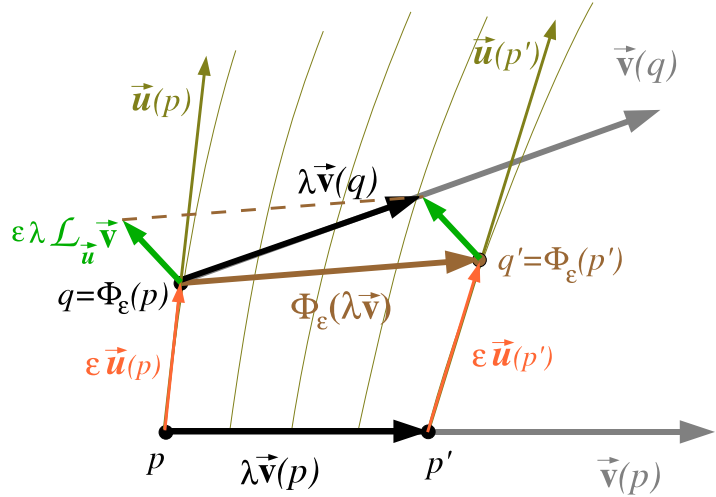}}
\caption[]{\label{f:lie:deriv} Geometrical construction of the Lie derivative of a
vector field: given a small parameter $\lambda$, each extremity of the arrow
$\lambda\w{v}$ is dragged by some small parameter $\varepsilon$ 
along $\w{u}$, to form
the vector denoted by $\Phi_\varepsilon(\lambda\w{v})$. The latter is then compared with
the actual value of $\lambda\w{v}$ at the point $q$, the difference (divided 
by $\lambda\varepsilon$) defining the Lie derivative $\Lie{u}\w{v}$.}
\end{figure}

Consider a vector field $\w{u}$ on $\M$, called hereafter the \emph{flow}.
Let $\w{v}$ be another vector field on $\M$, the variation of which is to be studied.
We can use the flow $\w{u}$ to transport the vector $\w{v}$ from one point $p$ to
a neighbouring one $q$ and then define rigorously the variation of $\w{v}$
as the difference between the actual value of $\w{v}$ at $q$ and the transported
value via $\w{u}$. More precisely the definition of the Lie derivative of 
$\w{v}$ with respect to $\w{u}$ is as follows (see Fig.~\ref{f:lie:deriv}).
We first define the image $\Phi_\varepsilon(p)$ of the point $p$ by the transport by an infinitesimal ``distance'' $\varepsilon$ along the field lines of $\w{u}$ as 
$\Phi_\varepsilon(p)=q$, where $q$ is the point close to $p$ such that
$\overrightarrow{pq}=\varepsilon\w{u}(p)$.
Besides, if we multiply the vector $\w{v}(p)$ by 
some infinitesimal parameter $\lambda$, it becomes an infinitesimal vector at $p$.
Then there exists a unique point $p'$ close to $p$ such that 
$\lambda\w{v}(p)=\overrightarrow{pp'}$.
We may transport the point $p'$ to a point $q'$ along the field lines of
$\w{u}$ by the same ``distance'' $\varepsilon$ as that used to transport
$p$ to $q$: $q'=\Phi_\varepsilon(p')$ (see Fig.~\ref{f:lie:deriv}). $\overrightarrow{qq'}$ is then an
infinitesimal vector at $q$ and we
define the transport by the distance $\varepsilon$ of the vector $\w{v}(p)$ 
along the field lines of $\w{u}$ according to
\be
	\Phi_\varepsilon(\w{v}(p)) := \frac{1}{\lambda} \, \overrightarrow{qq'}.
\ee
$\Phi_\varepsilon(\w{v}(p))$ is vector in $\T_q(\M)$. We may then subtract it from the
actual value of the field $\w{v}$ at $q$ and define the \defin{Lie derivative}
of $\w{v}$ along $\w{u}$ by
\be
	\Lie{u} \w{v} := \lim_{\varepsilon\rightarrow 0} \frac{1}{\varepsilon}
	\left[ \w{v}(q) - \Phi_\varepsilon(\w{v}(p)) \right] .
\ee

If we consider a coordinate system $(x^\alpha)$ adapted to the
field $\w{u}$ in the sense that $\w{u}=\w{e}_0$ where $\w{e}_0$ is the first
vector of the natural basis associated with the coordinates $(x^\alpha)$, then
the Lie derivative is simply given by the partial derivative of the vector components
with respect to $x^0$:
\be \label{e:Lie_adapted}
	\left( \Lie{u} \w{v} \right)^\alpha = \der{v^\alpha}{x^0} .
\ee
In an arbitrary coordinate system, this formula is generalized to 
\be
	\encadre{ \Liec{u} v^\alpha = u^\mu \der{v^\alpha}{x^\mu}
	- v^\mu \der{u^\alpha}{x^\mu} }, 
\ee
where use has been made of the standard notation 
$\Liec{u} v^\alpha := \left( \Lie{u} \w{v} \right)^\alpha$.
The above relation shows that the Lie derivative of a vector with respect to another
one is nothing but the commutator of these two vectors:
\be
	\encadre{ \Lie{u} \w{v} = [\w{u},\w{v}] } . 
\ee

\section{Generalization to any tensor field}

The Lie derivative is extended to any tensor field by (i) demanding that for
a scalar field $f$, $\Lie{u} f = \langle\dd f,\w{u}\rangle$ and (ii) using the Leibniz
rule. As a result, the Lie derivative $\Lie{u}\w{T}$ of a tensor field $\w{T}$ of type 
$\left({k \atop \ell}\right)$ is a tensor field of the same type, the components of which
with respect to a given coordinate system $(x^\alpha)$ are
\be
\Liec{u} T^{\alpha_1\ldots\alpha_k}_{\qquad\ \; \beta_1\ldots\beta_\ell}=
u^\mu \der{}{x^\mu} T^{\alpha_1\ldots\alpha_k}_{\qquad\ \; \beta_1\ldots\beta_\ell} 
- \sum_{i=1}^k T^{\alpha_1\ldots
\!{{{\scriptstyle i\atop\downarrow}\atop \scriptstyle\sigma}\atop\ }\!\!
\ldots\alpha_k}_{\qquad\ \ \ \  \  \  \beta_1\ldots\beta_\ell}
 \; \der{u^{\alpha_i}}{x^\sigma} 
+  \sum_{i=1}^\ell T^{\alpha_1\ldots\alpha_k}_{\qquad\ \; \beta_1\ldots
\!{\ \atop {\scriptstyle\sigma \atop {\uparrow\atop \scriptstyle i}} }\!\!
\ldots\beta_\ell} 
\; \der{u^{\sigma}}{x^{\beta_i}} . \label{e:Lie_der_comp}
\ee 
In particular, for a 1-form,
\be \label{e:Lie_der_1form}
	\Liec{u} \omega_\alpha = u^\mu \der{\omega_\alpha}{x^\mu}
	+ \omega_\mu \der{u^\mu}{x^\alpha} .
\ee
Notice that the partial derivatives in Eq.~(\ref{e:Lie_der_comp}) can be 
remplaced by any connection without torsion, such as the Levi-Civita connection
$\wnab$ associated with the metric $\w{g}$, yielding
\be
\Liec{u} T^{\alpha_1\ldots\alpha_k}_{\qquad\ \; \beta_1\ldots\beta_\ell}=
u^\mu \nabla_\mu T^{\alpha_1\ldots\alpha_k}_{\qquad\ \; \beta_1\ldots\beta_\ell} 
- \sum_{i=1}^k T^{\alpha_1\ldots
\!{{{\scriptstyle i\atop\downarrow}\atop \scriptstyle\sigma}\atop\ }\!\!
\ldots\alpha_k}_{\qquad\ \ \ \  \  \  \beta_1\ldots\beta_\ell}
 \; \nabla_\sigma u^{\alpha_i} 
+  \sum_{i=1}^\ell T^{\alpha_1\ldots\alpha_k}_{\qquad\ \; \beta_1\ldots
\!{\ \atop {\scriptstyle\sigma \atop {\uparrow\atop \scriptstyle i}} }\!\!
\ldots\beta_\ell} 
\; \nabla_{\beta_i} u^{\sigma} . \label{e:lie:der_comp_nab}
\ee

%  
%    Annexe : 
%
% $Date: 2007-03-05 22:39:07 +0100 (lun, 05 mar 2007) $
% $Rev: 182 $
% $Author: e_gourgoulhon $
%%%%%%%%%%%%%%%%%%%%%%%%%%%%%

\chapter{Conformal Killing operator and conformal vector Laplacian} \label{s:cko}

%\verb$Date: 2007-03-05 22:39:07 +0100 (lun, 05 mar 2007) $

\minitoc
\vspace{1cm}

%%%%%%%%%%%%%%%%%%%%%%%%%%%%%%%%%%%%%%%%%%%%%%%%%%%%%%%%%%%%%%%%%%%%%%%%%%%%

In this Appendix, we investigate the main properties of two important
vectorial operators on Riemannian manifolds: the \emph{conformal Killing operator}
and the associated \emph{conformal vector Laplacian}. 
The framework is that of a single three-dimensional manifold $\Sigma$,
endowed with a positive definite metric (i.e. a Riemannian metric). 
In practice, $\Sigma$ is embedded in some spacetime $(\M,\w{g})$, 
as being part of a 3+1 foliation $(\Sigma_t)_{t\in\R}$, but we shall
not make such an assumption here. For concreteness, we shall denote
$\Sigma$'s Riemannian metric by $\wtgm$, because in most
applications of the 3+1 formalism, the conformal Killing operator
appears for the metric $\wtgm$ conformally related to the physical
metric $\wgm$ and introduced in Chap.~\ref{s:cfd}. 
But again, we shall not use the hypothesis that $\wtgm$ is derived from
some ``physical'' metric $\wgm$.
So in all what follows, $\wtgm$ can be replaced by the 
physical metric $\wgm$ or any other Riemannian metric, as for
instance the background metric $\w{f}$ introduced in Chap.~\ref{s:cfd}
and \ref{s:glo}. 

\section{Conformal Killing operator}

\subsection{Definition}

The \defin{conformal Killing operator} $\w{\tilde L}$ associated with 
the metric $\wtgm$ is the linear mapping from 
the space $\T(\Sigma)$ of vector fields on $\Sigma$ to the space 
of symmetric tensor fields of type $\left({2\atop 0}\right)$
defined by
\be \label{e:cko:conf_Killing_def}
	\encadre{\forall \w{v}\in\T(\Sigma),\quad
		 (\tilde L v)^{ij} := \tD^i v^j + \tD^j v^i 
			- \frac{2}{3} \tD_k v^k \, \tgm^{ij} } ,
\ee
where $\wtD$ is the Levi-Civita connection associated with $\wtgm$
and $\tD^i := \tgm^{ij} \tD_j$.
An immediate property of $\w{\tilde L}$ is to be traceless with respect
to $\wtgm$, thanks to the $-2/3$ factor: for any vector $\w{v}$,
\be
	\tgm_{ij} (\tilde L v)^{ij} = 0 . 
\ee

\subsection{Behavior under conformal transformations}

An important property of $\w{\tilde L}$ is to be invariant, except for some scale
factor, with respect to conformal transformations. 
Indeed let us consider a metric $\wgm$ conformally related 
to $\wtgm$:
\be \label{e:cko:gm_Psi_tgm}
	\wgm = \Psi^4 \wtgm . 
\ee
In practice $\wgm$ will be the metric induced on $\Sigma$ by the spacetime
metric $\w{g}$  and $\Psi$ the conformal factor defined in Chap.~\ref{s:cfd}, but
we shall not employ this here. So $\wgm$ and $\wtgm$ are any two 
Riemannian metrics on $\Sigma$
that are conformally related (we could have called them $\wgm_1$ and $\wgm_2$)
and $\Psi$ is simply the conformal factor between them. 
We can employ the formul\ae\ derived in Chap.~\ref{s:cfd} to relate the
conformal Killing operator of $\wtgm$, $\w{\tilde L}$, with that of
$\wgm$, $\w{L}$ say. Formula (\ref{e:cfd:der_vector}) gives
\bea
	D^j v^i & = & \gm^{jk} D_k v^i
		= \Psi^{-4} \tgm^{jk}
			\left[ \tD_k v^i + 2 \left( v^l \tD_l \ln\Psi \, \delta^i_{\ \, k}
		+ v^i \tD_k\ln\Psi - \tD^i\ln\Psi\, \tgm_{kl} v^l \right) 
			\right] \nonumber \\
	& = & \Psi^{-4} \left[ \tD^j v^i + 2 \left( v^k \tD_k \ln\Psi \, \tgm^{ij}
		+ v^i \tD^j \ln\Psi - v^j \tD^i\ln\Psi \right) 
			\right] .
\eea
Hence 
\be
	D^i v^j + D^j v^i = \Psi^{-4} \left( \tD^i v^j + \tD^j v^i
	+ 4  v^k \tD_k \ln\Psi \, \tgm^{ij} \right) 
\ee
Besides, from Eq.~(\ref{e:cfd:divergence_conf0}),
\be
	- \frac{2}{3} D_k v^k \, \gm^{ij}
		= - \frac{2}{3} \left( \tD_k v^k + 6 v^k \tD_k \ln \Psi
			\right) \Psi^{-4} \tgm^{ij} . 
\ee
Adding the above two equations, we get the simple relation
\be
	\encadre{ (L v)^{ij} = \Psi^{-4} (\tilde L v)^{ij} } . 
\ee
Hence the conformal Killing operator is invariant, up to the scale
factor $\Psi^{-4}$, under a conformal transformation. 

\subsection{Conformal Killing vectors} \label{s:cko:ckv}

Let us examine the kernel  of the conformal Killing operator, 
i.e. the subspace $\mathrm{ker}\, \w{\tilde L}$ of $\T(\Sigma)$
constituted by vectors $\w{v}$ satisfying
\be \label{e:cko:ckv_def}
	(\tilde L v)^{ij} = 0 . 
\ee
A vector field which obeys Eq.~(\ref{e:cko:ckv_def}) is called
a \defin{conformal Killing vector}. It is the generator of 
some conformal isometry of $(\Sigma,\wtgm)$.
A \defin{conformal isometry} is a diffeomorphism 
$\Phi: \Sigma \rightarrow \Sigma$ for which there exists some scalar field 
$\Omega$ such that $\Phi_* \wtgm = \Omega^2 \wtgm$. Notice that any isometry
is a conformal isometry (corresponding to $\Omega=1$), which means that
every Killing vector is a conformal Killing vector. The latter property is
obvious from the definition (\ref{e:cko:conf_Killing_def}) of the conformal
Killing operator. Notice also that any conformal isometry of $(\Sigma,\wtgm)$
is a conformal isometry of $(\Sigma,\wgm)$, where $\wgm$ is a metric 
conformally related to $\wtgm$ [cf. Eq.~(\ref{e:cko:gm_Psi_tgm})]. 
Of course, $(\Sigma,\wtgm)$ may not admit any conformal isometry at all,
yielding $\mathrm{ker}\, \w{\tilde L}=\{0\}$. The maximum dimension of
$\mathrm{ker}\, \w{\tilde L}$ is 10 (taking into account that $\Sigma$
has dimension 3). If $(\Sigma,\wtgm)$ is the Euclidean space $(\R^3,\w{f})$,
the conformal isometries are constituted by the isometries (translations, rotations)
augmented by the homotheties. 

%%%%%%%%%%%%%%%%%%%%%%%%%%%%%%%%%%%%%%%%%%%%%%%%%%%%%%%%%%%%%%%%%%%%%%%%%%%%%%%%%%%%%%%

\section{Conformal vector Laplacian} 

\subsection{Definition} \label{s:cko:def_cvlap}

The \defin{conformal vector Laplacian} associated with the metric $\wtgm$
is the endomorphism $\w{\tilde\Delta}_L$ of the space $\T(\Sigma)$
of vector fields on $\Sigma$ defined by taking the divergence 
of the conformal Killing operator:
\be
	\encadre{ \forall \w{v}\in\T(\Sigma),\quad
		\tilde\Delta_L \, v^i := \tD_j (\tilde L v)^{ij} } .
\ee
From Eq.~(\ref{e:cko:conf_Killing_def}),
\bea
	\tilde\Delta_L \, v^i & = & \tD_j \tD^i v^j 
	+ \tD_j \tD^j v^i - \frac{2}{3} \tD^i \tD_k v^k \nonumber \\
	& = &  \tD^i \tD_j v^j + {\tilde R}^i_{\ \, j} v^j 
	+ \tD_j \tD^j v^i - \frac{2}{3} \tD^i \tD_j v^j \nonumber \\
 	& = & \tD_j \tD^j v^i + \frac{1}{3} \tD^i \tD_j v^j
	+ {\tilde R}^i_{\ \, j} v^j  ,		\label{e:cko:DeltaL_DD}
\eea
where we have used the contracted Ricci identity (\ref{e:cfd:contr_Ricci_tD})
to get the second line. 
Hence $\tilde\Delta_L \, v^i$ is a second order operator acting on the vector
$\w{v}$, which is the sum of the vector Laplacian $\tD_j \tD^j v^i$, one third of the
gradient of divergence $ \tD^i \tD_j v^j$ and the curvature term 
${\tilde R}^i_{\ \, j} v^j$:
\be \label{e:cko:tDelta_DD_Ricci}
	\encadre{ \tilde\Delta_L \, v^i 
	= \tD_j \tD^j v^i + \frac{1}{3} \tD^i \tD_j v^j
	+ {\tilde R}^i_{\ \, j} v^j } 
\ee
The conformal vector Laplacian plays an important role
in 3+1 general relativity, 
for solving the constraint equations (Chap.~\ref{s:ini}), 
but also for the time evolution
problem (Sec.~\ref{s:evo:min_distort}).
The main properties of $\w{\tilde\Delta}_L$ have been first investigated
by York \cite{York73,York74}. 

\subsection{Elliptic character} \label{s:cko:ellip_char}

Given $p\in\Sigma$ and a linear form $\w{\xi}\in\T_p^*(\Sigma)$,
the \defin{principal symbol} of $\w{\tilde\Delta}_L$ with respect to $p$
and $\w{\xi}$ is the linear map $\w{P}_{(p,\w{\xi})}:\T_p(\Sigma) \rightarrow \T_p(\Sigma)$
defined as follows (see e.g. \cite{Dain06}). 
Keep only the terms involving the highest 
derivatives in $\w{\tilde\Delta}_L$ (i.e. the second order ones):
in terms of components, the operator is then reduced to
\be
	v^i \longmapsto \tgm^{jk} \der{}{x^j} \der{}{x^k} v^i
	+ \frac{1}{3} \tgm^{ik} \der{}{x^k} \der{}{x^j} v^j
\ee
Replace each occurrence of $\dert{}{x^j}$ by the component $\xi_j$ of
the linear form $\w{\xi}$, thereby obtaining a mapping which is no longer differential, i.e. that involves only values of the fields at the point $p$; 
this is the principal symbol of $\w{\tilde\Delta}_L$ at $p$ with respect
to $\w{\xi}$:
\be
	\begin{array}{rrcl}
	\w{P}_{(p,\w{\xi})} : &  \T_p(\Sigma) & \longrightarrow & \T_p(\Sigma) \\
	   &  \w{v} = (v^i) & \longmapsto & \displaystyle
	 \w{P}_{(p,\w{\xi})}(\w{v}) = \left( \tgm^{jk}(p)  \xi_j \xi_k \, v^i + \frac{1}{3}
		 \tgm^{ik}(p) \xi_k \xi_j v^j \right), 
	\end{array}
\ee
The differential operator $\w{\tilde\Delta}_L$ is said to be 
\defin{elliptic} on $\Sigma$
iff the principal symbol $\w{P}_{(p,\w{\xi})}$ is an isomorphism 
for every $p\in\Sigma$ and every non-vanishing linear form 
$\w{\xi}\in\T_p^*(\Sigma)$. It is said to be \defin{strongly elliptic}
if all the eigenvalues of $\w{P}_{(p,\w{\xi})}$ are non-vanishing 
and have the same sign. To check whether it is the case, let
us consider the bilinear form $\w{\tilde P}_{(p,\w{\xi})}$ associated to the endomorphism
$\w{P}_{(p,\w{\xi})}$ by the conformal metric:
\be
	\forall (\w{v},\w{w})\in \T_p(\Sigma)^2,\quad
	\w{\tilde P}_{(p,\w{\xi})}(\w{v},\w{w}) = \wtgm\left( 
	\w{v},\, \w{P}_{(p,\w{\xi})}(\w{w}) \right) . 
\ee
Its matrix $\tilde P_{ij}$ is deduced from the matrix $P^i_{\ \, j}$ of $\w{P}_{(p,\w{\xi})}$ by
lowering the index $i$ with $\wtgm(p)$. We get
\be
	\tilde P_{ij} = \tgm^{kl}(p) \xi_k \xi_l \, \tgm_{ij}(p)
	+ \frac{1}{3} \xi_i \xi_j . 
\ee
Hence $\w{\tilde P}_{(p,\w{\xi})}$ is clearly a symmetric bilinear form. 
Moreover it is positive definite for $\w{\xi}\not=0$: 
for any vector $\w{v}\in\T_p(\Sigma)$ such that $\w{v}\not=0$, we have
\be
    	\w{\tilde P}_{(p,\w{\xi})} (\w{v},\w{v}) =
	\tgm^{kl}(p) \xi_k \xi_l \, \tgm_{ij}(p) v^i v^j + 
	\frac{1}{3} (\xi_i v^i)^2 > 0 ,
\ee
where the $>0$ follows from the positive definite character of $\wtgm$.
$\w{\tilde P}_{(p,\w{\xi})}$ being positive definite symmetric bilinear form, 
we conclude that $\w{P}_{(p,\w{\xi})}$ is an isomorphism and that
all its eigenvalues are real and strictly positive. Therefore 
$\w{\tilde\Delta}_L$ is a strongly elliptic operator. 

\subsection{Kernel}

Let us now determine the kernel of $\w{\tilde\Delta}_L$.
Clearly this kernel contains the kernel of the conformal Killing
operator $\w{\tilde L}$. Actually it is not larger than that kernel:
\be \label{e:cko:ker_Delta}
	\encadre{ 
	\mathrm{ker} \, \w{\tilde\Delta}_L = \mathrm{ker}\, \w{\tilde L} } .
\ee
Let us establish this property. 
For any vector field $\w{v}\in\T(\Sigma)$, we have
\bea
	\int_{\Sigma} \tgm_{ij} v^i \tilde\Delta_L \, v^j \, 
	\sqrt{\tgm} \, d^3 x & = & \int_{\Sigma} \tgm_{ij} v^i \tD_l (\tilde L v)^{jl}
		\sqrt{\tgm} \, d^3 x \nonumber \\
  & = &  \int_{\Sigma} \left\{ \tD_l \left[ \tgm_{ij} v^i (\tilde L v)^{jl} \right]
  	- \tgm_{ij} \tD_l v^i \, (\tilde L v)^{jl} \right\} \sqrt{\tgm} \, d^3 x
		\nonumber \\
  & = & \oint_{\partial\Sigma} \tgm_{ij} v^i (\tilde L v)^{jl}  {\tilde s}_l
  	 \sqrt{\tilde q} \, d^2 y
	 - \int_{\Sigma} 
	 \tgm_{ij} \tD_l v^i \, (\tilde L v)^{jl} \sqrt{\tgm} \, d^3 x	,
	 	\nonumber \\
		\label{e:cko:int_vDelta1}
\eea
where the Gauss-Ostrogradsky theorem has been used to get the last line. 
We shall consider two situations for $(\Sigma, \wgm)$:
\begin{itemize}
\item $\Sigma$ is a \emph{closed manifold}, i.e. is compact without boundary;
\item $(\Sigma,\wtgm)$ is an \emph{asymptotically flat manifold}, in the sense
made precise in Sec.~\ref{s:glo:asymp_flat}. 
\end{itemize}
In the former case the lack of boundary of $\Sigma$ implies that the first
integral in the right-hand side of Eq.~(\ref{e:cko:int_vDelta1}) is zero.
In the latter case, we will restrict our attention to vectors $\w{v}$
which decay at spatial infinity according to (cf. Sec.~\ref{s:glo:asymp_flat})
\bea
	v^i & = & O(r^{-1})  \label{e:cko:v_decay1} \\
	\der{v^i}{x^j} & = & O(r^{-2})  \label{e:cko:v_decay2} ,
\eea
where the components are to be taken with respect to the asymptotically 
Cartesian coordinate system $(x^i)$ introduced in Sec.~\ref{s:glo:asymp_flat}.
The behavior (\ref{e:cko:v_decay1})-(\ref{e:cko:v_decay2}) implies
\be
	v^i (\tilde L v)^{jl} = O(r^{-3}) , 
\ee
so that the surface integral in Eq.~(\ref{e:cko:int_vDelta1}) vanishes. 
So for both cases of $\Sigma$ closed or asymptotically flat, Eq.~(\ref{e:cko:int_vDelta1})
reduces to 
\be \label{e:cko:int_vDelta2}
	\int_{\Sigma} \tgm_{ij} v^i \tilde\Delta_L \, v^j \, 
	\sqrt{\tgm} \, d^3 x = - \int_{\Sigma} 
	 \tgm_{ij} \tD_l v^i \, (\tilde L v)^{jl} \sqrt{\tgm} \, d^3 x .
\ee
In view of the right-hand side integrand, let us evaluate
\bea
	\tgm_{ij}  \tgm_{kl}  (\tilde L v)^{ik}(\tilde L v)^{jl}
	& = & \tgm_{ij}  \tgm_{kl} (\tD^i v^k + \tD^k v^i) (\tilde L v)^{jl}
			- \frac{2}{3} \tD_m v^m \, 
			\underbrace{\tgm^{ik} \tgm_{ij}}_{=\delta^k_{\ \, j}} 
				\tgm_{kl} (\tilde L v)^{jl} \nonumber \\
	& = & \left( \tgm_{kl} \tD_j v^k + \tgm_{ij} \tD_l v^i \right) 
	(\tilde L v)^{jl} - \frac{2}{3} \tD_m v^m \, 
		 	\underbrace{ \tgm_{jl} (\tilde L v)^{jl} }_{=0} \nonumber \\
	& = & 2 \tgm_{ij} \tD_l v^i \, (\tilde L v)^{jl} , 
\eea
where we have used the symmetry and the traceless property of $(\tilde L v)^{jl}$
to get the last line. 
Hence Eq.~(\ref{e:cko:int_vDelta2}) becomes
\be
	\int_{\Sigma} \tgm_{ij} v^i \tilde\Delta_L \, v^j \, 
	\sqrt{\tgm} \, d^3 x = - \frac{1}{2}
		\int_{\Sigma}  \tgm_{ij}  \tgm_{kl}  (\tilde L v)^{ik}
	(\tilde L v)^{jl}
		\sqrt{\tgm} \, d^3 x .	
\ee
Let us assume now that $\w{v}\in \mathrm{ker} \, \w{\tilde\Delta}_L$:
$\tilde\Delta_L \, v^j =0$. Then the left-hand side of the above equation
vanishes, leaving
\be
	\int_{\Sigma}  \tgm_{ij}  \tgm_{kl}  (\tilde L v)^{ik}(\tilde L v)^{jl}
		\sqrt{\tgm} \, d^3 x = 0 . 
\ee
Since $\wtgm$ is a positive definite metric, we conclude that 
$(\tilde L v)^{ij} = 0$, i.e. that $\w{v}\in \mathrm{ker} \, \w{\tilde L}$. 
This demonstrates property (\ref{e:cko:ker_Delta}). 
Hence the ``harmonic functions'' of the conformal vector Laplacian 
$\w{\tilde\Delta}_L$ are nothing but the conformal Killing vectors
(one should add ``which vanish at spatial infinity as (\ref{e:cko:v_decay1})-(\ref{e:cko:v_decay2})'' in the case of 
an asymptotically flat space). 

\subsection{Solutions to the conformal vector Poisson equation} \label{s:cko:Poisson}

Let now discuss the existence and uniqueness of solutions to the 
conformal vector Poisson equation
\be \label{e:cko:DeltaLv_S}
	\encadre{ \tilde\Delta_L \, v^i = S^i }, 
\ee
where the vector field $\w{S}$ is given (the source). 
Again, we shall distinguish two cases: the closed manifold case and
the asymptotically flat one. 
When $\Sigma$ is a closed manifold, we notice first
that a necessary condition for the solution to exist
is that the source must be orthogonal to any vector field in the kernel,
in the sense that 
\be  \label{e:cko:intCS}
	\forall \w{C}\in\mathrm{ker}\, \w{\tilde L},\quad 
	 \int_{\Sigma}  \tgm_{ij}  C^i S^j \sqrt{\tgm} \, d^3 x = 0  . 
\ee
This is easily established by replacing $S^j$ by $\tilde\Delta_L \, v^i$ 
and performing the same integration by part as above to get 
\be
	\int_{\Sigma}  \tgm_{ij}  C^i S^j \sqrt{\tgm} \, d^3 x = 
	- \frac{1}{2}
	\int_{\Sigma}  \tgm_{ij}  \tgm_{kl}  (\tilde L C)^{ik}(\tilde L v)^{jl}
		\sqrt{\tgm} \, d^3 x .
\ee
Since, by definition $(\tilde L C)^{ik}=0$, Eq.~(\ref{e:cko:intCS}) follows. 
If condition (\ref{e:cko:intCS}) is fulfilled (it may be trivial since
the metric $\wtgm$ may not admit any conformal Killing vector at all), 
it can be shown that Eq.~(\ref{e:cko:DeltaLv_S}) admits a solution 
and that this solution is unique up to the addition of a conformal 
Killing vector. 

In the asymptotically flat case, we assume that, in terms of the
asymptotically Cartesian coordinates $(x^i)$ introduced 
in Sec.~\ref{s:glo:asymp_flat}
\be \label{e:cko:decay_source_Poisson}
	S^i = O(r^{-3}) .
\ee
Moreover, because of the presence of the Ricci
tensor in $\w{\tilde\Delta}_L$, one must add the decay condition
\be
	\dderp{\tgm_{ij}}{x^k}{x^l} = O(r^{-3})	\label{e:cko:aflat_extra}
\ee
to the asymptotic flatness conditions introduced in 
Sec.~\ref{s:glo:asymp_flat} [Eqs.~(\ref{e:glob:aflat1}) to 
(\ref{e:glob:aflat4})]. Indeed Eq.~(\ref{e:cko:aflat_extra})
along with Eqs.~(\ref{e:glob:aflat1})-(\ref{e:glob:aflat2})
guarantees that 
\be \label{e:cko:decay_Ricci}
	{\tilde R}_{ij} = O(r^{-3}) . 
\ee
Then a general theorem by Cantor (1979) \cite{Canto79} 
on elliptic operators on asymptotically flat manifolds can be invoked
(see Appendix~B of Ref.~\cite{SmarrY78a} as well as Ref.~\cite{ChoquIY00}) to
conclude that the solution of Eq.~(\ref{e:cko:DeltaLv_S}) 
with the boundary condition
\be
	v^i = 0 \qquad \mbox{when} \ r\rightarrow 0
\ee
exists and is unique.
The possibility to add a conformal Killing vector to the solution, 
as in the compact case, does no longer exist because there is
no conformal Killing vector which vanishes at spatial infinity
on asymptotically flat Riemannian manifolds. 

Regarding numerical techniques to solve the conformal vector Poisson
equation (\ref{e:cko:DeltaLv_S}), let us mention that a very accurate
spectral method has been developed by Grandcl\'ement et al. (2001)
\cite{GrandBGM01} in the case of the Euclidean space:
$(\Sigma,\wtgm)=(\R^3,\w{f})$. It is based on the use of Cartesian components
of vector fields altogether with spherical coordinates.
An alternative technique, using both spherical components and spherical 
coordinates is presented in Ref.~\cite{BonazGGN04}.

%\backmatter
% 
%    References
%
% $Date: 2007-03-06 11:59:03 +0100 (mar, 06 mar 2007) $
% $Rev: 183 $
% $Author: e_gourgoulhon $
%%%%%%%%%%%%%%%%%%%%%%%%

\printindex

\end{document}